\documentclass[twocolumn,rmp,aps,nofootinbib,amsmath,amssymb,floatfix]{revtex4}

\usepackage{graphics}
\usepackage{epsfig}
\usepackage{longtable}
\usepackage{bm}
\usepackage{color}
\definecolor{Blue}{rgb}{0.3,0.3,0.9}
\definecolor{Red}{rgb}{1,0.1,0.1}

\begin{document}
\title{Nuclear spin physics in quantum dots: an optical investigation}

\author{Bernhard Urbaszek, Xavier Marie and Thierry Amand}
\affiliation{Universit\'e de
Toulouse, INSA-CNRS-UPS, LPCNO, 135 Av. Rangueil, 31077 Toulouse,
France}

\author{Olivier Krebs and Paul Voisin}
\affiliation{Laboratoire de Photonique
et Nanostructures CNRS, route de Nozay, 91460 Marcoussis, France}

\author{Patrick Maletinsky}
\affiliation{Department of Physics, University of Basel, Klingelbergstrasse 82, Basel CH-4056, Switzerland}

\author{Alexander H\"ogele}
\affiliation{Ludwig-Maximilians-Universit\"at, Fakult\"at f\"ur
Physik and CeNS, Geschwister-Scholl-Platz 1, D-80539 M\"unchen,
Germany}

\author{Atac Imamoglu}
\affiliation{Institute of Quantum Electronics, ETH-Zurich,
CH-8093, Zurich, Switzerland}

\begin{abstract}
The mesoscopic spin system formed by the $10^4-10^6$ nuclear spins
in a semiconductor quantum dot offers a unique setting for
the study of many-body spin physics in the condensed matter. The
dynamics of this system and its coupling to electron spins is
fundamentally different from its bulk counter-part as well as that
of atoms due to increased fluctuations that result from reduced
dimensions. In recent years, the interest in studying quantum dot
nuclear spin systems and their coupling to confined electron spins
has been fueled by its direct implication for possible applications
of such systems in quantum information processing as well as by the
fascinating nonlinear (quantum-)dynamics of the coupled
electron-nuclear spin system. In this article, we review experimental
work performed over the last decades in studying this mesoscopic,
coupled electron-nuclear spin system and discuss how optical
addressing of electron spins can be exploited to manipulate and
read-out quantum dot nuclei. We discuss how such techniques have
been applied in quantum dots to efficiently establish a non-zero
mean nuclear spin polarization and, most recently, were used to
reduce fluctuations of the average quantum dot nuclear spin
orientation. Both results in turn have important implications for
the preservation of electron spin coherence in quantum dots, which
we discuss. We conclude by speculating how this recently gained
understanding of the quantum dot nuclear spin system could in the
future enable experimental observation of quantum-mechanical
signatures or possible collective behavior of mesoscopic nuclear
spin ensembles.

\end{abstract}

\maketitle
\tableofcontents

\section{INTRODUCTION}
\label{sec:intro}

Electronic spins in most semiconductors are relatively well
decoupled from orbital or charge degrees of freedom. As a
consequence, electronic spin coherence is not hindered by the
prevalent charge decoherence, rendering spins good candidates for
the realization of novel devices whose functionalities rely on
quantum coherence. The isolation of spins from adverse effects of
fluctuating charge environments is particularly effective in quantum
dots (QDs) where electronic motion is quantum confined in all
directions to length scales on the order of 10 nanometers.
In such atom-like structures, hyperfine coupling is the dominant interaction for both the spin of the electron confined to the QD and the nuclear spins, making this system a nearly ideal realization of the central spin model. 


In this Article, we review recent work literally shedding light on
this unique coupled spin system. The basic principle of optical
manipulation and measurement of QD spins we describe has its roots
in the use of strong spin-orbit interaction of valence band states
which allow for correlating the optically excited electron spin with
the polarization of the excitation laser. This is in fact the same
physics used in optical pumping experiments carried out in atomic
vapors: in 1952 Kastler, Brossel and Winter investigated Mercury
atoms in a weak magnetic field which splits the electron states into
Zeeman sublevels. By irradiation of the atoms with circularly
polarized light the authors could selectively populate one of the
electron Zeeman levels \cite{Kastler:1952a}. Subsequent
optical pumping experiments on atoms with nonzero nuclear spin
resulted in direct preparation of correlated electron-nuclear spin
states in atoms.

In pioneering work in the solid-state, Knight observed that
polarized electrons lead to a shift in nuclear magnetic resonance
frequency \cite{Knight:1949a}. In 1953 Overhauser proposed to
polarize nuclear spins by transferring spin polarization from
electrons to the nuclear spin system. In this original proposal, a
net electron spin polarization was created simply by allowing
thermalisation in an applied longitudinal field
\cite{Overhauser:1953a}. Soon afterwards the ideas of
Overhauser and Brossel \textit{et al} were combined in the first
work on optical preparation of  electron spins and the resulting
interaction of these spin polarized electrons with the nuclear spin
system in a semiconductor \cite{Lampel:1968a}. In this
experiment performed with silicon the initial pumping of spin
oriented conduction electrons induced by polarized light leads to a
polarization of the nuclear spins of the atoms of the silicon
lattice via the hyperfine interaction (Overhauser effect). This is
based on the angular momentum transfer between photons and electrons
and subsequently between electrons and nuclei. The nuclear
polarization was detected by Lampel through the enhancement of the
nuclear magnetic resonance signal. A detailed review of the nuclear
spin effects in bulk semiconductor optics can be found in
\textcite{Meier:1984a}, where the key ingredients for strong
hyperfine effects in solids were clearly identified: localization of
the carrier wave function around a finite number of nuclei and
temporal fluctuations in the electron spin system i.e. a short correlation time of
the hyperfine interaction. As a result strong nuclear effects
imprinted on the polarization of the emitted photons were observed in
n-doped bulk semiconductors which show strong localisation of
carriers around donors \cite{Dzhioev:2002a}.

\begin{figure}
\epsfxsize=3.5in
\epsfbox{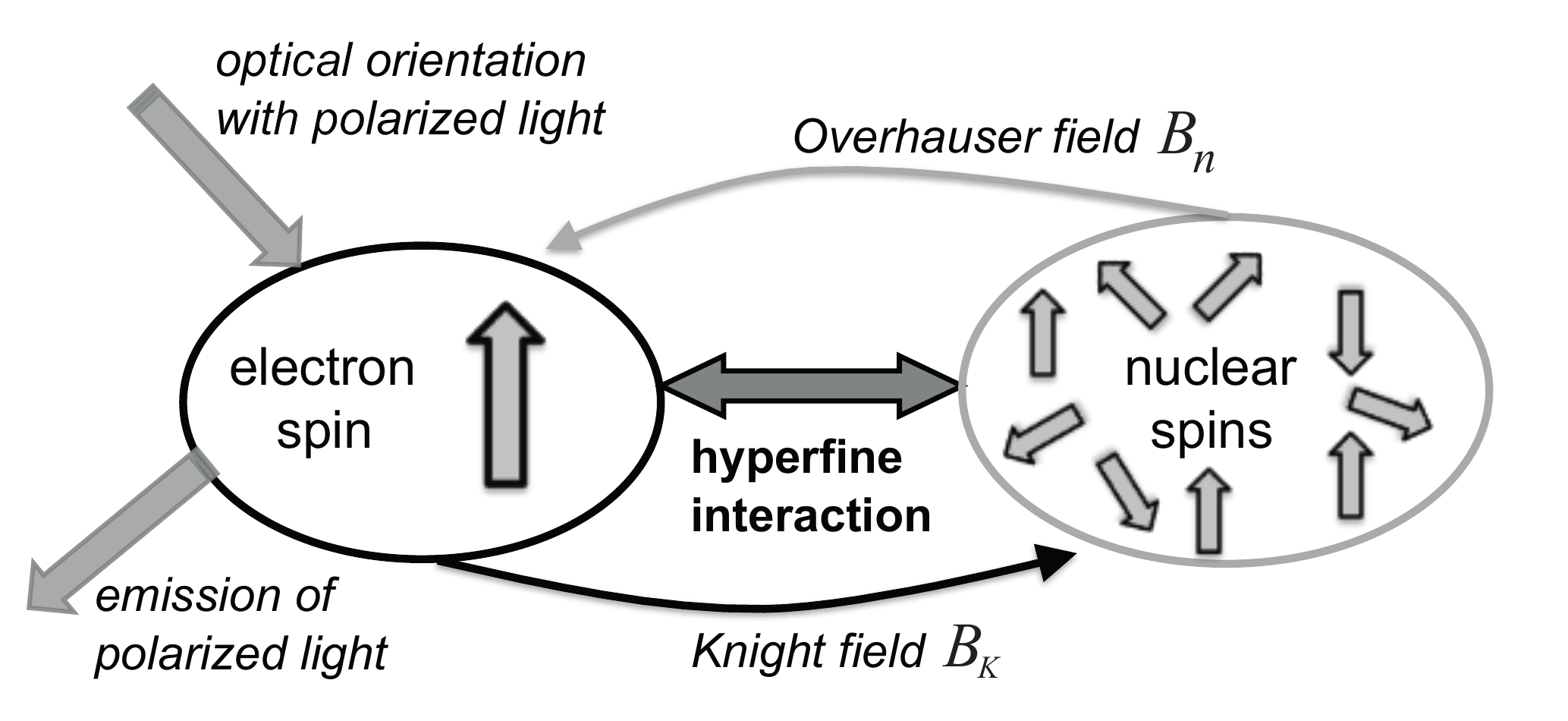}
\caption{Reciprocal interaction between electron and nuclear spins. The electron spin state is initialised through optical pumping.
 \label{intro0}}
\end{figure}

Due to the strong localization of the carrier wave function in a QD,
the role of hyperfine interactions in spin dynamics is drastically
enhanced; this is a direct consequence of enhanced fluctuations in
the effective magnetic field seen by the electron spin (Overhauser field) 
due to its interactions with randomly oriented QD nuclei. Similarly,
the effective magnetic field seen by each nucleus (Knight field) is more
susceptible to fluctuations in the electron spin. Soon after the
first observation of emission from single QDs
\cite{Marzin:1994a} it became clear that studies of the
electron spin system cannot be done without taking nuclear effects
into account. Conversely, ultra-narrow QD optical transition
linewidths allow for a direct measurement of the nuclear field,
greatly enhancing the possibilities for investigating nuclear spin
dynamics using optical spectroscopy. This is shown in pioneering
work on optically detected nuclear magnetic resonance ODNMR in GaAs
dots in AlGaAs \cite{Gammon:1997a, Brown:1998a}.

Initialization of an individual electron or hole  spin with a laser
pulse is possible due to angular momentum transfer between photons
and electrons, enabled by spin-orbit interaction and ensuing optical
selection rules. Once initialized, the prospects for controlled,
coherent manipulations of spins in QDs are very good as the
main spin relaxation mechanisms known from experiments in bulk or 2D
semiconductors do not apply to localized carriers in dots
\cite{Khaetskii:2000a,Paillard:2001a,Kroutvar:2004a,Pines:1957a}. However,
it had been pointed out early
\cite{Burkard:1999a,Merkulov:2002a,Dyakonov:1973a,Dyakonov:1974a}
that interactions with fluctuating, arbitrarily aligned nuclear
spins of the atoms that form the QD might severely limit
the electron spin coherence time.  This prediction has indeed been
confirmed independently for electrons in transport measurements and
in optical spectroscopy \cite{Petta:2005a,Braun:2005a}.
Extending the carrier spin coherence time for controlled quantum
state (qubit) manipulation was one of the strong motivations that
led to increased interest in nuclear spin physics in QDs
\cite{Bluhm:2010b}. Many fascinating experiments have been
reported confirming the strong, reciprocal interaction between the
spin systems. For example, the magnitude and direction of the
Overhauser field created via optical pumping can be tuned by
adjusting laser power and polarization
\cite{Bracker:2005a,Eble:2006a,Tartakovskii:2007a,Maletinsky:2009a}.
The nuclear spin system can be stable up to several hours under
certain conditions, which is interesting for information storage
schemes \cite{Taylor:2003a}. The hyperfine
interaction allows for tuning the exact energy of the electronic
states and for controlling the polarization of the emitted light.
This is particularly true for experiments in the absence of magnetic
fields \cite{Lai:2006a,Belhadj:2009a,Larsson:2011a} and could
become important for applications in photonics. For example, knowing
the exact polarization basis is crucial when evaluating the degree
of entanglement of a source of photon pairs based on optical
transitions from the conduction to valence state in a single quantum
dot \cite{Akopian:2006a,Dousse:2010a,Stevenson:2011a}.

Hyperfine effects in QDs can have other  spectacular
consequences, such as locking of a QD transition to a
resonant pump laser
\cite{Latta:2009a,Xu:2009a,Chekhovich:2010a}, bistability of
the nuclear spin system
\cite{Braun:2006a,Maletinsky:2007a,Tartakovskii:2007a,Kaji:2008a}
depending not only on the experimental parameters at the time the
measurement was performed but also on the history of the experiment
(non-Markovian behaviour). The mesoscopic nuclear spin system may
enable observation of physical phenomena such as Levy flights
\cite{Issler:2010a}, spin-squeezed states
\cite{Rudner:2011a} and dissipative quantum phase transitions
\cite{Kessler:2010a,Kessler:2012a}. Studies of hyperfine effects in dots are
also relevant for other systems with localised carriers, such as
nitrogen vacancy centres in diamond
\cite{Childress:2006a,Balasubramanian:2009a}.

\section{BASICS OF SEMICONDUCTOR QUANTUM DOTS}
\label{sec:basics}

Semiconductor QDs are nanometer sized objects that contain
typically several thousand atoms of a semi-conducting compound
resulting in a quantum confinement of the carriers in the
three spatial directions. As a consequence, the energy levels in
semiconductor QDs are discrete. Micro-photoluminescence experiments
\cite{Marzin:1994a}, photon correlation
measurements \cite{Michler:2000a} and resonant laser scattering
\cite{Hogele:2004a} have established the atom like character of the
interband transitions. This motivated many research groups to probe
and manipulate charge and spin states of individual carriers.
These experiments test the possibility of using these QD
states as qubits for quantum information processing \cite{Henneberger:2008a}.

\begin{figure}
\epsfysize=3.5in
\epsfbox{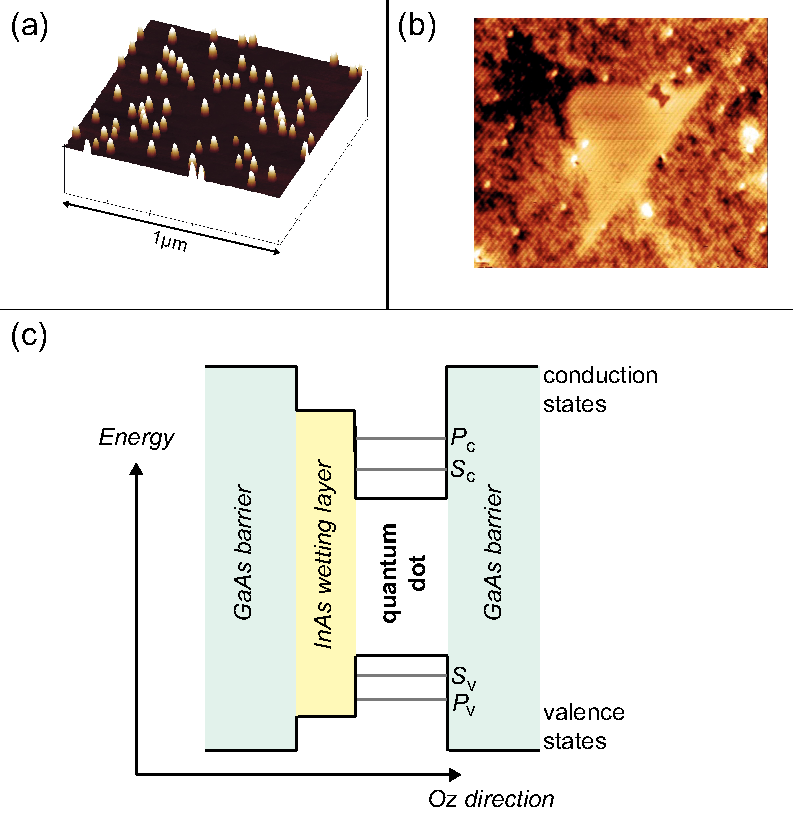}
\caption{(a) 1~$\mu$m $\times$ 1~$\mu$m Atomic force microscopy image of InAs dots on GaAs (b) 40~nm $\times$ 34~nm cross-sectional scanning tunneling microscopy image of a GaAs dot in AlGaAs \cite{Keizer:2010a}. (c) schematic energy level diagram for an InAs QD in GaAs, where the growth axis is along the $Oz$ direction.
 \label{intro1}}
\end{figure}

\subsection{Growth and sample structures}
\label{sec:growth} Semiconductor QDs can be synthesized by
a large variety of methods based on colloidal chemistry, molecular
beam epitaxy (MBE) or metalorganic chemical vapor deposition
(MOCVD). QDs can be formed at interface steps of thin
quantum wells
\cite{Gammon:1996a,Gammon:1997a,Hours:2005a,Besombes:2000a} or by
self assembly in the Stransky-Krastanov growth mode during molecular
beam epitaxy \cite{Goldstein:1985a,Leonard:1994a}. The latter
process is driven by the strain resulting from the smaller lattice
parameter of the matrix (barrier) compared to that of the dots, for
example 7\% for InAs dots in GaAs. The QDs obtained in this
well studied system are typically 20~nm in diameter and 5~nm in
height (see Fig.\,\ref{intro1}(a)) and are formed on a thin InAs
quantum well called \textit{wetting layer}, as can be seen in STM
measurements \cite{Offermans:2005a}. Samples used for optical
spectroscopy are then covered again by the barrier material. In
realistic samples InAs dots contain a significant fraction $x$ of Ga, leading to the formation of In$_{1-x}$Ga$_{x}$As dots.
The Stransky-Krastanov growth mode is applied to a large variety of
III-V and II-VI compounds. An interesting alternative for
fabricating GaAs or InAs QDs is provided by a technique
which is not strain driven, called molecular droplet epitaxy
\cite{Koguchi:1991a}, see Fig.\,\ref{intro1}(b) for a
cross-sectional scanning tunneling microscopy image of a GaAs dot in
AlGaAs \cite{Keizer:2010a}. The recently achieved high optical
quality of GaAs droplet dots has allowed first investigations of
carrier and nuclear spin dynamics \cite{Belhadj:2008a,Sallen:2011a}.
Due to carrier confinement potentials between tens and hundreds of
meV, the samples elaborated with the above techniques are suitable
for optical spin manipulation often carried out at a temperature of
4K, with the possibility for detailed spectroscopy up to few tens of
Kelvin.

This review will concentrate on the \textit{optical}
manipulation of spin states. A very high degree of control
over carrier spin states and the mesoscopic nuclear spin system is
also achieved in QDs defined by electrostatic potentials as
summarized in the detailed review by \textcite{Hanson:2007a}. The
electron (not hole) spin physics probed in these transport
measurements at very low temperature (100mK) provide a powerful,
complementary approach
\cite{Bluhm:2010b,Petta:2005a,Takahashi:2011a} to optical
spectroscopy.

\subsection{Addressing individual charge states}
\label{sec:charge}
\begin{figure}
\epsfxsize=3.5in
\epsfbox{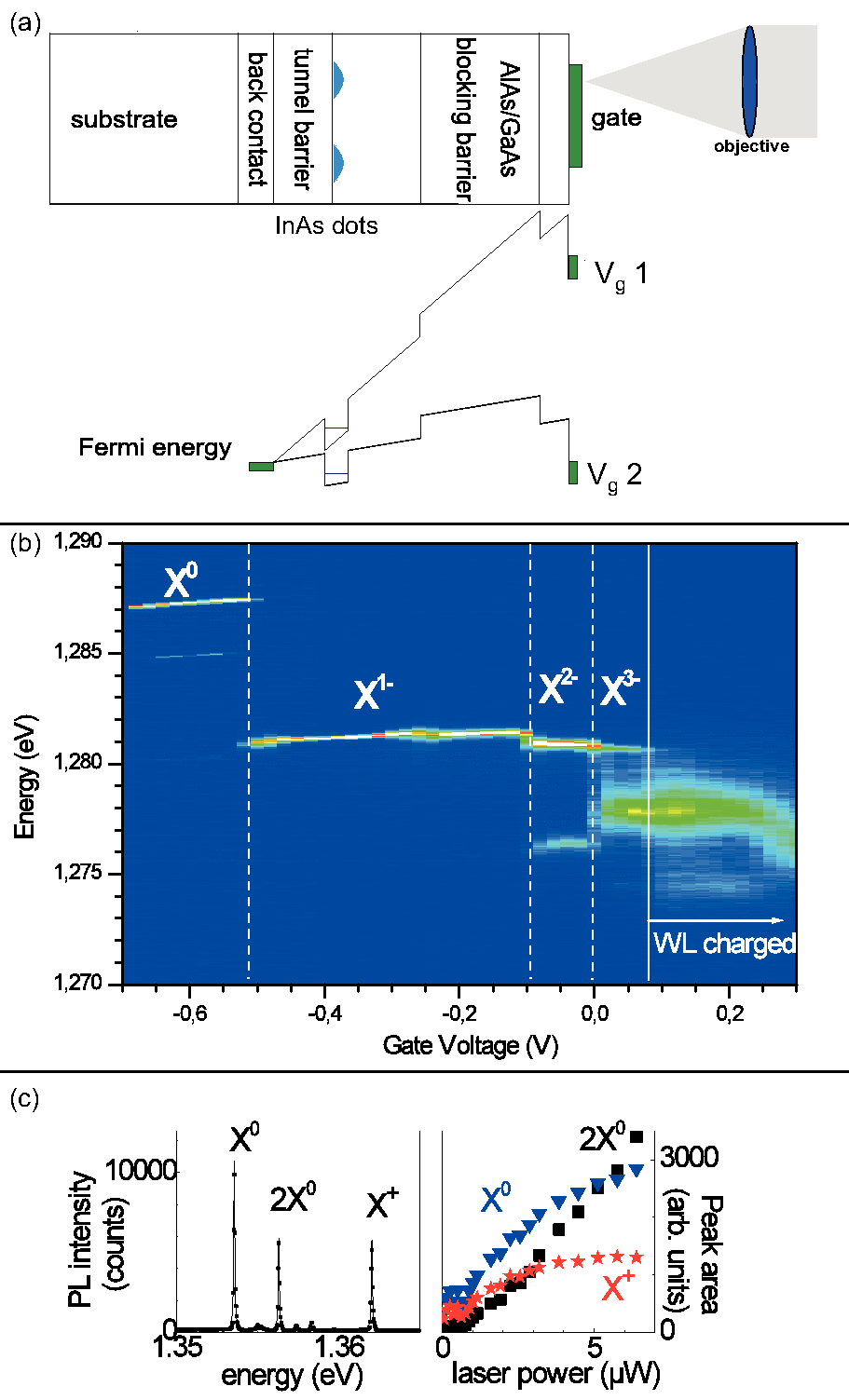}
\caption{\textbf{Sample A} (a) Scheme of InAs QDs embedded into a charge tunable device as in \textcite{Warburton:2000a}, where for a voltage V$_{g1}$ applied to the top gate the electronic level of the dot is above the Fermi energy of the highly n-doped back contact. The QD contains no conduction electron. For a gate voltage V$_{g2}$ the electronic level of the dot is now below the Fermi sea and an electron can tunnel into the dot. (b) The charging of a single InAs QD with electrons is accompanied by discrete jumps in the emission energy when going from the neutral exciton X$^0$ (1 electron, 1 hole) to the charged exciton X$^-$ (2 electrons, 1 hole) etc until the wetting layer (WL) is charged. \textbf{Sample B} (c) Left: charge fluctuations (a doping hole or electron tunnel into and out of the dot) in non-intentionally doped dots allow the observation of neutral excitons $X^0$, charged excitons $X^+$  and biexcitons $2X^0$ in photoluminescence (PL) spectra that are integrated over seconds i.e. over times much longer than the charge fluctuation times \cite{Belhadj:2009a} Right: In addition to the fine structure, the emission intensity of each transition as a function of optical excitation power allows to distinguish between different exciton complexes containing two, three or four optically generated charge carriers.
 \label{chargestates}}
\end{figure}

Controlling the charge state of QDs relies on the remarkable possibility of
doping semiconductor materials with $n$-type or $p$-type impurities. In some cases the non-intentional residual doping is sufficient
to obtain singly-charged QDs~\cite{Belhadj:2009a,Akimov:2002a}, see Fig.\,\ref{chargestates}(c), but usually a delta-doped layer is grown a
few nanometers  below the QD layer with a density adjusted to reach  the desired average QD
charge~\cite{Cortez:2002a,Laurent:2006a,Greilich:2006a}. This modulation doping technique can be significantly
improved by controlling the chemical potential of the QD electrons with an electric
voltage applied between the doped layer and a semitransparent top contact \cite{Drexler:1994a}.
In these charge-tuneable structures a given QD is coupled to a reservoir of free carriers (a heavily doped layer) through a tunnel barrier as in Fig.\,\ref{chargestates}(a).
The energy levels of the QD can be adjusted with respect to the Fermi level in the highly doped barrier, to vary deterministically the charge state with the precision of a single elementary charge due to Coulomb blockade. This effect is clearly observed in micro-photoluminescence (PL) spectra by abrupt jumps of the (charged) exciton emission energy when the gate voltage is varied (see Fig.\,\ref{chargestates}(b)) as a result of changes of the strong few particle direct Coulomb terms \cite{Warburton:2000a}

\subsection{Electronic states, Optical selection rules and Carrier Coulomb exchange interactions}
\label{sec:states}
The electronic structure of QDs can be analysed by techniques
such as capacitance-voltage measurements, scanning-tunneling mircoscopy \cite{Girard:2009a}, electron-spin resonance, photo-current spectroscopy and a large variety of
optical spectroscopy experiments. The latter allow a detailed study of the optically
active electronic states and their symmetry by analysing the energy and polarization of
absorbed or emitted photons. These experiments probe the interplay between carrier
confinement, direct and exchange Coulomb terms and the hyperfine interaction.
The orders of magnitude of the different effects that determine
the optical and spin properties are given in table\,\ref{tab:ordermag} for the model system of InAs dots in GaAs.\\

\begin{table}
\caption{\label{tab:ordermag} Typical transition and interaction energies for a standard \textbf{InAs} QD in \textbf{GaAs} grown along the [001] axis, measured at a temperature of 4K, see for example \textcite{Warburton:1998a, Bayer:2002a,Urbaszek:2003a}, are listed. It is important to note that all properties which are linked to the QD size and shape and hence the exact confinement potential can vary considerably from dot to dot, this table merely indicates typical values for confinement energies, Coulomb interactions and Zeeman energies to establish the relative strength of the different interactions.}
\begin{ruledtabular}
\begin{tabular}{p{5cm} p{2.5cm}}  
interaction & energy in eV \\
\hline
GaAs barrier & 1.519  \\
Electron to heavy hole transition in wetting layer & 1.44 \\
InAs dot electron to heavy hole transition & 1.3 \\
electron confinement energy & 50 $\cdot 10^{-3}$ \\
heavy hole confinement energy & 25 $\cdot 10^{-3}$ \\
direct Coulomb interaction between two S electrons & 20 $\cdot 10^{-3}$ \\
exchange Coulomb interaction between an S and a P electron & 5 $\cdot 10^{-3}$ \\
fine structure splitting between J=2 and J=1 $X^0$ due to isotropic e-h Coulomb exchange interaction $\delta_0$ & 100...500 $\cdot 10^{-6}$ \\
fine structure splitting of J=1 $X^0$ due to anisotropic e-h Coulomb exchange interaction $|\delta_1|$ & 0...150 $\cdot 10^{-6}$ \\
electron Zeeman splitting $\hbar\omega^e_Z$ at $B_z=~1$ T & 30 $\cdot 10^{-6}$ \\
nuclear Zeeman splitting $\hbar\omega^n_Z$ at $B_z=~1$ T & 30 $\cdot 10^{-9}$ \\

\end{tabular}
\end{ruledtabular}
\end{table}

QDs can be populated by valence holes and conduction electrons through optical
excitation and/or through controlled tunneling in charge tuneable structures
\cite{Warburton:1998a}. For a simplified calculation of optical transition energies
between conduction band electron states and valence band hole states the single particle
energies are determined by treating the electron-hole confinement potential within the harmonic approximation. For self-assembled as
well as interface fluctuation dots the vertical confinement energies (along the growth axis $z$) are almost an order
of magnitude larger than the lateral confinement energies in the $xy$ plane. The quantization energies of
both electrons and holes are larger than the Coulomb energies. The Coulomb effects can
therefore be treated as perturbations to the single particle structure
\cite{Warburton:2000a}. At zero magnetic field the lowest lying conduction (valence)
level $S_c$($S_v$) is twofold degenerate and the adjacent $P_c$($P_v$) level is fourfold
degenerate in the case of axial symmetry, as in an ideal two-dimensional harmonic potential \cite{Warburton:1998a}, see Fig.\,\ref{intro1} for the energy level diagram. 
Here $S$ and $P$ refer to the symmetry of the envelope part of the Bloch function of the carrier state. 
For brevity, a Coulomb correlated electron-hole pair trapped inside a dot by the confinement potential will be
called \emph{exciton} in the following.\\
\indent The electric-dipole interaction of an electromagnetic wave with carriers in a semiconductor is governed by strict optical selection rules \cite{Meier:1984a}. Energy and angular momentum are conserved for transitions between the valence and conduction band of a typical zincblende semiconductor like GaAs. The periodic part of the Bloch function of the conduction states is s-like, so the electron angular momentum is simply $m_s=\pm 1/2$ in units of $\hbar$ ($\uparrow$ or $\downarrow$). The p-like valence states are determined by spin-orbit coupling and we consider here only the states with total angular momentum of $J=3/2$ as the split-off states $J=1/2$ are very far in energy (hundreds of meV in GaAs based samples) and can usually be neglected.

\begin{figure}
\epsfxsize=3in
\epsfbox{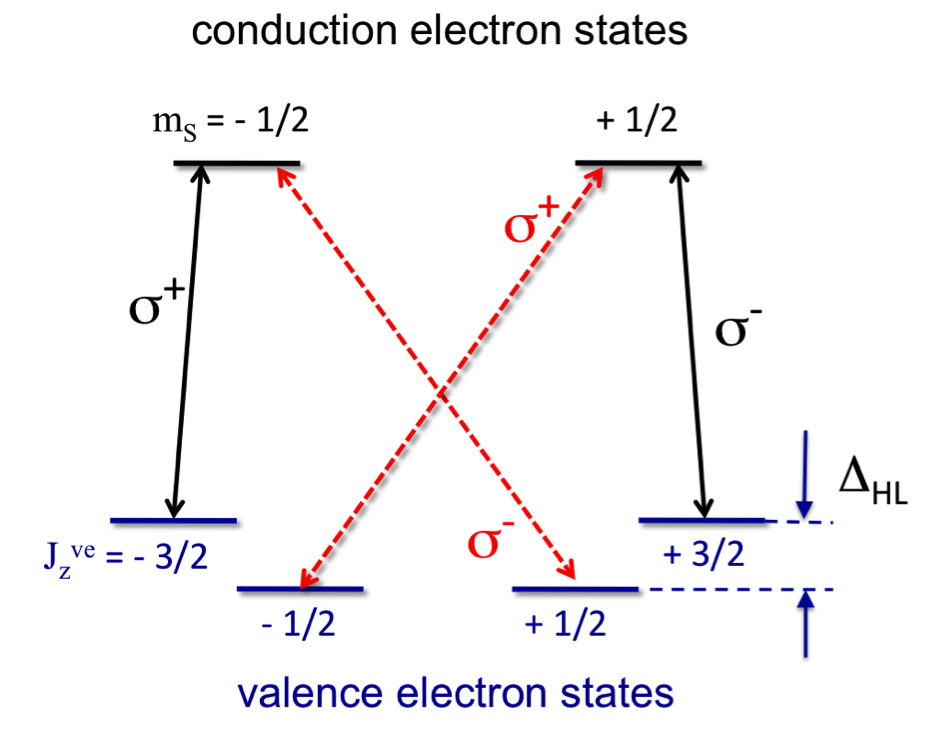}
\caption{Optical selection rules for inter-band transitions involving valence band electrons with angular momentum $J^{ve}=3/2$ in units of $\hbar$, the corresponding photon polarization $\sigma^+$ or $\sigma^-$ is indicated. Valence states with $J^{ve}=1/2$ are well separated in energy and are not shown here. Absorption of a $\sigma^+$ polarized photon by a $J_z^{ve}=-3/2$ valence electron, results in the promotion of this electron to the conduction state $m_s=-1/2$ and a hole heavy hole $J_z=+3/2$ is left behind in the valence band. 
The energy separation between valence heavy and light hole states  $\Delta_{HL}$ is typically several tens of meV in InAs dots in GaAs
 \label{states}}
\end{figure}

The quantization axis $z$ is chosen perpendicular to the QD plane and in most experiments $z$ is also parallel to the excitation light propagation direction. 
Following absorption of a photon of suitable energy, an electron is promoted from a valence state to a  conduction state. The absorption of a photon can increase the electron angular momentum by $1$ for a $\sigma^+ $ polarized photon or lower it by $1$ for a $\sigma^- $ polarized photon,  see Fig.\,\ref{states} for all possible transitions between valence and conduction states in a simple picture. The selection rules for photon absorption and emission are identical. The unoccupied valence state left behind due to the promotion of the electron to the conduction state is called hole.  The states with a projection of $J_z=\pm 3/2$ ($\Uparrow$ or $\Downarrow$) are called heavy holes, $J_z=\pm 1/2$ are called light holes. 

The heavy and light hole valence states are separated by an energy $\Delta_{HL}$ of typically several tens of meV due to quantum confinement and/or strain. For most of the experiments the light hole states can safely be ignored and optical exciton spin state preparation is straightforward. In practice however, strain, interface rotational symmetry breaking \cite{Grundmann:1995a,Bester:2005a,Krebs:1996a} and shape anisotropy introduce heavy to light hole coupling which make all the transitions between the states indicated in Fig.\,\ref{states} possible, yet with very different probabilities \cite{Bayer:2002a,Belhadj:2010a,Koudinov:2004a,Leger:2007a}. \\
\indent Optical excitation of an empty dot  with a suitable energy results
in a transition from a valence to a conduction state and in the
formation of a neutral exciton $X^0$, which allows to study carrier
spin dynamics during the radiative lifetime of typically hundreds of
picoseconds \cite{Paillard:2000a}. For studies on longer time scales the
spin information can be transferred to resident carriers in doped
dots. In this review we focus on the three most relevant
configurations: a conduction electron-valence hole pair $X^0$ with
two optically active \textit{bright} states ($\Uparrow \downarrow$
or $\Downarrow \uparrow$) and two \textit{dark} states ($\Uparrow
\uparrow$ or $\Downarrow \downarrow$), the negatively charged
exciton (trion) X$^-$ ($\Uparrow \uparrow \downarrow$ or $\Downarrow
\uparrow \downarrow$) and the positively charged exciton X$^+$
($\Uparrow \Downarrow \uparrow$ or $\Uparrow \Downarrow
\downarrow$). Here $\uparrow$ ($\downarrow$)  and $\Uparrow$
($\Downarrow$) represent the conduction electron spin and hole
pseudo-spins, and for example $\Downarrow \uparrow \downarrow$
stands for
$\frac{1}{\sqrt{2}}(\uparrow\downarrow-\downarrow\uparrow)\otimes\Downarrow$,
where the antisymmetrization of the conduction states is more
explicit. Due to strong localization of the carrier wave
function, direct and exchange Coulomb, as well as correlation
effects are very strong in dots. For the trions the direct and
exchange Coulomb interaction lead to a renormalization of the
transitions energies in the meV range but no fine structure
splitting due to Kramers degeneracy
\cite{Hogele:2004a,Bayer:2002a,Belhadj:2008a}.

For neutral excitons (in zero magnetic field and in the  absence of
strong nuclear polarization) selection rules are affected by the
electron hole Coulomb exchange interaction. This interaction includes an
anisotropic contribution  \cite{Bayer:2002a,Tong:2011a} due to
deviation of the real QD shape from a perfectly circular
shape, see microscopy images in Fig.\,\ref{intro1}, and/or due to
the dot-semiconductor matrix interface anisotropy. Due to
anisotropic exchange, X$^0$ recombination results in a doublet of
\emph{linearly-polarized} transitions, separated by an energy
$\delta_1$ that varies from a few to a few tens of $\mu$eV from dot
to dot in InAs/GaAs samples, see table \ref{tab:ordermag}.

\subsection{Optical spectroscopy techniques}
\label{sec:spectro}

To investigate the spin dynamics of carriers and nuclei a large
variety of optical spectroscopy techniques have been developed, each
adapted to the time scales relevant for the experiment. The typical
radiative lifetime of a  neutral or charged exciton is hundreds of
picoseconds \cite{Paillard:2000a,Paillard:2001a}, electron spin
coherence times can be prolonged up to 200~$\mu$s
\cite{Greilich:2006a,Bluhm:2010b}. Efficient collection of single
dot photoluminescence following non-resonant excitation
\cite{Maletinsky:2007b} and resonant fluorescence \cite{Lu:2010a} result in signal
integration times well below the millisecond range, which
provides the time resolution necessary to measure for example the
nuclear polarization build-up time.

The discreteness of the QD energy states was demonstrated
in optical spectroscopy experiments as early as 1994
\cite{Marzin:1994a}. Reducing the detection spot size in
optical experiments to an area that contains only one nano-object
permits studying directly the optical properties of an individual dot. 
A simple and powerful tool is
non-resonant photoluminescence (PL), where carriers are optically
excited in the surrounding semiconductor matrix by a laser tuned
above the QD resonance energy i.e. either into wetting
layer or barrier states. The carriers are subsequently trapped by
the QD confinement potential and, following energy
relaxation, recombine radiatively at the ground state energy, see
$S_c$ to $S_v$ transition in Fig.\,\ref{intro1}(c). More recently
resonant fluorescence experiments where the excitation laser is
resonant with the energy necessary for absorption from the highest
lying valence level to the lowest lying conduction level
\cite{Muller:2007a} have shown beautiful analogies to atomic
physics \cite{Vamivakas:2009a}. Two closely related,
powerful techniques developed in charge tuneable structures are
differential transmission and reflectivity, which also allow
resonant probing of QD states \cite{Hogele:2004a,Alen:2006a}. These experiments are carried out with
pulsed or cw excitation. The challenge is to detect a very weak optical signal stemming from only one photon per recombination process.
In practice efficient cw detection with Si-based CCD cameras and avalanche diodes are adapted to single dot measurements.  For time
resolved measurements and also to observe the spin physics in
several thousand dots simultaneously experiments on QD
ensembles are useful, that allowed important discoveries in the
field, in resonant PL \cite{Paillard:2001a}, Kerr and
Faraday rotation probing the real part of the refractive index
\cite{Greilich:2006a} and photoinduced circular dichroism
\cite{Eble:2009a} probing the imaginary part of the
refractive index. An interesting alternative to conventional pump-probe techniques
is to passively detect the spectrum of
intrinsic random spin fluctuations of carriers in thermal equilibrium
(i.e., without optical pumping or initialization). This technique labelled \textit{spin noise spectroscopy} has been successfully applied to 
electron \cite{Crooker:2010a} and hole spins \cite{Li:2012a,Dahbashi:2012a}, respectively, interacting with nuclear spins.

\subsection{Electron spin orientation mechanisms}
\label{sec:orientation}

The target of this subsection is to explain how carrier spin states
in QDs can be initialised in optical experiments. Two very
different scenarios have to be distinguished: non-resonant and
resonant optical excitation. The technical advantage of non-resonant
optical excitation is the possibility to avoid blinding of the
detector by the excitation laser light, thanks to spectral
filtering. Also, contrary to atomic physics, non-resonant excitation
is very efficient due to the high absorption probability of the
thick barrier layer or 2D wetting layer. In the case of non-resonant
excitation, the carriers have to relax in energy towards the quantum
dot ground state $S_v$ for holes and $S_c$ for electrons. The
average carrier spin $\left< S_z \right>$ that can be initialised in
the QD ground state is the result of the (i) spin
initialisation upon photon absorption in the barrier or in the
wetting layer according to the optical selection rules (ii) spin
relaxation of the carrier during its presence in the barrier or
wetting layer (iii) spin and energy relaxation during capture
into the QD. \\
In spite of the expected phonon
bottleneck\footnote{At first sight energy relaxation from the continuous barrier states to the quantum dot ground states via discrete states separated in energy by tens of meV seems very unlikely if the relevant energy level spacing does not exactly match the energy of lattice phonons. The anticipated slowing down of the relaxation via phonon emission is termed \textit{phonon bottleneck}, but is rarely observed in practice as, for example, the involvement of polarons assures energy conservation during relaxation \cite{Verzelen:2002a}.},
it is observed that the carriers in the majority of samples relax
on a picosecond time-scale towards the dot ground state
\cite{Verzelen:2002a}. As a general rule, hole spin relaxation is efficient in bulk
semiconductors and quantum wells \cite{Dyakonov:2008a,Damen:1991a} i.e. during relaxation,
whereas the electron keeps its spin orientation for longer and can
to a high degree preserve its spin state during capture
\cite{Braun:2005a,Kalevich:2001a}.

Excitation in the GaAs barrier for InAs dots (or the AlGaAs barrier
for GaAs dots) involves both  light and heavy hole
transitions. As a result, a circularly polarized excitation creates
both up and down electron spins (see Fig \ref{states} for selection rules). 
The heavy-hole transition has a roughly 3 times larger oscillator
strength than the light hole transition. As a result, under $\sigma^-$ excitation
for 3 spin $\uparrow$ electrons only 1 spin $\downarrow$ electron is created in a conduction state. 
This corresponds to an optical spin initialization of $\frac{n_{\uparrow}-n_{\downarrow}}{n_{\uparrow}+n_{\downarrow}}=\frac{3-1}{3+1}=50\%$. 
To increase the optically generated average
spin, excitation into the bi-dimensional wetting layer (if present in the
sample) allows in principle injection of 100\% spin polarized
electrons when driving heavy hole transitions, which are separated
in energy from the light hole transitions in the wetting layer due to confinement and/or strain.\\
\indent \textit{The neutral exciton $X^0$.}--- For an empty QD, ground state electron and hole form an $X^0$. 
Due to the strong overlap of the carrier wavefunctions, Coulomb correlations are important and the anisotropic part of the exchange interaction results in two linearly polarized exciton eigenstates that are separated in energy by $\delta_1$ \cite{Gammon:1996a}. 
Assume that the QD is excited with a pulsed laser (temporal pulse length $\tau_L$) that is $\sigma^+$ polarized and for which $\hbar/\tau_L>\delta_1$ holds. Due to anisotropic exchange, the created exciton is \textit{not} in an $X^0$ eigenstate, but in a superposition of the linearly polarized eigenstates, so during the radiative $X^0$ lifetime $\tau_r$ quantum beats in the $\sigma^{+/-}$ basis are observed, for a detailed discussion see \cite{Senes:2005a}. If the beat period $\approx \hbar/\delta_1\ll\tau_r$, 
then the time averaged circular polarization degree $\frac{I_{\sigma^+}-I_{\sigma^+}}{I_{\sigma^+}+I_{\sigma^+}}$,
where $I_{\sigma^+}$ ($I_{\sigma^-}$) are the $\sigma^+$ ($\sigma^-$) polarized emission intensities, will tend to zero. 
In general $\hbar/\delta_1$ and $\tau_r$ can be of similar magnitude, 
which leads in cw experiments to a decrease of exciton pseudo-spin polarization from initially
$\rho_{c}^0$ down to $\rho_c$ during the radiative lifetime as $\rho_c = \rho_{c}^0(1+\omega ^2 \tau_r ^2)^{-1}$ with
$\hbar\omega=\delta_1$.\\
\indent \textit{The positively charged exciton $X^+$.}--- In the case of the $X^+$ exciton, a doping hole is present before
the optically generated electron and hole are captured. The incoming
hole spin is random, so the resident hole (which has a given spin orientation) and the optically generated hole can form a hole pseudo spin
singlet. As a result, the subsequent evolution of the spin orientated
\textit{electron} can be monitored during the $X^+$ lifetime
\cite{Laurent:2005a,Krebs:2008a} in the absence of Coulomb exchange effects. Recording the $X^+$ emission from
an InAs dot, initialisation of electron spin polarization as high as
80\% ($\langle S_z \rangle = 0.4$) has been achieved through
non-resonant excitation into the wetting layer, about 100~meV above
the dot ground state \cite{Urbaszek:2007a}.\\
\indent \textit{The negatively charged exciton $X^-$.}--- Non-resonant excitation with a circularly polarized laser  of a dot
doped with a resident electron is in principle not expected to yield
polarized emission, as the incoming electron will form a spin
singlet with the resident electron (total spin $S=0$). The hole
spin, completely randomised, will determine the polarization of the
emitted photon after $X^-$ recombination. But surprisingly, this
prediction has not been confirmed in experiments; instead,
non-resonant  circularly-polarized  excitation of a QD
results in a partially polarized ground state emission with an
helicity opposite to that of the
excitation~\cite{Dzhioev:1998b,Cortez:2002a,Laurent:2006a,Oulton:2007a,Shabaev:2009a}.
The origin of this \textit{negative polarization} has been ascribed
to exchange related electron-hole spin flip-flop processes during
carrier energy relaxation \cite{Laurent:2006a,Ware:2005a}. Another
possible scenario involving the accumulation of dark excitons in the
barriers, that are subsequently captured by the dots resulting in
negative polarization is likely to be applicable to GaAs interface
fluctuation dots \cite{Bracker:2005a}. Independent of its origin,
changes in the negative polarization degree observed for $X^-$
initialisation and subsequent recombination, can be used as a
sensitive probe for nuclear spin effects \cite{Auer:2009a}.

\begin{figure}
\epsfxsize=3.5in
\epsfbox{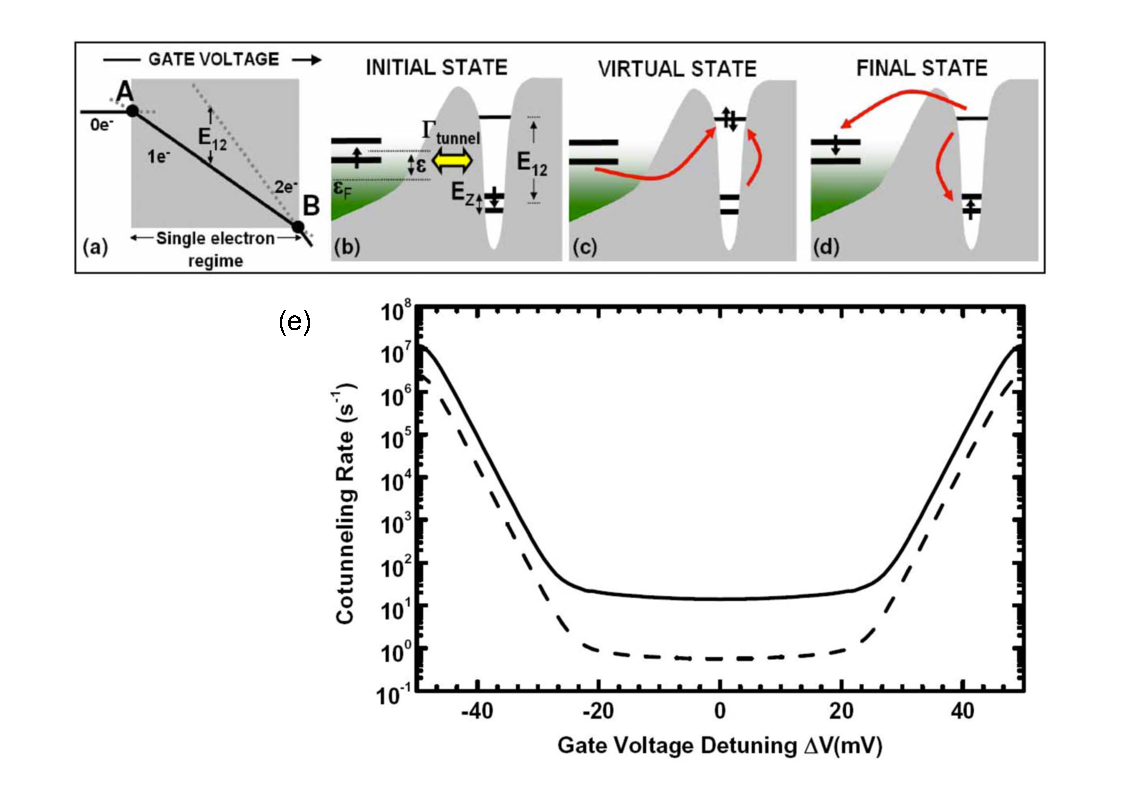}
\caption{(a) to (d) a QD exchanges an electron with the reservoir via a virtual two electron state and (e) calculation of the co-tunneling rate as a function of gate voltage after \textcite{Dreiser:2008a} for tunneling rates $0.1$~ns$^{-1}$ (solid curve) and $0.02$~ns$^{-1}$ (dashed curve).
 \label{cotunnel}}
\end{figure}

\indent \textit{Strictly resonant and quasi-resonant optical excitation.}--- Although experimentally simple,  non-resonant excitation has several
disadvantages: (i) Initialisation of the QD in a well
defined coherent superposition of polarization states is not
possible, as coherence is lost during relaxation. (ii) As carriers
with well defined spin orientation have been injected into the
barrier and/or wetting layer material, the electron can interact
with the nuclear spins during its presence in these layers. If the
QD emission shows that nuclear spins in the dot are
polarized, in the case of non-resonant excitation one cannot be
100\% sure that this polarization originates from nuclear spins in
the QD only, or, if the QD is simply a nanoscopic
probe of a macroscopic nuclear spin polarization created inside the
sample through spin diffusion \cite{Paget:1982a}. A first step to
circumvent these problems is to use what is termed quasi-resonant
excitation, for example 1LO phonon energy above the S$_v$ - S$_c$
transition or directly the  P$_v$ - P$_c$ transition. Reaching the
ground state via emission of a single LO phonon partially preserves
coherence \cite{Flissikowski:2001a,Scheibner:2003a,Senes:2005a}. In
addition, as in these experiments the photon is absorbed directly by
the QD states, one can be sure that the nuclear spins in
the QD are the direct source of the detected nuclear
spin polarization. This is demonstrated, for example, by
illuminating the dot with a fixed laser polarization and changing
the sign of the nuclear polarization by varying the gate voltage applied to a charge tunable structure i.e. by
going from the X$^+$ to the X$^-$ emission
\cite{Lai:2006a,Eble:2006a}. Resonant experiments allow precise
control over the created spin state or superposition of states
\cite{Greilich:2006b}. Whereas in non-resonant experiments the
hyperfine interaction will have negligible influence on the photon
absorption probability, in resonant experiments under certain
conditions the hyperfine interaction determines the polarization and
energy of the preferentially absorbed photons
\cite{Latta:2009a,Kloeffel:2011a,Chekhovich:2010a,Gerardot:2008a,Klotz:2010a},
as discussed in Sec.\,\ref{sec:beyond}. Once carrier spin initialisation has been achieved, the spin  will
interact with its nuclear spin environment during its lifetime.\\
\indent \textit{Electron co-tunneling in charge tunable structures.}--- Our
main focus in this review is on the interaction with nuclear spins,
but before going into details an important spin interaction present
in charge tunable structures has to be mentioned. An electron in a
dot embedded in a charge tuneable structure is coupled to the
continuum of delocalised electron states in the n-doped layer (Fermi
sea) via the tunnel barrier. The physical problem itself of a single
spin coupled coherently to the Fermi sea has parallels to the Kondo
effect \cite{Smith:2005a,Dreiser:2008a,Latta:2011b}. Essentially,
each charge state corresponds to the ground state for a given gate voltage range $\Delta
V=V_\text{start}-V_\text{end}$ over typically several tens of mV. Close to
$V_\text{start}$ and $V_\text{end}$ the exchange coupling to the
electron reservoir is strong; in between it is weak (charging
plateau), see Fig.\,\ref{cotunnel}e. When the coupling is strong the
electron spin can flip via an intermediate virtual transition either
to an empty or doubly-occupied QD state, as shown in
Fig.\,\ref{cotunnel}(a)-(d) \cite{Dreiser:2008a}. The net result of
exchanging an electron with the reservoir is a spin flip of the
electron inside the dot characterized by a spin flip co-tunneling
rate. So for experiments where stable carrier spins are required,
the structures must be operated at a gate voltage close to the
centre of the charging plateau where the co-tunneling rate is low
and hence the spin state of the resident electron is long-lived.

\section{ELEMENTARY INTERACTIONS WITH NUCLEI IN QUANTUM DOTS}
\label{sec:enint}

\begin{figure}
\epsfxsize=2.5in
\epsfbox{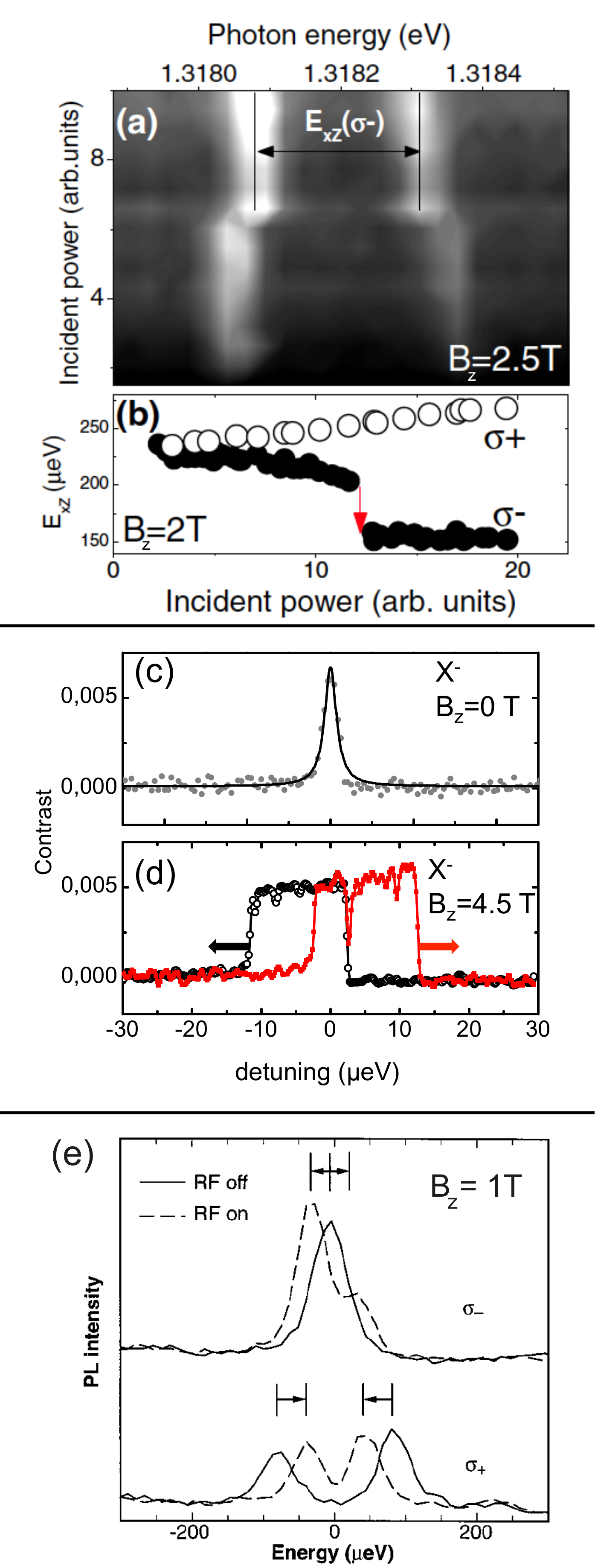}
\caption{(a) Gray-scale plot showing exciton PL spectra recorded for an individual InGaAs dot. The spectra are
recorded at $B_z=2.5$~T using unpolarized detection. (b) power dependences measured at $B_z=2$~T for $\sigma^+$and $\sigma^-$ excitation polarizations (after \textcite{Tartakovskii:2007a}) (c) trion $X^-$ absorption at zero magnetic field with Lorentzian fit of linewidth $2~\mu$eV and strong deviation from Lorentzian lineshape at $B_z=4.5$~T (d) after \textcite{Latta:2009a}. (e) Zeeman splitting for a single GaAs interface fluctuation QD. Due to dynamic nuclear polarization the Zeeman splitting for $\sigma^+$ and $\sigma^-$ polarized excitation are different. Randomizing the nuclear spin orientation with a radiofrequency field (RF on) results in the same Zeeman spliting for both laser polarizations, after \textcite{Gammon:1997a}.
 \label{introhyper}}
\end{figure}

Nuclear spin effects are important for experiments in QDs
that investigate optical carrier spin manipulation. Three striking
examples are shown in Fig.\,\ref{introhyper}: Figures (a) and (b)
show that the Zeeman splitting of an exciton in a longitudinal
magnetic field $B_z$ depends in a strongly non-linear fashion on the
laser excitation power as nuclear spins start to get polarized
\cite{Tartakovskii:2007a}, see Sec.\,\ref{sec:dnp}. These
measurements show that \textit{hyperfine} effects in III-V dots are
of the same order as the \textit{fine} structure of bright excitons,
see also Fig.\,\ref{introhyper}e and table \ref{tab:ordermag}.
Figure\,\ref{introhyper}(c) shows absorption of the charged exciton
$X^-$ line at zero Tesla, with a text book Lorentzian
lineshape. Figure\,\ref{introhyper}(d) represents a highly unusual
absorption spectrum at $B_z=4.5$~T, that is strongly broadened,
asymmetric and changes with laser scan direction
\cite{Latta:2009a}. In this experiment nuclear spin
polarization allows the QD transitions to be locked to the
driving laser field as it changes frequency, see section
\ref{sec:locking}. So nuclear spin effects have to be taken into
account even in a simple measurement of the transition energy of the
$X^-$ in an applied magnetic field. A direct proof that this type
of behaviour is related to nuclear spins comes from original ODNMR
measurements by \textcite{Gammon:1997a}, as detailed in section\ref{sec:odnmr}. The electron Zeeman splitting in a longitudinal field
of $B_z=1$~T for single GaAs/AlGaAs dots changes, as nuclear spins
are depolarized by a chirped radio-frequency source (RF on) scanning
the nuclear spin resonances for Ga and As, see Fig.\,\ref{introhyper}(e). The measurements show
that the dynamic nuclear polarization created through optical
pumping has an effect on the electrons that is comparable to the
applied magnetic field, see Sec.\,\ref{sec:dnp}.

Below we highlight the basics of the magnetic and electrostatic coupling between electrons and nuclei
that will allow us in the following sections to interpret quantitatively, whenever possible, the fascinating nuclear spin
effects observed in optical spectroscopy experiments in quantum dots for a wide range of experimental conditions.

\subsection{Magnetic coupling of electrons to nuclei: Hyperfine interaction}
\label{sec:hyperfine}

The strength of the hyperfine  interaction in QDs is
enhanced compared to semiconductor bulk or quantum well structures
due to the strong localization of the electron wavefunction over
typically only $10^5$ lattice sites. This number is too small for
efficient cancellation of the total nuclear spin by averaging
\cite{Burkard:1999a,Merkulov:2002a}, yet too large to address
each nuclear spin state individually. In III-V QDs like
GaAs, InP and InAs, 100\% of the lattice sites have a non-zero
nuclear spin and these materials are taken here as model systems.
Even for solids with very few isotopes carrying a nuclear spin like
diamond \cite{Childress:2006a}, ZnO \cite{Liu:2007a} or
CdSe \cite{Feng:2007a} hyperfine effects still
play a key role in the carrier spin state evolution.

There are two main contributions to the hyperfine interaction
\cite{Abragam:1961a}: (i) The Fermi contact interaction is efficient
when there is a physical overlap of the carrier wavefunction with
the lattice site. This type of interaction is dominant for
\textit{s}-type wave functions  (periodic part of the Bloch
function) of conduction electrons. (ii) The dipole-dipole
interaction is effective for p-type (non-zero orbital angular
momentum) wave functions. This term is therefore dominant for
valence-band states (holes). It is about one order of magnitude
weaker than the Fermi contact interaction for conduction electrons
\cite{Fischer:2008a,Eble:2009a,Chekhovich:2011a,Fallahi:2010a,Desfonds:2010a}. In
the bulk of this paper, namely sections \ref{sec:elecdynamics}, \ref{sec:dnp}, \ref{sec:tdnp} we concentrate on the interaction of a conduction
electron with nuclear spins, the interaction between valence holes
and nuclear spins will be discussed separately in Sec.\,\ref{sec:hole}.

\begin{table}
\caption{\label{tab:hyperconst} Hyperfine constants in GaAs, InAs, InP and CdTe for a cell containing two atoms, see \textcite{Testelin:2009a} and references therein. Please note that an average is quoted for Ga and In for which two stable isotopes exist.}
\begin{ruledtabular}
\begin{tabular}{p{1.2cm} p{1.2cm} p{1.2cm} p{1.5cm}}  
isotope & nuclear spin I & abundance (\%) & hyperfine constant A in $\mu$eV   \\
\hline
In & 9/2 & 100 & 56 \\
Ga & 3/2 & 100 & 42  \\
As & 3/2 & 100 & 46   \\
P & 1/2 & 100 & 44 \\
\hline
Cd & 1/2 & 25 & -30 \\
Te & 1/2 & 8 & -45

\end{tabular}
\end{ruledtabular}
\end{table}

\begin{table}
\caption{\label{tab:ohs} Electron Overhauser splitting in $\mu$eV  for 100 \% nuclear polarization $\hbar\omega^e_{OS}= I^\text{Ga} A^\text{Ga}+I^\text{As} A^\text{As}$ for GaAs and InAs quantum dots, using the nuclear spin and hyperfine constant values from table \ref{tab:hyperconst}}
\begin{ruledtabular}
\begin{tabular}{p{2cm} p{2cm}}  

 InAs & 315  \\
In$_{0.5}$Ga$_{0.5}$As & 230  \\
GaAs & 135 \\
\end{tabular}
\end{ruledtabular}
\end{table}

To introduce the orders of magnitude of the energy shifts due to the
hyperfine interaction between electron and nuclear spins, a
comparison with the Zeeman splitting of the spin levels in an
external magnetic field $\bm{B}=(0,0,B_z)$ is helpful
\cite{Abragam:1961a,Dyakonov:2008a}: The Zeeman energy of an
electron spin with $\hat{S}^e_z=\frac{1}{2}\hat{\sigma}^e$ is
\begin{equation}
\label{eq:HZe} \hat{H}_{Ze}= \mu_B g_e B_z \hat{S}_z^e=\hbar
\omega_Z^e \hat{S}_z^e
\end{equation}
where $g_e$ is the longitudinal electron g-factor and $\mu_B=9.27
\times 10^{-24}$~J/T$=58~\mu$eV/T. The Zeeman energy of a system of nuclear spins
$I^j$ is given by:
\begin{equation}
\label{eq:HZN}
\hat{H}_{ZN}= -\mu_N \sum_{j} g_{Nj} B_z \hat{I}_z^j
\end{equation}
summing over all nuclei $j$ in the system. Here $g_N$ is the nuclear
g-factor and $\mu_N \simeq \frac{\mu_B}{2000}$ is the nuclear
magneton. For an order of magnitude calculation, we take the example
of Indium and an electron g-factor of 0.6, and find
$(g_e\mu_B)/(g_N\mu_N)\simeq 1000$. The energy separation between
the nuclear spin states is therefore negligible compared to that of
the electron spins.

The Fermi contact ($fc$) hyperfine interaction in a QD between an electron spin
and the $N$ nuclei of the atoms forming the dot is
\cite{Abragam:1961a,Gammon:2001a}:
\begin{eqnarray}
\label{eq:eqHf}
\hat{H}_{hf}^{fc} = \frac{\nu_0}{2}\sum_{j}A^j
\vert\psi(\bm{r}_j)\vert^2 \left(2\hat{I}_z^j\hat{S}_z^e+
[\hat{I}_+^j\hat{S}_-^e+\hat{I}_-^j\hat{S}_+^e]\right)
\end{eqnarray}
where $\nu_0$ is the two atom cell volume, $\bm{r}_j$ is the
position of the nuclei $j$ with spin $\hat{I}^j$ and
$\psi(\bm{r}_j)$ is the normalized electron envelope function. The
nuclear spin is 3/2 for Ga and As, 5/2 for Al, 9/2 for In in units
of $\hbar$. $A^j$ is the constant of the hyperfine interaction with
the electron in the order or $50~\mu$eV for In, Ga and As, see
table \ref{tab:hyperconst}.

As an electron interacts simultaneously  with about $10^5$ lattice
sites, one can consider in the mean field approach that the electron
spin is affected by a mean nuclear spin polarization $\langle
\hat{\bm{I}}^j\rangle$ acting like an effective magnetic field $B_n$
(\textbf{Overhauser field}) :
\begin{equation}
\label{eq:Bn}
\bm{B}_n=\frac{\nu_0 \sum_{j}A^j \vert\psi(\bm{r}_j)\vert^2 \langle \hat{\bm{I}}^j\rangle }{g_e \mu_B}
\end{equation}
For uniform nuclear polarization, the field $B_n$ is independent of
the electron localization volume and is in the order of  $B_n^\text{max}
\simeq 5$~T for fully polarized nuclei in GaAs \cite{Paget:1977a},
as the maximum Overhauser shift is simply $g_e \mu_B
B_n^\text{max}=I^{Ga} A^{Ga}+I^{As} A^{As}=135\mu$eV.

The hyperfine interaction is reciprocal,  see scheme in
Fig.\,\ref{intro0}, so also the nuclei are effected by the average
electron spin polarization acting like an effective magnetic field
$B_K$ (\textbf{Knight field}). The time averaged Knight field acting
on one specific nucleus $j$ is given by:

\begin{equation}
\label{eq:BK}
\bm{B}_{Kj}=f_e \frac{\nu_0 A^j}{g_N \mu_N} \vert\psi(\bm{r}_j)\vert^2 \langle \hat{\bm{S}}^e \rangle
\end{equation}
where $f_e$ is the filling factor $\in[0,1]$ characterizing the
occupation of the dot by electrons, underlining that the Knight
field is zero in the absence of electrons. The maximum Knight field
can be estimated as $B_K^\text{max}\simeq\frac{B_n^\text{max}}{N} \frac{g_e
\mu_B}{g_N \mu_N}$, so for $N\simeq 10^5$ results in $B_K^\text{max}$ in
the tens of mT range. The amplitude of the Knight field for a
nucleus situated in the centre of the dot (where electron occupation
probability is strongest) will be higher than for a nucleus in the
dot periphery. The Knight field experienced by the nuclei leads to
frequency shifts in ODNMR spectra of individual QDs \cite{Brown:1998a}.

Introducing $\tilde{A}$ as the  average of the hyperfine constants
$A^j$ and assuming a strongly simplified, uniform electron
wavefunction $\psi(\bm{r})=\sqrt{2/(N\nu_0)}$ over the involved
nuclei, Eq.\,\ref{eq:eqHf} simplifies to:
\begin{equation}
\label{eq:eqHf1}
\hat{H}_{hf}^{fc} = \frac{2\tilde{A}}{N} \left(\hat{I}_z\hat{S}_z^e+ \frac{\hat{I}_+\hat{S}_-^e+\hat{I}_-\hat{S}_+^e}{2}\right)
\end{equation}
where $\hat{I}=\sum_{j=1}^{N}\hat{I}^j$.

The energy level splittings between the  different nuclear and
electron spin states are determined by the hyperfine interaction in
combination with the applied magnetic field $B_z$. $\hbar
\omega_{OS}=2\tilde{A}\langle\hat{I}_z\rangle/N=\mu_B g_e B_n$
relates the Overhauser shift $\hbar \omega_{OS}$ to the average
nuclear polarisation. We can therefore access the average nuclear
polarisation by measuring $\hbar \omega_{OS}$ in single dot
spectroscopy, as in Fig.\,\ref{introhyper}e.  For example, when the
nuclear spins are polarized (i.e. the RF source is off) the
\textit{total} electron Zeeman splitting $\hbar\omega_e$ in
Fig.\,\ref{introhyper}e is given by
$\hbar\omega_e=\hbar(\omega_Z^e+\omega_{OS}^e)$. When the RF source
is on, the nuclei are depolarized, the Overhauser field $B_n$ is
vanishingly small and $\hbar\omega_e=\hbar\omega_Z^e$. The
difference between the two cases allows to measure the Overhauser
shift $\hbar\omega_{OS}^e$.

The hyperfine interaction is  time dependent since the electron
lifetime is finite and its spin may also relax during its lifetime.
The time dependence of the second term in Eq.\,\ref{eq:eqHf1} can be
explicitly written as:
$\hat{H}_1(t)=\frac{\tilde{A}}{N}(\hat{I}_+\hat{S}_-^e+\hat{I}_-\hat{S}_+^e)h_1(t)$.
This term allows for spin transfer via simultaneous spin flips
(flip-flop) of a carrier and nuclear spin. As the nuclear Zeeman
splitting is negligible, the electron Zeeman splitting plays a
crucial role in determining the probability of these spin
flip-flops, as already pointed out in the original paper by
\textcite{Overhauser:1953a}. It should be emphasized that
while the term $\propto \hat{I}_z\hat{S}_z^eh_1(t)$ also fluctuates in
time, it does not directly induce any spin flips. Depending on the exact
experimental conditions, the electron-nuclear spin flip-flop term
can lead to electron spin dephasing \cite{Braun:2005a}, dynamic
nuclear polarization \cite{Gammon:1997a} or nuclear spin dephasing
\cite{Abragam:1961a,Merkulov:2002a}.

$\hat{H}_1(t)$ can be visualized as a random perturbation between
states split in energy by $\hbar\omega_e$. The function $h_1(t)$ is
characterized by its mean value $\overline{h_1(t)}=f_e$ and a
simple, auto-correlation function
$\overline{h_1(t)h_1^*(t+\tau)}=\exp(-\frac{\vert\tau\vert}{\tau_c^e})$
with a correlation time $\tau_c^e$. The fraction of time the quantum
dot contains an electron $f_e$ takes values between 0 and 1.
The rate of nuclear polarisation will depend on the splitting $\hbar\omega_e$ and the level broadening $\hbar/\tau_c^e$, \cite{Eble:2006a,Urbaszek:2007a}, as discussed in Sec.\,\ref{sec:dnp}.\\

For commonly achieved nuclear  spin polarization values well below
100\%, the nuclear field fluctuates around a mean value $\langle B_n
\rangle$. The fluctuations (root mean square deviation) can be
written as an effective field $\delta B_{n} = \sqrt{\langle B_n^2
\rangle - \langle B_n \rangle^2} $. Several theoretical
studies have predicted that the dominant mechanism of electron spin
relaxation in QDs at low temperature and zero external
magnetic field is due to the hyperfine interaction with these
nuclear field fluctuations $\delta B_{n}$
\cite{Burkard:1999a,Khaetskii:2002a,Merkulov:2002a,Semenov:2003a}.
The reason for the non-negligible $\delta B_{n}$ lies in the finite
number of nuclei within the dot: The mesoscopic nuclear spin system of a QD is described by the nuclear spin operators $
\hat{I}_x $,$ \hat{I}_y $,$ \hat{I}_z $. These operators do not commute, it is therefore
impossible to determine the $x$,$y$ and $z$ components of the
nuclear spin system with equal precision i.e. they can not all be exactly zero.
In the absence of DNP
\textit{repeated} measurements of the expectation value of $B_n$ at
time intervals longer than the nuclear spin correlation time of the
order of $10^{-4}$~s give an average of $\langle B_n \rangle=0$. 
But, employing a useful qualitative physical picture \footnote{The electron really interacts with a quantum field
of indeterminate magnitude and direction at any time scale for $B=0$.}, 
an electron spin will interact during its lifetime (about 1ns in InAs QDs) with a field of
typical magnitude $\delta B_n$ and random orientation during about
$10^{-4}$~s; this is referred to as the \textit{frozen
fluctuation model} \cite{Merkulov:2002a} and is detailed in
Sec.\,\ref{sec:elecdynamics}.

An important interaction between nuclear spins is the dipole-dipole
interaction that allows for example nuclear spin diffusion in bulk
GaAs samples \cite{Paget:1982a} with spatially inhomogeneous nuclear
polarization. The dipole-dipole interaction of a nucleus $n$ with
the other nuclei $n'$ separated by the translation vector
$\bm{r}_{nn'}$ can be written as \cite{Abragam:1961a}:

\begin{equation}
\label{eq:dipdip}
\hat{H}_{dd}= \frac{\mu_N^2}{2}\sum_{n\neq n'}\frac{g_ng_{n'}}{r_{nn'}^3}\left( \hat{\bm{I}}^n \hat{\bm{I}}^{n'} - 3\frac{(\hat{\bm{I}}^n \bm{r}_{nn'})(\hat{\bm{I}}^{n'} \bm{r}_{nn'})}{r_{nn'}^2}\right)
\end{equation}

As a result of the dipole-dipole interaction each nucleus experiences a fluctuating local effective magnetic field $\delta B_L$, where $\delta B_L \simeq 0.15$~mT in GaAs, created by the other nuclei. Via the non-secular (non spin conserving) part of the dipole-dipole interaction nuclear spin is transferred to the crystal as a whole and is not conserved, see Ch. VIII.E of \textcite{Abragam:1961a} were secular and non-secular parts of the dipole-dipole interaction are detailed.
The precession of the nuclear spins around $B_L$ is one of the reasons why dynamic nuclear polarization in GaAs bulk in the absence of any applied magnetic field is not possible \cite{Meier:1984a}. In QDs two interactions, namely the Knight field $B_K$ and the nuclear quadrupole interaction can in principle dominate $B_L$ already at zero field, as discussed in detail in Sec.\,\ref{sec:quadintro} and \ref{sec:zerofield}.

\subsection{Electrostatic coupling: Nuclear quadrupole effects}
\label{sec:quadintro}

\begin{figure}[t]
\centering
\includegraphics*[width=0.48 \textwidth,angle=0]{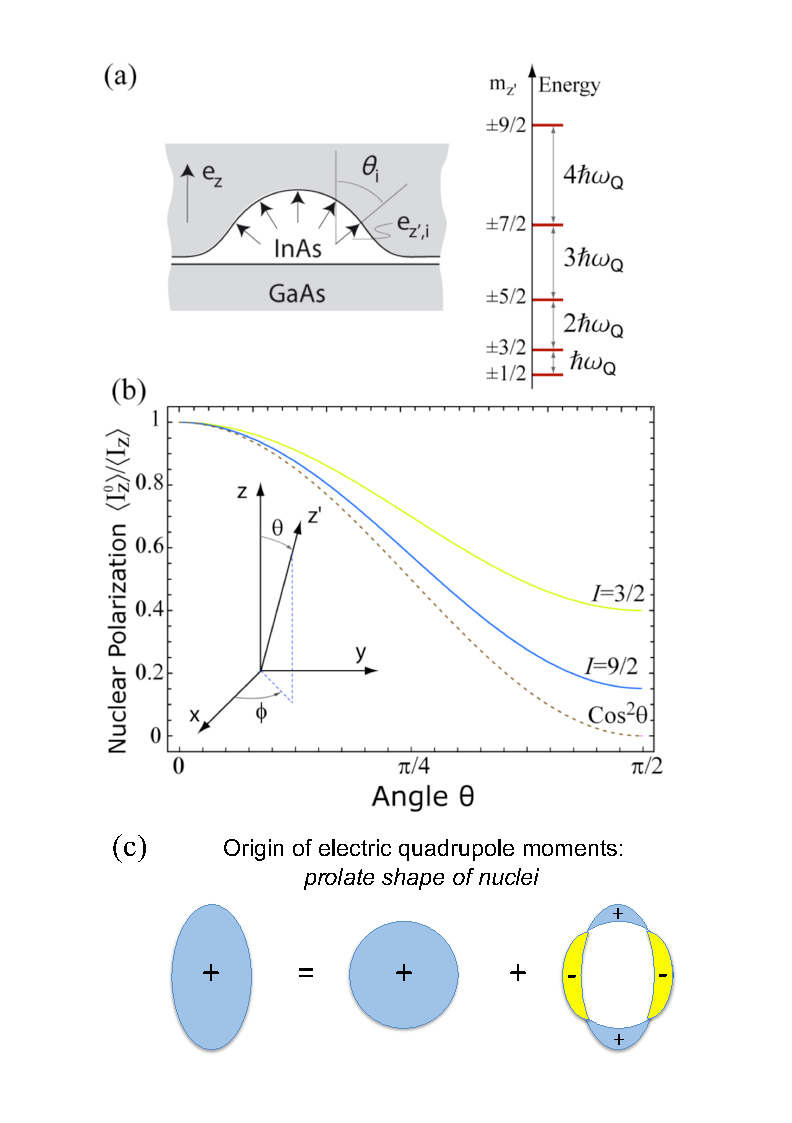}
\caption{ (a) Sketch of the  strain distribution in an InAs/GaAs
QD and splitting of energy levels in zero field due to
axially symmetric quadrupolar interaction for a nuclear spin
$I=9/2$. (b) Reduction factor of nuclear polarization along $z$ due
to the inclination $\theta$ of the quadrupolar main axis.  In the
approximation of high spin temperature, it is given by  $\langle
I_z\rangle/\langle I_z^0\rangle=
(I+3/4)(I-1/2)\cos^2\theta/(I(I+1))+3 (I+1/2)/(4I(I+1))$ (c) A nucleus with non-spherical, prolate charge distribution is equivalent to a spherically symmetric charge distribution plus some positive charge shared between the two polar regions and a band of equal negative charge added around the equator. This addition has  no dipole moment but does have electric quadrupole moment, after \textcite{Williams:1991a}.}\label{LPN-QIintro}\end{figure}

Due to lattice strain and atomic inter-diffusion electric quadrupolar effects are strong for nuclei in QDs  compared to the influence of alloy disorder in unstrained bulk samples and hence play a central role in nuclear spin dynamics in QDs \cite{Dzhioev:2008a}, see \textcite{Bulutay:2012a} for a detailed discussion. Quadrupolar effects are at the heart of many of the surprising effects that go beyond the nuclear spin physics known from bulk and quantum well systems, such as for example zero field DNP \cite{Lai:2006a,Oulton:2007a} (see Sec,\,\ref{sec:zerofield}), strongly suppressed spin diffusion \cite{Maletinsky:2009a} (see Sec\,\ref{sec:demag}), the anomalous Hanle effect \cite{Krebs:2010a} (see Sec\,\ref{sec:ahanle}) and the locking of quantum-dot resonances to an incident laser \cite{Latta:2009a,Hogele:2012a} (see Sec\,\ref{sec:locking}).\\
\indent Nuclei have no electric dipole moment and are thus insensitive to homogeneous electric fields \cite{Abragam:1961a}.
But the non-spherical (prolate) charge distribution of  atomic nuclei with
spin $I>1/2$ presents an electric \textit{quadrupolar} moment, as sketched in Fig.\,\ref{LPN-QIintro}c,  which can couple to inhomogeneous electric fields produced by electron clouds, expressed as an
electric field gradient $\partial^2 V/\partial x_\alpha\partial
x_\beta$ where $V$ is the electrical potential due to local charge distribution.
If the nuclear environment has cubic symmetry, the electric field gradient vanishes  and so
does the quadrupolar coupling \cite{Abragam:1961a,Slichter:1990a}. This situation
prevails in  bulk GaAs, but the cubic symmetry breaks down in
self-assembled QDs like InAs/GaAs because of large biaxial strain
associated  with the $\sim$7\% lattice mismatch between InAs and GaAs.
Also,  inter-diffusion  of In and Ga atoms during QD growth
results in a substantial fraction of As atoms for which all  first
neighbours  are no longer identical. The local tetrahedral symmetry
is  then lost and an electric field gradient arises along one of the
crystallographic  directions $\langle111\rangle$ or
$\langle100\rangle$ \cite{Meier:1984a}.

This electrostatic coupling of the electronic system with the nuclear spin system can manifest itself in principle in the analysis of either system.
For the conduction electrons most relevant for the physics described here the quadrupolar interaction vanishes (for $s$-electrons i.e. $l=0$) \cite{Abragam:1961a} and in what follows we only consider the effect of the electric field gradients on the nuclear spin system.
For simplicity we assume that the electric field gradients in the dot have cylindrical (axial) symmetry i.e. the strength of the interaction does not depend on the angle $\phi$ in the $x-y$ plane, defined in Fig.\,\ref{LPN-QIintro}b with respect to the growth axis $z$. The influence of the electric field gradients oriented along an axis $z'$ on a nucleus of spin
$I$ can then be described by \cite{Abragam:1961a}:
\begin{equation}
\hat{H}_\text{Q} = \frac{\hbar \omega_\text{Q}}{2}\left(\hat{I}_{z'}^2-\frac{I(I+1)}{3}\right)
\label{eq:HQ}
\end{equation}
where  $\hbar \omega_\text{Q}$ is the quadrupolar splitting
proportional to the nuclear quadrupolar moment and to the
electric field gradient, and  $\hat{I}_{z'}$ is the angular momentum
projection on the principal axis $z'$ of the electric field
gradient. Without loss of generality one can assume
$\hat{I}_{z'}=\hat{I}_z\cos\theta+\hat{I}_x\sin\theta$ to analyze
the effect of the Eq.\,\eqref{eq:HQ} on the nuclear polarization:

\begin{eqnarray}
\hat{H}_\text{Q} &=& \frac{\hbar \omega_\text{Q}}{2} [\hat{I}_{z}^2 \cos^2\theta -\frac{I(I+1)}{3} \nonumber \\
&+& (\hat{I}_{z}\hat{I}_{x}+\hat{I}_{x}\hat{I}_{z})\sin \theta \cos \theta + \hat{I}_{x}^2 \sin^2\theta]
\label{eqn:HQlong}
\end{eqnarray}

In zero magnetic field, the $(2I+1)$ spin levels are split according to
the square of their angular momentum projection $m_{z'}$ onto  $z'$.
For a half-integer spin $I$, this leads to pairs of levels
$m_{z'}=\pm1/2, \pm 3/2\ldots\pm I$, which are separated from each
other by $1, 2, \ldots (I-1/2)\times\hbar \omega_\text{Q}$, see
Fig.\,\ref{LPN-QIintro}(a).

Although the quadrupolar interaction is qualitatively
different from a magnetic field, for the purpose of quantitative
comparison it is useful to express $\hbar\omega_\text{Q}$ as an effective field
$B_\text{Q}$, where $g_N \mu_N B_\text{Q}\equiv \hbar\omega_\text{Q}$. In self-assembled InAs QDs estimated values of
$B_\text{Q}$ fall in the 100~mT range.
\cite{Dzhioev:2008a,Maletinsky:2009a,Krebs:2008a}. As a result,
the dipolar coupling between nuclear states of angular momentum
difference $|\Delta m_{z'}|=1$ or $|\Delta m_{z'}|=2$ is strongly
inhibited. Only the states $m_{z'}=\pm1/2$ still experience the
local small fluctuating field due to surrounding nuclear spins in the order of $\delta B_L\approx 0.15~$mT in GaAs.
The polarization relaxation induced by the non-secular part of the dipolar interaction is then essentially
suppressed for the levels $|m_{z'}|>1/2$ in agreement with the
substantial Overhauser shift observed in QDs for zero external field, as discussed in section \ref{sec:zerofield}.
The quadrupolar shifts also lead to an energy mismatch between nuclear spin levels of atoms inside the dot compared to atoms in the surrounding barrier material which leads to a strong suppression of nuclear spin diffusion from the dot towards the barrier.
As a result the mesoscopic spin system of $\approx 10^5$ nuclear spins in a highly strained QD like InAs in GaAs is well isolated from its surroundings.\\
The inclination $\theta$ of the quadrupolar axis
induces oscillations of the nuclear polarization component
perpendicular to $z'$,  while keeping constant the longitudinal
projection along $z'$. Under cw optical excitations this transverse
part   vanishes and the nuclear polarization created in a quantum
dot along the external field direction  $z$ is reduced by $\sim
\cos^2\theta$, as shown in Fig.\,\ref{LPN-QIintro}(b) for nuclear spins
$I=3/2$ and $I=9/2$. This effect which has to be averaged over the
angle dispersion of $z'$ may contribute to the enhancement of the
effective nuclear spin relaxation observed experimentally in low magnetic fields, see Sec.\,\ref{sec:DNPvsBz}.

\section{DYNAMICS OF ELECTRON SPINS COUPLED TO A FLUCTUATING NUCLEAR FIELD}
\label{sec:elecdynamics}

In bulk semiconductor or quantum well structures,  the electron spin
dephasing induced by the interaction with nuclear spins is usually
much weaker than the well-known mechanisms originating from
spin-orbit interactions, well documented in
\textcite{Meier:1984a} and \textcite{Dyakonov:2008a}. Due to the absence of
translational carrier motion in semiconductor QDs, the
discrete energy levels due to carrier localization and the
corresponding lack of energy dispersion lead to a strong suppression
of these well-known electron spin relaxation processes
\cite{Burkard:1999a,Khaetskii:2000a,Paillard:2001a}. The spin
relaxation time due to hyperfine interaction with lattice nuclei was
first derived by Dyakonov and Perel for donor-bound electrons
\cite{Dyakonov:1973a,Dyakonov:1974a,Dzhioev:2002a,Paget:1977a} and subsequently in great detail for electrons confined to QDs
\cite{Burkard:1999a,Khaetskii:2002a,Merkulov:2002a,Semenov:2003a}.
We review in this section the experimental and theoretical work on
this topic. Throughout this section the mean nuclear spin
polarization $\langle B_n \rangle$ is taken to be zero, i.e. \underline{no}
dynamic nuclear polarization (DNP) is created, see sections \ref{sec:dnp} and \ref{sec:tdnp} 
for a detailed discussion about build-up, manipulation and decay of nuclear spin polarization.

\subsection{The Merkulov-Efros-Rosen Model}
\label{sec:merk}

\begin{figure}
\epsfysize=2.4in
\epsfbox{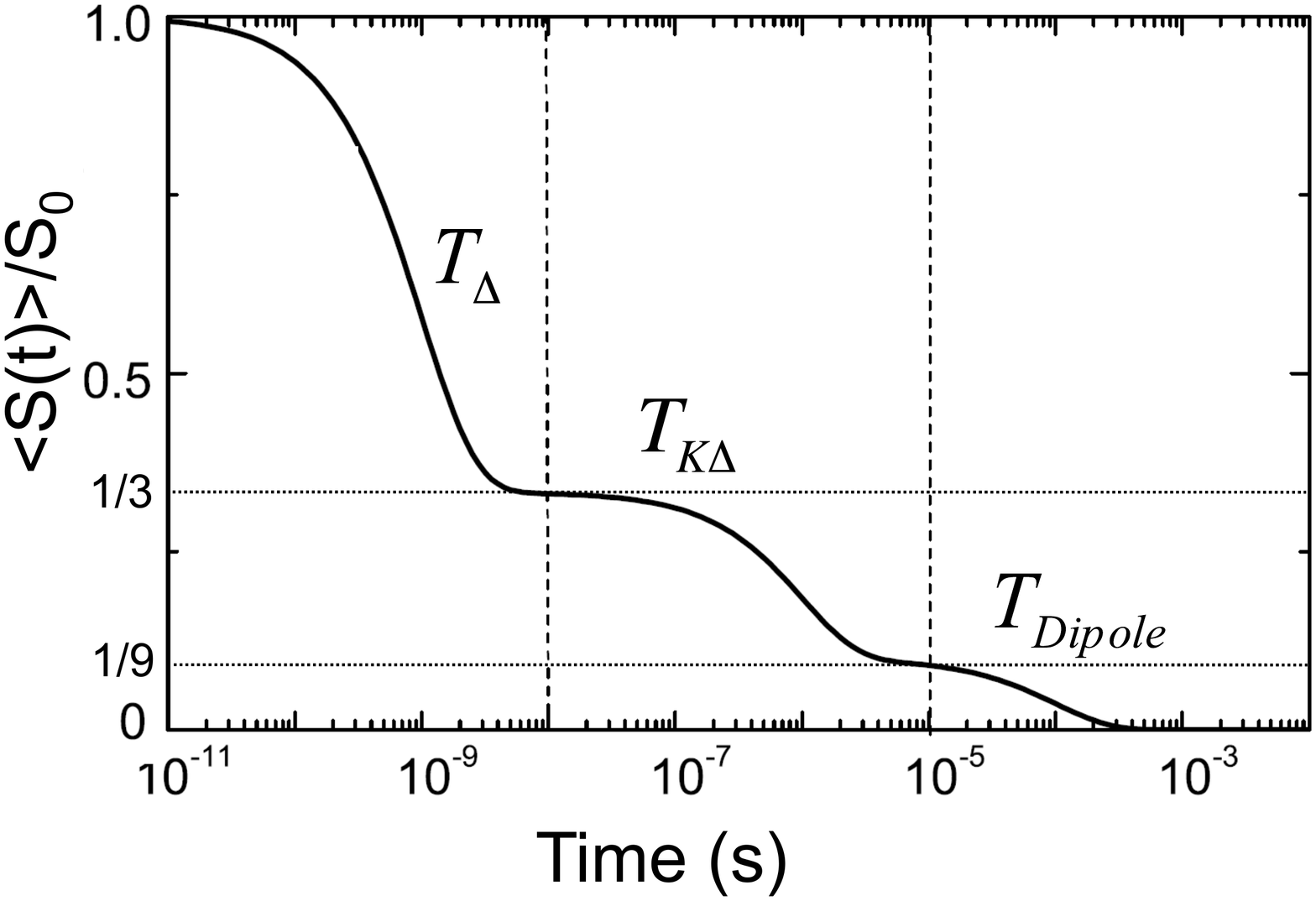}
\caption{Electron spin relaxation induced by nuclei in a QD. The calculation is done for $N=10^5$ nuclei per dot (after Chapt. 11 in \textcite{Dyakonov:2008a}).
 \label{xmfig1}}
\end{figure}

Three distinct time scales are relevant for describing the
electron-nuclei  spin system evolution in a QD according to
the Merkulov-Efros-Rosen (MER) model \cite{Merkulov:2002a}:\\
\indent (1) the first time corresponds to the electron-spin precession around the
frozen nuclear field fluctuations given by $\delta \bm{B_n}$
\cite{Burkard:1999a,Khaetskii:2002a,Semenov:2003a}: the typical
dephasing time is of the order of $T_\Delta\approx\frac{\hbar}{g \mu_B \delta B_n}\sim 1$~ns for InAs QDs
containing $10^5$  nuclei  (see Fig.\,\ref{xmfig1}). \\
\indent (2) The second time
is controlled by nuclear-spin precession in the inhomogeneous hyperfine field of
the localized electron (Knight field $\bm{B_K}$): the typical time is
given by $T_{K\Delta}\simeq\sqrt{N} T_{\Delta}$ which results for
$N=10^5$ in $T_{K\Delta}\sim 1$~$\mu$s. \\
\indent (3) The third time is given by
the nuclear-spin relaxation due to dipole-dipole interaction with
nuclei in the vicinity of the QDs : its order of magnitude is given
by the average precession time of a nuclear spin in the local field
fluctuation $\delta \bm{B_L}$,   occurring on a typical timescale
$T_\text{Dipole}\sim 100$~$\mu$s.

During the first two stages, the total angular magnetic moment  of
an electron and the nuclei interacting with this particular electron
is conserved. Thus the global coherence of the electron-nuclear spin
system is preserved, while during the last stage it is not, since
the dipolar interaction does not conserve the total angular magnetic moment (see Ch.VIII.E of \textcite{Abragam:1961a}).

\begin{figure}
\epsfxsize=3.5in
\epsfbox{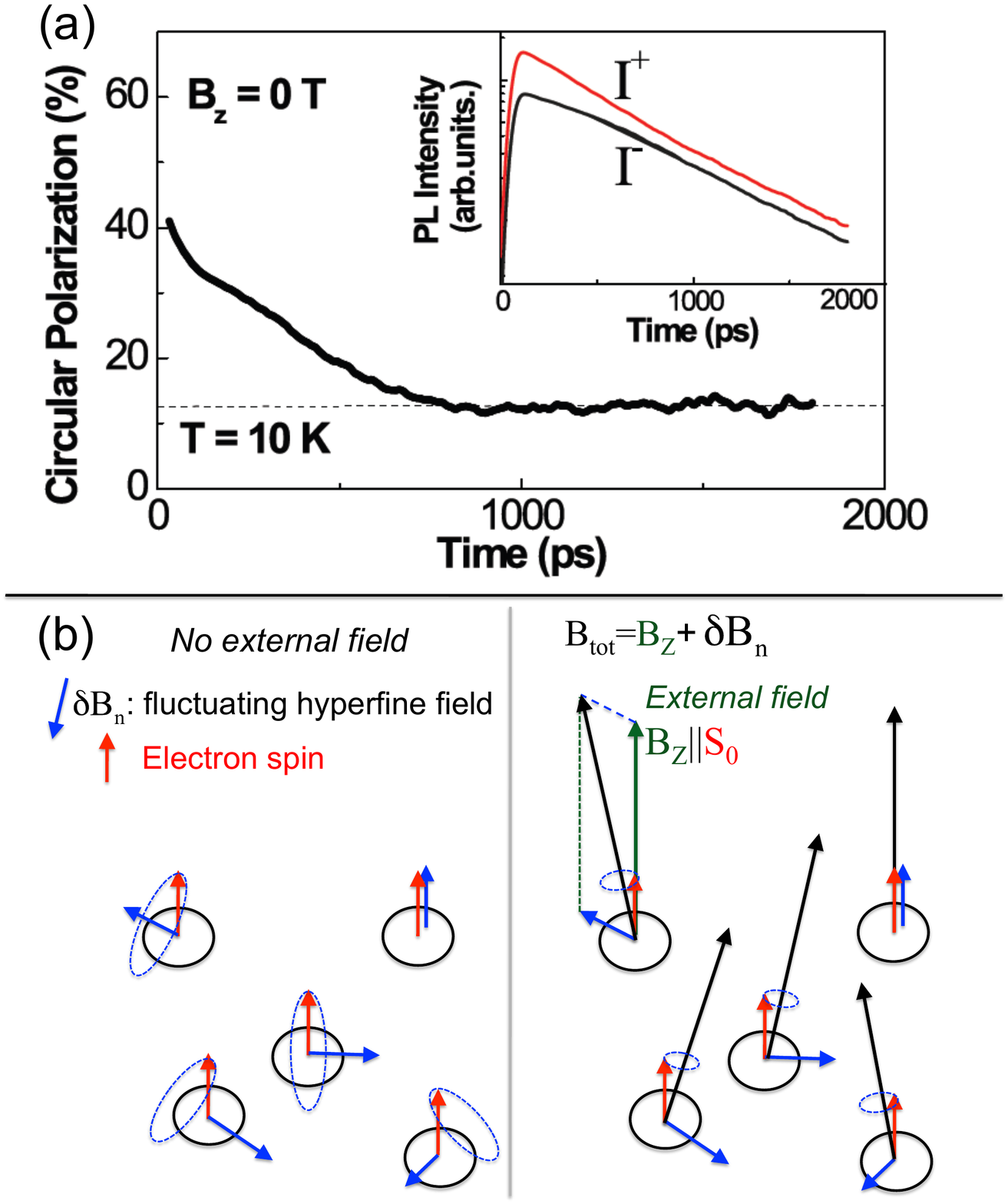}
\caption{(a) Circular polarization dynamics of the positively charged exciton $X^+$ luminescence for $B = 0$ for an ensemble of InAs dots. $X^+$ is formed by a photocreated electron-hole pair and a resident hole. Since unpolarized holes form a zero spin singlet, the spin-polarized electron is not coupled to the holes by the exchange interaction, which can dominate the hyperfine interaction \cite{Braun:2005a,Dyakonov:2008a}. The inset displays the $I^+$ and $I^-$ luminescence dynamics.
(b) Schematics of the electron spin precession around (left) the effective nuclear field $\delta B_n$  and (right) the total field $B_{tot}=B_z+ \delta B_n$ if an external magnetic field $B_z$ is applied.
 \label{xmfig2}}
\end{figure}

Let us focus first on the shortest decay time corresponding to the
electron-spin precession  around the frozen nuclear field
fluctuations. An electron spin in a QD interacts with a
large but finite number of nuclei for instance $N\sim 10^5$ in an
InAs QD. In the frozen fluctuation model, the sum over the
interacting nuclear spins gives rise to an average local effective hyperfine
field $\bm{B_n}$ with a fluctuation characterized by $\delta \bm{B_n}$. The
dispersion of the nuclear hyperfine field $\bm{B_{n}}$ in the absence of
dynamic nuclear polarization can be described  by a Gaussian
distribution $W(B_n)\propto \exp(-3B_n^2/2\delta B_{n}^2)$. $W(B_n)$
has a spherical symmetry indicating that $\bm{B_n}$ has no preferred spatial
orientation. To estimate the magnitude of $\delta B_{n}$, consider a
QD made of $N$ identical nuclear spins $I$. The average
amplitude of the fluctuating hyperfine field reads $\delta
B_{n}=\frac{1}{g_{e}\mu_{B}}\frac{2\tilde{A}}{\sqrt{N}}\sqrt{I(I+1)}$.
The maximum Overhauser field corresponding to a 100$\%$ nuclear spin
polarization is $B_{n}^\text{max}=\frac{1}{g_{e}\mu_{B}}2\tilde{A}I$ in
the order of several Tesla in InAs and GaAs. As a result $\delta
B_{n}\sim B_{n}^\text{max}/ \sqrt{N}$ which corresponds to an effective
field in InAs dots of typically 30~mT which arbitrarily changes
orientation on  a time scale of $10^{-4}$~s
\cite{Merkulov:2002a}.
The exact value of $\delta B_{n}$ extracted from measurements in InAs/GaAs QDs varies typically from 20 to 40~mT. This variation has two main origins: \\
(i) The exact value of the hyperfine constant $\tilde{A}$ and the relevant nuclear spin $I$ depends on the exact chemical composition of the dot influenced, for example, by Gallium inter diffusion into nominally pure InAs dots, resulting in reality in In$_{1-x}$Ga$_{x}$As dot formation. \\
(ii) The exact QD dimensions and hence the number of nuclei N interacting with the electron spin vary from dot to dot even within the same sample wafer.\\
\indent Although an electron spin precesses
coherently  around $\delta \bm{B_n}$ in a given dot,  the amplitude and
the direction of this effective nuclear field vary strongly from dot
to dot. The average electron spin $\left\langle
\textbf{S}(t)\right\rangle$ in an ensemble of dots will thus decay during this \textbf{first stage}
as a consequence of the random distribution of the local nuclear
effective field \cite{Merkulov:2002a,Braun:2005a,Khaetskii:2002a}:
\begin{equation}
\label{eq:A} \left\langle
\textbf{S}(t)\right\rangle=\frac{\textbf{S}_0}{3}\left\{1+2\left[1-2\left(\frac{t}{2T_\Delta}
\right)^2\right]\exp\left[-\left(\frac{t}{2T_\Delta}\right)^2\right]\right\}
\end{equation}
where $S_0$ is the initial spin and
\begin{equation}
\label{eq:tdelta}
T_\Delta=\hbar\left( \frac{3N}{2n\sum_{j=1}^n I^j(I^j+1)(A^j)^2}\right)^{1/2}
\end{equation}
is the dephasing time due to the random electron precession
frequencies in the randomly distributed frozen fluctuation of the
nuclear hyperfine field; here $n$ is the number of atoms per unit
cell of the lattice and the index $j$ runs over the nuclei of a unit
cell. The fluctuating field $\delta \bm{B_n}$ is assumed to be isotropic
for electron spin dynamics. If the electron spin is initially
orientated along the z axis, only the components of  $\delta \bm{B_n}$ in
the x and y directions will contribute to spin dephasing (see
Fig.\,\ref{xmfig2}(b)). Hence only 2/3 of the initial electron spin
polarization is lost.
Figure\,\ref{xmfig1} shows that the calculated  average electron
spin $\left\langle \textbf{S}(t)\right\rangle$ drops down to about
1/3 of its initial value on a characteristic time $T_\Delta\sim 1~$ns.

The second stage of spin relaxation, occurring on a
characteristic time $T_{K\Delta}$, is due to the Larmor precession
of the nuclear spins in the \textit{inhomogeneous} Knight field due
to the electron spin (see Eq.\,\eqref{eq:BK}). This precession
results in a new configuration of the random nuclear field. During
this \textbf{second stage}:
\begin{equation}
\label{eq:st}
\langle \bm{S} (t) \rangle = \frac{\bm{S}_0}{3} \langle \bm{B}_n (t) \cdot \bm{B}_n (0) \rangle _N/\langle \bm{B}_n^2 (0) \rangle _N
\end{equation}
where the time origin is set here at a delay of the order of
$T_{\Delta}$. For $T_{\Delta}\ll t < T_\text{Dipole}$ one finds $\langle
\bm{B}_n (t) \cdot \bm{B}_n (0) \rangle _N/\langle \bm{B}_n^2 (0)
\rangle _N\sim 1/3$, since only the components $\langle
\hat{I}_x\rangle$ and $\langle \hat{I}_y\rangle$ of the nuclear
field precess about $\langle S_z^e\rangle \vec{e}_z$. The average
electron spin evolves like the nuclear field correlation function
\cite{Merkulov:2002a}. The time $T_{K\Delta}$ is much longer than
$T_\Delta$ because the interaction of an electron spin with a given
nucleus is $\sqrt{N}$ times weaker compared to the interaction
with the effective magnetic field of the nuclear fluctuations:
$T_{K\Delta}\sim T_\Delta \cdot\sqrt{N}\sim1$~$\mu$s. During this second
stage the electron spin feels a slow variation of the effective
nuclear field yielding again a decrease of the average electron spin
down to $1/9$ of its initial value $\langle \bm{S} (t) \rangle =
\frac{\bm{S}_0}{9}$, as indicated in Fig.\,\ref{xmfig1}.

Finally the \textbf{third stage} of electron spin  relaxation, labelled
$T_\text{Dipole}\sim100~\mu$s, is due to the dipole-dipole interaction
of nuclear spins. In contrast to the two first stages, this
dipole-dipole interaction does not conserve the total nuclear spin
and is thus the upper limit for the electron spin coherence $T_2$.
Whereas the electron spin dephasing time in the fluctuating nuclear
field ($T_{\Delta}$) has been measured by different groups (see the
next section), the two other stages occurring on much longer
timescales are still under discussion \cite{Fras:2011a}.  Using a terminology
associated to quantum coherence, the $T_{\Delta}$ time is sometimes
referred to as an inhomogeneous electron spin dephasing time.\\
It is important to note that the nuclear spin dynamics predicted during the second and third stage of the Merkulov model will be strongly influenced by inhomogeneous \textit{quadrupolar coupling} of the nuclear spins with local electric
field gradients, present in InAs QDs due to local anisotropic strain, as introduced earlier in section \ref{sec:quadintro}.
In practice, the strong quadrupolar effects in InAs quantum dots \cite{Maletinsky:2009a,Flisinski:2010a} will prevent the nuclei from precessing around the Knight field and also suppress dipolar relaxation for $B_z<B_Q$. As a result, for QD systems with strong nuclear quadrupolar coupling the description by the Merkulov model of the first stage of the electron spin relaxation in the \textit{frozen} average nuclear field fluctuations on a timescale $T_{\Delta}$ is by far the most relevant of the three.

\subsection{Experimental studies of electron spin dephasing in zero external magnetic field}
\label{sec:dephzero}
In optical experiments performed on undoped QDs the photogenerated electron feels a strong effective magnetic field due to the exchange interaction with the photogenerated hole in the neutral exciton $X^0$ \cite{Bayer:2002a,Senes:2005a}. This exchange field (with a characteristic energy $\delta_0$ of typically hundreds of $\mu$eV) is much stronger than the effective field due to the fluctuating nuclear field $\delta B_n$ of the nuclei, which thus plays a negligible role  for the spin dynamics of the electron in $X^0$ \cite{Erlingsson:2001a} in most experiments (see \textcite{Stevenson:2011a} for experiments investigating electron dephasing due to $\delta B_n$ in neutral QDs). The positively charged excitons $X^+$ , consisting of one electron and two holes forming a spin singlet, is the ideal configuration to probe the electron spin relaxation mediated by nuclei in QDs with optical experiments. Due to Kramer's theorem the anisotropic contribution to the exchange interaction between the electron and the two holes does not lead to any fine structure splitting of the $X^+$ ground state. Thus the analysis of the circular polarization $\rho_c$ of the $X^+$ luminescence in p-doped QDs following a circularly polarized laser excitation probes directly the spin polarization of the electron as
$\left\langle S_z^e\right\rangle =-\rho_c/2$.

Figure\,\ref{xmfig2}(a) displays the circular polarization dynamics of the $X^+$ photoluminescence from an ensemble of InAs/GaAs QDs \cite{Braun:2005a}. The inset shows the time evolution of the polarised luminescence intensity components.
The circular polarization dynamics in Fig.\,\ref{xmfig2}(a) presents two regimes: The polarization decays within the first 800ps down to about 1/3 of its initial value; then it remains stable with no measurable decay on the radiative life-time scale. The observed electron spin relaxation is due to the hyperfine interaction with the nuclei \cite{Merkulov:2002a} : from Eq.\,\eqref{eq:tdelta}, we calculate $T_{\Delta}\sim0.5$~ns, in agreement with the observed decay time in Fig.\,\ref{xmfig2}(a). This corresponds to a dispersion of the nuclear hyperfine field distribution $\delta B_n\sim 45$~mT. The subsequent electron spin dephasing $T_{K\Delta}$, which is the result of the variations of the random nuclear field direction, occurs on a time
scale typically 100 times longer than $T_{\Delta}$. Thus it cannot be observed on the $X^+$ radiative lifetime scale ($\sim$1~ns) accessible in PL measurements.

\subsection{Electron spin dephasing in a longitudinal magnetic field: Faraday geometry }
\label{sec:dephfara}

\begin{figure}
\epsfxsize=3.5in
\epsfbox{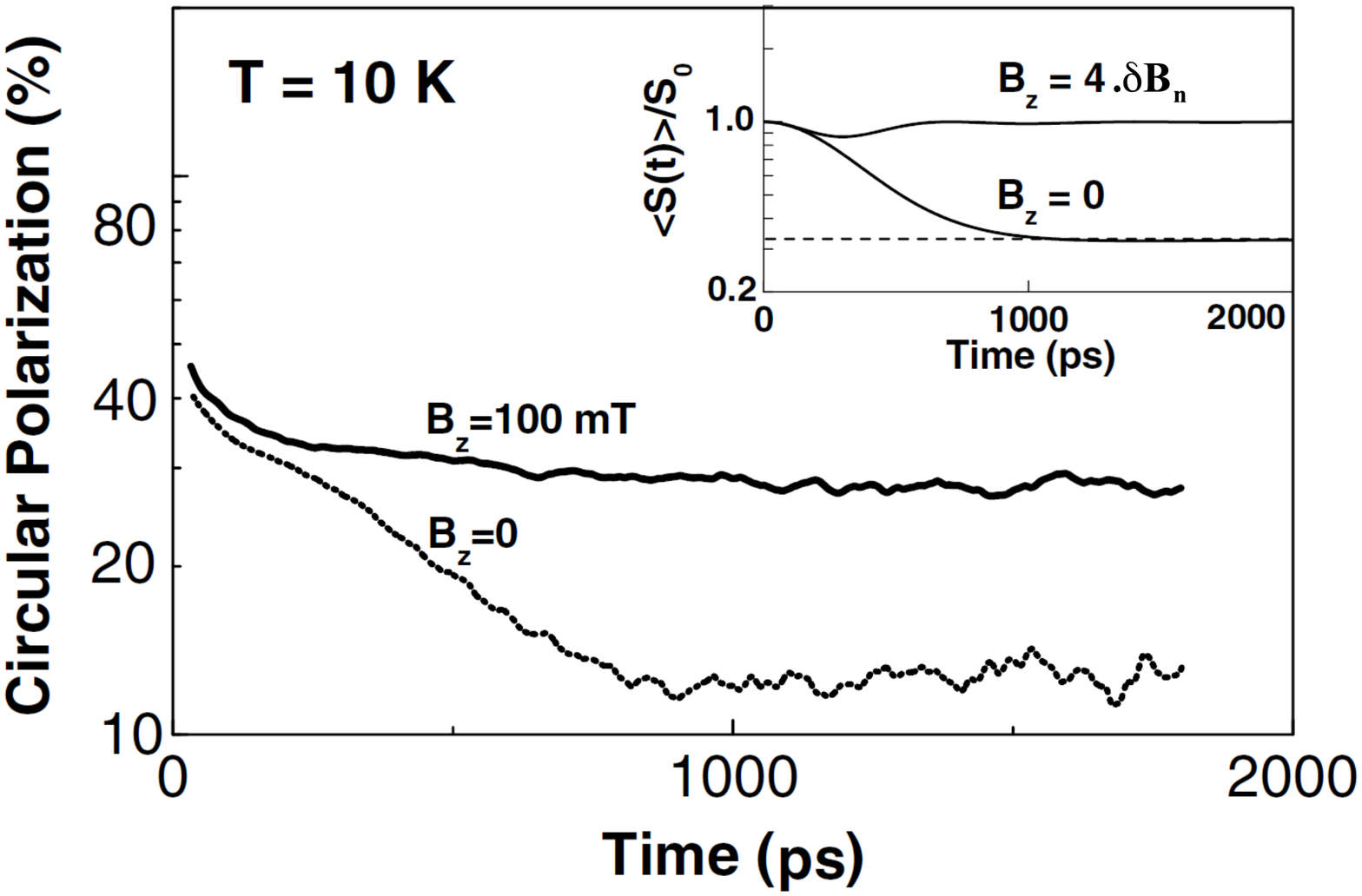}
\caption{Circular polarization dynamics of the $X^+$ luminescence in an ensemble of InAs dots for $B_z = 0, B_z = 100$~mT. The inset displays the calculated time dependence of the average electron spin $\left\langle \textbf{S}(t)\right\rangle/S_0$ \cite{Merkulov:2002a,Braun:2005a}.
 \label{xmfig3}}
\end{figure}

An external magnetic field applied along the $z$ growth axis
(Faraday configuration) which adds to the nuclear field fluctuations
$\delta B_n$, can stabilize the electron spin, which will then
precess about the resulting total field $B_\text{tot}=B_z+\delta B_n$.
This effect is sometimes referred to as the \textit{screening} of
$\delta B_n$  by the external field. $B_z$ must be larger than
$\delta B_n$, to ensure that the Zeeman interaction of the electron
spin with the magnetic field is stronger than the interaction with
the nuclei \cite{Merkulov:2002a}. Figure\,\ref{xmfig3} displays the
circular polarization dynamics of the $X^+$ luminescence with a
magnetic field $B_z=100$~mT; the dynamics for $B_z = 0$ is also
presented for comparison \cite{Braun:2005a}. Note that the Zeeman
splitting energy of the electron in this weak magnetic field is
about 100 times smaller than $k_BT$. By applying a field of $B_z =
100$~mT, the initial decay is suppressed since the total field
affecting the electron spin becomes almost parallel to the initial
spin direction as in Fig.\,\ref{xmfig2}(b) (right). Let us emphasize
that this does not mean the nuclear field fluctuations disappeared:
they still strongly affect the $S_x$ and $S_y$ electron spin
components which is a key obstacle for use of electron spin states in
quantum information schemes, as shown in Sec.\,\ref{sec:dephvoigt}.
This pronounced effect of the small external magnetic field observed
in Fig.\,\ref{xmfig3} agrees well with the predicted influence of
the external magnetic field on the electron spin relaxation by
nuclei in InAs QDs (see inset of Fig.\,\ref{xmfig3}, \textcite{Merkulov:2002a}, \textcite{Semenov:2003a}) or in InP
QDs \cite{Pal:2007a}. The effect observed here is similar
to the suppression of the nuclear hyperfine interaction effects
measured for localized  electrons in lightly doped bulk n-GaAs
\cite{Dzhioev:2002a,Colton:2004a}. For larger external magnetic
fields the relaxation of the $z$ component of the electron  spin is
no longer governed by hyperfine interaction effects but by spin-orbit
mechanisms mediated by phonon coupling \cite{Khaetskii:2000a}.
Electron spin relaxation times of from milliseconds to seconds can then
be measured for magnetic fields of the order of a few Tesla
\cite{Amasha:2008a,Elzerman:2004a,Kroutvar:2004a}.

\begin{figure}
\epsfxsize=3.6in
\epsfbox{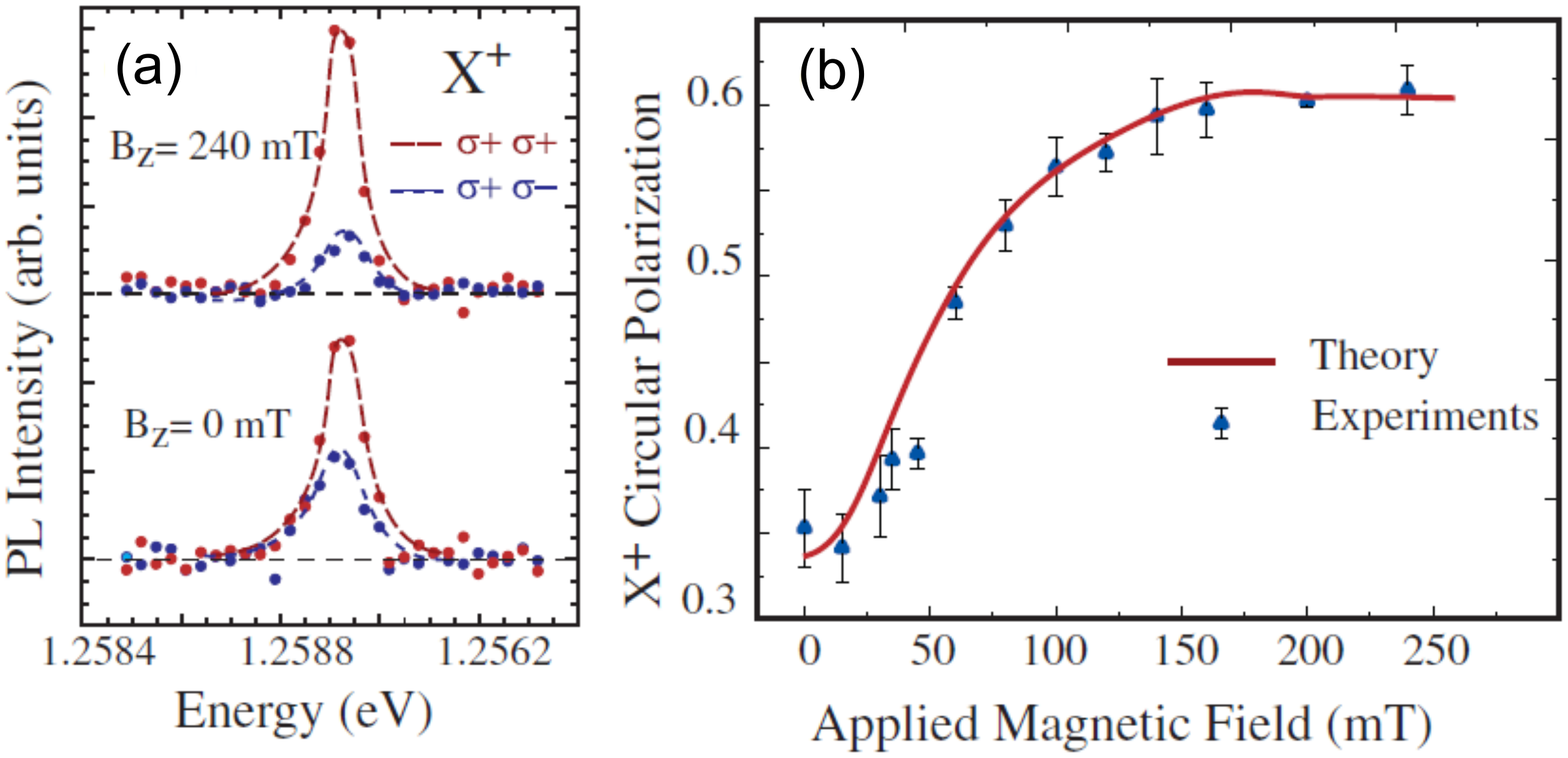}
\caption{Single InAs dot measurement (a) Polarization-resolved spectra of $X^+$ luminescence for two different magnetic fields. (b) Circular polarization of $X^+$ line versus external magnetic field $B_z$. The excitation polarization is provided by a 50 kHz-photo-elastic modulator to avoid the build up of a nuclear polarization through optical pumping \cite{Braun:2006b}.
 \label{xmfig4}}
\end{figure}

For repeated measurements on a single QD the  hyperfine
interaction has the same effect as for an ensemble of dots : $B_n$
will change orientation from one measurement to another since the
signal integration times are commonly much longer than $T_\text{Dipole}
\sim 100$~$\mu$s so that the average is taken over a large number of
uncorrelated nuclear spin configurations, as nicely demonstrated by
\textcite{Dou:2011a}. Figure\,\ref{xmfig4} presents the measurements
of circular polarization on the $X^+$ luminescence in a single InAs
dot following a right circularly polarized excitation light
\cite{Braun:2006b}. Here, the time-integrated circular polarization
appears to be limited to about 35\% for $B_z=0$ as a result of the
electron spin dephasing induced by the fluctuating nuclear field. At
$B_z=240$~mT the circular polarization increases significantly up to
$\sim 60\%$. In Fig.\,\ref{xmfig4}(b) the experimental data show a
very good agreement with the theoretical field dependence according
to the MER model  \cite{Merkulov:2002a} which has been calculated
here with $T_\Delta=470$~ps and an initial photogenerated   electron
spin polarization of the $X^+$ state of 62\%. Note that  the
excitation polarization in this experiment is modulated at 50 kHz
between $\sigma^+$  and $\sigma^-$ to prevent the dynamic
polarization of nuclear spins. In the presence of DNP, the resulting
mean nuclear field $\langle \bm{B_n} \rangle$ would screen its own
fluctuations allowing for a strong $X^+$ circular polarization even
for very weak (or even zero) external magnetic fields
\cite{Krebs:2008a}.

A spin dephasing time $T_\Delta\sim 16$~ns for a resident electron in
a single GaAs QD defined by monolayer fluctuations in a
GaAs/AlGaAs quantum well  has also been deduced from cw
photominescence Hanle experiments \cite{Bracker:2005a} ; this
corresponds to $N\sim 5 \cdot 10^6$ interacting nuclei, which is
consistent with the larger size of these GaAs dots (diameter $\sim
170$~nm) compared to the InAs  QDs (diameter $\sim 20$~nm).

\begin{figure}
\epsfysize=2in
\epsfbox{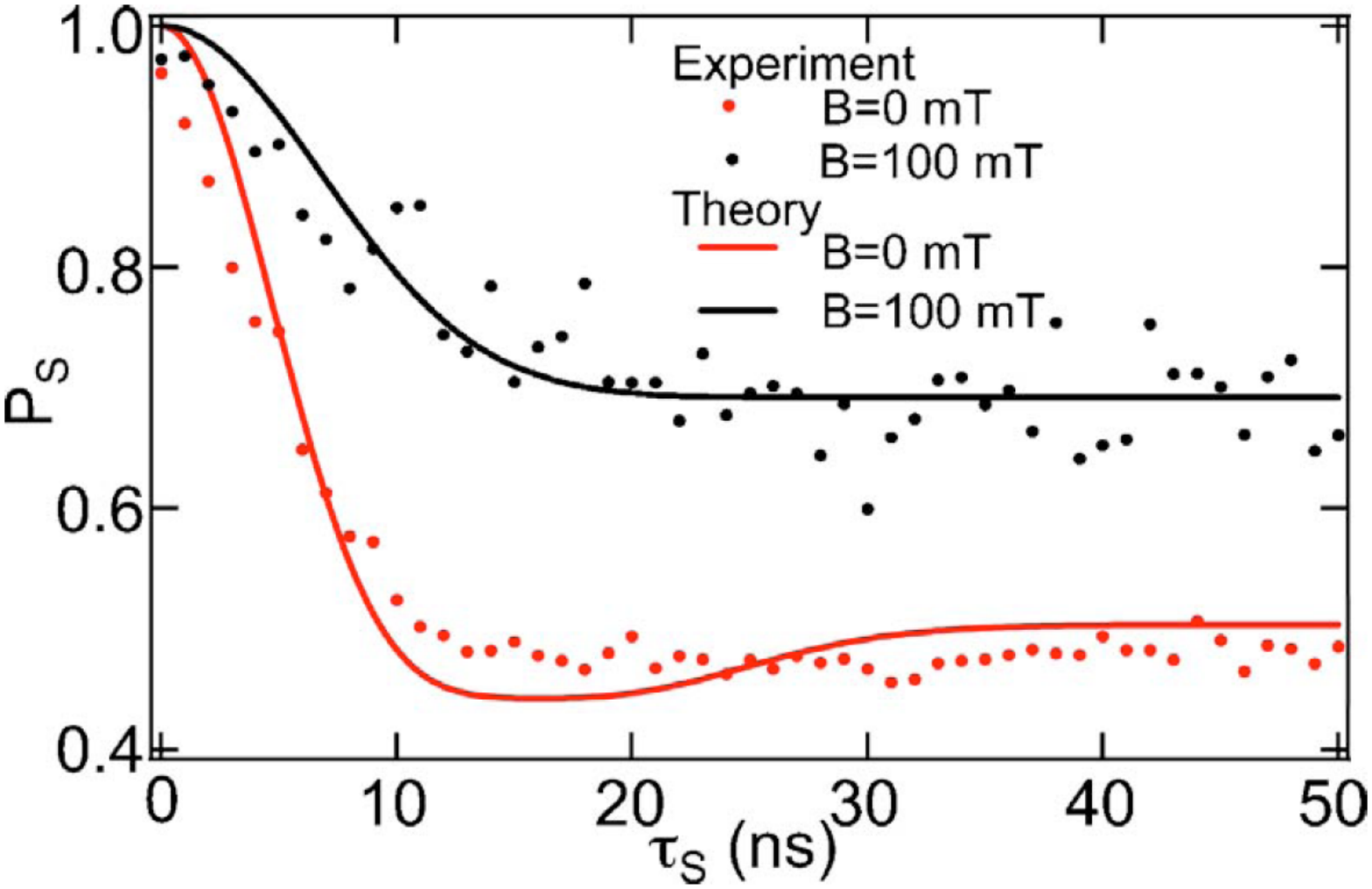}
\caption{Time dependence of  the singlet (two-electron-spin) return probability in an electrically-defined  GaAs double-QD. The data are fitted using a semi-classical model \cite{Schulten:1978a} which is similar to the Merkulov-Efros-Rosen approach. The fits correspond to $\delta B_n =2.3~$mT and a corresponding $T_2^* =10~$ns. \cite{Petta:2005a}.
 \label{xmfig5}}
\end{figure}

The electron spin dephasing induced by nuclei was also intensively investigated
in transport experiments in gate-defined GaAs double QDs \cite{Hanson:2007a} at very
low temperature ($\sim$100~mK). Rapid electrical control of the
exchange interaction in gate-defined double QD devices
allow the measurement of the single electron spin dynamics with an
average value that decays on a characteristic time $T_\Delta$. The measurements show that the separated electron spins
in the two QDs loose coherence in $T_\Delta\sim 10$~ns (see
Fig.\,\ref{xmfig5}). The increase of the long time saturation value of the average electron spin in a
weak external magnetic field ($\sim$100~mT) is also clearly observed
\cite{Petta:2005a}.

\begin{figure}
\epsfysize=4in
\epsfbox{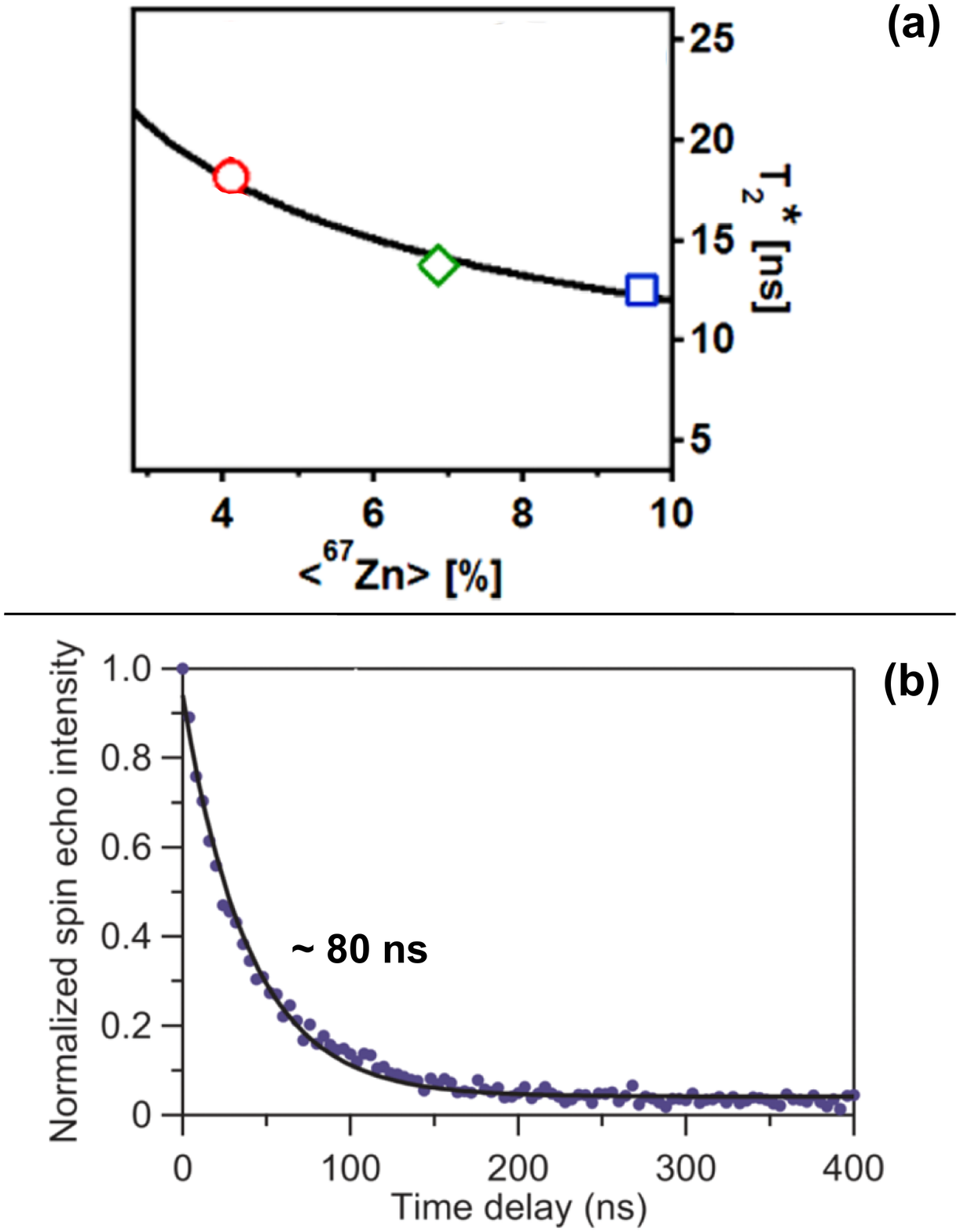}
\caption{(a) Electron spin dephasing time $T_2^*=T_{\Delta}$ in colloidal n-type ZnO nanocrystals as a function $^{67}$Zn: (circle) $4.1$\% natural abundance, (diamond) $~6.8$\%, and (square) $9.6$\% $^{67}$Zn. \cite{Liu:2007a} ; (b) Spin-echo decay curve measured by Electron
Paramagnetic Resonance spectroscopy of ZnO nanocrystals ($d\sim 4.0~$nm, $^{67}$Zn$=4.1$\%) at $5~$K \cite{Whitaker:2010a}.
 \label{xmfig6}}
\end{figure}

The electron spin dephasing induced by the nuclear field
fluctuations has also  been clearly evidenced in materials where
only a fraction of the nuclei has non-zero nuclear spins (in InAs or
GaAs, all the nuclei have a non-zero spin) such as  ZnO
\cite{Whitaker:2010a,Liu:2007a}, diamond
\cite{Childress:2006a,Balasubramanian:2009a} and CdSe
\cite{Akimov:2006a}. The electron spin relaxation dynamics in
colloidal n-type ZnO QDs has been studied using electron
paramagnetic resonance spectroscopy. In ZnO, only the isotope
$^{67}$Zn has a non zero nuclear spin (I=5/2) with a natural
abundance of 4.1 \%. The nuclear spin contents in the ZnO QDs can be
controlled chemically by preparing nanocrystals from precursors
containing different concentrations of $^{67}$Zn (see
Fig.\,\ref{xmfig6}(a)). As expected, the electron spin dephasing time
decreases as the $^{67}$Zn concentration increases. Note that this
dependence was observed at room temperature.

As $T_{\Delta}$ is governed by inhomogeneous contributions  of the
fluctuating nuclear field, spin echo measurements should yield much
longer spin coherence times. Indeed single pulse Hahn-echo techniques were successfully
applied to transport measurements in a gate defined GaAs
QD-molecule with a measured coherence time on the order of $30~\mu$s
\cite{Bluhm:2010b} and also in optical studies of single
InAs QDs \cite{Press:2010a}. However this time does not correspond
to a true $T_2$ time, since the slow time evolution of the nuclear
field ($T_{K\Delta}$, see Sec.\,\ref{sec:merk})  leads to an incomplete recurrence. One origin of the inhomogeneity comes from the different
precession period of the nuclear spins due to the different nuclear
species in the GaAs dot ( $^{69}$Ga, $^{71}$Ga,$ ^{75}$As). By
implementing a multiple-pulse echo sequence and taking into account
these different precession periods, a record coherence time of
200~$\mu$s has been measured \cite{Bluhm:2010b}, see also an interesting theoretical discussion in this context \cite{Cywinski:2009a}. In
ZnO, pulsed electron
paramagnetic resonance measurements also yield a longer spin coherence time
(spin echo decay time $\sim 80~$ns for $^{67}$Zn$=4\%$ compared to
$T_2^*\sim20$~ns) as displayed in Fig.\,\ref{xmfig6}(b). In practice, long spin
coherence times reaching $T_2=3.0$~$\mu$s for individual dots in an ensemble of InAs dots, 
were also demonstrated for resident electrons using an original mode-locking technique \cite{Greilich:2006a} 
which can be described in the framework of narrowed state dynamics, see \cite{Yao:2006a} for discussion.

\subsection{Electron spin dephasing in a transverse magnetic field: Voigt geometry }
\label{sec:dephvoigt}

\begin{figure}
\epsfysize=2in
\epsfbox{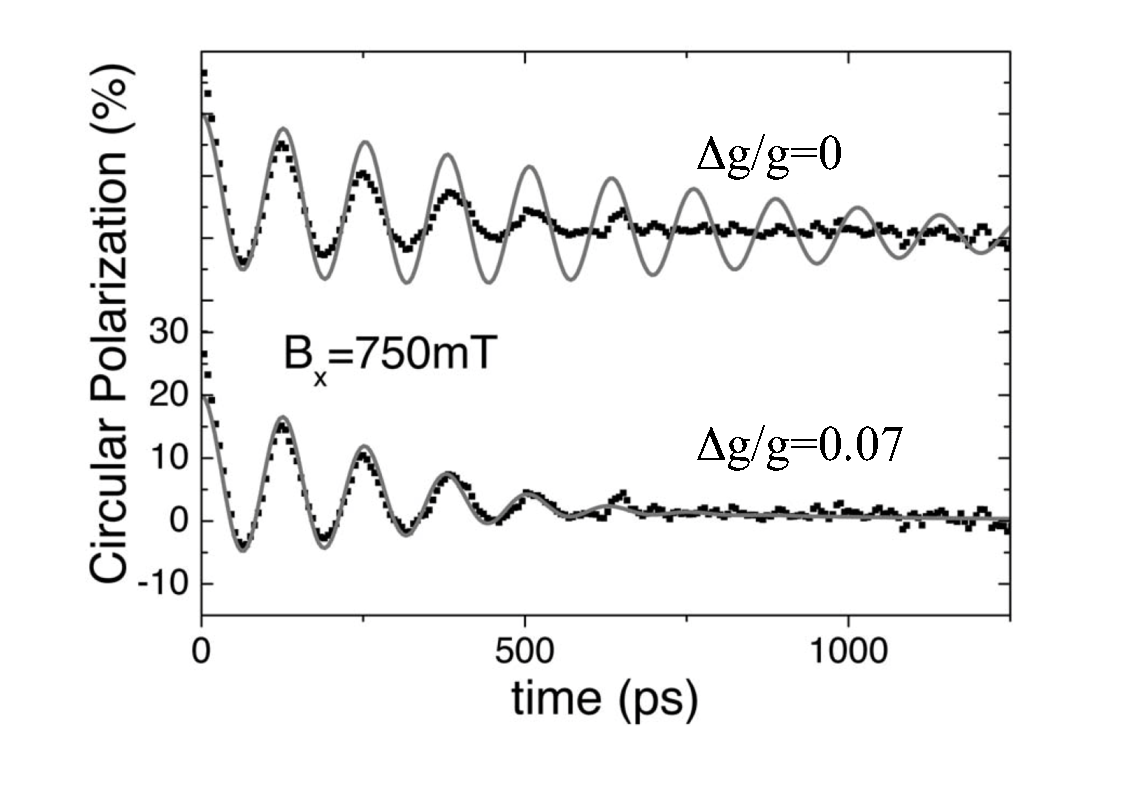}
\caption{Comparison of experimental (dotted line) and theoretical curves of time-resolved  photoluminescence circular polarization of $X^+$ for a transverse magnetic field $B_x=750$~mT. The damping reflects the electron spin dephasing in the ensemble of InAs dots. The theoretical curves (gray line) is given with $\Delta g/g=0.07$ or without $\Delta g/g=0$ electron g-factor fluctuations \cite{Lombez:2007a}.
 \label{xmfig7}}
\end{figure}

When an external magnetic field $B_x$ is applied perpendicular to the initial electron spin orientation which corresponds to the excitation light propagation axis z (Voigt geometry), the electron spin will precess coherently around the external magnetic field axis. The damping of these oscillations reflects directly the spin dephasing time \cite{Dutt:2005a}.
As shown in Sec.\,\ref{sec:dephzero}, the time- and polarization resolved $X^+$ photoluminescence signal following a circularly polarized pulsed excitation in an ensemble of p-doped InAs QDs can directly probe the electron spin dynamics during the charged exciton $X^+$ radiative lifetime. Figure\,\ref{xmfig7} shows the damping of the PL circular polarization oscillations in a transverse magnetic field $B_x=750$~mT. The observed damping of the oscillations due to electron spin dephasing has two origins.
The first one is due to the $\delta B_n$ induced spin dephasing, with a characteristic dephasing time $T_\Delta\sim500$~ps \cite{Lombez:2007a}. This is the same value obtained in the absence of an external magnetic field and confirms that prolonging the macroscopic coherence time can not simply be achieved through application of a magnetic field. A true narrowing of the nuclear spin distribution i.e. lowering of $\delta B_n$ is necessary, for example via dynamic nuclear polarization, discussed in detail in Sec.\,\ref{sec:dnp}. Yet, the characteristic spin dephasing time $T_\Delta\sim500$~ps is not enough to explain all experimental observations, see Fig.\,\ref{xmfig7} (top curve). At the origin of the second contribution lies the dispersion of the transverse electron Lande g factor, due to the inherent inhomogeneity of the system. This magnetic-field-dependent damping arises simply from the variations of the electron g factor over the QD ensemble \cite{Dutt:2005a,Greilich:2006b}, resulting in a spreading of the Larmor frequencies with increasing $B_x$ \cite{Yugova:2007a}. A typical fluctuation of $\Delta g/g=0.07$ explains the observed damping of the electron spin oscillations (see Fig.\,\ref{xmfig7}, bottom curve) which follows the simple law:
\begin{equation}
\label{eq:ttwostar}
T_2^*=T_{\Delta}/\sqrt{1+2\left( \frac{ \Delta g}{g}\frac{B}{\delta B_n}\right)^2}
\end{equation}
where $\delta B_n=\hbar/g \mu_B T_\Delta $.

\begin{figure}
\epsfysize=4.5in
\epsfbox{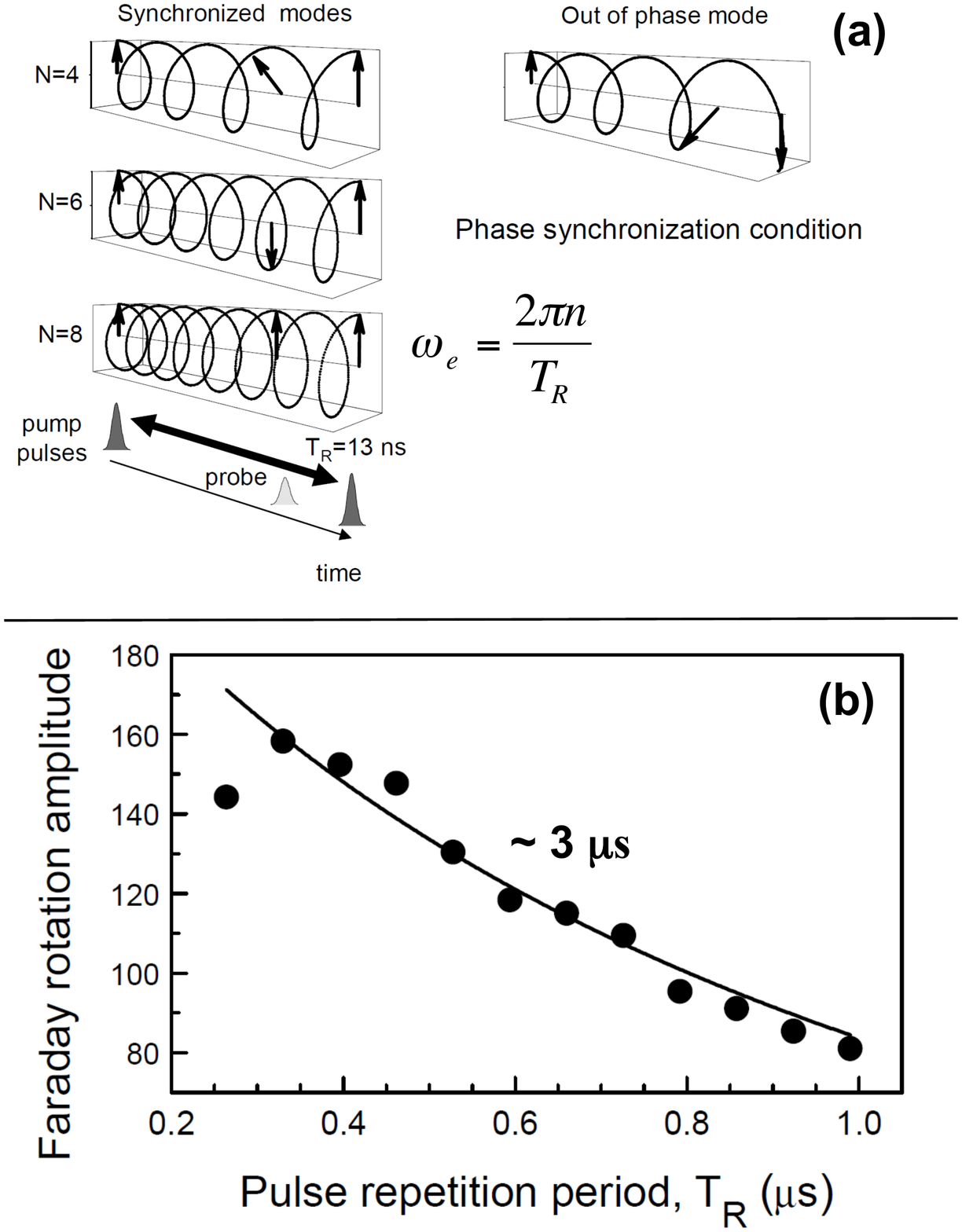}
\caption{(a) Schematics of the mode-locking of electron spins in an ensemble of spins precessing around a transverse magnetic field at different frequencies due to the g factor fluctuations from dot to dot (see text). (b) Faraday rotation amplitude at negative delay in an ensemble of n-doped InAs QDs as a function of the time interval between subsequent pump pulses measured at $B_x = 6$~T and T = 6 K \cite{Greilich:2006a,Dyakonov:2008a}.
 \label{xmfig8}}
\end{figure}

The ensemble spin dephasing observed in Fig.\,\ref{xmfig7} does not lead to a destruction of the individual spin coherence, but masks it due to phase differences among different spins in the different dots. An elegant technique, called mode-locking of electron spins, can be used to measure the \textit{single} electron coherence time with the measurements performed in an \textit{ensemble} of dots \cite{Greilich:2006a}. The principle is to excite an ensemble of n-doped QDs (containing a single resident electron) with a periodic train of polarized picosecond laser pulses resonant with the $X^-$ state and then probe the spin dynamics by Faraday rotation technique. This excitation will spin polarize the resident electrons which will precess around the transverse magnetic field $B_x$. The key point is that the pulse train will yield a synchronization of the electron spin precession. If the pulse period, $T_R$, is equal to an integer number $n$ times the electron spin precession period, the action of such pulses leads to almost complete electron spin alignment along $z$ at each pulse arrival time (see Fig.\,\ref{xmfig8}(a)).

In the ensemble of excited QDs, the electrons do not precess with the same frequency because of the fluctuations of the electron g factors. In this ensemble, some QDs will have a precession frequency which fulfils this synchronization relation with the laser, termed phase synchronization condition (PSC). When a given pump pulse excites the sample the spin coherence generated by the previous pulse has the same orientation as the one which the subsequent pulse induces for these dots. In other words the contributions of all pulses in the train are constructive. In contrast, in the non-PSC dots the contributions have arbitrary orientations (see right panel of Fig.\,\ref{xmfig8}(a)) and for these dots the degree of spin synchronization will vanish. 
The fraction of dots in the ensemble that fulfil the PSC will increase when nuclear spin induced frequency focussing starts to be efficient, as discussed in the follow-up work by \textcite{Greilich:2007a}, see discussion in section \ref{sec:Bayer} and references therein. In the experiment a PSC dot makes a stronger contribution to the Faraday rotation signal than a non-PSC dot. These latter dots do not contribute to the average electron spin polarization $S_z(t)$ at times $t \gg T_2^*$, due to dephasing.
The sum of oscillating terms from all synchronized subsets leads to a constructive interference of their contributions to the Faraday rotation signal around the times of pump pulse arrival.
By measuring the Faraday signal at such delays, the synchronized spin dynamics  which  move on a background of dephased electrons, can be measured. When the pump pulse period becomes comparable with the electron spin coherence time, the amplitude of the signal decreases, Fig.\,\ref{xmfig8}(b).

This yields the measurement of a spin coherence time of the order of a few $\mu s$ whereas the ensemble spin dephasing time $T_{\Delta} = 0.4$~ns at $B_x = 6$~T.
This measured value of $T_2^*$ with respect to $T_{\Delta}$ is however not limited by the spin relaxation time within one particular dot.
The nuclear field fluctuations may help to achieve the phase synchronisation condition, by using $\frac{1}{\hbar}\left( g_e \mu_B (B_n+\delta B_n) T_R \right)=2\pi n$. However, this condition cannot hold at times longer than the correlation time of the nuclear field fluctuations. As a consequence, the nuclear field fluctuations contribute towards limiting the spin coherence time $T_2^*$.

\subsection{Influence of the fluctuating nuclear field on electron and hole spin pumping processes}
\label{sec:spinpumping}

\begin{figure}
\epsfxsize=3.5in
\epsfbox{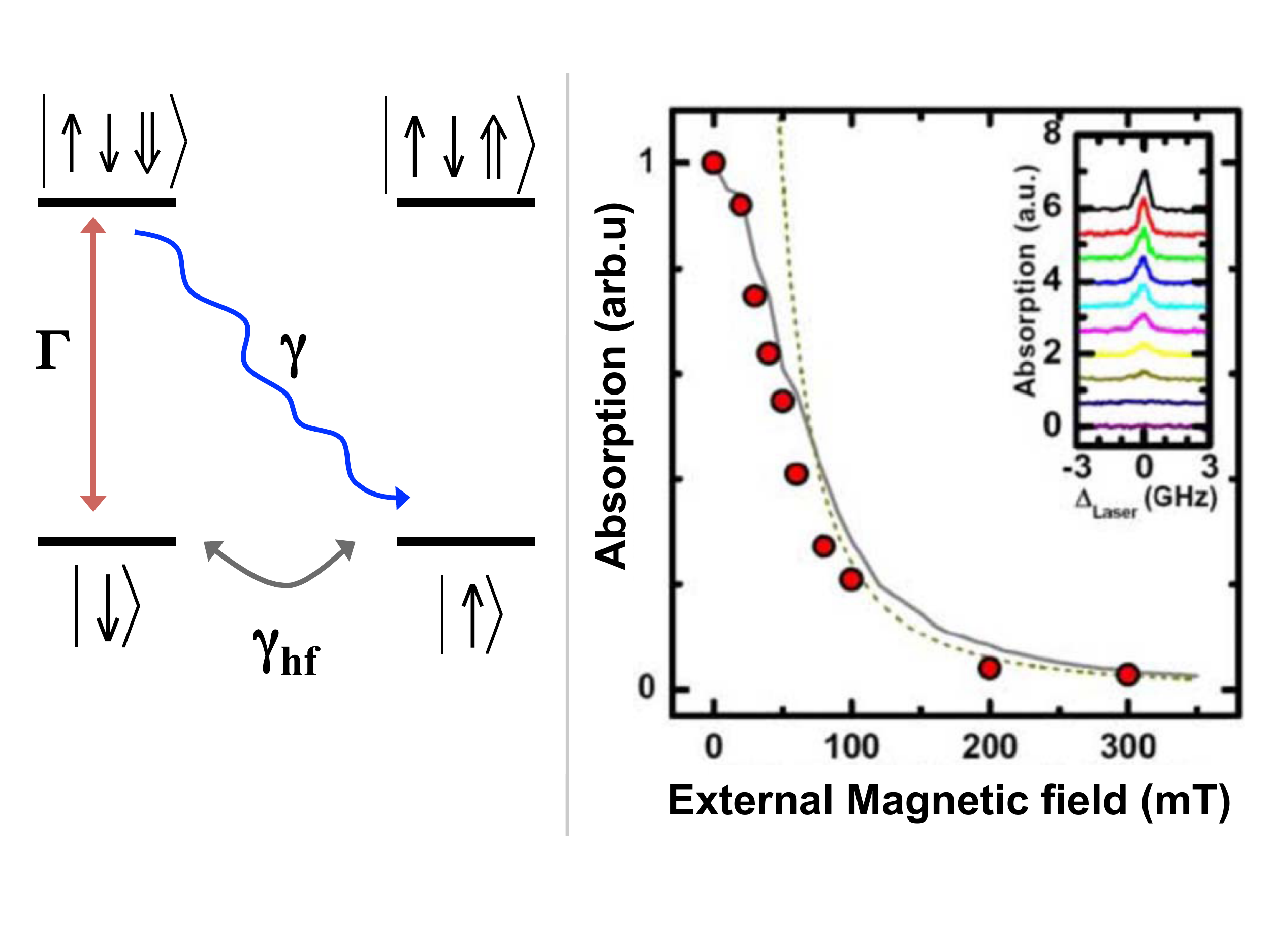}
\caption{Electron spin pumping in a single dot. (Left) Four-level system describing the singly charged QD. The fluctuations of the hyperfine field lead
to a slowly varying coherent coupling  of the spin ground states with a rate $\gamma _{hf}$ and the heavy hole - light hole mixing yields a diagonal transition with a rate $\gamma$ (Right) Absorption of the single dot as a function of the magnetic field $B_z$. The absorption decreases with increasing $B_z$ due to Optical Spin Pumping which becomes more efficient since the efficieny of the hyperfine field coupling of the ground state decreases.The inset
shows the corresponding raw laser scans from 0 T (top) to 300 mT (bottom) \cite{Atature:2006a}.
 \label{xmfig9}}
\end{figure}

The interaction of the localized electron spin in a quantum
dot with the surrounding fluctuating nuclear spins plays  an
important role for the QD spin state preparation, a key
issue from the perspective of quantum information processing
\cite{Imamoglu:1999a}. High-fidelity preparation of a
QD spin state via laser cooling, also referred to as optical spin
pumping, has been convincingly demonstrated for electrons \cite{Atature:2006a} and holes \cite{Gerardot:2008a}.
This is at first sight surprising, given the strong influence of fluctuating nuclear fields on electron spin dephasing.
But as shown below, the presence of a fluctuating nuclear field can be exploited to \textit{enable} efficient carrier spin pumping processes.\\
\indent \textit{Electron spin pumping.}--- In the experiment of \,\textcite{Atature:2006a} optical coupling of electronic spin states was
achieved using resonant excitation of the negatively charged exciton
(trion) transition X$^-$. A singly-charged QD is described as a
four-state system consisting of twofold degenerate  ground and
excited states, coupled by two vertical optical transitions, as
illustrated in Fig.\,\ref{xmfig9}(left).  The ground state $\left|
\uparrow \right> (\left| \downarrow \right>)$ is coupled to the
trion state consisting of two electrons in a singlet and a heavy
hole $\Uparrow (\Downarrow)$ according to the optical selection
rules for $\sigma^+(\sigma^-)$ optical transition. The diagonal
transition between the trion state $\left|
\uparrow\downarrow\Downarrow \right>$ and the electron $\left|
\uparrow \right>$ is  forbidden for a heavy hole with pure $\pm$3/2
angular momentum. 
In reality the diagonal transition characterized by a rate
$\gamma$ is permitted and the efficiency  of electron spin pumping
depends on the relative magnitude of $\gamma$ and
$\gamma_\text{hf}$. There are three physically distinct contributions that determine the strength of $\gamma$:
 (i) Valence band mixing due to the reduced
symmetry of the QD (elongation, strain anisotropy, interfaces), leading to a small light hole admixture of the
heavy hole state \cite{Calarco:2003a,Krizhanovskii:2005a,Belhadj:2010a}.  (ii) The (unintentional)
application of a transverse magnetic field \cite{Xu:2007a}, arising from a slight tilt in the sample holder. (iii) In-plane component of the fluctuating Overhauser field.\\
In a small applied longitudinal field $B_z \le \delta B_n$ no efficient
electron spin state preparation can be achieved using resonant
excitation since the strong hyperfine interaction of the resident
electron spin with the QD nuclear spin ensemble leads to random
spin-flip events at rate $\gamma _\text{hf}$ \cite{Merkulov:2002a}, see
Sec.\,\ref{sec:dephzero}.

\begin{figure}
\epsfxsize=3.5in
\epsfbox{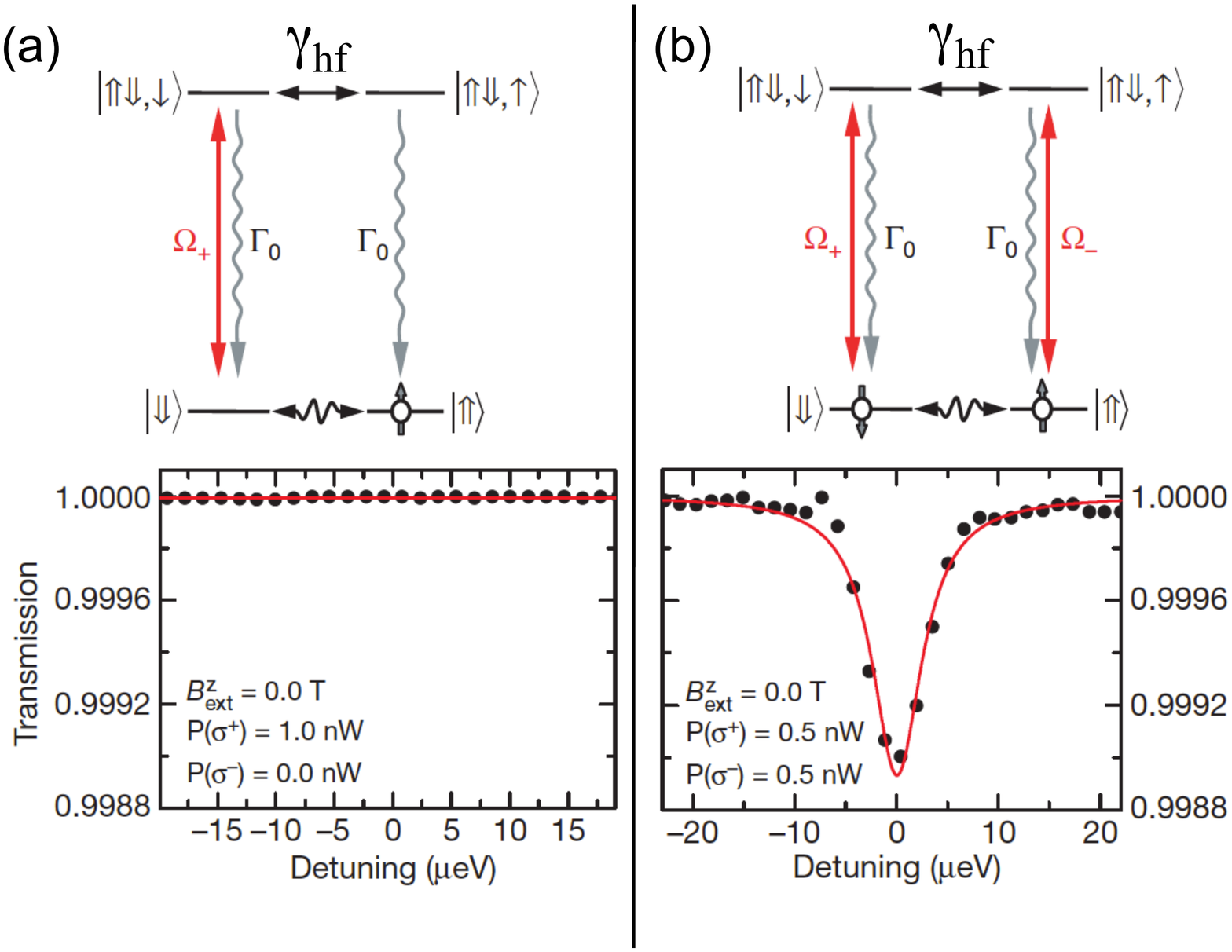}
\caption{Hole spin pumping in a single dot. (a) A laser with $\sigma^+$ polarization drives the $ \left| \Downarrow \right> \longleftrightarrow \left| \Uparrow \Downarrow, \downarrow \right>$ transition. No transmission dip is observed demonstrating that the hole population is shelved into state $\left| \Uparrow \right>$. (b) simultaneous excitation with both $\sigma^+$ and $\sigma^-$ at the same frequency. A large transmission dip is observed, as the additional $\sigma_-$ excitation leads to a re-pumping of the electron spin and thereby suppresses electron spin pumping \cite{Gerardot:2008a}.
 \label{xmfig10}}
\end{figure}

As shown in Sec.\,\ref{sec:dephfara}, a longitudinal  magnetic
field exceeding $\delta B_n$ suppresses the effect of this
interaction. As a consequence, the $\left|\uparrow\right>$ state can be
prepared and an absorption drop in differential transmission
experiments  is clearly observed in Fig.\,\ref{xmfig9} when an
external magnetic field is applied. With increasing $B_z$, the QD
becomes transparent which is a signature of optical electron-spin
pumping into the spin $\left|\uparrow\right>$ state due to the weak
recombination path ($\gamma$) that dominates over the
hyperfine-induced bidirectional spin-flip process with a rate
$\gamma _\text{hf}(B_z)$ which is decreasing as a function of the applied
magnetic field  (see Fig.\,\ref{xmfig9}). For $B_z=200$~mT, it has
been shown that the electron spin can be prepared with an average
polarization as high as  $98.5$\% \cite{Dreiser:2008a}. In summary,
the hyperfine interaction of the resident electron with the
fluctuating nuclear spins prevents the realization of spin-state
preparation with high fidelity in negatively charged QDs. The
application of an external magnetic field is essential to achieve
this spin state preparation. \\
\indent \textit{Hole spin pumping.}--- In a positively charged QD (i.e a QD containing a resident hole),  the
situation is reversed as compared to the $X^-$ case: the interaction between the electron in the
positively charged exciton $X^+$ with the nuclear spins allows
coherent coupling between the two electron spin states, i.e. between $|\Uparrow\Downarrow,\downarrow\rangle$ and $|\Uparrow\Downarrow,\uparrow\rangle$ yielding an
efficient hole spin cooling \cite{Gerardot:2008a,Eble:2009a}, as
schematically shown in Fig.\,\ref{xmfig10}(a). The hole spin cooling
is efficient since the rate $\gamma _\text{hf}$ is faster than the
spontaneous radiative recombination rate $\Gamma_0$. If a
longitudinal magnetic field larger than $\delta B_n$ is applied, the
coherent coupling between the two electron spin states within the
trion is diminished and as a consequence the hole spin preparation
fidelity decreases \cite{Gerardot:2008a}. Although hole spin pumping is very efficient for $B_z=0$,
it need not be absent for $B_z>0$.
When an exciting laser strictly resonant with the $X^+$ transition  in a single QD has circular polarization, either $\sigma^+$ (see Fig.\,\ref{xmfig10}(a)) or $\sigma^-$, no absorption is observed, demonstrating the efficient hole spin cooling. However by pumping with two lasers with identical wavelength and with the same total power, one with $\sigma^+$ and one with $\sigma^-$ polarization,  a clear absorption appears (Fig.\,\ref{xmfig10}(b)). This arises because spin pumping with $\sigma^+$ polarization is frustrated by  the $\sigma^-$ excitation, and vice-versa.\\

A comparison of hole and electron spin  pumping is shown in
\textcite{Desfonds:2010a} which highlights that at zero applied
magnetic field hole spin pumping is possible, whereas electron spin
pumping is not. Both the hyperfine mediated electron
and hole spin flips as well as heavy hole - light hole mixing are limiting processes for electron spin pumping, as discussed in the
context of a \textit{quantum-dot-spin single-photon interface} in
\textcite{Lu:2010a} and \textcite{Yilmaz:2010a}.

\subsection{Beyond the nuclear mean field approach}

\label{sec:beyondmean} In order to go beyond the nuclear mean field
approach developed above, powerful quantum models have been developed \cite{Khaetskii:2003a,Zhang:2006a}. 
G. G. Kozlov proposed a full quantum model for describing the time evolution of a single electron spin $S
= 1/2$ interacting with an even number of nuclei 2N (N$\approx
10^3$) in a single QD \cite{Kozlov:2007a}. The main
assumptions of this model are: (i) the electron envelope function is
considered as constant within the dot; (ii) The nuclear spin
interacting with the electron are identical and possess an angular
momentum $I^ j = 1/2$. The first assumption allows the use of the
total angular momentum of the QD nuclei to classify their spin
states. At zero external magnetic field, the solution of the coupled
system evolution problem is analytical, and the time dependence of
the electron average spin $\langle \hat{S}_z(t) \rangle$ is then identical
to the one obtained in the Merkulov model, see Eq.~\,\eqref{eq:st} .
It thus reproduces quantitatively the fast relaxation of the
electron spin observed experimentally just after the pulsed laser excitation of an
\textit{individual} dot \cite{Dou:2011a}. This is in contrast to measurements on \textit{ensembles}, where this initial dip
is usually masked due to variations of this particular electron spin relaxation time from dot to dot.\\
Some fundamental differences exist between the quantum and
the effective field model \cite{Petrov:2009a}. In the mean field approach, coherences
between the electron spin states arise if one considers a single dot
excited with a single pulse i.e. an electron spin precesses
coherently around $\delta B_n$. These coherences will disappear in
the averaging process of repeated laser pulses, when a sufficiently
high number of different nuclear field configurations have been
probed within the measurement process. On the other hand, in the
microscopic model, coherences arise only within the coupled
electron-nuclear spin system, which becomes fully entangled as time
increases. As a consequence, it can be shown that no spin coherences
can develop in the electron subsystem \cite{Petrov:2009a}. The spin populations are then
evaluated through repeated measurements.

%
\section{OPTICAL PUMPING OF NUCLEAR SPINS}
\label{sec:dnp}

\begin{figure}
\epsfxsize=3.5in
\includegraphics[width=0.45 \textwidth,angle=0]{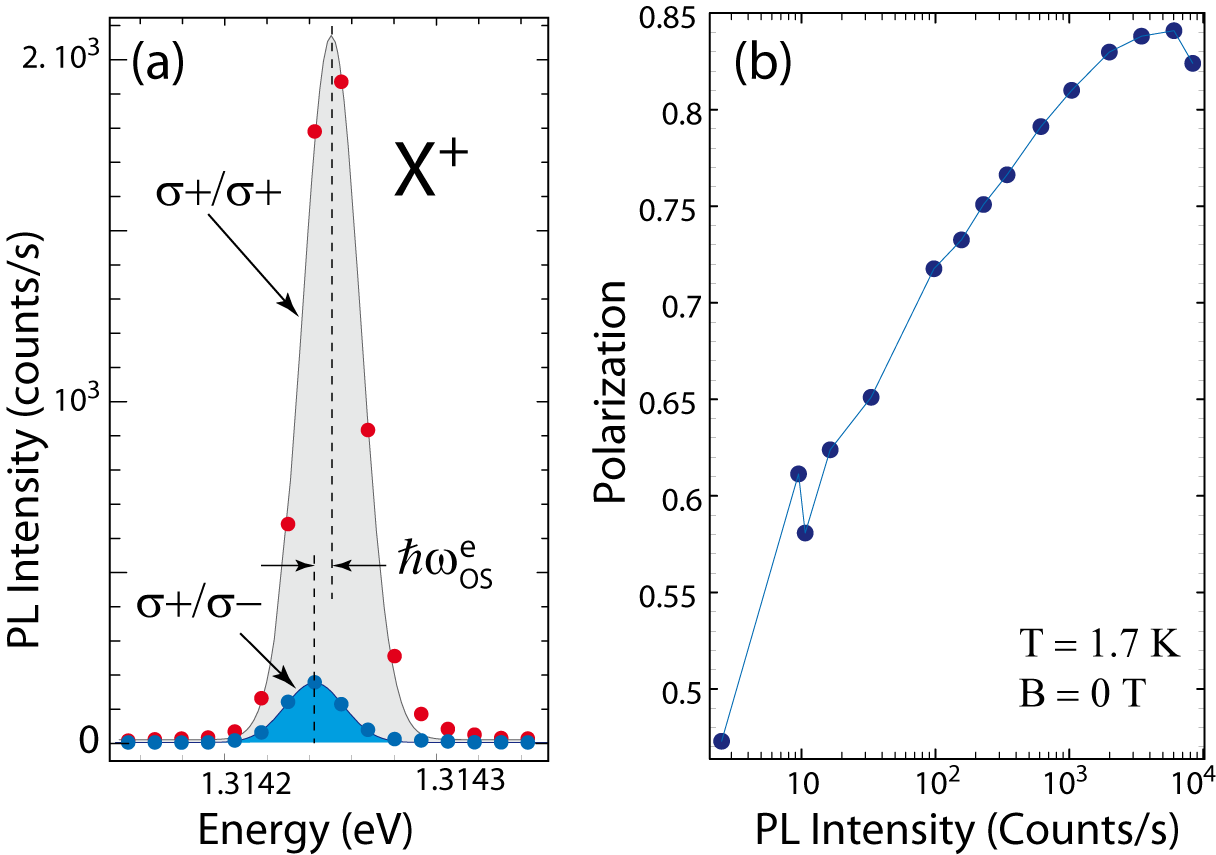}
\caption{(a) PL spectra of a single $X^+$ line in zero magnetic field measured under  co-polarized ($\sigma^+/\sigma^+$) and cross-polarized ($\sigma^+/\sigma^-$) excitation/detection configuration. (b) Evolution of the corresponding circular polarization as a function of the total PL intensity for an incident laser excitation varied from 150~nW to 250~$\mu$W.}\label{LPN-PolarvsPower}
\end{figure}

Starting in sec.\,\ref{sec:orientation} we have described experiments that allow preparing an electron spin state with a suitable laser pulse. In this section we explore under which conditions the electron spin polarization can be transferred to the nuclear spin ensemble in the QD. \\
As discussed in Sec.\,\ref{sec:merk} the hyperfine  interaction
leads to electron  spin dephasing  within a characteristic time
$T_\Delta\sim1$~ns due to the statistical distribution of
nuclear Overhauser field $\bm{B}_n$ in a QD
ensemble~\cite{Merkulov:2002a}. This effect is directly  evidenced
in the time domain for an ensemble of positively charged QDs by
monitoring the  PL polarization decay of $X^+$
trions shown in Fig.\,\ref{xmfig2}~\cite{Braun:2005a}. Since  the correlation time $T_2$ of
$\bm{B}_n$ amounts to $\approx$10$^{-4}$s,
this spin dephasing also manifests itself when repeatedly recording the emission stemming from a single QD for a signal
integration time $\tau_i\gg T_2$. Under such conditions, the
time-integrated circular polarization of a single $X^+$ line excited
by circularly polarized light reads:
\begin{equation}
\rho_{c}=2\Gamma\int\langle S_z(t)\rangle\exp(-\Gamma t)dt\label{Eq:AveragePolar}
\end{equation}
where $\langle S_z(t)\rangle$ is the electron spin evolution
averaged over the distribution of random nuclear fields and $\Gamma=1/\tau_r$
is the $X^+$  recombination rate. Using Eq.\,\ref{eq:st} from Sec.\,\ref{sec:merk} for $\langle S_z(t)\rangle$ with a nuclear field fluctuation of $\delta B_n\approx
30$~mT, the maximum degree of polarization  for an initially
photo-created spin $S_z(0)=1/2$  is estimated  to be
$\rho_c^\text{max}\sim$50\%. However, this limit  is commonly exceeded
in experiments measuring the  PL polarization of single $X^+$ line
under cw excitation for signal integration times
$\tau_i\sim$1~s~\cite{Eble:2006a,Lai:2006a,Ebbens:2005a}. This is
illustrated in Fig.\,\ref{LPN-PolarvsPower}(a) and (b) showing that the
circular polarization of a single $X^+$ line reaches more than
$80\%$ under quasi-resonant excitation.  This apparent discrepancy
turns out to be the manifestation of efficient DNP acting back on
the electron  spin dynamics.  The power-dependence of the circular
polarization shown in Fig.\,\ref{LPN-PolarvsPower}(b)  supports this
interpretation: at very low excitation density the $X^+$ circular
polarization reduces to $\approx$50\% because the average nuclear
spin polarization vanishes.
The experimental approach is often as follows: The electron spin is initialised via optical pumping that induces DNP. The back action of the nuclear spin system on the electron is then monitored by recording the resulting changes in electron spin orientation and transitions energy, revealing the surprising bistability, memory and line dragging effects detailed below.

\subsection{Dynamic Nuclear Polarization : general features}
\label{sec:dnpgeneral} The dynamic nuclear polarization in
semiconductor quantum  dots results from the scalar form
$\hat{\bm{S}}^e\cdot\hat{\bm{I}}$ of the hyperfine interaction which conserves
the total spin. When an electron spin relaxes its initial
orientation via this interaction,  its spin angular momentum is
transferred to the nuclear spins. This corresponds to the
electron-nuclei flip-flop term in Eq.\,\eqref{eq:eqHf1}. To achieve
a significant nuclear polarization  via the Overhauser effect
\cite{Overhauser:1953a} it is required to maintain the  spin
polarization of unpaired electrons far from their  equilibrium value
$\langle S^e_z\rangle_0$ which is usually determined by the lattice
temperature $T$ and the external magnetic field $B_z$. The trend of
electrons to return to their thermal equilibrium populations leads to an increase of the
nuclear spin polarization via the spin flip channel provided by the
hyperfine interaction. Since the  nuclei are themselves much less
coupled to the lattice, their polarization relaxes on a much longer timescale than electron polarization and a
large DNP degree can be reached. The relation between the DNP build-up time
and its relaxation time will be discussed more closely in the
following sections.\\
\indent In many experiments the out-of-equilibrium electron spin polarization is achieved by saturating the electron spin resonance by an RF
field to achieve equal  populations of spin $\uparrow$ and
$\downarrow$ in a non-zero magnetic field. Yet,  in semiconductors
and more specifically in QDs, a large electron spin
polarization can be created \emph{optically}, even in zero magnetic
field, due to the selection rules permitting excitation of
specific spin states with the light polarization as discussed in Sect.\,\ref{sec:orientation}. The very efficient
optical orientation of electrons in p-type InAs QDs is
therefore well suited for generation of a large DNP via the
Overhauser effect. Interestingly, the polarization-resolved PL
spectroscopy of such individual QDs provides an access to
both  the electron spin component $\langle S^e_z\rangle$ via the
circular polarization, and the average nuclear field $B_{n,z}$ via
the energy splitting of the $\sigma^+$ and $\sigma^-$ components. In
Fig.\,\ref{LPN-PolarvsPower}(a) for $B_z=0$, the $\sigma^+$ PL line is clearly
blue-shifted from the $\sigma^-$ line by $\hbar\omega^e_{OS}=7~\mu$eV. This corresponds to an average  nuclear field
$B_{n,z}=\hbar\omega^e_{OS}/ (g_e \mu_\text{B})=-200$~mT  which is
notably larger than its own fluctuations $\delta B_n$. In this
experiment, a stationary regime is thus established,  where the
relaxation by the  hyperfine interaction is efficiently quenched
due to the nuclear field it has \textit{itself}  established during some
initial DNP transient.

The  spin flip-flops  permitted by the hyperfine interaction are
however not energy-conserving in a finite external magnetic field.
Electron and nuclear spins have indeed very different Land\'{e}
factors yielding a ratio of their  Larmor frequencies
$\omega^n_Z/\omega^e_Z\sim 10^{-3}$, so that the electron Zeeman
energy $\hbar\omega^e_Z$ has to be exchanged with some other degrees
of freedom.  This is the reason why the coupling of the electron
to a reservoir is in general important for the generation of a
sizeable Overhauser field: it provides a finite width (given by a finite correlation time $\tau_c^e$) for the
electronic levels necessary to account for the energy cost of
flip-flops, see \cite{Baugh:2007a} for a related discussion \footnote{There are several physical mechanisms that will influence the correlation time,  for example any physical process that limits the electron spin lifetime. Following quasi-resonant or non-resonant optical excitation, the spin flip-flop with a nucleus can take place while the electron is in an excited state in the QD, leading to $\tau_c^e$ in the tens of ps range, or for an electron in the QD ground state, so $\tau_c^e$ is limited by the radiative lifetime $\tau_r\simeq 1$~ns. The electron spin lifetime could also be limited by co-tunneling events in charge tunable structures. In addition, the hyperfine induced flip-flop events themselves  will limit the correlation time.}.  
In QDs, this issue of energy conservation is
essential for explaining the emergence of strong non-linearities of
the electron-nuclei system that will be discussed in the next
sections.

A different mechanism for DNP called 'solid-state effect' was first
observed in  crystals with paramagnetic impurities
\cite{Abragam:1961a}. Here the energy necessary for a spin flip-flop
is provided by an external RF source. In close analogy in optical
pumping experiments the energy necessary for a spin flip-flop can be
provided by a driving laser, in experiments similar to the
solid-state effect carried out by \textcite{Chekhovich:2010a}, where
a resonant laser drives an optical transitions that is only weakly
allowed due to hyperfine coupling, see Sec.\,\ref{sec:beyond}.

In general, the specific conditions of QD optical excitation (e.g.
strictly- or quasi-resonant, with our without applied magnetic
fields)  determine the DNP mechanism that dominates and hence
controls the magnitude and sign of the nuclear field experienced by
electrons in the QD volume. If the electron spin coherence time
resulting from optical excitation is short, then the flip-flop term
$\propto (\hat{I}_+^j\hat{S}_-^e+\hat{I}_-^j\hat{S}_+^e)$ of
Eq.\,\ref{eq:eqHf1} introduced in Sec.\,\ref{sec:hyperfine} will be the dominant nuclear spin pumping
mechanism; this is typically the case for quasi-resonant or non-resonant
excitation. In strong magnetic fields and under resonant excitation
of the fundamental exciton or trion resonances on the other hand,
the electron spin flip is too costly in energy and nuclear spin
flips can be induced by two additional forms of electron-nuclear
spin coupling, that do not require a simultaneous electron spin
flip: \\
\textbf{(i)} The nuclear spin dynamics in strained QDs will also be
influenced by strong nuclear quadrupole effects, introduced in
Sec.\,\ref{sec:quadintro}, giving rise to a non-collinear coupling
term, which we will use in the following simplified form
\cite{Huang:2010a}:

\begin{equation}
    \hat{H}_{hf}^{nc} =\sum_i A_\mathrm{nc}^i \hat{I}_\mathrm{x}^i \hat{S}_\mathrm{z}^e
    \label{eq:nc-hyperfine} \;.
\end{equation}

\noindent with a non-collinear hyperfine interaction constant
$A_\mathrm{nc}$ that is typically $A_\mathrm{nc}\approx 0.01 A$.
The physical origins of this coupling are detailed in
Sec.\,\ref{sec:locking}.
$\hat{H}_{hf}^{nc}$ can induce under certain conditions nuclear spin relaxation, as detailed in Sec.\,\ref{sec:quadlong}. Interestingly, this term is also at the origin of bi-directional DNP and explains the experimentally observed locking of a QD transition to a resonant driving laser ('dragging') \cite{Yang:2010a,Hogele:2012a} described in Sec.\,\ref{sec:locking}. \\
\textbf{(ii) } Even as direct electron-nuclear spin-flip processes
are forbidden by energy conservation at high magnetic fields,
elimination of the flip-flop terms of Eq.\,\ref{eq:eqHf1} using a
Schrieffer-Wolff transformation \cite{Schrieffer:1966a} shows that the QD electron can still
strongly influence the nuclear dynamics: in addition to the nuclear
dipole-dipole interaction given by Eq.\,\ref{eq:dipdip}, nuclear
spins can also be coupled to each other by an indirect interaction
mediated by the electron in the dot which has the form
\cite{Klauser:2006a,Abragam:1961a}:
\begin{equation}
\hat{H}_{ind}=\sum_{i,j}\frac{A^i
A^j}{\omega^e_Z}\hat{S}_z\hat{I}^i_+\hat{I}^j_- \label{eq:hfind}
\;.
\end{equation}
This Hamiltonian ensures the conservation of the total nuclear spin polarization but leads to nuclear spin diffusion \cite{Latta:2011a}, as discussed in Sec.\,\ref{sec:DecmK}. Indirect coupling of nuclear spins causing spin relaxation is also induced in charge tunable structures \cite{Warburton:2000a} through electron co-tunneling \cite{Smith:2005a,Dreiser:2008a} introduced in Sec.\,\ref{sec:orientation}.\\
\indent It is the intricate balance between these different
nuclear spin pumping and depolarization mechanisms that gives rise
to the intriguing experimental findings detailed below.

\subsection{Dynamic nuclear polarization in single quantum dot optics}
    \label{sec:dnpqdoptics}

\subsubsection{Nonlinearity of the dynamic nuclear polarization}
\label{sec:nonlinear}

The very first observations of  nuclear polarization in InAs quantum
dots
\cite{Eble:2006a,Maletinsky:2007a,Braun:2006a,Tartakovskii:2007a}
have revealed a striking asymmetry of the magnitude of the generated
Overhauser field $\bm{B}_n$ as a function of the light helicity
$\sigma^+$ or $\sigma^-$ used for excitation when a magnetic field
$\bm{B}$ of only  a few 100~mT is applied parallel to the optical
axis (Faraday configuration)\footnote{This asymmetry was not
observed in the well characterised GaAs interface fluctuation dots
\cite{Gammon:2001a}.}. The general expression for $\bm{B}_n$ due to
the Overhauser effect in bulk semiconductors is however essentially
symmetrical with respect to the electron spin direction
\cite{Meier:1984a} :
\begin{equation}
\bm{B}_n=b_n\frac{\bm{B}(\bm{B}\cdot\bm{S}^e)}{\bm{B}^2+\xi B_L^2}
\label{eq:BnBulk}
\end{equation}
Here $b_n$ is a proportionality constant and $B_L\sim 0.15$~mT is
the  small local effective  magnetic field experienced by the nuclei due
to their mutual dipole-dipole interaction and $\xi$ is a coefficient
close to unity. Note that this expression assumes that a nuclear
spin temperature exists\footnote{The validity of the spin temperature concept in InAs/GaAs QDs in the presence of nuclear quadrupole effects is investigated in detail in section \ref{sec:demag}.}. This expression is valid for moderate
fields where the equilibrium spin polarization $\langle
S_z^e\rangle_0=1/2\tanh(g_e \mu_\text{B}B_z/k_\text{B}T)$ can be
neglected. For magnetic fields below $\sim$10~mT, the Knight field
$\bm{B}_K\propto\langle\bm{S}^e\rangle $ should also be added to
$\bm{B}$ in Eq.\,\eqref{eq:BnBulk}. This introduces an asymmetry
which however vanishes for fields above 100~mT.
Also the strong nuclear quadrupole effects in strained QDs introduced in Sec.\,\ref{sec:quadintro} need to be included when analysing the magnitude and orientation of $\bm{B}_n$.

To establish an expression similar to Eq.\,\eqref{eq:BnBulk} valid
in the case of QDs where large nuclear fields are generated, the
dependence of the electron-nuclei flip-flop rate (hidden in $b_n$)
on the total magnetic field $\bm{B}+\bm{B}_n$  has to be taken into
account explicitly. This will account for the non-energy-conserving
character of the hyperfine interaction. Assuming a uniform electron
wavefunction $\psi(\mathbf{r})=\sqrt{2/N\nu_0}$ spanning over  $N$
nuclei in the QD of volume $\nu_0 N/2$, one can first
derive the electron-induced relaxation rate  $T_{1e}^{-1}$ of
nuclear spins due to the temporal fluctuations of the flip-flop term
$\propto (\hat{S}^e_-\hat{I}_+ +\hat{S}^e_+\hat{I}_-)h_1(t)$ of the
hyperfine coupling~\cite{Abragam:1961a,Meier:1984a,Eble:2006a}:
\begin{equation}
T_{1e}^{-1}=\left(\frac{\tilde{A}}{N\hbar}\right)^2\frac{2f_e\tau_c^e}{1+(\omega^e\tau_c^e)^2}
\label{eq:Te}
\end{equation}
where $\hbar\omega^e=g_e\mu_\text{B}(B_z+B_n)$ is the electron  spin
splitting in the \emph{total} field, $\tau_c^e$ the  correlation
time of the hyperfine perturbation $\hat{H}_1(t)$, and $f_e$ is the
fraction of time that the QD contains an unpaired electron. The
average constant $\tilde{A}$ of the hyperfine interaction in the QD
is used to obtain an expression for $T_{1e}^{-1}$ independent
of the nuclear species. Let us recall that the $A^j$'s  vary
indeed only slightly  for different nuclear spins $I^j$ in
InAs/GaAs QDs, see table \ref{tab:hyperconst}.
In this limit of homogeneous coupling, the approximation of the existence of a high nuclear spin temperature is valid ~\cite{Abragam:1961a} and allows one to derive a simple
differential equation for the dynamic polarization of the average
nuclear spin $z$~component per nucleus $\langle \tilde{I}_z
\rangle=\frac{1}{N}\sum_j\langle \hat{I}^{j}_{z}\rangle$ in an
external magnetic field along $z$ \cite{Dyakonov:1974a}:
\begin{equation}\label{eq:DNP}
\frac{d\langle \tilde{I}_z\rangle}{dt}=-\frac{1}{T_{1e}}\left[\langle \tilde{I}_z\rangle-\widetilde{Q}\left(\langle \hat{S}_z^e\rangle-\langle \hat{S}_z^e\rangle_0\right)\right] - \frac{\langle \tilde{I}_z \rangle}{T_d}
\end{equation}
The first term in the right-hand side of Eq.\,\eqref{eq:DNP} is the
DNP source driven by the  departure of the electron spin from its
thermal equilibrium $(\langle \hat{S}_z^e\rangle-\langle
\hat{S}_z^e\rangle_0)$. It is strictly zero when the electron-nuclear spin system is in thermal equilibrium.
Note that the equilibrium value of nuclear spin $\langle I_z\rangle_0$  is assumed
to be zero in the usual experimental conditions\footnote{$\langle I_z\rangle_0$ on the order of $10^{-5}$ to $10^{-4}$ at several Tesla is negligible compared to $\langle I_z\rangle$ on the order of up to about 50\% achieved via optical pumping \cite{Urbaszek:2007a}.}.
The factor
$\widetilde{Q}=\sum_jI_j(I_j+1)/(NS(S+1))$  is a numerical constant
which amounts to $\sim15$ for realistic In$_{1-x}$Ga$_x$As QDs
containing a fraction $x\sim0.5$ of Gallium.

The second term in Eq.\,\eqref{eq:DNP} accounts for the return to equilibrium of the
average nuclear spin following an exponential decay with the  time
constant $T_d$. This term is necessary because the stationary
solution of Eq.\,\eqref{eq:DNP} without nuclear spin relaxation
would lead to  nuclear polarization much higher than what has
been observed experimentally. Different mechanisms may contribute
to this relaxation: (i) the dipolar interaction between nuclei
responsible for fast depolarization in a very weak field and for a
slower field-independent spin diffusion. (ii) Because of local
anisotropic strain, the quadrupolar coupling with local electric
field gradients introduced in Sec.\,\ref{sec:quadintro} could also produce an important reduction of the
nuclear polarization $z$ component in magnetic fields, depending on
the exact angle $\theta$ defined in Fig.\,\ref{LPN-QIintro}(b) of the principal axis given by the electric field
gradients with respect to the quantization axis~$z$
\cite{Huang:2010a}. This effect will be discussed in more detail in the next Sec.\,\ref{sec:quadlong}. For InAs
QDs, the experimental evidence suggests that the quadrupolar fields
not only render point (ii) important in many experiments, but also
suppress the contribution of argument (i) to nuclear spin relaxation. It is
worth noting that in a non-zero magnetic field, say above a
few 10~mT, the nuclear polarization can survive for minutes or even
hours as long as the QD is neither excited (i.e. does not contain any
charge carrier) nor coupled to a Fermi
sea~\cite{Maletinsky:2009a,Chekhovich:2010b}. The decay of  nuclear
polarization  must therefore contain a term associated to the
presence of the electron in the QD and proportional to
$f_e$.  The associated relaxation mechanism originates most likely
from the temporal fluctuations of the Knight field
$h_1(t)(\tilde{A}/N)(\hat{S}_z^e(t)-\langle
\hat{S}_z^e\rangle)\hat{I}_z$ coupled to quasi-static non-diagonal
perturbations experienced by the nuclei like the
dipole-dipole~\cite{Gammon:2001a} or quadrupolar
interactions ~\cite{Huang:2010a} as detailed in the next Sec.\,\ref{sec:quadlong}.
The time-dependent quadrupolar
interaction associated to the  fluctuations of electrical \textit{gradients}
induced by the creation or annihilation of an exciton could also
directly contribute to the relaxation as estimated for donors in
GaAs~\cite{Paget:2008a}.

Because of the dependence of $T_{1e}^{-1}$ on the generated  nuclear
field  $B_{n,z}\equiv2\tilde{A}\sum_j\langle
\hat{I}^{j}_{z}\rangle/(g_e \mu_\text{B})$,  Eq.\,\eqref{eq:DNP}
acquires a non-linear character. Its steady state solution, that
would be equivalent  to Eq.\,\eqref{eq:BnBulk} in the case of a
constant $T_{1e}^{-1}$ and  $T_d\propto B_z^2$, gives rise to an
implicit equation for $B_{n,z}$ reading \cite{Eble:2006a}:
\begin{equation}\label{eq:BnQD}
B_{n,z}=\frac{2\tilde{A}\widetilde{Q}}{g_e \mu_\text{B}}\frac{\langle \hat{S}_z^e\rangle-\langle \hat{S}_z^e\rangle_0}{1+\varsigma\left[1+\Big(g_e \mu_\text{B} (B_z+B_{n,z}) \tau_c^e/\hbar\Big)^2\right]}
\end{equation}
where $\varsigma=\left(\frac{N\hbar}{\tilde{A}}\right)^2/(2\tau_c^e
f_e T_d)$ is a constant corresponding to the ratio $T_{1e}/T_d$ when
the total magnetic field vanishes (i.e.   $\vert
B_z+B_{n,z}\vert\ll\hbar/(g_e \mu_\text{B}\tau_c^e)$). It determines
the minimal value $(1+\varsigma)^{-1}$ of the leakage factor
limiting the magnitude of the  nuclear polarization.
Equation\,\eqref{eq:BnQD} shows that the effect of $\varsigma$ is
amplified by  $\left[1+\left(g_e \mu_\text{B}
(B_z+B_{n,z})\tau_c^e/\hbar\right)^2\right]$ which represents the
influence of the  electron spin splitting  on the flip-flop rate.
Clearly, this introduces a dependence on the relative signs of
$B_{n,z}$ and $B_z$. The build-up of  $B_{n,z}$ is favored when
both fields point in opposite directions (i.e. partially compensate
each other),  whereas DNP tends to be inhibited when the fields are
parallel,  as confirmed experimentally \cite{Eble:2006a}. Moreover
this simple formalism predicts possible regimes of nuclear field
bistability since Eq.\,\eqref{eq:BnQD} is actually  a 3$^{rd}$ order
polynomial equation that may accept three real solutions (2 stable,
1 unstable) determined by its coefficients i.e. the experimental conditions
\cite{Maletinsky:2007a,Braun:2006a}.


\subsubsection{Dynamic nuclear polarization in the presence of nuclear quadrupole effects}
\label{sec:quadlong}

\begin{figure}[t]
\centering
\includegraphics*[width=0.48 \textwidth,angle=0]{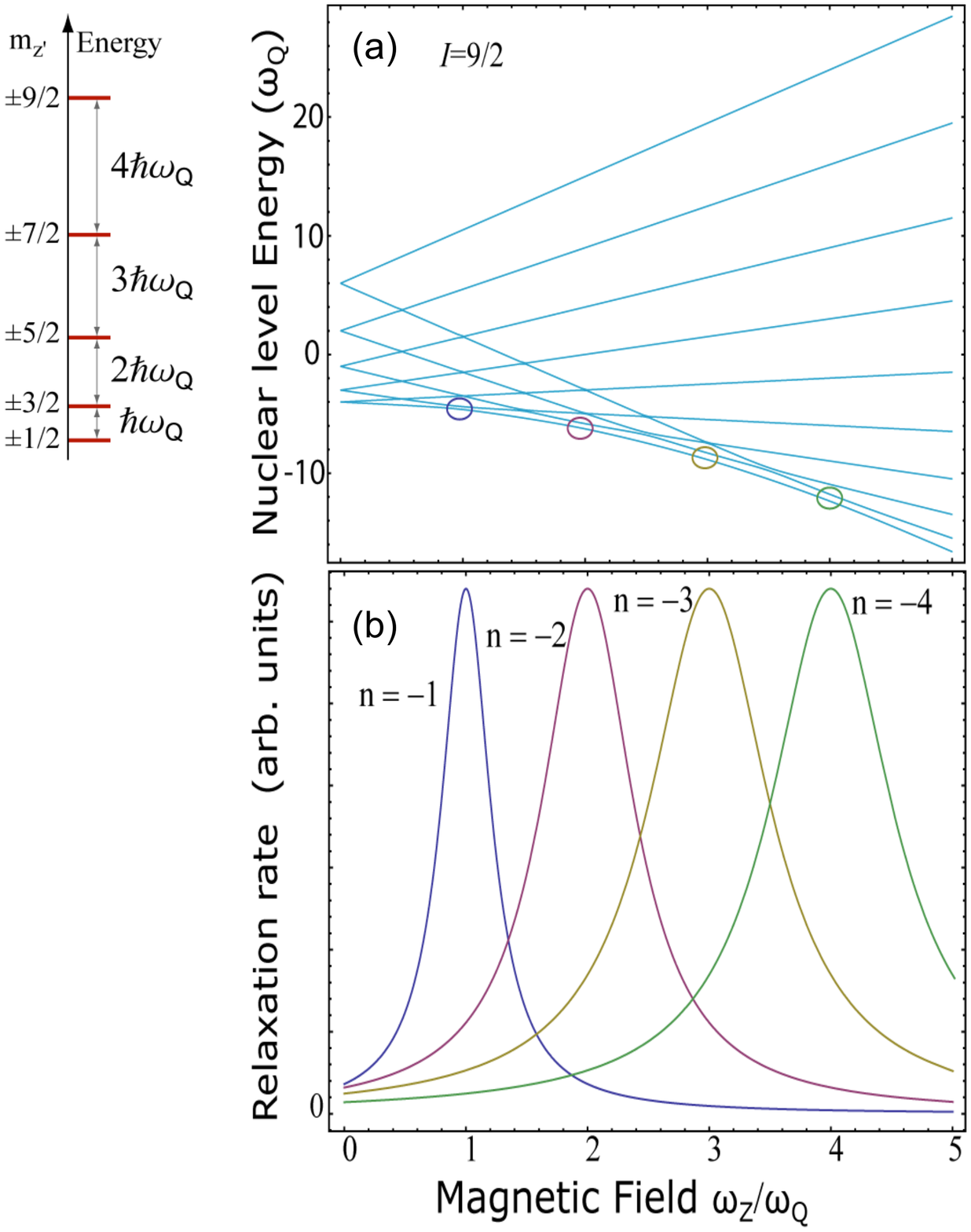}
\caption{(a), (b) Energy levels and relaxation rates according to Eq.(\,\ref{eq:TeQ}) for a spin $I=9/2$ and $\theta=0.05$. The anti-crossings are marked by circles of different color in (a), the corresponding relaxation rates are shown in (b) in the same color.}\label{LPN-QI}\end{figure}

A quantitative description of DNP achievable in optical pumping experiments in QDs has to take into account the nuclear quadrupole interaction introduced in Sec.\,\ref{sec:quadintro} induced by strain and alloy disorder \cite{Bulutay:2012a}.
In addition to zero order effects, the quadrupolar
interaction is  also responsible for  a specific  mechanism of
nuclear spin relaxation which arises even for small $\theta$ and
which is induced by the longitudinal part $\propto
\hat{S}^e_z\hat{I}_z$ of the hyperfine interaction fluctuations. If
the quadrupolar axis is tilted by an angle $\theta$ the momentum
operator $\hat{I}_z$ couples  the nuclear eigenstates of angular
momentum difference $|\Delta m_{z'}|=1$ (except the pair of levels
$m_{z'}=\pm1/2$) which tends to equalize the populations of these
states and therefore cancel their respective  polarization. In the
interaction representation and for small $\theta$, the
time-dependent coupling is proportional to $(h_1(t)
A/N)\theta\omega_\text{Q}/\Delta\omega_n$ where the relevant energy detuning is the sum of external field and contributions from the nuclear quadrupole effects $\Delta\omega_n=(\gamma_n B_z+n\,\omega_\text{Q})$ where
$n=m_{z'}+1/2$ corresponds to the energy splitting between the
states $|m_{z'}\rangle$ and $|m_{z'}+1\rangle$. Remarkably, in
contrast to the flip-flop relaxation induced by the transverse part
of the hyperfine interaction,  no electron spin-flip is required
if we take into account the non-collinear hyperfine coupling $ \hat{H}_{\rm hf}^{\rm nc}$ given by Eq.\,\ref{eq:nc-hyperfine}. This makes the associated relaxation independent of the
electron spin splitting. Besides, since  the nuclear splitting
$\Delta\omega_n\sim 10$~neV is small in usual experimental
conditions compared to the inverse correlation time
$(\tau_c^e)^{-1}\sim 10~\mu$eV  determined in
Sec.\,\ref{sec:DNPvsBz}, the issue of energy conservation does not
require special consideration and the spectral density of $h_1(t)$
reads simply $2\tau_c^e/(1+
(\Delta\omega_n\,\tau_c^e)^2)\approx2\tau_c^e$.  Because of the
non-harmonicity  produced by the quadrupolar splitting
($\Delta\omega_n$ depends on  $n$) it is not possible to derive a
single relaxation time for the nuclear polarization, as done  for
the flip-flop term in Sec.\,\ref{sec:DNPvsBz}. However, the
transition rate between  the specific  levels  $|n\pm1/2\rangle$
where  $n\in\{\!-\!I+1/2,\ldots,I\!-\!1/2\}$  can still be
calculated perturbatively according to Redfield's theory
\cite{Abragam:1961a,Slichter:1990a} or from the Fermi's golden rule
assuming a Lorentzian broadening $\hbar/\tau_c^e$ of the level
spectral density \cite{Huang:2010a}.  To first order in $\theta$
we obtain \footnote{Please note that as a consequence of keeping only first order terms in $\theta$ for small angles we can neglect transitions with $\Delta I_z=2$.}:
\begin{equation}
\frac{1}{T_\text{e-Q}^{I,n}}= \left(\frac{A_{I}}{2\hbar N }\right)^2
\frac{f_e
\tau_c^e}{1+(\Delta\omega_n/c_{I,n}\theta\omega_\text{Q})^2}
\label{eq:TeQ}
\end{equation}
where $c_{I,n}=n\sqrt{(I+1/2)^2-n^2}$. For each pair of levels the
relaxation rate is maximum at  the magnetic field $B_z=-n B_\text{Q}$
corresponding to a minimum of $\Delta\omega_n$ and follows a
Lorentzian dependence with a typical width given by
$2c_{I,n}\theta\omega_\text{Q}$. Figure\,\ref{LPN-QI}(b)   shows the
relative evolution of the different rates for  a spin $I=9/2$ (case
of Indium) and assuming a small angle $\theta=0.05$.
This magnetic field dependent nuclear spin relaxation mechanism will be considered in the following section that details the evolution of DNP in measurements at variable magnetic fields.


\subsubsection{Nuclear field bistability in magnetic field sweeps}
\label{sec:DNPvsBz}

\begin{figure}
\epsfxsize=3.5in
\includegraphics[width=0.48 \textwidth,angle=0]{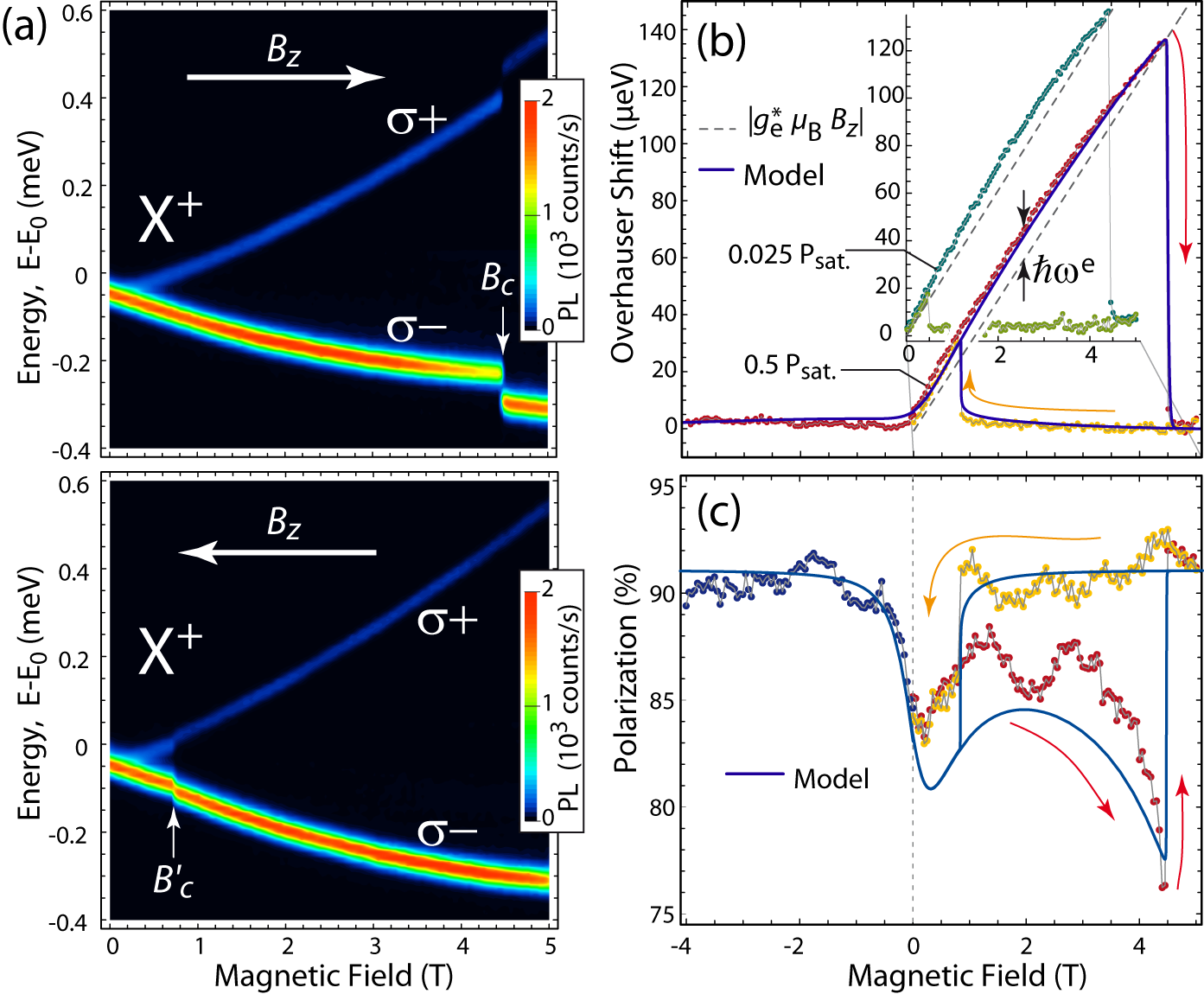}
\caption{(a) Density plots of $X^+$ PL  intensity from a single  QD at T=1.7~K as a function of  an increasing or decreasing magnetic field $B_z$,  detected  over a 1~meV energy  range around $E_0=1.3143$~eV. The  quasi-resonant $\sigma^-$ polarized excitation is $\sim60$~meV above the trion line in zero field.  (b) Absolute Overhauser shift of the Zeeman splitting for two excitation powers, and (c) circular polarization as a function of $B_z$. The measurement at 0.025$P_\text{sat}$ has been shifted and slightly reduced in size for clarity. adapted from \textcite{Krebs:2008a}}\label{LPN-OHSvsFieldXp}
\end{figure}

For a given set of experimental parameters there exist several stable nuclear spin configurations, see Eq.\,\ref{eq:BnQD}. 
The experimentally achieved nuclear spin polarization degree will in addition depend on the history of the experiment (non-Markovian behaviour).
Different bistability regimes of DNP  were observed in single QD optics by varying continuously different
parameters in the experiment like the polarization of the optical excitation
\cite{Braun:2006a}, the laser excitation power
\cite{Tartakovskii:2007a,Kaji:2008a,Belhadj:2008a,Skiba-Szymanska:2008a}
or more commonly the external magnetic field
\cite{Maletinsky:2007a,Braun:2006a,Krebs:2008a,Kaji:2008a}.
Figure\,\ref{LPN-OHSvsFieldXp}(a)  shows the typical evolution of
the PL from an $X^+$ trion as a function of a the applied
longitudinal magnetic field when the QD is excited with a
quasi resonant, $\sigma^-$ polarized laser. The Zeeman splitting measured between
the $\sigma^+$ and $\sigma^-$ components
increases steadily up to a critical field $B_c\approx$~4.5~T where
it undergoes an abrupt increase. This provides direct
evidence that up to  $B_c$, a nuclear field $B_n$ is created in the quantum
dot in the direction opposite to $B_z$. When the magnetic field is
swept back, a similar abrupt reduction of the Zeeman  splitting
takes place but at a lower critical field $B'_c\approx 1$~T.

The Overhauser shift $\hbar\omega^e_{OS}\equiv
\vert g_e \mu_\text{B} B_{n,z}\vert$ due to the nuclear field can be
extracted as shown in Fig.\,\ref{LPN-OHSvsFieldXp}(b) by subtracting the Zeeman splitting of the trion $(g_h-g_e)
\mu_\text{B} B_z$ exclusively due to the external field, where $g_h\simeq2.4$ is the  $g$-factor of the
heavy-hole pseudo-spin 1/2 for this particular QD and
$g_e\simeq-0.6$.  Note that the
Zeeman splitting used for reference is precisely  determined from a
measurement where $B_n=0$: under linearly polarized excitation in moderate external
fields (below $\sim$3~T)  since in this case $\langle
\hat{S}_z^e\rangle=0$ due to optical pumping, $\langle \hat{S}_z^e\rangle_0\approx
0$ due to thermalisation result in zero nuclear polarization. The asymmetrical  dependence of $B_n$on
$B_z$ and the bistability regimes of DNP are clearly evidenced in
Fig.\,\ref{LPN-OHSvsFieldXp}(b). Remarkably, the Overhauser shift
grows almost linearly with the positive  $B_z$  (for $\sigma^-$
excitation) in order to  remain above the dashed line which
represents the electron Zeeman splitting. As long as this condition
is fulfilled, the nuclear field $B_n$ fully compensates the applied field
(even slightly overcompensates) so that the amplification of
$\varsigma$ in Eq.\,\eqref{eq:BnQD} remains moderate of the order of
unity. When  $B_z$ approaches  the critical field  $B_c$, the
electron spin splitting $\hbar\omega^e$ vanishes, indicating that the
nuclear field has reached its maximum in the conditions of the
experiment. Beyond this point a further increase of $B_z$ yields a
reduction of $B_{n,z}$. This provides a negative feedback ($\varsigma$ in Eq.\,\eqref{eq:BnQD} is amplified)
and the Overhauser shift suddenly drops as the applied field is slightly increased.
The  maximum  Overhauser shift of 137~$\mu$eV achieved here would
amount  to an average nuclear polarization of  43\% for a pure InAs
QD, but it is certainly closer to  60\%  for a more realistic
In$_{0.5}$Ga$_{0.5}$As intermixed QD, using the values for 100\% nuclear polarization in table \ref{tab:ohs} as references.
By adjusting the parameters of
the model it is possible to reproduce the experimental
magnetic field dependence of $\hbar\omega^e_{OS}$ reasonably well using the numerical solution
of Eq.\,\eqref{eq:DNP}. In practice only $\tau_c^e$, $g_e$ and the
product $f_e T_d$  have to be varied for the fit, whereas other parameters can
be fixed  to the  values expected for usual InAs QDs ($N=5 \cdot 10^4$,
$\tilde{Q}=13$, $T=1.7$~K).

To improve the agreement around
zero field where the experimental slope
$(\partial\omega^e_{OS}/\partial B_z)$ presents a kind of
discontinuity, it is necessary to include a field-dependent
nuclear relaxation rate $T_d^{-1}$.
One plausible approach is to include the gradual inhibition of nuclear spin relaxation
due to the quadrupolar coupling as the longitudinal magnetic field increases~\cite{Meier:1984a,Abragam:1961a,Huang:2010a}.
In realistic QDs we have to average over the  distribution of
quadrupolar interactions both in magnitude and direction, and hence the
relaxation resonances in Eq.\,\eqref{eq:TeQ} are drastically broadened compared to the estimations shown in Fig.\,\ref{LPN-QI}. Also, the applied longitudinal field will restore the eigenaxis along $z$.
This justifies the incorporation of a simplified magnetic field dependence as we assume in addition to a constant term a Lorentzian part describing qualitatively the
slowdown of relaxation at high fields:
\begin{equation}\label{eq:TdvsBz}
T_d^{-1}=T_{d_\infty}^{-1}+\frac{T_{d_0}^{-1}}{1+(B_z/B_{Q'})^2}
\end{equation}

A manual  fit of the model to the
experimental points is plotted in Fig.\,\ref{LPN-OHSvsFieldXp}(b)
(solid line) with $\tau_c^e=75$~ps, $g_e=-0.6$ for the electron g-factor\footnote{It was
assumed that  the hyperfine interaction with the hole reduces the
theoretical Overhauser shift $\vert 2 \tilde{A} \langle
I_z\rangle\vert$ by about 10\%, see Sec.\,\ref{sec:holeover}. This
enables us to use the  nominal value expected for the electron
$g$-factor instead of the reduced effective  factor $g_e^\star=0.9
g_e$  determined from the maximum Overhauser shift
$\hbar\omega_{OS}^e$ at $B_c$ and corresponding to the dashed line
in Fig.\,\ref{LPN-OHSvsFieldXp}(b). This is  not a critical
assumption since it  improves only slightly the agreement with the
experimental data in weak fields.} , $f_e
T_{d_0}=65\,\mu$s, $f_e T_{d_\infty}=5$~ms and $B_{Q'}=0.4$~T, where $B_{Q'}$ is a measure of the strength of the nuclear quadrupole effects, typically on the order of 100~mT. The inclusion of this magnetic field dependent nuclear spin relaxation rate improves the fit considerably for $B_z<1$~T \cite{Krebs:2008a}.

In the present simple model the fraction of time of QD occupation $f_e$
and the relaxation time $T_d$ can not be  distinguished since only
their product appears in  $\varsigma$. Furthermore, when the
laser excitation power (proportional to $f_e$ below the saturation power
$P_\text{sat}$) is reduced from $P_\text{sat}/2$ to
$P_\text{sat}/40$, the generation of the nuclear field is almost not
affected as shown in Fig.\,\ref{LPN-OHSvsFieldXp}(b): despite a
smaller initial value in zero field, the Overhauser shift still
follows the increase of $B_z$  up to  about the same critical field
$B_c$ around 4.5~T, reaching the same maximum Overhauser shift
$\simeq135~\mu$eV with $P_\text{sat}/40$ as achieved with $P_\text{sat}/2$. This is a clear
indication that while the power  \textit{decreases} the relaxation
time  $T_d$ \textit{increases}. To  reach the same $B_c$  in  the
simulations  an almost  constant product $f_e T_{d}$ is required. It
can therefore be  inferred that $T_d\propto f_e^{-1}$  in
qualitative agreement with the very slow nuclear relaxation observed
when the sample is left in the dark i.e. does not contain any charge carriers (see Sec.\,\ref{sec:demag}).
This observation points towards a depolarization mechanism due to
Knight field fluctuations  coupled to  the quasi-static local
dipolar field~\cite{Gammon:2001a} or to the quadrupolar interaction
\cite{Huang:2010a}  as the dominant cause of nuclear spin relaxation
in optically excited QDs.

The $X^+$ is created through quasi-resonant optical excitation. The electron-nuclear spin flip-flop can either occur during
the trion lifetime $\tau_r\sim$1~ns, or alternatively, during the shorter stage of energy relaxation from the initial
excited level since the correlation time extracted from the fits $\tau_c^e\ll\tau_r$. A reduction of $\langle
S^e_z \rangle=-\rho_{c}/2$ due to this spin flip-flop is expected and we shall consider if this decrease is observable in realistic experiments:
From Eq.\,\eqref{eq:DNP} one can infer the spin-flip time $T_{1n}$
experienced by the electron due to the nuclei   to be $T_{1n}=T_{1e}
f_e/(N \tilde{Q})$. From the fitting parameters used above, it
typically amounts to $\sim$10~ns  in the magnetic field range where a non-zero
Overhauser shift is created. The probability of an electron spin-flip
during its lifetime is therefore not completely negligible ($\tau_r/T_{1n}\approx 0.1$) and the polarization of the photons emitted by
$X^+$ recombination  should  be affected by the preceding electron-nuclear spin flip-flop process.
Also, when the electron splitting $\hbar\omega^e$ becomes of the order of $\tilde{A}(\tilde{Q}/N)^{1/2}\sim 1 \mu$eV\footnote{The secular approximation used to derive  the expression of $T_{1n}$ to the second order in perturbation is no longer valid.}, the first order effect of spin precession in the nuclear field fluctuations (as discussed in Sec.\,\ref{sec:merk}) becomes predominant and should appreciably enhance the electron spin relaxation because $T_{\Delta}/T_{1n}\sim\tilde{A}\tau_c^e/\hbar (\tilde{Q}/N)^{1/2}\sim 0.1$ \cite{Krebs:2008a}.
This is confirmed experimentally as a clear correlation between the measured $X^+$
polarization and the achieved Overhauser shift is found, see
Fig.\,\ref{LPN-OHSvsFieldXp}(c). The polarization is reduced by
$\sim$5\% in the conditions where DNP is built up efficiently i.e. when $T_{1n}$ is short. More specifically
the $X^+$ polarization exhibits a pronounced dip down to 76\% when approaching the
critical field $B_c$, immediately  followed by an abrupt increase to
a roughly constant level of 91\%. For a qualitative comparison  the
theoretical  reduction of the polarization $\rho_c$ of the $X^+$ emission is plotted in
Fig.\,\ref{LPN-OHSvsFieldXp}(c) using
$\rho_c=\rho_c^\text{max}/(1+\tau_r/T_{1n})$ with
$\rho_c^\text{max}=0.91$ and $\tau_r=0.7$~ns. This formula is
only approximate because it assumes that electron-nuclei
flip-flops take place for the whole radiative lifetime of $X^+$
although the correlation time used in the model  is much shorter,
but on the other hand it neglects the faster  spin relaxation rate
$\sim T_\Delta^{-1}$  near $B_c$.  The general idea is captured: the smaller the splitting in energy
$\vert\hbar\omega^e\vert$ between electron spin up and spin down states, the shorter the spin lifetime and
consequently the smaller the measured polarization, irrespective of the actual
spin-flip mechanism.  The dip observed at $B_c$ is clearly related to the vanishing $\vert\hbar\omega^e\vert$. It is remarkable that the Zeeman splitting of the electron is zero in an applied field $B_z$ of more than 4~Tesla, and an electron depolarization due to the Merkulov effect is observed, in analogy to the observed decrease in electron polarization around zero magnetic field \cite{Merkulov:2002a,Braun:2005a}.

\begin{figure}
\epsfxsize=3.5in
\includegraphics[width=0.48 \textwidth,angle=0]{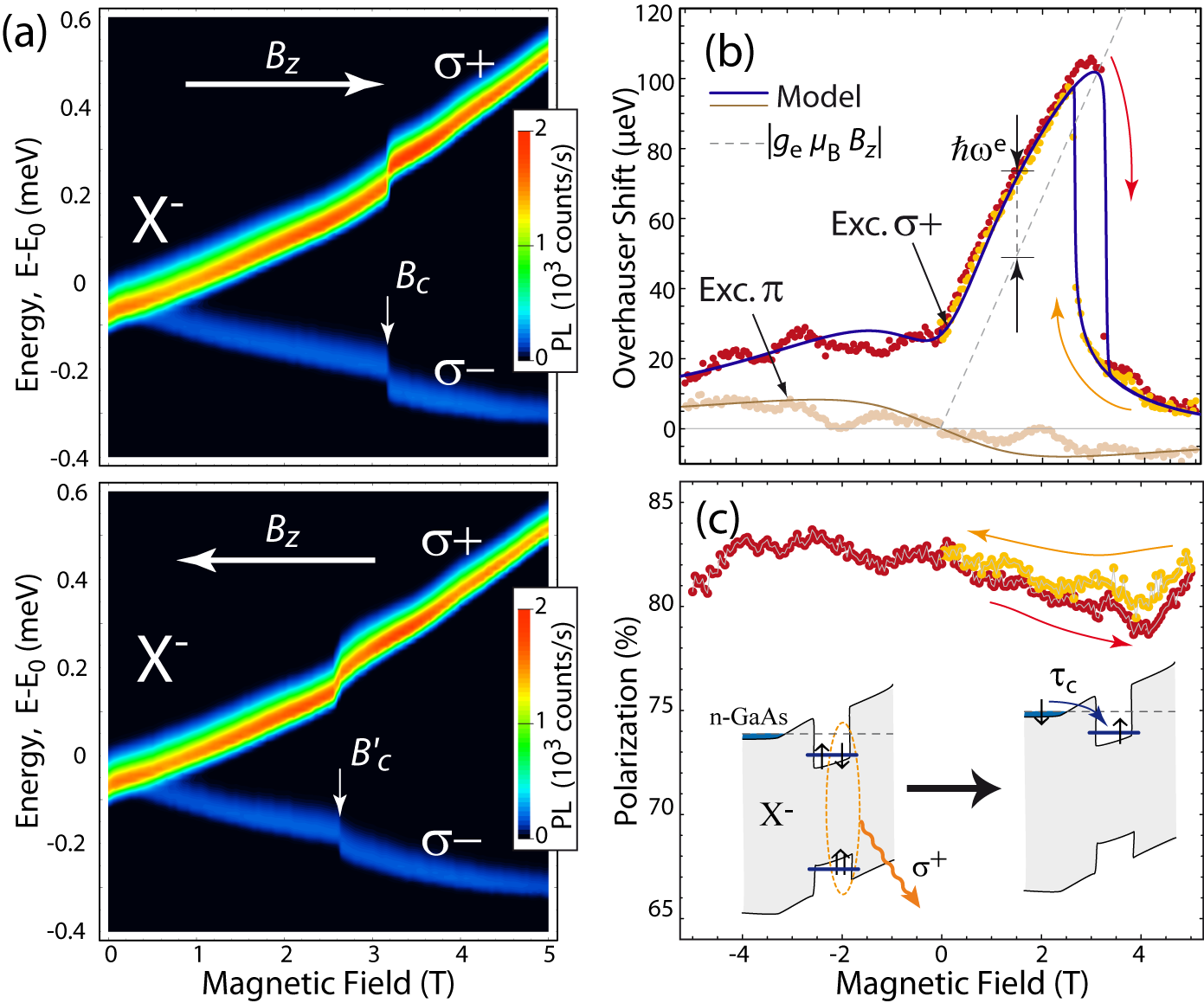}
\caption{(a) Density plots of $X^-$ PL  intensity from the same  QD of Fig.\,\ref{LPN-OHSvsFieldXp} at T=1.7~K as a function of  an increasing or decreasing magnetic field $B_z$, detected  around $E_0=1.305$~eV. The  quasi-resonant $\sigma^+$ or $\pi$ polarized excitation is $\sim40$~meV above the trion line in zero field.  (b) Overhauser shifts of the Zeeman splitting for both excitation polarizations with superimposed   fits of the model using the same parameters (solid lines). The measurements under  $\pi$ polarization, performed in positive fields,  have been duplicated anti-symmetrically in negative fields. (c) Circular polarization under $\sigma^+$ excitation as a function of $B_z$ showing no correlation to the Overhauser shift in contrast to $X^+$. Inset illustrates schematically $X^-$  emission, followed by the capture of a second resident electron defining the correlation time $\tau_c^e\sim20$~ps in the model. adapted from \textcite{Krebs:2008a}}\label{LPN-OHSvsFieldXm}
\end{figure}

Figure\,\ref{LPN-OHSvsFieldXm} shows for comparison similar
measurements for the same QD but in the negative charge
state giving rise to the PL emission of an $X^-$ line located about
10~meV below  the $X^+$ line. This charge state is achieved by
increasing the gate voltage applied to the $n$-Schottky device  by
+0.8~V, in a regime where the QD is actually charged by two
electrons. The quasi-resonant optical excitation creates a hole
confined in the QD valence ground state and an  electron in an excited
state which escapes from the dot in a few picoseconds. This leads to  $X^-$
formation in a  polarization state  determined by the hole spin
\cite{Laurent:2005a}.  Non-linear DNP is again observed when a
longitudinal magnetic field is applied, see
Fig.\,\ref{LPN-OHSvsFieldXm}(a),(b), with some notable differences compared
to the $X^+$ case:. \\
\textbf{(i)} the Overhauser shift  develops   when the
$\sigma^+$ branch of the trion   (i.e. the Zeeman  line which
blue-shifts with the applied magnetic field)  is  predominantly
pumped by the $\sigma^+$ excitation. This inversion of the required
polarization with respect to $X^+$ case   results from the  symmetry
between $X^+$  and $X^-$ \cite{Eble:2006a}: As shown in
Fig.\,\ref{LPN-OHSvsFieldXm}(c) inset, a $\sigma^+$ polarized $X^-$
does not itself polarize the nuclei because the conduction electrons
are in a singlet state, but  after recombination with the $\Uparrow$
hole, it yields an unpaired electron with $\uparrow$ spin exactly as
for a  $\sigma^-$ photo-created $X^+$ trion. The same regime of DNP
can therefore take place in a positive magnetic field which is still
opposite to the generated nuclear field $B_n$.\\
\textbf{(ii)} In contrast to $X^+$, the measured PL polarization
in Fig.\,\ref{LPN-OHSvsFieldXm}(c) shows no correlation with the
measured Overhauser shift, because it is determined by the optically
created hole spin  polarization. \\
\textbf{(iii)} The effective  electronic splitting $|\hbar\omega^e|\simeq25\,
\mu$eV is significantly larger, the critical   magnetic fields
$B_c\simeq$3.3~T and $B'_c\simeq$2.8~T associated with the abrupt
decrease and recovery of nuclear polarization are respectively
reduced and enhanced, which results in a narrower  domain of
bistability, and eventually there is only a partial decrease of
$\vert\hbar\omega^e_\text{OS}\vert$ in negative fields leading to a
residual shift of $\sim$15~$\mu$eV at $B_z=-5$~T. \\
 All these changes can
still be reproduced by the above model as shown by the simulation
plotted in Fig.\,\ref{LPN-OHSvsFieldXm}(b)  (solid line) assuming
$\tau_c^e=18$~ps, $g_e=-0.62$, $f_e T_{d_0}=1$~ms, $f_e
T_{d_\infty}=8.7$~ms and $B_{Q'}=0.7$~T.  The  reduction of the
non-linearity  quantified by $(B_c-B'_c)/(B_c+B'_c)$ results
essentially from the  shortening of  $\tau_c^e$ which leads to a
smoothing of  the Lorentzian  factor in the right-hand side of
Eq.\,\eqref{eq:BnQD} and makes the condition $\hbar\omega^e \approx
0$ less critical. This explains also why in  negative fields as well
as for $B_z> B_c$, a finite Overhauser shift  can still be
generated. There is however a clear asymmetry  between the measured degree of DNP at $B_z=+5~$T and $-5~$T
indicating a favored  DNP in large negative fields.
In this situation  the electron splitting $\hbar\omega^e$ is dominated by
the Zeeman part so that the flip-flop rate $T_{1e}^{-1}$ should be
essentially symmetrical. The reason for the  asymmetry comes now
from the increase of the  equilibrium spin polarization $2\langle
S_z^e\rangle_0$  which amounts to $\pm$53\% at $T=~$1.7~K and
$B_z=\pm5$~T. The departure from equilibrium in
Eq.\,\eqref{eq:BnQD} is therefore drastically altered according to
the field direction. The role played  by this equilibrium spin
$\langle S_z^e\rangle_0$ is confirmed in  measurements with  a
linearly polarized ($\pi$) excitation which does not directly
create any spin polarization ($\langle S_z^e\rangle=0$), but still
show a finite Overhauser shift in high fields, see
Fig.\,\ref{LPN-OHSvsFieldXm}(b).

\begin{figure}
\epsfxsize=3.5in
\includegraphics[width=0.45 \textwidth,angle=0]{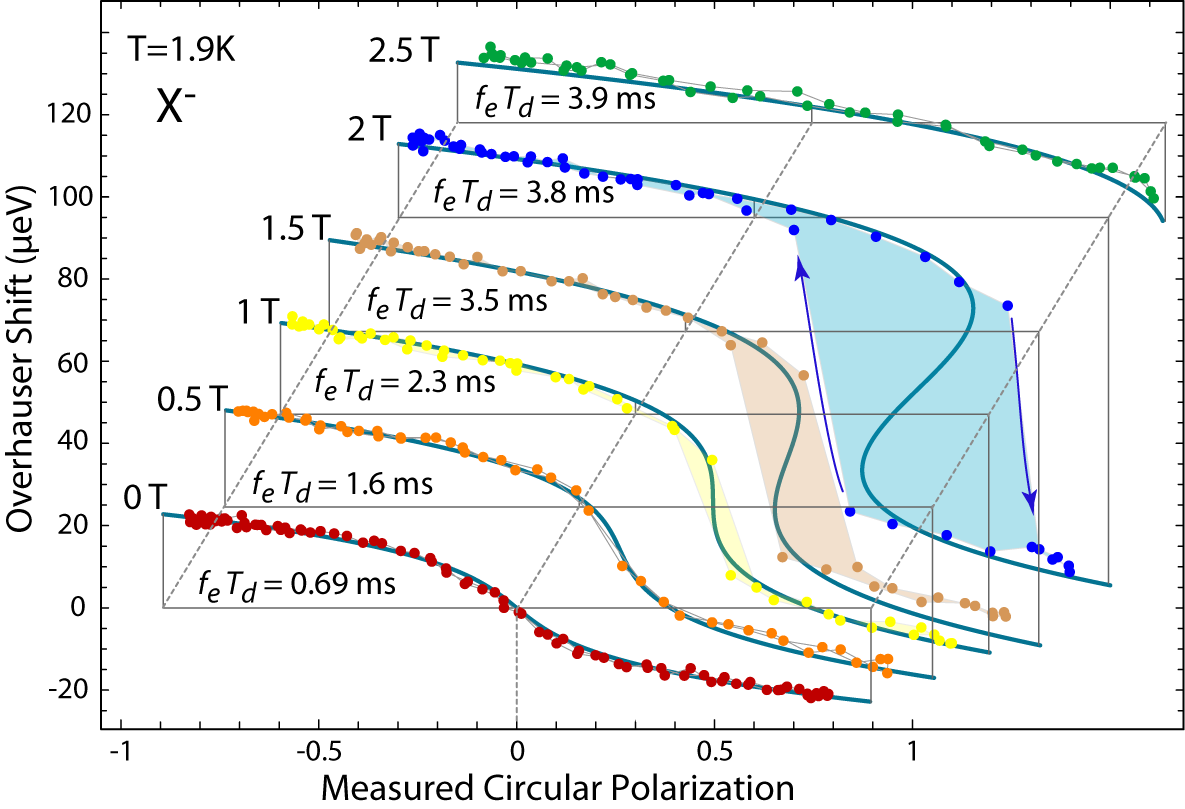}
\caption{Overhauser shift $\propto \langle \hat{I}_z \rangle$ in a single InAs/GaAs QD as a function of the \emph{measured} circular polarization from the $X^-$ PL line (essentially $\propto \langle S_z^e \rangle$ of the electron left behind) around $E_0=1.265$~eV at different longitudinal magnetic fields.   The  polarization of the quasi-resonant excitation is varied  by stepwise rotating  a quarter-wave plate over a total angle of $\pi$, which achieves  the complete polarization cycle $\pi\rightarrow\sigma^+\rightarrow\pi\rightarrow\sigma^-\rightarrow\pi$. Solid lines represent fits of Eq.\,\eqref{eq:BnQD} with $\tau_c^e=31\pm 2$~ps, $g_e^\star=0.54$ and the product $f_e T_d$ adjusted for each field as indicated. The curves and experimental points at different magnetic fields have been translated for clarity.}\label{LPN-OHSvsSz}
\end{figure}
\subsubsection{Nonlinearity of the nuclear polarization as a function of excitation laser polarization}
\label{sec:DNPvsSz}

The case of single $X^-$ trions is particularly interesting because
the  circular polarization of the corresponding PL line allows
for a direct determination of the unpaired electron spin left in the
QD just after the optical recombination and responsible for
the subsequent DNP. The validity of Eq.\,\eqref{eq:BnQD} which
relates the Overhauser shift $\propto \langle \hat{I}_z \rangle$ to the average electron spin $\langle
S^e_z\rangle$ can thus be tested in a fixed magnetic field without the need
of Eq.\,\eqref{eq:TdvsBz} \cite{Eble:2006a}.
Figure\,\ref{LPN-OHSvsSz} shows a series of such measurements where
the polarization of the excitation light is changed stepwise by
rotating  the quarter-wave plate through which the  linearly
polarized laser passes, allowing essentially to plot $\langle \hat{I}_z \rangle$ as a function of $\langle
S^e_z\rangle$. Along a $\pi$  rotation the incident
polarization covers the cycle
$\pi\rightarrow\sigma^+\rightarrow\pi\rightarrow\sigma^-\rightarrow\pi$,
as a result, the recorded circular polarization degree of the PL varies between
$-0.8$ and $+0.8$. For each excitation polarization the $X^-$ PL emission is measured in
$\sigma^+$ and $\sigma^-$ polarization separately, typically for a
1~s integration time which is  long enough to reach the steady state
regime described by Eq.\,\eqref{eq:BnQD}.
Fig.\,\ref{LPN-OHSvsSz} clearly shows that the DNP achieved via optical pumping depends in a highly non-linear fashion on the average injected electron spin $\langle S^e_z\rangle$  for magnetic fields below the critical field $B_c$ which amounts to $\simeq $2.2~T for the $X^-$ of this QD.
This non-linearity manifests itself as an asymmetry of the
Overhauser shift as  $\langle
S^e_z\rangle$ changes from positive to negative, in perfect agreement with the asymmetry observed when then magnetic field changes sign, as discussed earlier for Fig.\,\ref{LPN-OHSvsFieldXm}. \\
In zero applied field the Overhauser shift shown in Fig.\,\ref{LPN-OHSvsSz} is as
expected exactly anti-symmetrical with respect to a change in sign of $\langle S^e_z\rangle$. Interestingly, it still exhibits a non-linear dependence on $\langle S^e_z\rangle$ indicating that the nuclear field which develops is strong enough to restrain the DNP rate
$T_{1e}^{-1}$: a negative feedback develops as any further spin flip-flop will increase $\hbar \omega^e=\hbar \omega^e_\text{OS}$ and hence make the next spin flip-flop less likely than the previous one.
At higher positive fields, this drastic feedback
shifts the non-linear response of the nuclear system towards
positive values of $\langle S^e_z\rangle$ and enhances its magnitude
to the point that a bistability region shows up, as evidenced by the
measurements  at 2~T in Fig.\,\ref{LPN-OHSvsSz}
\cite{Braun:2006a,Maletinsky:2007a}. When the external field exceeds
$B_c$, the level of high nuclear polarization can no longer be
reached by increasing the electron spin polarization. The system
remains in the  low nuclear polarization state which now depends
almost linearly on $\langle S^e_z\rangle$ because the role of
$B_{n,z}$ in $T_{1e}^{-1}$ is negligibly small
($|B_{n,z}|\ll|B_z|$). Note that there is yet a residual Overhauser
shift  of a few $\mu$eV at $\langle S^e_z\rangle=0$  due the
finite equilibrium spin $\langle S^e_z\rangle_0$ in
Eq.\,\eqref{eq:BnQD}.  As shown in Fig.\,\ref{LPN-OHSvsSz}, the
agreement with the model is very good. From Eq.\,\eqref{eq:BnQD} one
can express $\langle S^e_z\rangle$ as a function of the nuclear
field $B_{n,z}$ and therefore  determine $\tau_c^e$, $f_e T_d$ and
$g_e$   from a least-square fit to the experimental points. All the
theoretical curves in Fig.\,\ref{LPN-OHSvsSz} are obtained with the
same set of parameters except for   $\tau_c^e$ and   $f_e T_d$. The
estimated values of the correlation time which presumably should not
depend on the magnetic field varies indeed very moderately
($\tau_c^e=31\pm 2$~ps),  while in contrast   the product  $f_e T_d$
increases appreciably with the field  in qualitative agreement with
the  Lorentzian dependence assumed in Eq.\,\eqref{eq:TdvsBz}.

\begin{figure}
\epsfxsize=3.5in
\includegraphics[width=0.48 \textwidth,angle=0]{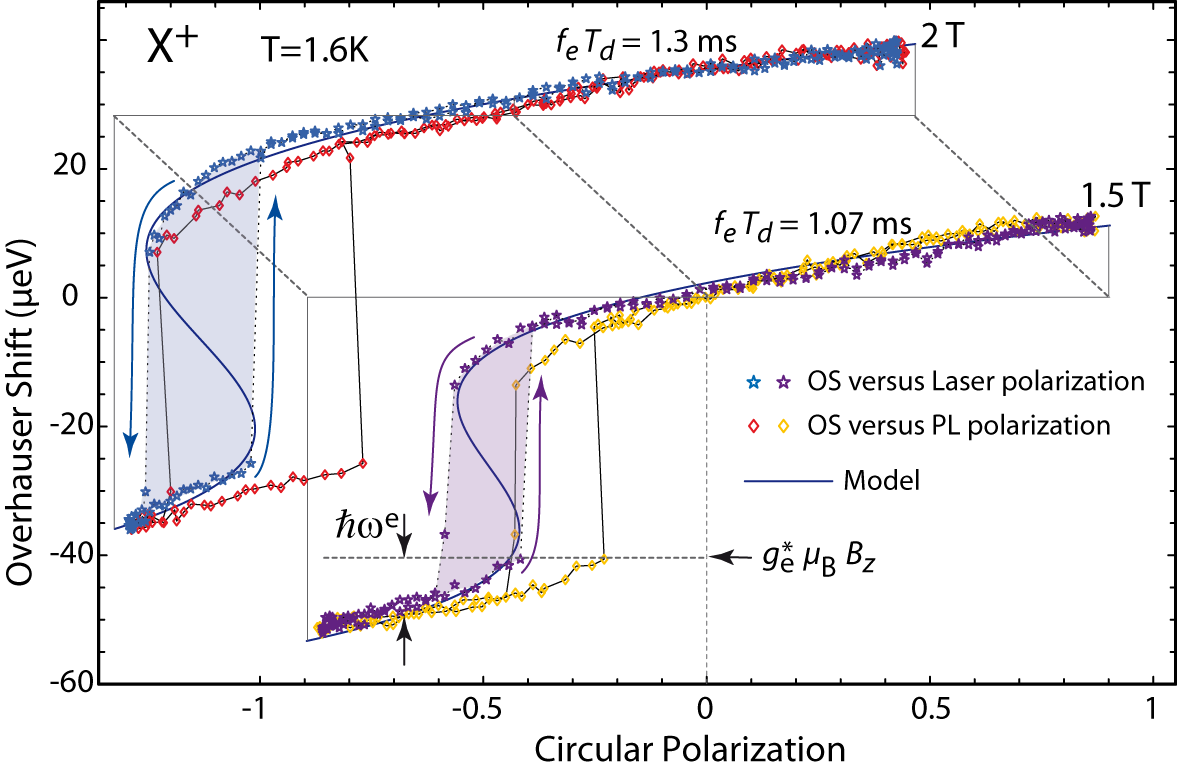}
\caption{ Overhauser shift in a single InAs/GaAs QD as a function of the Laser or PL circular polarization  for an   $X^+$ PL line around $E_0=1.348$~eV at two longitudinal magnetic fields.   The  polarization of the quasi-resonant excitation is changed  by stepwise rotating  a quarter-wave plate as in Fig.\,\ref{LPN-OHSvsSz}. Solid blue lines represent fits of Eq.\,\eqref{eq:BnQD} with $\tau_c^e=31\pm 2$~ps, $\vert g_e \vert = 0.5$, $\vert g_e^\star \vert =0.45$ and the product  $f_e T_d$  as indicated. The curves and experimental points for both  magnetic fields have been translated for clarity.}\label{LPN-OHSvsSz-Xp}
\end{figure}
\indent Figure\,\ref{LPN-OHSvsSz-Xp} shows the analogous evolution of the Overhauser shift for an $X^+$ trion when the excitation polarization varies from  $\sigma^+$ to  $\sigma^-$ and back. The global behavior is very similar, with a bistability region in magnetic  fields below $B_c$. The main difference, besides the symmetry change between $X^+$ and $X^-$ case already discussed above, is that the Overhauser shift can no longer be satisfactorily correlated  to the measured PL circular polarization according to  Eq.\,\eqref{eq:BnQD}.

\begin{figure}
\epsfxsize=3.5in
\includegraphics[width=0.45 \textwidth,angle=0]{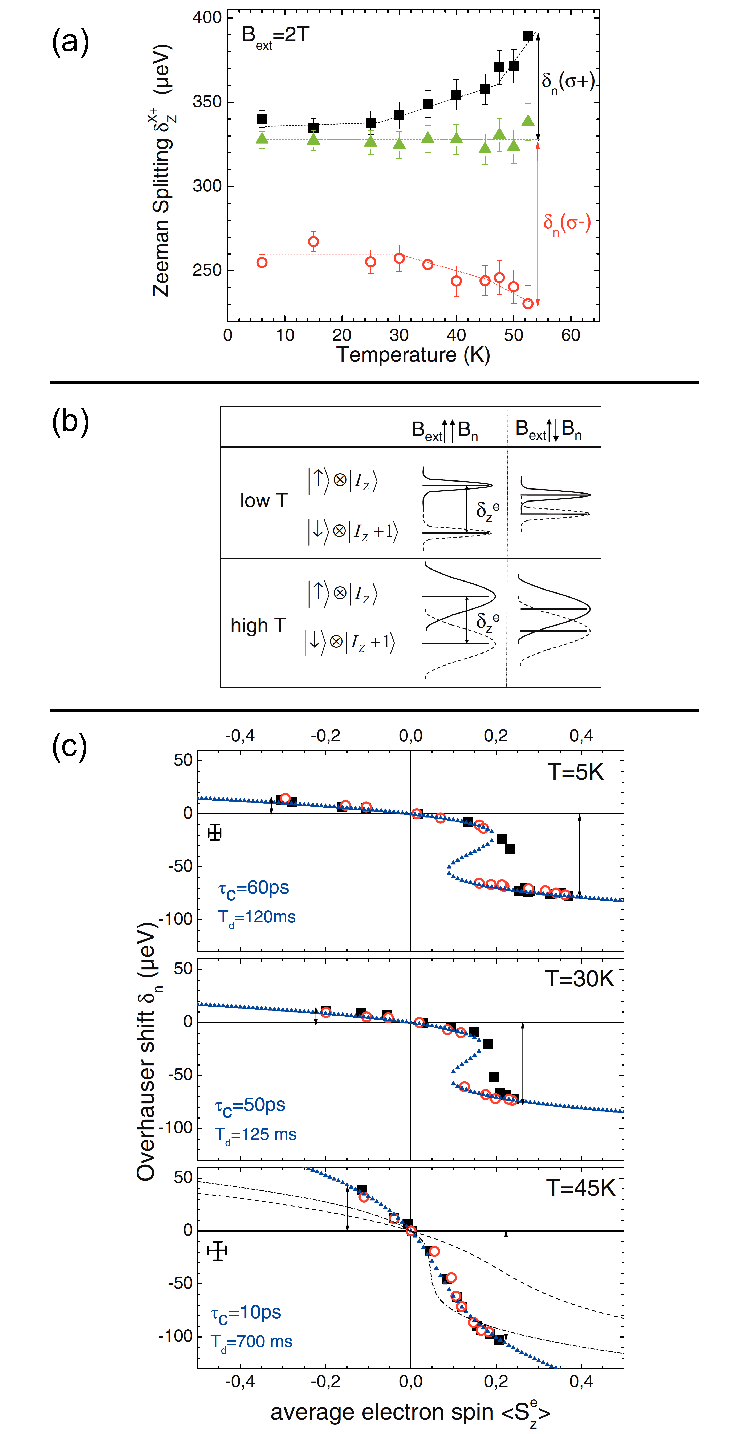}
\caption{(a)  Spin splitting in a single InAs/GaAs QD  for an   $X^+$ PL line as a function of temperature.  The black (red) symbols are measured following $\sigma^+$ ($\sigma^-$) polarized laser excitation and the green symbols following linear laser excitation, for which DNP is absent. (b) Schematics of the electron-nuclei level splitting and the change of broadening due to temperature raising. (c) Evolution of the Overhauser shift as a function of $\langle S_z^e\rangle$ deduced from the PL circular polarization ($\langle S_z^e\rangle=-\rho_c/2$) for different temperatures. Solid lines are fit of the model to the experimental data with the  corresponding correlation time $\tau_c^e$ and relaxation time $T_d$ as indicated (after \textcite{Urbaszek:2007a}). }\label{LPN-OHSvsT-Xp}
\end{figure}

The difference between the laser polarization ($\propto \langle S^e_z\rangle_{t=0}$) and the $X^+$ PL polarization ($\propto \langle S^e_z\rangle_{t=\tau_r}$) is a measure of the impact of the electron-nuclei spin flip-flops on the average electron polarization $\langle S^e_z\rangle$. As can be expected, the difference between excitation polarization and emitted polarization in Fig.\,\ref{LPN-OHSvsSz-Xp} is largest when nuclear polarization is strongest, in close analogy to the dip at $B_c$ in Fig.\,\ref{LPN-OHSvsFieldXp}(c). Good agreement with Eq.\,\eqref{eq:BnQD} can still be obtained by assuming  $\langle S^e_z\rangle$ proportional to the nominal \textit{laser} polarization, see fits to the data in Fig.\,\ref{LPN-OHSvsSz-Xp}.\\
\indent Investigating the non-linearity of DNP as a function of the electron or laser circular polarization gives access to the key  parameters $\tau_c^e$ and $f_e T_d$  without causing their simultaneous variation, as  is the case when varying the magnetic field or excitation power. As seen in Fig.\,\ref{LPN-OHSvsSz} this can be used to evidence and possibly measure the evolution of the relaxation time  $T_d$ in magnetic field. In \textcite{Urbaszek:2007a}, this method was employed to study the effect of temperature on the magnitude of the optically generated nuclear polarization. Figure\,\ref{LPN-OHSvsT-Xp}(a) shows that the Overhauser shift in a magnetic field $B_z=2$~T appreciably increases with the temperature in the range  30-55~K, for both   $\sigma^+$ and $\sigma^-$ excitation polarizations. This general trend could seem rather counterintuitive because both the electron and nuclear spin lifetimes decrease at higher temperatures when they are measured independently \cite{Senes:2004a,Lu:2006a}. However, the  sensitivity of the DNP  on the energy cost of electron-nuclei  flip-flops  can lead to  a specific dependence. Indeed, when the temperature is increased one may naturally expect a shortening of the correlation time $\tau_c^e$. As depicted in Fig.\,\ref{LPN-OHSvsT-Xp}(b), the competition between the broadening $\hbar/\tau_c^e$ of the levels involved in the flip-flop and the energy cost $\hbar\omega^e$ reaches a new  equilibrium  corresponding roughly to $|\omega^e\tau_c^e|\sim1$ obtained  for a larger nuclear field.\\

\indent This qualitative interpretation is confirmed by the determination of $\tau_c^e$ at different temperatures from the fitting of the model to the non-linearity of  the Overhauser shift $\delta_n=\hbar\omega^e_{OS}$ as a function of $\langle S_z^e\rangle$, see Fig.\,\ref{LPN-OHSvsT-Xp}(c). $\tau_c^e$ has to be reduced from 60~ps at 5~K to  only 10~ps at 45~K to neatly reproduce the experimental data, but interestingly this also requires to increase the relaxation time of the nuclei $T_d$ (note that in Fig.\,\ref{LPN-OHSvsT-Xp}(c), $f_e$ is taken as an independent constant). Strikingly,  the relative increase of $T_d$ is approximately the same as the relative reduction of $\tau_c^e$. This behavior is expected if the nuclear relaxation is dominated by the Knight field fluctuations produced by the electron in the QD. In this case it can be shown (see \textcite{Huang:2010a}) that $T_d^{-1}\propto f_e \tau_c^e/(1+(\tilde{\omega}^n\tau_c^e)^2)$ where $\tilde{\omega}^n$, the typical  nuclear spin-splitting  due to external field and quadrupolar interaction, is a small correction that can be neglected since $|\tilde{\omega}^n\tau_c^e|\ll1$ for $\tau_c^e\sim50$~ps.   As a result, the effect of the temperature  in the range 2-55~K is   essentially  governed by the reduction of the correlation time  $\tau_c^e$. On the one hand it produces an increase of both  relaxation times  $T_{1e}$ and $T_{d}$, equivalent to  the phenomenon of motional narrowing. However, these changes  compensate each other and therefore do not affect the magnitude of nuclear polarization. Note that  the parameter $\varsigma$ in  Eq.\,\eqref{eq:BnQD} is indeed a constant independent of $\tau_c^e$ when $T_d^{-1}\propto \tau_c^e$.  On the other hand, the  flip-flop probability is directly favored by the level broadening, which finally gives rise to higher DNP. The reduction of $\langle S^e_z\rangle$ with the temperature due to concurrent relaxation mechanisms  \cite{Senes:2004a} still limits the effective DNP magnitude that can be reached in this way.\\

\clearpage
\subsubsection{Overhauser effect in zero magnetic field : role of the Knight field }
        \label{sec:zerofield}

\begin{figure}
\epsfxsize=3.5in
\includegraphics[width=0.4 \textwidth,angle=0]{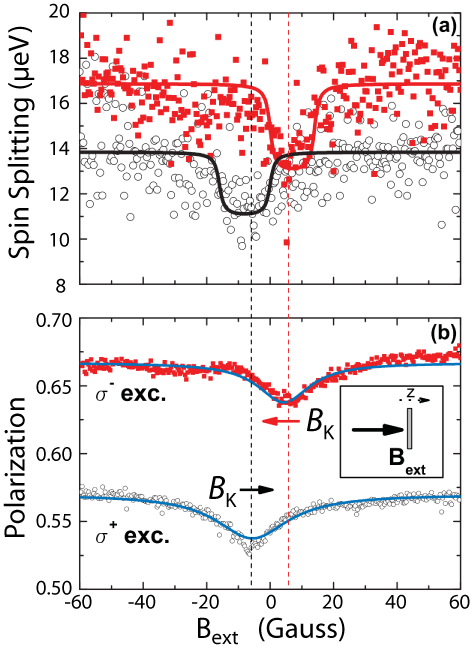}
\caption{(a)  Overhauser shift and (b) circular polarization of  a single $X^-$ PL line from an InAs/GaAs QD  as a function of a small longitudinal magnetic field for $\sigma^+$ and $\sigma^-$ circularly polarized excitation.    (after \textcite{Lai:2006a}). }\label{LPN-KnightField}
\end{figure}

A very remarkable observation in InAs QDs is the occurrence of substantial nuclear spin
polarization in zero external magnetic field \cite{Lai:2006a}. In contrast,
in n-type bulk semiconductor samples this requires a small  external  field
of a few millitesla to suppress the nuclear spin relaxation
induced by the dipolar interaction \cite{Meier:1984a}. For each
nucleus this spin-spin interaction  indeed amounts to a randomly
oriented local effective magnetic field $B_\text{L}\sim$0.15~mT with
an associated spin precession period of $T_2\sim$100~$\mu$s
determining the timescale of both the  nuclear spin relaxation and
the reorientation of $B_\text{L}$. In comparison, the intrinsic
nuclear spin decay time $T_d$ is in the 1-100~ms range, from an
analysis of the non-linearity (assuming e.g.  $f_e=0.1$) or from
direct measurements (see below Sec.\,\ref{sec:tdnp}). The fact that
$T_d\gg T_2$, is definitely a remarkable feature of InAs QDs. In
\textcite{Lai:2006a} it was proposed that the electron spin $\langle
S^e_z \rangle$  could produce a  Knight field
$\bm{B}_\text{K}\propto\langle \bm{S}^e \rangle$ (see
Eq.\,\eqref{eq:BK} in Sec.\,\ref{sec:hyperfine}), strong  enough
compared to $B_\text{L}$ to stabilize the nuclear polarization along
$z$ in zero external field. As shown in
Fig.\,\ref{LPN-KnightField}(a) for a QD charged by a single
electron,  the Overhauser shift is  almost constant  over the
$\pm$6~mT range around zero field, with still a  reduction at
$\sim\pm$0.6~mT for  $\sigma^\mp$ circular polarization of
excitation. These dips would result from  the compensation of the  Knight field by the
external field, which in turn  yields a faster
decay of nuclear spin polarization and a reduction of Overhauser
shift. They are  also observed on the circular polarization of $X^-$
PL, see Fig.\,\ref{LPN-KnightField}(b), although in this case the
polarization should solely reflect   the hole spin polarization.
This  can be interpreted as (i) the  partial inhibition by the nuclear
field of the   hole spin relaxation due to anisotropic electron-hole
exchange during  $X^-$ formation or alternatively (ii) the hole spin itself becoming more sensitive to the nuclear field fluctuations,  as discussed in section \ref{sec:holedeco}. To discuss  the Knight field
hypothesis more quantitatively,  one  can use Eq.\,\eqref{eq:BnBulk}
where the field $\bm{B}$ experienced by the nuclei is the sum
$\bm{B}=\bm{B}_\text{ext}+\bm{B}_\text{K}$, since the non-linearity
due to the field-dependence of Eq.\,\eqref{eq:Te} is negligible in
this  very narrow field range. Accordingly, the Overhauser shift
should  indeed be reduced at $\bm{B}_\text{ext}=-\bm{B}_\text{K}$.
Furthermore, the non-uniform nature of the Knight field allows  to
explain very satisfactorily both the partial reduction of the
Overhauser shift, whereas Eq.\,\eqref{eq:BnBulk} predicts complete
cancellation, and the enhancement of the corresponding  magnetic
width compared to $B_\text{L}$.

However, the exact role played by the Knight field in stabilizing
the nuclear field needs to be clarified in the light of the recently discovered importance of the nuclear quadrupole interaction introduced in Sec.\,\ref{sec:quadintro}, revealed for instance in experiments
carried out in transverse magnetic fields \cite{Krebs:2010a}: they
show that an external field perpendicular to $z$ in the 10-100~mT
range, i.e. a few orders of magnitude larger than $B_\text{L}$ and
$B_\text{K}$,  does not destroy the nuclear polarization along $z$.
In \textcite{Dzhioev:2008a} it was suggested that  the
nuclear quadrupole interaction is more likely to be responsible for allowing DNP at (near) zero applied magnetic field,
as depolarization of the nuclei via the dipole-dipole interaction is quenched. This
alternative explanation will be reviewed in more detail in
Sec.\,\ref{sec:ahanle} together with the striking observation of the
anomalous Hanle effect. Still, the assignment of the dips observed
in Fig.\,\ref{LPN-KnightField} to the Knight field compensation is the most
likely explanation. Since the polarization of the $\pm1/2$ nuclear
spin sublevels is not protected from $B_\text{L}$ by the quadrupolar
interaction, their sensitivity to any magnetic field of the order of
$B_\text{L}$ should still reflect the contribution of the Knight
field to the stabilization of the total nuclear field.

\subsubsection{Nuclear field versus electron-hole exchange : the case of neutral excitons}
\label{sec:OSwithX0}

\begin{figure}
\epsfxsize=4.5in
\includegraphics[width=0.45 \textwidth,angle=0]{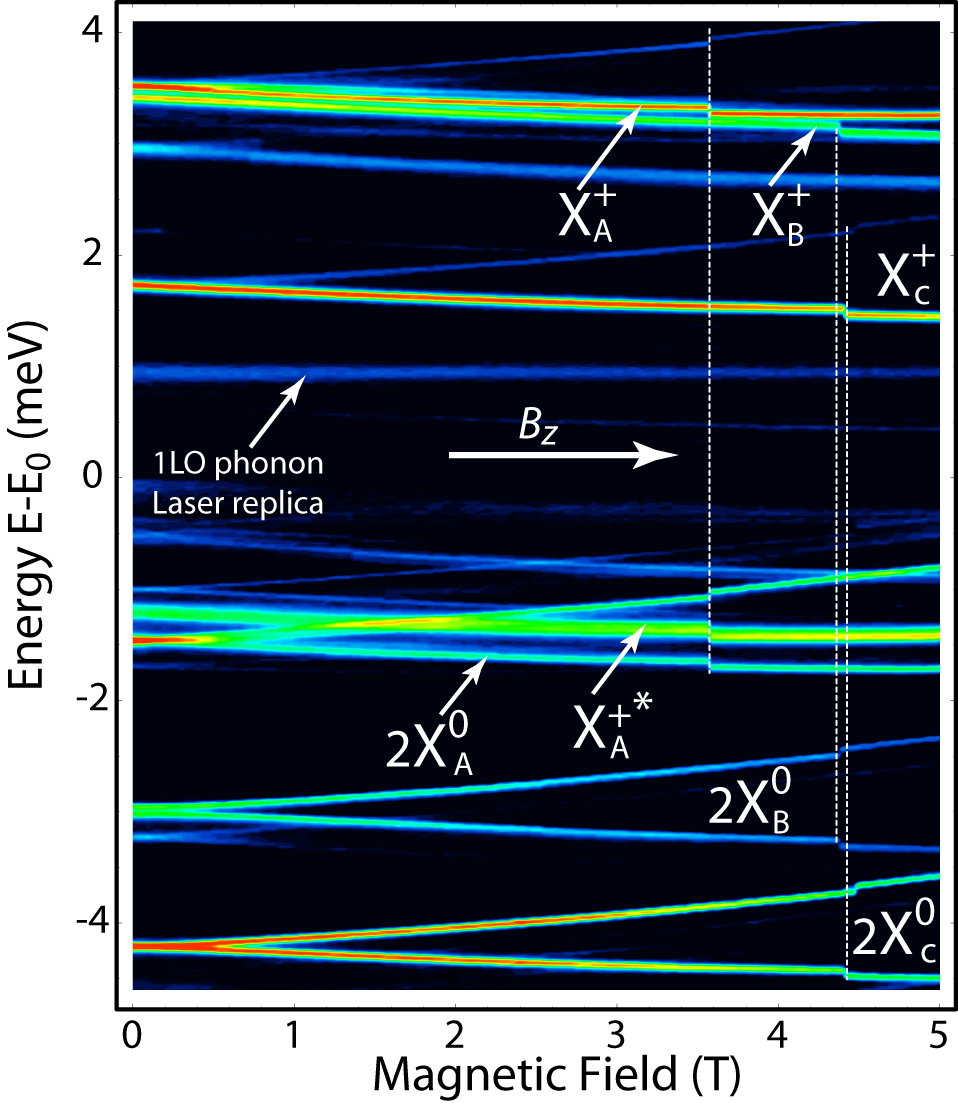}
\caption{Density plot of PL spectra of three different QDs (labelled A, B  and C) around $E_0$=1.36~eV as a function of a longitudinal magnetic field under $\sigma^-$ excitation at about 1LO phonon energy (37 meV) above $E_0$. The specific critical fields at which the nuclear polarization collapses vary from dot to dot and allow to recognize \textit{different} excitonic lines originating from the \textit{same} QD. }\label{LPN-MultiQD-DNP}
\end{figure}

The nuclear field generated in QDs  under quasi-resonant
excitation with circularly polarized light  is definitely a local
property:  on the one hand, distinct dots illuminated under the
same laser spot of about 1~$\mu$m diameter exhibit distinctive
nuclear field effects;  on the other hand all excitonic states in a
given QD which involve the same conduction electron in its
ground or excited state experience the same nuclear field.  This
property is clearly evidenced by the abrupt cancellation of the
nuclear field $B_n$ in magnetic field sweeps which  take place
accordingly at different or identical critical fields, see
Fig.\,\ref{LPN-MultiQD-DNP}. Note that the coexistence of  different
excitonic states  in the same QD in a PL spectrum results from the
random capture of  excitons and individual  charges under
quasi-resonant excitation conditions. In the measurement of
Fig.\,\ref{LPN-MultiQD-DNP} the spin-polarized positive trions $X^+$
attract a photo-created electron during their lifetimes   and
transform into  neutral  biexcitons $2X^0$ with no polarization. For
QDs inserted in a Schottky structure such effect can be favored by
tuning the gate voltage near critical values separating different
charge states.

One could expect  the coexistence of biexcitons and trions to affect
the DNP magnitude. Biexcitons indeed recombine to an exciton state
with no specific circular polarization so that the average electron
spin  driving the nuclear field in Eq.\,\eqref{eq:BnQD} should be
reduced. However, the electron-nucleus flip-flop turns out to be
largely inhibited for neutral excitons because it couples bright ($\uparrow\Downarrow$ or $\downarrow\Uparrow$) to
dark excitons ($\uparrow\Uparrow$ or $\downarrow\Downarrow$) which are separated by the electron-hole exchange
energy $\delta_0$. For InAs/GaAs QDs  where $\delta_0\simeq
0.2-0.5$~meV \cite{Urbaszek:2003a} this would correspond to a magnetic field of
about 15~T to be compensated. The possible contribution of excitons
to DNP is therefore quite negligible compared to that of trions
under usual experimental conditions. However, in interfacial GaAs
QDs for which $\delta_0\sim100\,\mu$eV the observation of
large nuclear polarization in weak fields around 100~mT  could still be assigned to flip-flops between neutral exciton states
\cite{Gammon:2001a}. The low  DNP rate $T_{1e}^{-1}\propto
\delta_0^{-2}$ estimated to 2.5~s$^{-1}$ was compensated in this
case by a vanishing nuclear spin relaxation assumed to be only due
to  dipolar coupling. More recently, neutral excitons have been
proposed to be at the origin of a strong Overhauser effect measured
for InP/InGaP QDs at \textit{low} optical excitation power
in a strong magnetic field $B_z=$6~T  \cite{Chekhovich:2011b} near
the dark-bright exciton crossing point.

\begin{figure}
\epsfxsize=3.5in
\includegraphics[width=0.45 \textwidth,angle=0]{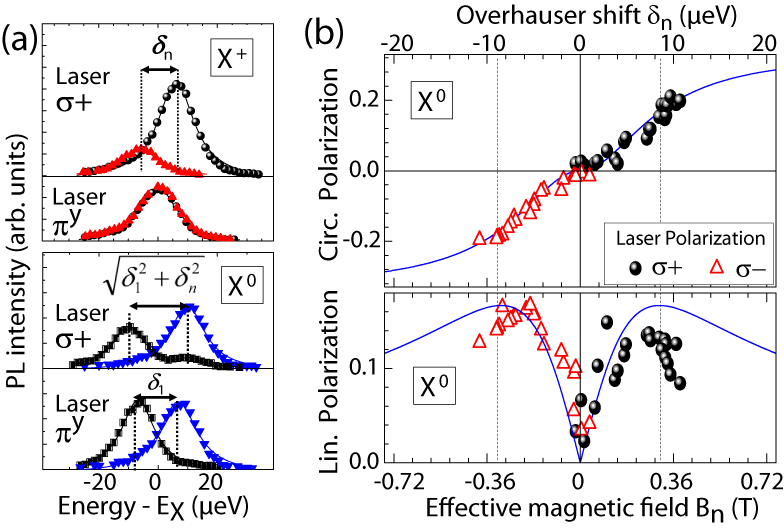}
\caption{(a) PL spectra of  $X^0$ and $X^+$ from the same InAs/GaAs QD in zero external magnetic field under the same experimental conditions, yet measured in $(\pi^x,\pi^y)$ and $(\sigma^+,\sigma^-)$ basis respectively. The same Overhauser shift $\delta_n\equiv\hbar\omega^e_{OS}$ generated under $\sigma^+$ circular polarization is evidenced for both $X^0$ and $X^+$. (b) Evolution of circular (top) and linear (bottom) polarization of $X^0$ PL as a function of the measured Overhauser shift  which increases in absolute value  when the power of the $\sigma^\pm$ excitation  increases. Solid lines follow the simple theoretical analysis given in \textcite{Belhadj:2009a}) }\label{LPN-X0-DNP}
\end{figure}

Even though neutral excitons do not necessarily contribute to DNP
they still experience the  nuclear field produced by trions
\cite{Eble:2006a}. To some extent this effect can be used to control
the intrinsic polarization of neutral excitons in zero magnetic
field \cite{Belhadj:2009a,Larsson:2011a,Moskalenko:2009a}. The
optical orientation of neutral excitons $X^0$ in InAs/GaAs QDs is
usually not possible because of the anisotropic electron-hole
exchange $\delta_1\sim 30\,\mu$eV which splits the levels into
linearly-polarized states. An external field substantially larger
than $\delta_1/g_{X^0}\mu_\text{B}\sim$150~mT would be required to
restore circularly-polarized eigenstates. As shown in
Fig.\,\ref{LPN-X0-DNP}, this can be partially achieved  when $X^+$
trions are formed alternatively   with $X^0$  excitons under
circularly-polarized non-resonant optical pumping, since the former
can generate a nuclear field $B_n$ reaching several 100~mT. Note
however that $B_n$ is the  effective magnetic field experienced by
the electron only and deduced from its Overhauser shift  according
to $\hbar\omega^e_{OS}/g_e\mu_\text{B}$. The  $X^0$  eigenstates are
actually determined by the ratio
$\eta=|\delta_1/\hbar\omega^e_{OS}|$.  When an exciton is created
with an initial circular polarization $\rho_c^0$, it experiences
quantum beats  during its lifetime $\tau_r$ which, if $\tau_r$ is
long enough, results in an average circular polarization
$\rho_c^0(1+\eta^2)^{-1}$ and  a linear polarization
$\rho_c^0\eta(1+\eta^2)^{-1}$ of the emitted
PL~\cite{Dzhioev:1998a}. This  effect is demonstrated in
Fig.\,\ref{LPN-X0-DNP}(b), with notably a substantial conversion of
the initial circular polarization into  a linear polarization when
the magnitude of the Overhauser shift increases with the excitation
power and becomes comparable to $\delta_1$.

{\subsection{Nuclear spin polarization under resonant optical excitation}
        \label{sec:beyond}

To investigate the coupled electron-nuclear spin dynamics
under resonant excitation sophisticated  pump probe
techniques~\cite{Greilich:2007a,Chekhovich:2010a}, photocurrent
experiments~\cite{Klotz:2010a} or schemes with spectator
states~\cite{Kloeffel:2011a} have been developed. These experiments
have uncovered original nuclear pumping cycles, not accessible
under non-resonant excitation conditions. In Sec.\,\ref{sec:OHSres}
an efficient electron spin initialization scheme is presented, that
results in DNP via the Overhauser effect. In
Sec.\,\ref{sec:forbidden} we will describe how a resonant laser can
drive an optically forbidden transition, where photon absorption is
accompanied by a simultaneous spin flip of an electron and a nuclear
spin. The DNP  achieved in these original schemes is based on the
flip-flop term  $\propto (\hat{I}_+\hat{S}_-^e+\hat{I}_-\hat{S}_+^e) $
in Eq.\,\eqref{eq:eqHf1}. In Sec.\,\ref{sec:locking} resonant
experiments will be presented that involve nuclear polarization
build-up that does \textbf{not} require the simultaneous spin-flip
of a carrier spin, i.e. that involves non-collinear terms of the form
$\propto \hat{I}_x\hat{S}^e_z$, paving the way for novel nuclear
spin control schemes.
\subsubsection{Overhauser effects under resonant optical excitation}
        \label{sec:OHSres}

An original cycle for efficient DNP generation  through resonant
optical pumping has been developed by \textcite{Kloeffel:2011a} in
small applied longitudinal fields $B_z=0.5$~T. In this scheme the
driving laser polarization and more surprisingly it's
frequency will define which electron spin state ($\uparrow$
or $\downarrow$) is initialised,  determining the direction of the
nuclear spin polarization created via the Overhauser effect. Here an
InAs QD is placed in a charge tunable structure with a highly n-doped layer, separated from
the dot by a 25nm thick GaAs tunnel barrier. In the experiment the
bias voltage is set just between the X$^0$ and X$^{-}$ plateaus. The
efficient cycle for electron spin initialisation, see
Fig.\,\ref{Figresonant2}, starts with a dot that is unoccupied
before laser excitation, as the first electronic state is just above
the Fermi sea of electrons. In the experiment the
laser polarisation is set to $\sigma^+$. Absorption of a photon
leads to the formation of neutral exciton X$^0$ state
$\left|\Uparrow\downarrow\right>$. The Coulomb attraction of the
hole lowers the electronic level just below the Fermi sea, so that after
a time $\tau_\text{in}$ of typically tens of picoseconds an electron tunnels into
the dot. Pauli exclusion determines the incoming electron to be in
the $\left|\uparrow\right>$ state and the charged exciton X$^-$
state $\left|\Uparrow\downarrow\uparrow\right>$ is formed. The X$^-$
does not polarize the nuclei efficiently since the two electrons
form a singlet and the coupling of the unpaired hole to nuclear
spins is much weaker than for a single electron (see
Sec.\,\ref{sec:hole}) \footnote{It is experimentally confirmed that the Ising like term $\propto \hat{I}_z \hat{S}_z$ is one order of magnitude weaker for heavy holes than for conduction electrons. In addition, as discussed in Sec.\,\ref{sec:hole}, the interaction of hole spins with nuclei can be strongly anisotropic, see Eq.\,\ref{dipprecise} and discussion in \cite{Fischer:2008a,Testelin:2009a}. As a result, the amplitude of the terms responsible for hole spin - nuclear spin flip flops strongly depend on the degree of valence band mixing and can therefore be several orders of magnitude weaker than for electrons and eventually tend to zero for pure heavy hole states.}. 
This charged exciton exists for its
radiative lifetime $\tau_r \approx~1$ns, leaving behind a single
electron after photon emission. The intensity of the X$^-$ emission
is a measure for the X$^0$ absorption \cite{Simon:2011a}. The
unpaired electron $\left|\uparrow\right>$ left behind interacts with
the nuclear spin system. But this interaction is limited in time: in
the absence of the hole the electron energy is above the Fermi sea,
therefore the electron will tunnel out of the dot within $\tau_{\rm
out}$ of tens of picoseconds. This short tunneling time (i.e. broadened
Zeeman levels) limits the correlation time for the electron-nuclear
spin interaction, allowing flip-flops between electron and nuclear
spins without violating energy conservation. Once the electron
has tunneled out, the dot is  again unoccupied and another
$\sigma^+$ photon resonant with the X$^0$ state can be absorbed. DNP
builds up by going repeatedly around this cycle.

\begin{figure}
\epsfxsize=3.4in
\epsfbox{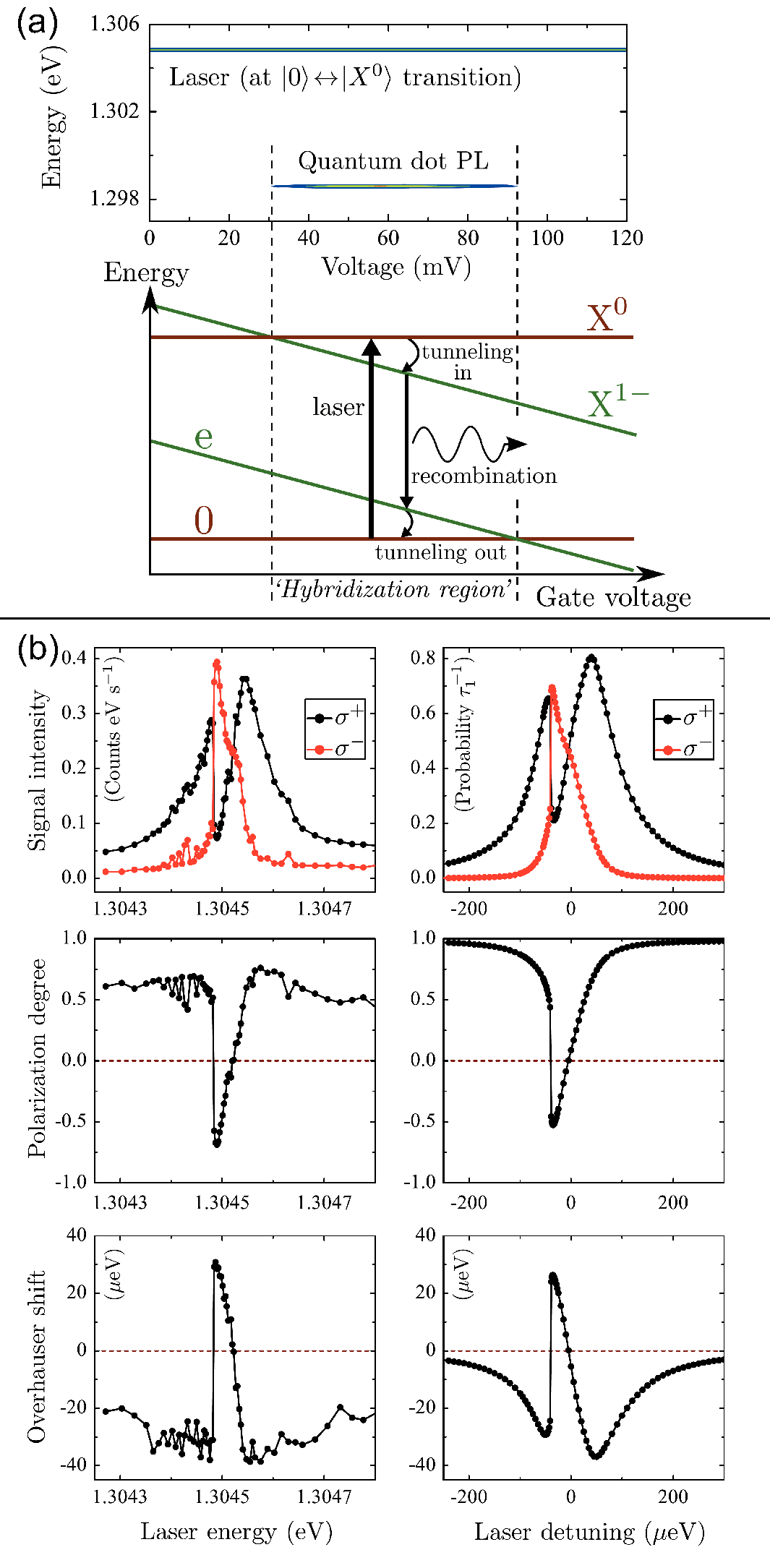}
\caption{{(a) Top: Photoluminescence (PL) at 4.2~K from a single QD versus applied bias driven with excitation at X$^0$ energy. X$^-$ PL appears in a narrow range of voltage. Bottom: Energy dependence versus bias for the QD vacuum state $|0\rangle$ and the single electron state $|e\rangle$, showing a crossing where the ground state changes. X$^0$ and X$^-$ cross at lower bias on account of the hole (Coulomb interaction). For the chosen bias region (\textit{Hybridization region}), automatic cycling takes place when a laser is tuned to the $|0\rangle\leftrightarrow | X^0\rangle$ transition.  (b) Left: Experimentally measured signal intensity, polarization degree and Overhauser shift versus laser energy (detuning) for a $\sigma^+$ pump and an external field of +0.5~T at fixed bias in the center of the hybridization region. Right: Comparison with theory from \textcite{Kloeffel:2011a}}
 \label{Figresonant2}}
\end{figure}

For a perfectly pure X$^0$ state $\left|\Uparrow\downarrow\right>$
one would  expect one sharp resonance in absorption. But very
surprisingly there is a second resonance in the absorption curve,
see Fig.\,\ref{Figresonant2}(b). This resonance corresponds to the
energy of the other X$^0$ bright state
$\left|\Downarrow\uparrow\right>$, separated from the
$\left|\Uparrow\downarrow\right>$ exciton, by an energy of
$\approx 40~\mu eV$. Both exciton states are coupled by the
anisotropic part of the Coulomb interaction, making the
$\left|0\right>\longrightarrow\left|\Downarrow\uparrow\right>$
transition weakly allowed under $\sigma^+$ excitation. Once
$\left|\Downarrow\uparrow\right>$ is created, the dot will receive
an additional electron from the Fermi sea, that has to be
$\left|\downarrow\right>$ due to Pauli exclusion, and the X$^-$
state $\left|\Downarrow\uparrow\downarrow\right>$ is formed.
Radiative recombination leads to $\sigma^-$ polarized emission, as
shown in the top part of Fig.\,\ref{Figresonant2}(b), and most
importantly the electron spin is now $\left|\downarrow\right>$
interacting with the nuclei, leading to nuclear spin polarization
that has the \textit{opposite} direction as compared to the previous
case. Therefore, the sign of the electron polarization and
hence the nuclear polarization can be tuned finely by varying the
excitation laser energy \cite{Kloeffel:2011a}, as shown in
Fig.\,\ref{Figresonant2}(b).  The DNP generation as a function of
detuning shows strong non-linearities as the transition energies
themselves change with each nuclear spin flip, driving the
transitions towards or away from the laser energy as described in
detail in the model developed by \textcite{Kloeffel:2011a}. Because
of this feedback, the QD transition can be 'pushed' or 'pulled' by
changing the laser frequency, a phenomenon even more pronounced in
the experiments described in the Sec.\,\ref{sec:locking}. The
presented resonant excitation cycle is an efficient way of creating
a net electron spin polarization that is subsequently transferred to
the nuclear spins via the Overhauser effect. The fact that the
carrier excitation is resonant has no direct consequences for the
DNP build-up process itself (here based on the standard Overhauser effect), in contrast to the experiments
described below.

\subsubsection{Pumping of Nuclear Spins by Optical Excitation of Spin-Forbidden Transitions}
        \label{sec:forbidden}

Experiments carried out on the X$^+$ transitions in InP dots in a
longitudinal magnetic field of 2.5T~\cite{Chekhovich:2010a} are
shown in Fig.\,\ref{Figresonant1}, where the nuclear spin
polarization is measured in non-resonant PL (probe) as a narrow excitation laser is swept
across the X$^+$ transitions (pump). In these experiments photon absorption
is accompanied by an electron-nuclear spin flip-flop. There is a
very important difference for the energy balance of the spin
flip-flop as compared to non-resonant excitation: the energy for the
electron spin flip is now directly provided by the excitation laser,
in close analogy to the  \textit{solid state effect} discussed by
\textcite{Abragam:1961a} and more recently in the context of QDs by
\textcite{Bracker:2008a} and \textcite{Korenev:2007a}.

Assuming initially 100\% pure optical selection rules, the hole
$\left|\Downarrow\right>$ state absorbs a $\sigma^+$ photon to
become $\left|\Downarrow\Uparrow\downarrow\right>$, the hole
$\left|\Uparrow\right>$ state absorbs a $\sigma^-$ photon to become
$\left|\Downarrow\Uparrow\uparrow\right>$ and as the electron and
hole g-factors are unequal ($|g_h|>|g_e|$) the energies of the
$\sigma^+$ and $\sigma^-$ transitions are different. In this case
one expects when scanning a $\sigma^+$ polarized laser across the
Zeeman branches one absorption line and no DNP, as no carrier spin
flip takes place. This is in stark contrast to the reported
experiments~\cite{Chekhovich:2010a}: Driving the dot with $\sigma^+$
laser light results as expected in absorption when in resonance with
the $\left|\Downarrow \right>\longrightarrow
\left|\Downarrow\Uparrow\downarrow\right>$ transition but, and this
is the first surprise, there is a second resonance at higher energy
for the normally spin forbidden transition $\left|\Downarrow\right>
\longrightarrow \left|\Downarrow\Uparrow\uparrow\right>$. The second
surprise is that DNP is created for both resonances, with the
normally spin-forbidden absorption scheme polarizing nuclear spins
even more efficiently.

\begin{figure}
\epsfxsize=3.5in
\epsfbox{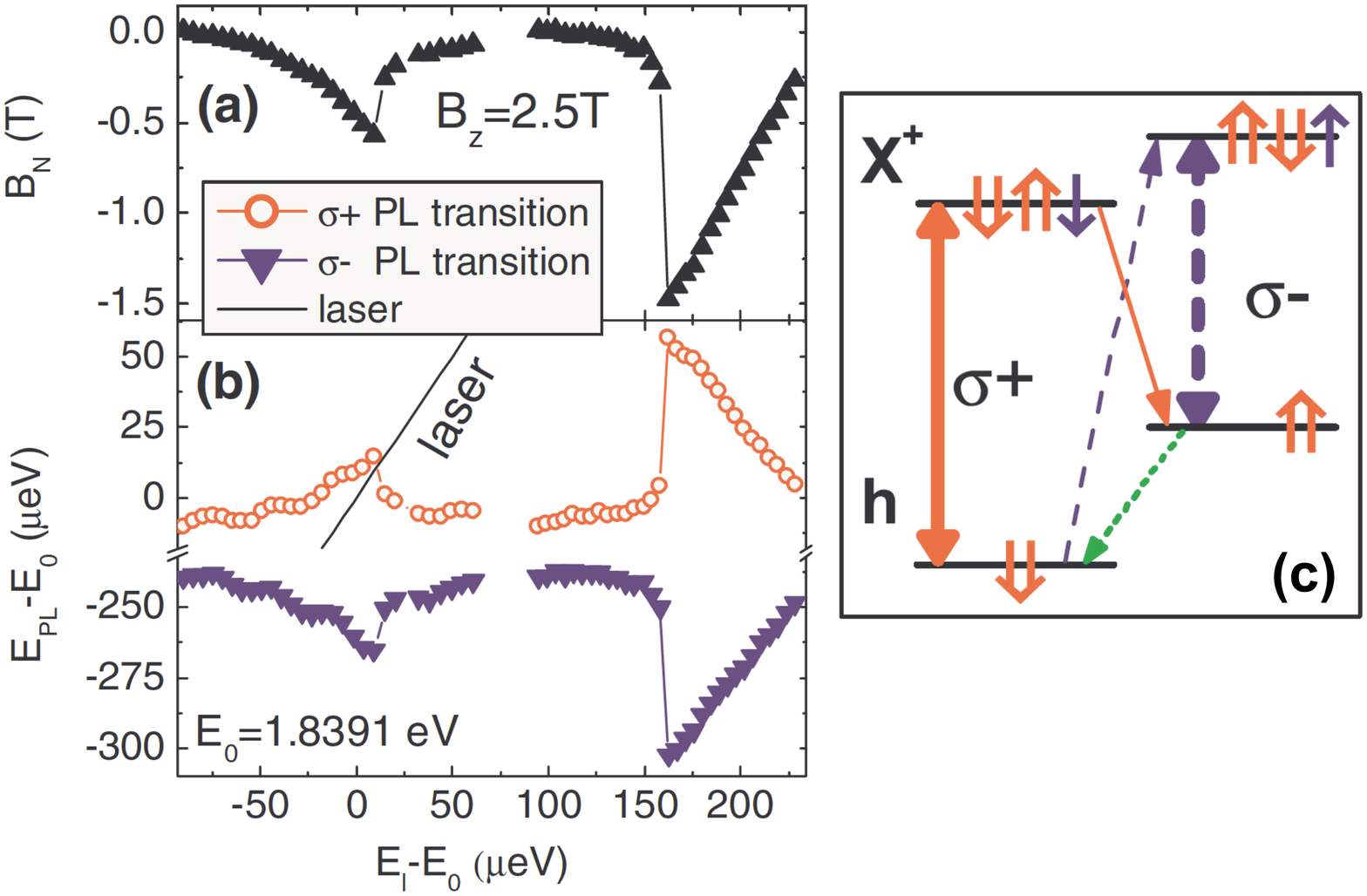}
\caption{{(a) Overhauser field $B_N$ in the dot for B$_z$=2.5T as a function of laser energy (b) PL transition energy as a function of laser energy (c) Energy level diagram of a positively charged dot in a magnetic field B$_z$. Long thick arrows show "allowed", thin arrows "forbidden" optical transition \cite{Chekhovich:2010a}.}
 \label{Figresonant1}}
\end{figure}

Optically assisted nuclear spin-flips can occur either in the
absorption or the ensuing spontaneous emission. The corresponding
absorption/emission cycles, as depicted in Fig.\,\ref{Figresonant1}(c) are:

(1) The $\sigma^+$ laser energy is tuned to the nominally forbidden
$\left|\Downarrow\right> \longrightarrow
\left|\Downarrow\Uparrow\uparrow\right>$ transition that is
difficult to saturate due to weak oscillator strength. In a second
order process, $\sigma^+$ photon absorption assisted by a flip of
the electron spin and a nuclear spin results in
$\left|\Downarrow\Uparrow\uparrow\right>$ formation. Subsequently
the spin allowed optical recombination
$\left|\Downarrow\Uparrow\uparrow\right>\longrightarrow
\left|\Uparrow\right>$  results in $\sigma^-$ photon emission. For
the pumping cycle to start again, the hole spin has to relax from
$\left|\Uparrow\right>\longrightarrow\left|\Downarrow\right>$, a
process that is fast enough for this dot system in strong magnetic
fields not to limit the efficiency of the spin pumping cycle
\cite{Chekhovich:2010a}.

(2) Starting again from $\left|\Downarrow\right>$ but tuning the
laser to the spin allowed $\left|\Downarrow \right>\longrightarrow
\left|\Downarrow\Uparrow\downarrow\right>$ transition results in
strong and easily saturable optical absorption. Re-emission of a
$\sigma^+$ photon at the same energy back to
$\left|\Downarrow\right>$ is the most likely course of events. But
due to the hyperfine interaction, the transition
$\left|\Downarrow\Uparrow\downarrow\right>\longrightarrow\left|\Uparrow\right>$
becomes weakly allowed, when assisted by a electron-nuclear spin
flip-flop. The final step of this cycle relies again on hole spin
relaxation
$\left|\Uparrow\right>\longrightarrow\left|\Downarrow\right>$.
Completing either one of these two cycles once results in lowering
the nuclear spin $z$ projection by $1$. To maximize the generated
DNP, the rates characterizing the optical absorption, emission and
the resident hole spin flip can be optimized by changing the applied
field B$_z$ and/or the laser intensity. Cycle (2) starting with an
allowed optical transition saturates already at low pumping power,
whereas for cycle (1) the absorption assisted by the hyperfine
interaction continues to grow in strength when increasing the laser
power. This results in more efficient nuclear spin pumping at high
laser powers using the cycle (1), as confirmed by the experiments as
well as the theoretical model detailed in
\textcite{Chekhovich:2010a}.

 \subsubsection{Locking of quantum-dot resonances to an incident laser}
        \label{sec:locking}
Several groups have
observed that the textbook-like Lorentzian absorption
line shape that one expects as a spectrally narrow single-mode laser field is
scanned across the QD resonance gets strongly modified at magnetic
fields $B_z$ exceeding $1$~Tesla. As the laser is tuned within
$\sim 2$ linewidths ($\Delta \nu$) of the QD resonance, the
absorption abruptly turns on. When the laser is further scanned
across the resonance (in either direction), the absorption
strength remains close to its maximum value for a laser detuning that
can exceed $\pm 10 \Delta \nu$, see Fig.\,\ref{figure:dragging:theory+exp}.

\begin{figure}[h] \centering
\includegraphics*[width=0.40 \textwidth,angle=0]{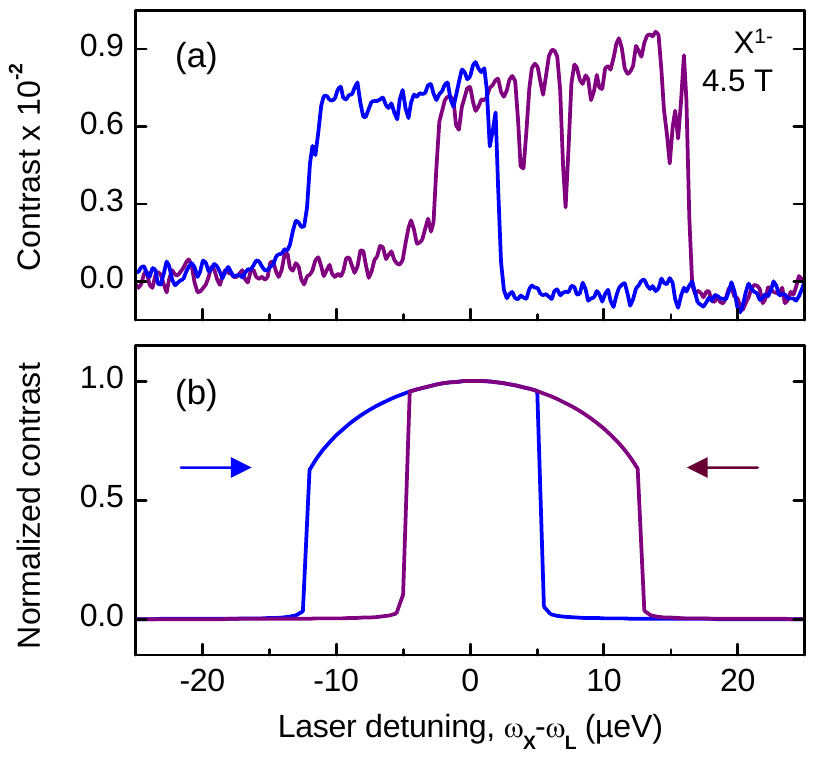}
\caption{Comparison of the dragging spectra of the blue-shifted
Zeeman branch for the X$^-$ trion as a function of laser detuning
for a sample with 25~nm thick tunnel barrier (a) and the
prediction of the rate equation \ref{eq:ratedrag} (b) for the following parameters:
$N = 3.2\cdot 10^4$, where $N$ is the number of nuclear spins,
$\tilde{A}^i= 120~\mu\mathrm{eV}/N$, $B_z = 4.5$~T,
$A_\mathrm{nc}^i=0.012\tilde{A}^i$, $\hbar\Gamma= 0.58~\mu$eV,
$\Omega_\mathrm{L}=0.7\,\Gamma$, step size $\Delta\delta =
0.23~\mu$eV and dwell time
$t_c=0.2$~s.(after \textcite{Hogele:2012a}).}\label{figure:dragging:theory+exp}
\end{figure}

\textcite{Latta:2009a} and \textcite{Hogele:2012a} have carried
out a detailed experimental study of this so-called {\sl dragging
effect} in charge tunable InAs/GaAs QDs \cite{Warburton:2000a}. The dependence on the laser intensity, laser
(or gate voltage) scan-speed and the electron spin-relaxation
(co-tunneling) rate have been mapped out. Their findings demonstrate that dragging is a
consequence of nuclear spin polarization that enables locking of
the QD resonance to the incident probe laser field frequency.
Remarkably, for any given (linear or circular) laser polarization,
nuclear spin polarization is bidirectional, allowing the combined
electron-nuclear spin system to track the changes in laser
frequency dynamically on both sides of the resonance. This latter
observation suggests that the underlying mechanism is \textit{not} related
to optical pumping of QD electron spin, that is used
to explain quasi-resonant DNP experiments for which the flip-flop term of the form $\propto (\hat{I}_+\hat{S}_-^e+\hat{I}_-\hat{S}_+^e)$ in $\hat{H}_\text{hf}^\text{fc}$ from Eq.\,\ref{eq:eqHf} dominates.

Experiments carried out on different charge-tunable structures
show that any QD transition that has an electron spin decay rate
$\lesssim 10^7 ~ s^{-1}$ will exhibit some degree of dragging. In
particular it has been shown that X$^0$, X$^-$ and X$^+$
transitions in self-assembled QD samples grown in different
laboratories under different conditions, exhibit dragging both in
Faraday and Voigt geometries. There is however, a striking
difference in the dragging profile of the blue-shifted and the
red-shifted Zeeman transition of a neutral (or singly charged) QD:
in contrast to the blue-Zeeman transition which exhibits the
symmetric flat-top response with maximal absorption described
earlier (shown in Fig.\,\ref{figure:dragging:theory+exp} and in
Fig.\,\ref{introhyper}(d)), nuclear spin polarization in the red
Zeeman transition ensures that the QD resonance is pushed away
from the incident laser frequency. In addition, the line shape is
asymmetric with respect to the laser (or gate voltage) scan
direction. Before we present an explanation of these features, we
highlight the general characteristics of dragging, determined by
examining the blue trion transition.

Figure\,\ref{FigDragPlateau}(a) shows the two-dimensional (2D) map
of resonant absorption (resonance fluorescence in
Fig.\,\ref{FigDragPlateau}(b)) of the blue-shifted trion
transition as a function of laser frequency and gate voltage
$V_{\rm{g}}$ for two different sample structures exhibiting
radically different ranges of the co-tunneling rate $\kappa$ (see section \,\ref{sec:orientation} for an introduction to co-tunneling). For
a QD that is separated from the Fermi sea by a 25~nm GaAs barrier
(Fig.\,\ref{FigDragPlateau}(a)), the co-tunneling rate ranges from
$> 10^{9} ~s^{-1}$ at the edge of the charging plateau to $\sim
10^6 ~s^{-1}$ in the center. Consequently, we observe that the
bi-directional dragging effect is strongest in the plateau center
and is completely suppressed at the edges ($V_{\rm{g}} \sim
160$~mV and $\sim 260$~mV in Fig.\,\ref{FigDragPlateau}(a), and
$V_{\rm{g}} \sim 220$~mV and $\sim 320$~mV in
Fig.\,\ref{FigDragPlateau}(b)), where the electron spin
orientation is completely randomized due to the spin flip assisted
co-tunneling introduced in Sec.\,\ref{sec:orientation}. In
Fig.\,\ref{FigDragPlateau}(a), each horizontal cut is obtained by
scanning the gate voltage for a fixed laser frequency; red (blue)
bars show data obtained by scanning the gate voltage such that the
detuning $\Delta \omega=\omega_{\mathrm{X}}-\omega_{\mathrm{L}}$
decreases (increases), where $\omega_{\mathrm{X}}$ is the
frequency of the QD transition and $\omega_{\mathrm{L}}$ is the
laser frequency. Figure\,\ref{FigDragPlateau}(b) on the other
hand, shows the resonance fluorescence map, obtained by scanning
the single-mode laser frequency at a fixed gate voltage, for a QD
separated from the Fermi sea by a 35~nm tunnel barrier. Due to
strong electron spin pumping in the center, dragging in this case
is restricted to the edges of the single-electron charging
plateau. The overall range for dragging is $\sim 30~\mu$eV for
$B_{z} = 5$~T, which has to be compared with the theoretical
absorption linewidth in the radiative limit $\hbar/\tau_r \approx
1~\mu$eV. The range of dragging depends on the laser Rabi
frequency $\Omega_{\rm{L}}$; while some degree of dragging is
observed for $\Omega_{\rm{L}}$ ranging from $\sim 0.3\, \Delta\nu$
to $3\, \Delta \nu$, the maximum width is obtained at saturation
intensity.

\begin{figure}[h] \centering
\includegraphics*[width=0.40 \textwidth,angle=0]{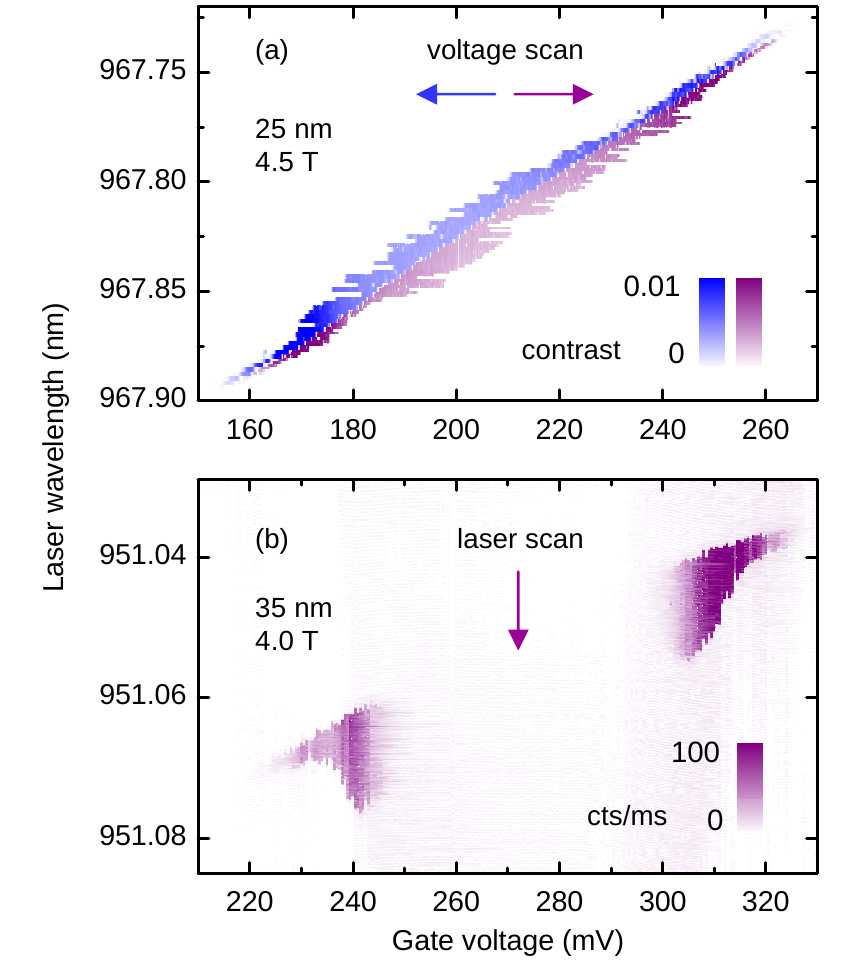}
\caption{Differential transmission as a function of gate voltage
showing a dragging plateau for a charge tunable sample with a
tunnel barrier thickness of  (a) 25 nm at $B_z=4.5$~T and (b) as
(a), but measured in resonance fluorescence for tunnel barrier
thickness 35~nm at $B_z=4.0$~T sample.}\label{FigDragPlateau}
\end{figure}

Experiments carried out on QDs in all samples showed that the
dragging width (in energy) increases sub-linearly with $B_{z}$
beginning at $\sim 1$~T \cite{Hogele:2012a}. This result is at
first sight unexpected, given that the maximum Overhauser field
that builds up as a result of dragging is a factor of 5 smaller
than the external field: therefore, a compensation of the electron
Zeeman splitting due to $B_z$ by the built-in Overhauser field
$B_n$ is not taking place. If the mechanism for DNP was based on
Fermi-contact hyperfine interaction $\hat{H}_{hf}^{fc}$ of
Eq.\,\ref{eq:eqHf}, we would have expected the polarization rate
at high $B_z$ to be suppressed due to energy non-conservation of the 
hyperfine induced spin flip-flop processes. We emphasize here that in contrast
to the quasi-resonant experiments, the electron spin coherence
time in resonant experiments is an order of magnitude longer.
Since the exchange interaction with the Fermi reservoir needs to
be suppressed to observe dragging, electron-nuclear flip-flop
processes accompanied by a co-tunneling event only lead to weak
directional Overhauser effect. Similar arguments apply to
spin-flip spontaneous Raman scattering leading to reverse
Overhauser effect \cite{Latta:2009a}.

The experimental observations therefore suggest that DNP based on
hyperfine flip-flop interaction and electron spin-pumping are
unlikely to explain the dragging effect. Noting this difficulty,
Yang and Sham \cite{Yang:2010a} proposed that the presence of a
small non-collinear hyperfine coupling of the form given in Eq.\,\ref{eq:nc-hyperfine} between the heavy-hole and
the nuclear spins would lead to symmetric bi-directional nuclear
spin polarization. Below it is argued that non-collinear hyperfine coupling is a key ingredient to explain dragging.
 The strength of the heavy-hole to nuclear spin
coupling does rely on heavy-hole light-hole mixing, see also
Sec.\,\ref{sec:hole} of this review. In the sample material used
by \textcite{Latta:2009a} and \textcite{Hogele:2012a}, however,
experimentally determined values of hole-hyperfine interaction
would suggest that the resulting nuclear spin-flip rates would be
an order of magnitude too small to explain the relative
insignificance of the directional reverse Overhauser effect. In
addition, the large apparent variation in heavy-light-hole mixing
in QDs as indicated by the measured in-plane hole $g$-factors
would suggest that dragging features would change appreciably from
QD to QD, if the mechanism was due to hole-hyperfine interaction,
described in detail in Sec.\,\ref{sec:hole}. This is in
contradiction with the experimental observations. Moreover, recent
experiments \cite{Latta:2011a} described in Sec.\,\ref{sec:DecmK}
demonstrate that non-collinear hyperfine interaction  between the
electron and the nuclei plays a significant role in determining QD
nuclear spin dynamics even in the absence of optically generated
holes. As an alternative origin to the proposed coupling of holes
to nuclei, this important non-collinear interaction with electrons
can be induced by large quadrupolar fields in strained
self-assembled QDs (introduced in Sec.\,\ref{sec:quadintro}) which
ensure that nuclear spin projection along $B_z$ is not a good
quantum number. We recall that an effective non-collinear
interaction between the electron in the dot and the nuclei of the
simple form
\begin{equation}
    \hat{H}_{hf}^{nc} =\sum_i A_\mathrm{nc}^i \hat{I}_{x}^i \hat{S}_{z}^e
    \tag{\ref{eq:nc-hyperfine}} \;
\end{equation}
for highly strained self-assembled QDs was introduced by Huang and
Hu \cite{Huang:2010a}. Here, $A_\mathrm{nc}^i=A^\mathrm{nc}/N$,
and $A^\mathrm{nc}$ is the non-collinear counterpart of
$2\tilde{A}$ in $\hat{H}_{hf}^{fc}$ of Eq.\,\eqref{eq:eqHf1}.

Our next target is to understand the physical origin of this effective non-collinear coupling.
Let us recall (see section \ref{sec:quadintro}) that the quadrupolar
interaction Hamiltonian for a nuclear spin with strain axis (i.e. the principal axis of the electric field gradient) tilted by an angle $\theta$
from the $z$-axis in the $xz$ plane in the limit of small angles (compare with Eq.\,\ref{eqn:HQlong}), can be written as:

\begin{eqnarray}
\hat{H}_\text{Q} &=& B_Q [\hat{I}_{z}^2 \cos^2\theta -\frac{I(I+1)}{3} \nonumber \\
&+& (\hat{I}_{z}\hat{I}_{x}+\hat{I}_{x}\hat{I}_{z})\sin \theta \cos \theta]
\label{eqn:quadrupolar} \;.
\end{eqnarray}


\noindent The first two terms are very small corrections to the
energy and will not be considered in the following. We focus our
attention on the term $\propto
(\hat{I}_{z}\hat{I}_{x}+\hat{I}_{x}\hat{I}_{z})$,
 which can be treated as a perturbation to the nuclear Zeeman energy under the experimental conditions
 $B_z=5~T\gg B_Q$, and apply a Schrieffer-Wolff (SW) transformation \cite{Schrieffer:1966a}.
When this transformation is applied to the $\sum_i A_i S_z I_z$ term
of the Fermi-contact hyperfine interaction Hamiltonian
$\hat{H}_{hf}^{fc}$, we obtain
\begin{eqnarray}
    \hat{H}_{hf-quad} &=& \sum_{i} \tilde{A}_\mathrm{nc}^i
    \hat{S}_{z}^e [\hat{I}_{x}^i \hat{I}_{z}^i + \hat{I}_{z}^i \hat{I}_{x}^i]
    \label{eqn:quadrupolar-hyperfine} \;,
\end{eqnarray}
with $\tilde{A}_\mathrm{nc}^i =
A^i\frac{B_Q^i}{\omega_Z^n}\sin 2\theta^i$. In
$\hat{H}_{hf-quad}$ we have only kept the terms that describe
processes which leave the electron spin-state unchanged, since
contributions that flip the electron spin will have a negligible
contribution at high external fields $B_z$ as they are
energetically forbidden.

The transformed hyperfine interaction Hamiltonian reads
$\hat{H}_{hf} = \hat{H}_{hf}^{fc} + \hat{H}_{hf-quad}$. A
calculation of $B_Q^i$ and $\theta_i$ for In$_{0.7}$Ga$_{0.3}$As
QDs has been carried out where atomistic strain and nuclear
quadrupolar distributions are extracted over a relaxed structure
\cite{Hogele:2012a}. These calculations yield an average value
$\tilde{A}_\mathrm{nc}^i \simeq 0.07 \tilde{A}^i$. Anionic As
nuclei dominate the non-collinear hyperfine coupling term by
contributing $75 \%$ to the average value; the residual $25\%$ are
due to the cations, where the contribution of In nuclei due to
their large nuclear spin of $9/2$ is a factor of 10 larger than
that of Ga nuclei.

We emphasize that $\hat{H}_{hf-quad}$ differs from the
non-collinear hyperfine term of Eq.\,\eqref{eq:nc-hyperfine} we
introduced earlier, since it does not allow for a coupling between
nuclear spin states with positive and negative spin-projection
along the $z$-axis (the difference compared to the $\hat{I}_x$
operator is that $\hat{I}_x \hat{I}_z+\hat{I}_z \hat{I}_x$ does
not couple the $\pm 1/2$ states to each other). It could be argued
however that even for large $B_z$, the dominant role of flip-flop
terms of Fermi-contact hyperfine interaction is to induce indirect
interaction between the QD nuclei \cite{Latta:2011a} as introduced
in Eq.\,\ref{eq:hfind}: the primary effect of this interaction, in
the presence of fast optical dephasing of the electronic spin
resonance, is to ensure that the nuclear spin population assumes a
thermal distribution on timescales fast compared to the
polarization timescale determined by $A_\mathrm{nc}^i$. In this
limit, the dynamics due to $\hat{H}_{hyp-quad}$ in
Eq.\,\eqref{eqn:quadrupolar-hyperfine} will be indistinguishable
from that described by $\hat{H}_{nc}$ in
Eq.\,\eqref{eq:nc-hyperfine}. Aiming at this stage for a
qualitative understanding of the dragging mechanism further
analysis is based on the reasonable assumption that the
quadrupolar fields in highly strained QDs can give rise to
non-collinear hyperfine interaction with effective dynamics of the
general form $\hat{I}_{x}^i \hat{S}_{z}^e$ provided we take
$\tilde{A}_\mathrm{nc}^i = A_\mathrm{nc}^i$.

The fact that $\hat{H}_{hf}^{nc}$ as defined in
Eq.\,\eqref{eq:nc-hyperfine} could explain dragging is at first
sight surprising since its dominant effect appears to be nuclear
spin diffusion. However, a careful inspection shows that the same
Hamiltonian also leads to a small polarization term whose
direction is determined by the sign of the optical detuning. To
explain this, we focus on X$^-$ and simplify the physical system
by considering only the blue-shifted trion transition. For this
system, we can write down an effective Hamiltonian:
\begin{eqnarray}
\hat{H}_{drag} &=&  \Delta \omega_\mathrm{L} \hat{\sigma}_{tt} + \omega_\mathrm{Z}^\mathrm{n} \hat{I}_{z} + \sum_i{2\tilde{A}^i\hat{I}_{z}^i\hat{S}_{z}^e} \nonumber\\
        &+& \Omega_\mathrm{L} (\hat{\sigma}_{t\uparrow}  +  \hat{\sigma}_{\uparrow t}) \nonumber\\
        &+& \sum_i A_\mathrm{nc}^i \hat{S}_{z}^e \hat{I}_x^i.
        \label{eqn:dragging}
\end{eqnarray}
with $\hat{\sigma}_{eg} = |e\rangle \langle g|$. Here,
$\left|t\right>$ and $\left|\uparrow\right>$ denote the excited
trion state and the single-electron spin ground states of X$^-$;
$\Delta \omega_\mathrm{L}=\omega_t-\omega_\mathrm{L}$ is the
optical detuning with respect to the bare trion resonance at
$\omega_t$ and $\Omega_\mathrm{L}$ is the Rabi frequency of the
laser field. The complete system dynamics is given by a master
equation for the reduced density operator $\rho$ involving
coherent dynamics due to $\hat{H}_{drag}$ and a dissipative
evolution given by the Liouvillian $\mathcal{L}(\rho) = 0.5 \Gamma
(2 \hat{\sigma}_{\uparrow t} \rho \hat{\sigma}_{t \uparrow} -
\hat{\sigma}_{t t} \rho - \rho \hat{\sigma}_{t t})$, with $\Gamma$
denoting the spontaneous emission rate of the trion state.

In Eq.\,\eqref{eqn:dragging}, we have neglected the flip-flop terms
of the Fermi-contact hyperfine interaction. To justify this
assumption, we consider the limit of a large external magnetic field
where $\omega_\mathrm{Z}^\mathrm{e} \gg \Omega_\mathrm{L} \sim
\Gamma$. If we in addition assume $\omega_\mathrm{Z}^\mathrm{n} \gg
\Omega_\mathrm{L} \sim \Gamma$, the nuclear spin-flip processes
described by the last term in Eq.\,\eqref{eqn:dragging} are
energetically forbidden to first order in perturbation theory.
Eliminating these terms by a SW transformation we arrive at the
following correction terms due to the non-collinear hyperfine interaction to the laser-trion coupling
\begin{equation}
\hat{H}_{nc-laser} =i \sum_i{\frac{\Omega_\mathrm{L}
A_\mathrm{nc}^i}{2(\omega_\mathrm{Z}^\mathrm{n}-\tilde{A}^i)}\left((\hat{\sigma}_{\uparrow
t}-\hat{\sigma}_{t\uparrow}) \hat{I}^i_y\right)}.
\label{eqn:nc-laser}
\end{equation}
Application of the same SW transformation to the Liouvillian term
leads to nuclear-spin-flip assisted spontaneous emission terms
with rate $\simeq  \Gamma
(A_\mathrm{nc}^i/4\omega_\mathrm{Z}^\mathrm{n})^2$. The
corresponding terms for spin-flip Raman scattering processes
arising from the $\hat{H}_{hf}^{fc}$ take place at a rate $\simeq
\Gamma (2\tilde{A}^i/4\omega_\mathrm{Z}^\mathrm{e})^2$; given that
$\omega_\mathrm{Z}^\mathrm{e} \simeq 1000
\omega_\mathrm{Z}^\mathrm{n}$ and $A_\mathrm{nc}^i \simeq 0.07
\tilde{A}^i$, we conclude that the latter processes will take
place at a rate that is $\sim 1000$ times slower.

\begin{figure}[h] \centering
\includegraphics*[width=0.48 \textwidth,angle=0]{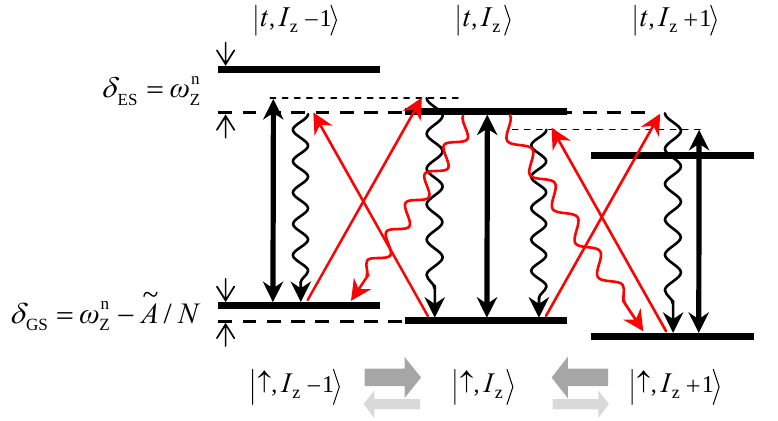}
\caption{Energy level diagram depicting the nuclear spin flip
assisted transitions for the $X^-$ trion higher energy Zeeman
branch in a finite magnetic field applied along the growth
direction $z$: a resonant laser field couples dipole allowed and
dipole forbidden transitions (straight and diagonal arrows,
respectively) of the trion-nuclear spin manifold. The lower states
of the manifold are electron spin-up states
$\left|\uparrow\right>$ split by the sum of nuclear Zeeman energy
and the Overhauser field,
$\delta_\text{GS}=\omega_{\mathrm{Z}}^{\mathrm{n}}-\tilde{A}/N$,
according to their nuclear spin projection $I_z$ along $z$. The
upper states are $X^-$ trion states (an individual hole and two
electrons in a spin-singlet state) split by the nuclear Zeeman
energy through
$\delta_\text{ES}=\omega_{\mathrm{Z}}^{\mathrm{n}}$. Nuclear spin
polarization occurs through spin-flip assisted diagonal
transitions (diagonal arrows) followed by spin preserving
radiative decay (wavy arrows). Finite laser detunings lead to an
imbalanced competition between the bidirectional nuclear spin
diffusion processes within the manifold (horizontal arrows): the
coupled trion-nuclear spin system reaches steady state by locking
to the laser resonance.(after \textcite{Hogele:2012a}).}\label{figure:dragging_ladderdiagram}
\end{figure}

To obtain a qualitative understanding of the coupled
electron-nuclear-optical dynamics, we consider the energy level
diagram in Fig.\,\ref{figure:dragging_ladderdiagram}, which shows
a ladder of two-level quantum systems coupled by nuclear spin-flip
processes. Here we adopt a mean-field description of the nuclear
spins by neglecting the quantum fluctuations in the Overhauser
field ($I_z=\langle\hat{I}_{z}\rangle$). We also make the
assumption that the electron couples equally to all nuclear spins
with a coupling strength $\tilde{A}^i = \tilde{A}/N$. For a given
nuclear spin polarization $I_z$ we can label the two level system
by the states $\left|\uparrow,I_z\right>$ and
$\left|t,I_z\right>$, where $\left|t\right>$ stands for the $X^-$
trion formed by  an individual hole and two electrons in a
spin-singlet state. The excited states $\left|t,I_z\right>$ do not
couple to the nuclear spins (we neglect the hole Overhauser field)
and differ by an energy $\omega_\mathrm{Z}^\mathrm{n}$. The ground
states which couple to the nuclear spins differ by an energy
$\omega_\mathrm{Z}^\mathrm{n}-\tilde{A}/N$. The strong direct
laser coupling and spontaneous emission processes are depicted in
black, whereas the hyperfine assisted processes are depicted in
red.

The diagonal spontaneous emission processes taking place at rate $\Gamma_+ =
\Gamma_- = \Gamma_\mathrm{diff} = \Gamma
(A_\mathrm{nc}^i/2\omega_\mathrm{Z}^\mathrm{n})^2$ cause
non-directional nuclear spin diffusion, limiting the range of
achievable DNP. The transition rate associated with
hyperfine-assisted laser coupling on the other hand is given by
\begin{eqnarray}
\label{eq:bluetrion} W_{\pm}(I_z) =&&
\left(\frac{\Omega_\mathrm{L}
\tilde{A}_\mathrm{nc}^i}{4(\omega_\mathrm{Z}^\mathrm{n}-\tilde{A}^i)}\right)^2\\
\nonumber && \frac{\Gamma}{4 (\Delta \omega_\mathrm{L}
-\tilde{A}^i(I_z\pm1) \mp \omega_\mathrm{Z}^\mathrm{n})^2 +
\Gamma^2}.
\end{eqnarray}
A remarkable feature of $W_{\pm}(I_z)$ is its dependence on the
sign of the laser detuning entering through the effective optical
detuning $\delta = \Delta \omega_\mathrm{L} -\tilde{A}^i(I_z\pm1)
\mp \omega_\mathrm{Z}^\mathrm{n}$: when the incident laser field
is red (blue) detuned, the transition rate $W_+(I_z)$ ($W_-(I_z)$)
dominates over $W_-(I_z)$ ($W_+(I_z)$) and ensures that the
Overhauser field increases (decreases). This directional nuclear
spin polarization will on the other hand result in a decrease of
$\delta$ from $\Delta \omega_\mathrm{L} - I_z$ to $\Delta
\omega_\mathrm{L} - I_z - \tilde{A}^i$ ($\Delta \omega_\mathrm{L}
- I_z + \tilde{A}^i$). If initially $I_z \ll N/2$, then DNP will
then continue until $\delta \simeq 0$, in consistence with the
experimental observations.

To obtain a more quantitative prediction, we consider the rate equation\footnote{For simplicity we consider a system of spin I=1/2 nulcei.}:
\begin{equation}
\label{eq:ratedrag}
\frac{dI_z}{dt} = W_+(I_z) (N/2 - I_z) - W_-(I_z)(N/2 + I_z) -
\Gamma_{\mathrm{diff}} I_z
\end{equation}
which exhibits bistability due to the nonlinear $I_z$ dependence
of the rates $W_{\pm}(I_z)$. A comparison between the predictions
of this rate equation and experiments is shown in
Fig.\,\ref{figure:dragging:theory+exp}.
Figure\,\ref{figure:dragging:theory+exp}(a) shows two laser scans
across the blue Zeeman line of $X^-$ at 4.5 T with opposite sweep
directions. For both scans, DNP was erased by waiting at a gate
voltage of strong co-tunneling for 10 s. A calculation of the
differential transmission signal is potted in
Fig.\,\ref{figure:dragging:theory+exp}(b).  The experimental
findings are qualitatively well reproduced by the theoretical
description based on the non-collinear hyperfine coupling.

So far, we have not discussed the red-shifted trion transition. While a
laser scan across the blue transition leads to a positive feedback
of the nuclear spins to ensure locking condition, a scan across
the red Zeeman line will cause an anti-dragging effect. To
understand this, we note that the transition rate associated with
hyperfine-assisted laser coupling in this case is given by
\begin{eqnarray}
R_{\pm}(I_z) = && \left(\frac{\Omega_\mathrm{L}
\tilde{A}_\mathrm{nc}^i}{4(\omega_\mathrm{Z}^\mathrm{n}-\tilde{A}^i)}\right)^2 \\
\nonumber && \frac{\Gamma}{4 (\Delta \omega_\mathrm{L}
+2\tilde{A}^i(I_z\pm1) \mp \omega_\mathrm{Z}^\mathrm{n})^2 +
\Gamma^2}.
\end{eqnarray}
The simple sign change in the effective optical detuning from
$\delta=\Delta \omega_\mathrm{L} -\tilde{A}^i(I_z\pm1) \mp
\omega_\mathrm{Z}^\mathrm{n}$ in Eq.\,\eqref{eq:bluetrion} for the
blue trion to $\delta=\Delta \omega_\mathrm{L}
+\tilde{A}^i(I_z\pm1) \mp \omega_\mathrm{Z}^\mathrm{n}$ renders
the exact resonance between the laser field and the trion
transition an unstable state. The DNP that ensues in the presence
of a small but non-zero $\delta$ will result in nuclear spin flip
processes that increase $|\delta|$ resulting in pushing the QD
trion transition away from the laser field.

These considerations allow to explain the dragging observed for
the blue- and red-shifted transitions \cite{Hogele:2012a},
suggesting that the dragging phenomenon in resonantly driven QD
transitions is due to non-collinear electron hyperfine
interaction. The main requirement for DNP via resonant laser
scattering is the presence of an unpaired electron spin with a
long spin-flip-time, either in the initial or the final state of
the optical transition. This condition is generically satisfied by
all fundamental QD transitions, i. e. the neutral exciton $X^0$ as
well as the singly negatively and positively charged excitons,
$X^-$ and $X^+$. In all cases, the Overhauser field experienced by
the unpaired electron facilitates the feedback that modifies the
QD transition energy. Whether or not this feedback leads to
resonance seeking or resonance avoiding excitations, as in the
blue and red Zeeman branch transitions, respectively, depends on
the spin orientation of the electron that couples to the incident
laser field.


\subsubsection{Preparation of nuclear spin states using periodic pulsed-laser excitation of quantum dot ensembles}
        \label{sec:Bayer}

Given the large inhomogeneous distribution of QD optical transition
energies, we would expect that nuclear spin manipulation in an
ensemble of QDs would only be possible either with non-resonant
excitation or by resonant short-pulse/broad-bandwidth laser pulses.
Even then, sizeable dot-to-dot variations in electron g-factor would
make it practically impossible to assess the degree of nuclear spin
polarization and it was far from obvious that there could be any
signatures of nuclear spin effects beyond electron-spin decoherence
in an ensemble setting.

Remarkably, Bayer and co-workers \cite{Greilich:2007a} have
demonstrated that hyperfine interaction effects are manifest in
electron spin dynamics when an ensemble of QDs is driven by
periodic short-laser pulse-trains. In their experiments (and as an extension to the experiments discussed in Sect.\,\ref{sec:dephvoigt}), electron
spin polarization is generated along the growth ($z$) direction
using circularly polarized laser pulses with a duration of $\tau_p =
1.5$~ps and a repetition rate of $75.6$~MHz, corresponding to an
inter-pulse separation of $T_R = 13.2$~ns. Even at relatively high
external (Voigt configuration) magnetic field strength of $B_x =
6$~T, the electron dynamics can be considered to be completely
frozen during the optical excitation. Following the
excitation/polarization, the electron spins undergo precession
around $B_x$ and dephase on a timescale given by $T_2^* < 1$~ns:
this anticipated dynamics, depicted in the top panel of
Fig.\,\ref{FigBayer}, is monitored using the Faraday rotation of a
second linearly-polarized probe laser pulse with a variable
time-delay with respect to the excitation pulse.

When a second pulse is applied at time $T_D$ with $T_D=T_R/k$ $(k\in \mathbb{Z})$, the authors have
observed a burst at a time delay of $2T_D$, indicating a revival of
the electron spin coherence in a manner reminiscent of spin-echo.
The fact that the observed Faraday signal cannot be explained as a
simple echo is evident because the spin polarization
also recovers before the arrival of the second laser pulse, as well
as at $nT_D$ with $n>2$. More strikingly, they observed that the
bursts continue to appear minutes after the second (delayed) pulse
excitation is turned off.

\begin{figure}
\epsfxsize=3.5in \epsfbox{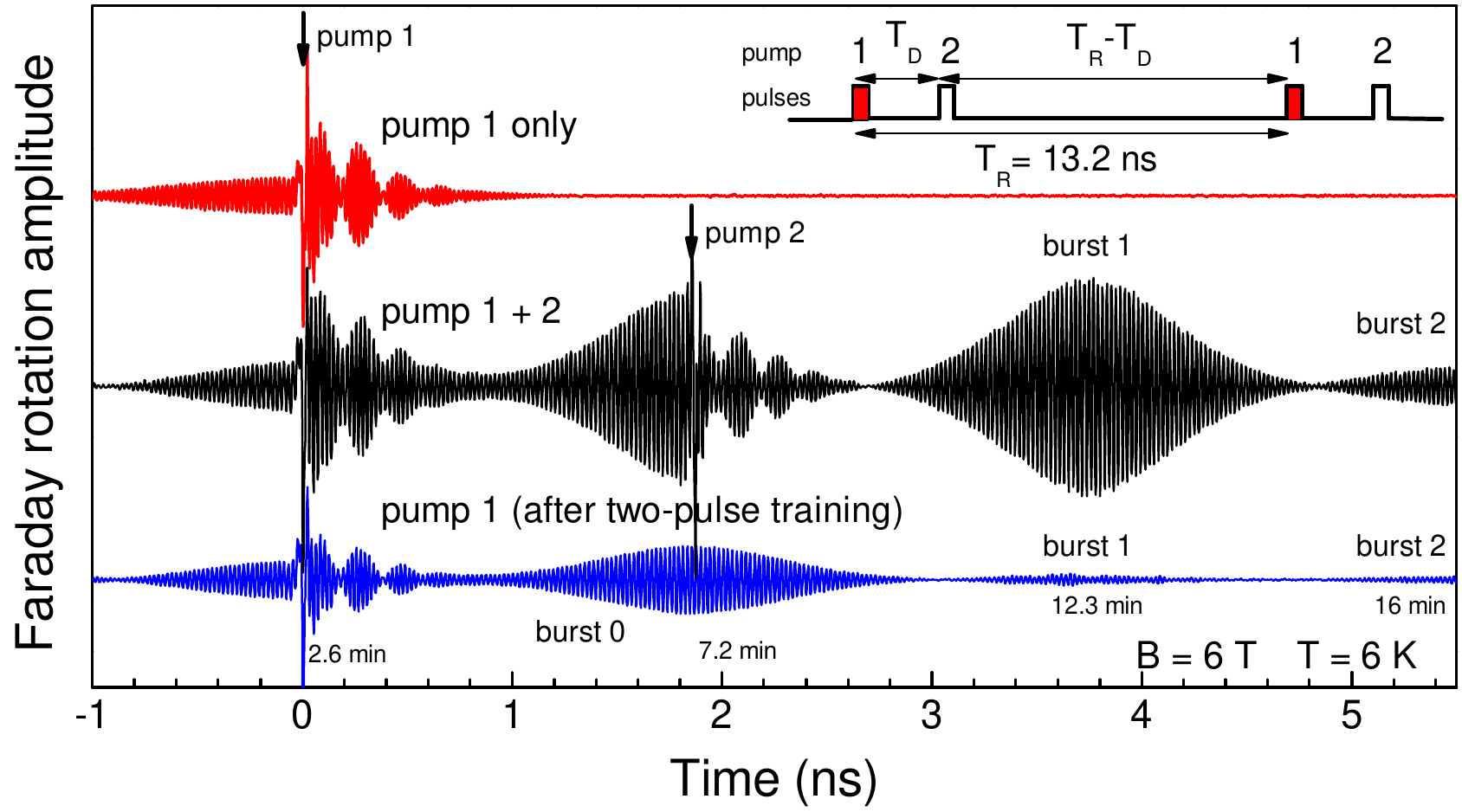} \caption{Faraday rotation
signal from an ensemble of single-electron-charged QDs
resonantly excited with a periodic laser pulse train. The data is
taken in Voigt geometry at $B_{x}=6$~T. The top panel shows the
Faraday signal when the ensemble is excited with a single pulse
train. Remarkably, excitation with a two-pulse train leads to bursts
in Faraday signal at time delays that are integer multiples of the
time-delay of the two laser pulses, where electron spins recover a
high degree of polarization. Even after the second laser pulse is
turned-off, the system continues to produce bursts, suggestive of
memory times exceeding minutes \cite{Greilich:2007a}.}
 \label{FigBayer}
\end{figure}

These stunning observations could be explained if the
pulsed-excitation protocol leads to a nuclear spin polarization in
each dot that ensures that the Larmor frequency is an integer
multiple of $2 \pi/T_D$ and hence of $2 \pi /T_R$. Even as one
starts from an ensemble of QDs, each with a different g-factor, a
quasi-continuous Overhauser field generated by nuclear spin
polarization and intra-dot diffusion could ensure that this
condition is satisfied. If the intrinsic nuclear spin lifetime is
$\tau_n$, then this polarization is maintained for times $\sim
\tau_n$ even after the pulses are turned off; since recent single QD
experiments have revealed nuclear spin polarization decay times
exceeding hours\,\cite{Maletinsky:2009a,Latta:2011a}, it is no longer surprising to see the memory of the
second pulse train survive for minutes time-scales.

To explain the observed locking effect, we introduce the concept of
time-dependent electronic dark-states: recently, it has been shown
theoretically that coherent population trapping (CPT) achieved using
continuous-wave laser fields satisfying two-photon resonance
condition in a single QD in Voigt geometry could lead to preparation
of Overhauser field eigenstates with uncertainty that can be as
small as the Overhauser field due to a single nuclear spin
\cite{Issler:2010a}. This nuclear spin cooling mechanism by Overhauser
field selective CPT relies on pure unbiased optical excitation
induced diffusion of QD nuclear spins. The counter-part of CPT
electron-spin dark states in the right-hand circularly-polarized
($\sigma_+$) pulsed-laser-train excitation setting is the
time-dependent superposition
\begin{equation}
|\psi_{dark-td}(t)\rangle = \frac{1}{\sqrt{2}} (|\uparrow_x\rangle -
e^{i \phi_p} e^{i (\omega^e_{x,j}-I_{x,j})t}
|\downarrow_x\rangle ,
\end{equation}
where the phase $\phi_p$ is set by the pulse-train; this phase
together with an Overhauser field $I_{x,j}$ specific to the $j^{th}$
QD ensure that at times satisfying $t = T_D$, the QD
electron spin is in state $\left|\downarrow_z\right>$, making it dark
against optical excitation by a $\sigma_+$ laser pulse. Provided
that the nuclear spin flips are only possible upon excitation of the
trion state, once this specific value of $I_{x,j}$ is attained by
optical excitation induced unbiased nuclear spin diffusion, nuclear
spin dynamics will freeze, ensuring that the dark state condition is
satisfied for times that are only limited by $\tau_n$.

Given the dominant role non-collinear hyperfine interaction
$\hat{H}_{\rm hf}^{\mathrm{nc}}$ plays in continuous-wave resonant laser
induced nuclear spin polarization discussed in
Sec.\,\ref{sec:locking}, it is natural to consider the origin of
the nuclear spin diffusion in the experiments of
\textcite{Greilich:2007a}. Having seen that at high external
magnetic fields, non-collinear hyperfine interaction leads to
nuclear spin-flip rates that are more than an order of magnitude
stronger than those induced by Fermi-contact flip-flop terms, we
argue that nuclear spin diffusion in self-assembled QDs under
pulsed-laser excitation should be primarily due to
$\hat{H}_{\rm hf}^{\mathrm{nc}}$. While the directional nuclear spin
polarization due to $\hat{H}_{\rm hf}^{\mathrm{nc}}$ during laser
excitation plays a key role in the dragging experiments
\cite{Hogele:2012a}, the corresponding processes are negligible in
the experiments reported by~\textcite{Greilich:2007a} due to the
short laser-pulse duration; simple considerations show that the
probability of a nuclear spin flip during optical excitation (induced
by $\hat{H}_{\rm hf}^{\mathrm{nc}}$) is smaller by a factor $\sim
(\omega_z^n \tau_p)^2$ as compared to those taking place during
radiative recombination from the trion state. We conclude therefore
that unlike continuous-wave excitation, the role of
$\hat{H}_{\rm hf}^{\mathrm{nc}}$ in pulsed-laser excitation is to induce
pure unbiased nuclear spin diffusion, allowing the QD
nuclei+electron system to find the time-dependent dark state and
thereby satisfying the observed phase synchronization condition
\cite{Greilich:2007a}.\\
The reader is referred to \textcite{Barnes:2011a} and \textcite{Glazov:2012a}
for interesting complementary models that help understanding the experimentally observed spin locking effects.


\section{NUCLEAR SPIN DYNAMICS IN QUANTUM DOTS} \label{sec:tdnp}

A key ingredient for the understanding of the coupled
electron-nuclear spin system is the knowledge of the relevant
timescales of the dynamics of nuclear spin polarization. This has
already become apparent in Sec.\,\ref{sec:nonlinear}, where
Eq.\,\eqref{eq:DNP} shows that the maximal nuclear spin polarization
in a QD is limited by the ratio of buildup and decay times of the
nuclear spins. Many other aspects like the respective roles of
nuclear spin diffusion, quadrupolar relaxation and trapped excess QD
charges can influence the dynamics of DNP. The buildup time of DNP
($\tau_{\rm buildup}$) typically depends on the experimental
conditions under which the nuclear spin system is addressed. For an
empty QD (free of any charge carriers) the DNP decay time
($\tau_{\rm decay}$) is an inherent property of the isolated nuclear
spin system. Furthermore, an experimental determination of
$\tau_{\rm decay}$ is essential for understanding the limits of
electron spin coherence in QDs\,\cite{Merkulov:2002a}, as $\tau_{\rm
decay}$ directly yields the correlation time of the fluctuations of
the Overhauser field along the axis in which the nuclei are
polarized (see Sect.\,\ref{sec:elecdynamics})

\subsection{Dynamics of nuclear spin polarization in low magnetic
fields} \label{sec:Dynamics}

The buildup and decay of DNP were studied by several groups
\cite{Maletinsky:2007b,Makhonin:2010a,Cherbunin:2009a,Nikolaenko:2009a,Chekhovich:2010b,Belhadj:2008a}
using optical ``pump-probe'' measurements. The principle of these
measurements is illustrated in Fig.\,\ref{FigUpDown}(a). Short
``pump'' and ``probe'' pulses of duration $\tau_{\rm pump}$ and
$\tau_{\rm probe}$ are used to polarize the nuclear spins and read out
the resulting degree of DNP after a waiting time $\tau_{\rm wait}$.
In order to measure the buildup (decay) time of DNP in these
experiments, $\tau_{\rm pump}$ ($\tau_{\rm wait}$) are varied,
respectively, while keeping all other experimental parameters fixed.

Figure\,\ref{FigUpDown}(b) and (c) show examples for buildup and
decay curves of DNP obtained with the above-mentioned technique at
zero magnetic field in a single InAs QD. In this regime, buildup and decay both follow
exponential curves with time-constants $\tau_{\rm buildup}=9.4~$ms
and $\tau_{\rm decay}=1.9~$ms, respectively.
The timescale of few tens of ms for optical pumping of nuclear spins
at low magnetic fields have additionally been confirmed in different
QD systems by \textcite{Nikolaenko:2009a} and
\textcite{Chekhovich:2010b}.

Compared to the rather slow timescales of nuclear spin buildup
suggested by\,\textcite{Gammon:2001a}
and\,\textcite{Maletinsky:2007a} the observed few-ms buildup times
are relatively fast. However, these early estimates were based on
high-field measurements of DNP, which leads to slower and, due to
the nonlinear electron-nuclear spin coupling, more complex dynamics
of DNP as will be shown in the next section. A further shortening of
$\tau_{\rm buildup}$ compared to earlier studies arises from the
strong localization of carriers in self-assembled QDs. Such
localization has been shown to be an important ingredient for
efficient nuclear spin polarization\,\cite{Malinowski:2001a} as it
increases the mean value of the Knight field (Eq.\,\eqref{eq:BK}) and
therefore the rate of electron-mediated nuclear spin relaxation in
QDs (see Eq.\,\eqref{eq:Te}).

\begin{figure}[t]
\centering
\includegraphics*[scale=1]{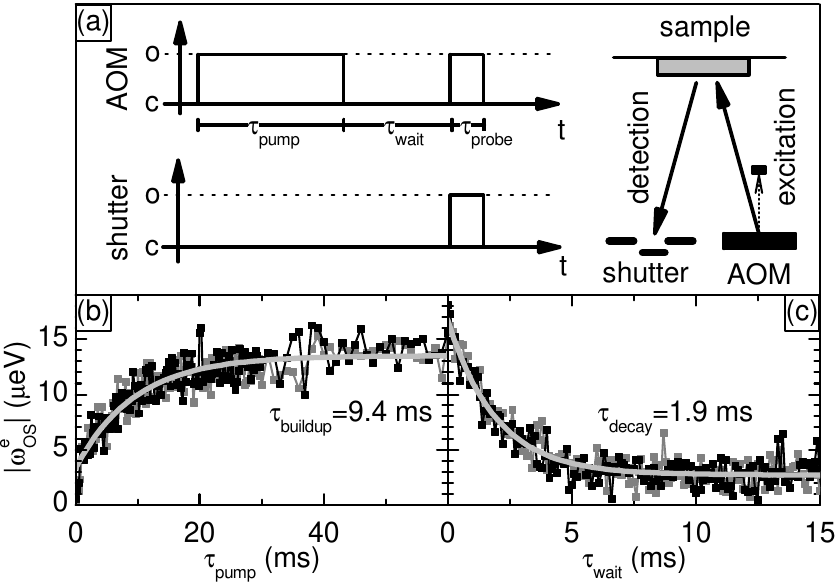}
\caption[Buildup and decay of QD nuclear spin
polarization]{\label{FigUpDown}(a) Schematic of the pulse sequences
used in the buildup and decay time measurements of DNP in zero external magnetic field. An optical
modulator is used as a fast switch for the PL excitation laser, where o and c denote the open and closed state, respectively. This creates pump
(probe) pulses of respective lengths $\tau_{\rm pump}$ ($\tau_{\rm
probe}$), separated by a waiting time $\tau_{\rm wait}$. A
mechanical shutter blocks the pump-induced luminescence from reaching the
spectrometer, while letting the probe-induced luminescence pass. (b) Buildup of
nuclear spin polarization measured by varying $\tau_{\rm pump}$ at
fixed $\tau_{\rm wait}$ ($0.5~$ms) and $\tau_{\rm probe}$ ($0.2~$ms)
for $\sigma^-$ ($\sigma^+$) excitation (black and grey,
respectively). (c) DNP decay curves obtained by varying $\tau_{\rm
wait}$ at fixed $\tau_{\rm pump}$ ($50~$ms) and $\tau_{\rm probe}$
($0.5~$ms). Exponential fits (light gray) yield time-constants
$\tau_{\rm buildup}=9.4~$ms and $\tau_{\rm decay}=1.9~$ms. Adapted
from\,\textcite{Maletinsky:2007b}}.
\end{figure}

\subsubsection{Nuclear spin relaxation in semiconductor quantum dots}
\label{sec:ElMediatedDecay}

While the buildup time of nuclear spin polarization depends on various experimental parameters (like optical
excitation intensity etc.), the dynamics of nuclear spin decay is a more
intrinsic property of the QD nuclear spin system. Three possible relaxation mechanisms for
nuclear spins in semiconductor QDs have been discussed in the past:

- \emph{nuclear spin diffusion}: Dipolar coupling of like-spins can
lead to diffusion of nuclear spin polarization to the surrounding,
unpolarized bulk\cite{Abragam:1961a}. The terms responsible for
nuclear spin diffusion are inter-nuclear flip-flop terms which are
energetically only allowed between nuclear spins exhibiting the same
level-spacing. In addition to the diffusion mediated by the
dipole-dipole interaction, the Fermi-contact hyperfine
interaction also leads to nuclear spin diffusion spatially limited
to the regions where the electron wave-function is
non-vanishing~\cite{Paget:1977a,Klauser:2006a,Latta:2011a}, see section \ref{sec:DecmK}.

- \emph{electron-mediated nuclear spin relaxation}: Fluctuations of
the spin of electrons occupying a QD typically lead to nuclear spin
relaxation through the same processes that allow for nuclear spin
pumping in the first place (i.e. electron-nuclear flip-flop
interactions). These spin fluctuations can either occur if the
charge-state of a given QD is fluctuating, or if a QD is occupied by
an odd number of (unpaired) electrons that are coupled to a nearby
charge reservoir, which induces electron spin relaxation through
spin-flip co-tunneling introduced in Sec.\,\ref{sec:orientation}.

- \emph{quadrupolar relaxation of nuclear spins}: As introduced in Sec.\,\ref{sec:quadintro},
nuclear spins couple to
electric field gradients in the lattice through their quadrupolar
moment. Modulation of electric field gradients at nuclear sites can
therefore lead to nuclear spin relaxation~\cite{Huang:2010a} described in detail in Sec.\,\ref{sec:quadlong}.
This modulation can
either occur in the form of lattice-vibrations (phonons) or through
QD charge-fluctuations. While phonon-induced quadrupolar relaxation
is typically suppressed for temperatures below
$20~$K\,\cite{McNeil:1976a}, quadrupolar relaxation of QD nuclei is
relevant even at low temperatures as QD charge fluctuations can be
induced either by optical excitation\,\cite{Paget:2008a} or by
charges tunneling in and out of a QD.

Which of these nuclear spin relaxation mechanisms has to be taken
into account to explain experimental observations, depends greatly
on the QD system under investigation and the experimental
parameters. As will be shown below, the more ``open'' nuclear spin
system of QDs formed by interface fluctuations in quantum
wells\,\cite{Gammon:1997a} is more likely to exhibit nuclear spin
diffusion. In contrast, the nuclear spin system in self-assembled
QDs, such as InAs QDs in GaAs, has been shown to be very well
isolated from its surrounding\,\cite{Maletinsky:2009a}; in
fact, the experiments reviewed in this Section demonstrate that
nuclear spin relaxation in these systems takes place only when the QD is occupied by an unpaired
resident electron.

The exceedingly high nuclear spin relaxation times reported
by\,\textcite{Greilich:2007a}, \textcite{Maletinsky:2009a} and
\textcite{Latta:2011a} for self-assembled QDs indicate that nuclear
spin diffusion from the QD to its surrounding is negligible in strained InAs/GaAs dots. This is in stark contrast to interfacial QD systems studied
by\,\textcite{Gammon:1996a} or nuclear spin pumping experiments
performed by \textcite{Paget:1982a} on shallow donors in GaAs. These
``open'' nuclear spin systems couple to their bulk surrounding
through spin diffusion and therefore exhibit nuclear spin relaxation
times on the order of few seconds. A recent study on interfacial QDs
however\,\cite{Nikolaenko:2009a} has shown that even in these
systems, nuclear spin diffusion can be inhibited under certain
experimental conditions. The exact physical mechanism leading to this
suppressed nuclear spin diffusion remains unknown up to now.
The different experiments suggest that the dipolar-induced spin diffusion is governed by the chemical contrast, as strongly reduced nuclear spin diffusion measured between two GaAs quantum wells separated by AlGaAs barriers seems to indicate \cite{Malinowski:2001a}.

In the following we will focus on nuclear spin dynamics in
self-assembled QDs, where nuclear spin diffusion from the QD nuclei towards the surrounding matrix is irrelevant and
electron-mediated nuclear spin relaxation as well as quadrupolar
relaxation can be studied in more detail.

The influence of the QD charge state on nuclear spin polarization
has been investigated by\,\textcite{Maletinsky:2007a}. The
experimental arrangement was analogous to the decay measurements
presented in Fig.\,\ref{FigUpDown}(a) with the addition that the QD
charge state during the decay time was varied using QD gate
electrodes  (cf. Sec.\,\ref{sec:charge}).
Figure\,\ref{FigVgSwitch}(b) shows a comparison between nuclear spin
decay for the same QD in the presence (gray line) and
absence (black lines) of a single electron. The  effect of the QD
charge state on DNP lifetime is striking. A further extension of the
decay interval (Fig.\,\ref{FigVgSwitch}(c)) shows that the nuclear
spin lifetime in these QD systems significantly exceeds one hour.
This observation is in accordance with earlier measurements of
nuclear spin memory time performed by\,\textcite{Greilich:2007a} on
ensembles of self assembled InAs QDs. This near-perfect isolation of
the QD nuclear spin system from its unpolarized environment is a
consequence of the strong inhomogeneous quadrupolar shifts of
the QD nuclei (c.f. Sec.\,\ref{sec:quadintro}) as well as the
material-mismatch between the QD and its surrounding, which taken
together almost completely suppress nuclear spin diffusion out of
the QD.

\begin{figure}[t]
\centering
\includegraphics*[scale=1]{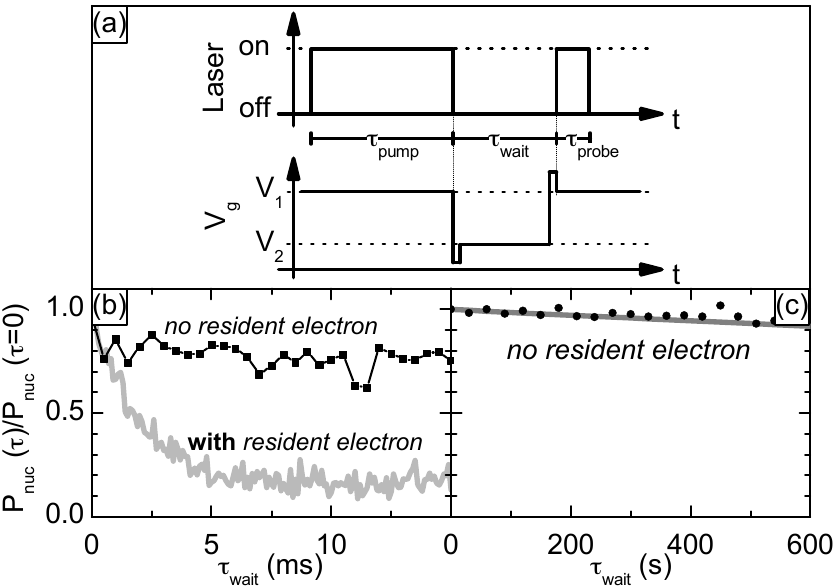}
\caption[Nuclear spin relaxation controlled by a QD
electron]{\label{FigVgSwitch} Measurements in zero external magnetic field.(a) Timing diagram for the gate voltage
switching experiment: DNP is established at a gate voltage $V_1$,
corresponding to the center of the $X^{-1}$ stability plateau.
During the period $\tau_{\rm wait}$, the QD gate voltage is switched
to a value $V_2$ and the nuclei are left to decay. (b) Measurement
of DNP decay, with $V_2$ corresponding to the center of the $X^0$
stability plateau. The gray (black) data points represent DNP decay
under \mbox{$\sigma^+$- ($\sigma^-$-)} excitation. For comparison, the
light gray curve shows the mean of the data presented in
Fig.\,\ref{FigUpDown}(b). (c) Same measurement as in (b), but over a
longer timescale. The exponential trace (light gray) indicates a decay
time constant of $1~$hour. Adapted
from\,\textcite{Maletinsky:2008a}}
\end{figure}

Two mechanisms could lead to the efficient decay of DNP due to the
residual QD electron. The first mechanism is caused by the
randomization of the electrons spin through co-tunneling to the
close-by electron reservoir. \,\textcite{Smith:2005a} showed that
co-tunneling occurs on a timescale of $\tau_{\rm
cot}\approx3~$ns in the center of the single-electron charging
plateau for a structure similar to the one studied in this work.
The resulting electron spin depolarization is mapped onto the
nuclear spin system via hyperfine flip-flop events. Taking into
account the detuning $\omega^e_Z$ of the two electron spin levels and assume $\omega^e_Z\gg1/\tau_\text{cot}$,
the nuclear spin depolarization rate can be estimated through
Eq.\,\eqref{eq:Te} as $T_{\rm
1e}^{-1}\approx(\tilde{A}/N\hbar\omega^e_Z)^2/\tau_{\rm
cot}$\,\cite{Meier:1984a}. In this expression, $\tau_\text{cot}$
plays the role of the hyperfine interaction correlation time in
Eq.\,\eqref{eq:Te}. With $\omega^e_Z$ assumed constant and equal to
half the maximum measured Overhauser shift for a rough estimate, one obtains a
nuclear spin depolarization time on the order of $10~$ms in
qualitative agreement with the measurement presented in
Fig.\,\ref{FigVgSwitch}(b).

A second possible mechanism is the indirect coupling of nuclear
spins due to the presence of a conduction electron in the QD \,\cite{Abragam:1961a} as given by Eq.\,\ref{eq:hfind}. Such indirect interactions result in
a long-range coupling of nuclear spins and can thereby significantly
enhance nuclear spin diffusion rates. While this process conserves
the total angular momentum of the nuclear spin system, it can lead
to a decay of the Overhauser shift by re-distributing the nuclear spin
polarization within the QD and by increasing the nuclear spin
diffusion rate out of the QD. The resulting decay rate for the
nuclear field has been estimated \cite{Klauser:2006a} to be on the order of $T_{\rm
ind}^{-1}\simeq \tilde{A}^2/\hbar^2 N^{3/2}\omega^e_Z$, assuming a weakly polarized nuclear spin
system and $\hbar\omega^e_Z\gg \tilde{A}$. It is important however
that $T_{\rm ind}$ gives the timescale over which coherent evolution
due to the indirect interaction takes place and as such, only marks
an upper limit to the nuclear spin relaxation rate due to indirect
interactions\,\cite{Klauser:2006a}. An estimate of nuclear spin
diffusion induced by indirect interactions is given in
Sect.\,\ref{sec:DecmK}. Finally, indirect interaction would only
lead to a partial relaxation of DNP, in contrast to the full decay
shown in Fig.\,\ref{FigUpDown}(b).

\begin{figure}[t]
\centering
\includegraphics*[scale=1]{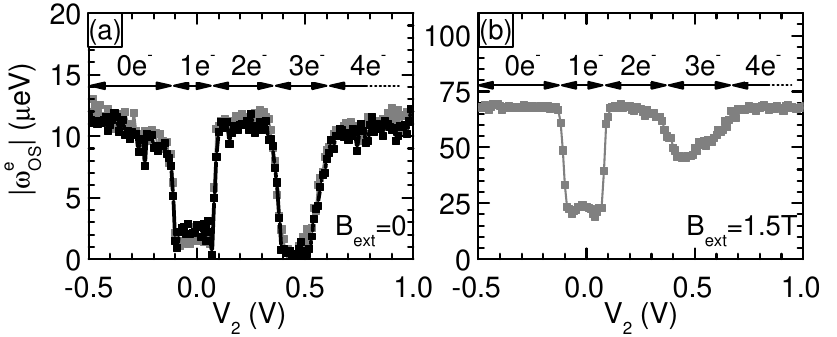}
\caption[Nuclear spin relaxation for different QD electron
occupation numbers]{\label{FigVgSwitchII} (a) Nuclear spin
polarization after free evolution of the nuclear spins during $\tau_{\rm wait}=20$~ms at a gate
voltage $V_2$ for (a) $B_{ext}=0$ and (b) $B_{ext}=1.5$~T.  For odd numbers of QD electrons, DNP decays in a
ms-timescale, while the decay takes seconds for an even number of
electrons (gray (black) data points correspond to QD excitation with
$\sigma^+$- ($\sigma^-$-) polarized light. Adapted
from\,\textcite{Maletinsky:2008a}
}\end{figure}

Electron-mediated nuclear spin relaxation is not limited to the
regime where QDs are occupied by a single electron.
Figure\,\ref{FigVgSwitchII} shows a comparison of nuclear spin
relaxation for QD charges between zero and four electrons. For this,
the QD nuclei were first polarized at a  QD gate voltage $V_1$ and then left to evolve for a
fixed period of $\tau_{\rm wait}=20~$ms at a variable gate voltage
$V_2$, before the final degree of DNP was measured with a short
probe laser pulse. Figure\,\ref{FigVgSwitchII} shows the resulting Overhauser shift
as a function of the gate voltage $V_2$. Regions of fast and slow
DNP decay can clearly be distinguished and correspond to QD
occupations with an odd or even number of electrons,
respectively\footnote{The QD charge state for a given gate voltage
was independently identified through corresponding features in the
QD PL spectra, as discussed by\,\textcite{Urbaszek:2003a}}.
When two electrons occupy the QD orbital ground state, the Pauli
principle ensures that they form a spin singlet, which does not
couple to the QD nuclei and therefore results in a prolonged
lifetime of DNP on the order of seconds. The third QD electron
occupies the next QD orbital (the $p$-shell), where it can again
interact with the QD nuclei and lead to a fast depolarization of
nuclear spins on a ms timescale. Injecting a fourth electron into
the QD at $V_g\approx0.6~$V increases the DNP lifetime again, indicating
that the two  $p$-shell electrons form a spin-singlet state.

This last observation is at first surprising, since Hunds rule states that the first two electrons occupying the $p$-shell of a QD should form a spin-triplet at ${B_{\rm ext}=0}$\,\cite{Warburton:1998a} and should therefore still couple to the QD nuclei. 
While the corresponding singlet-triplet splitting is typically on
the order of  $1~$meV\,\cite{Urbaszek:2003a,Karrai:2004a}, interactions breaking the
rotational symmetry of the QD such as QD
deformations\,\cite{Schulhauser:2004a} or lattice
effects\,\cite{Bester:2007a} can lower the singlet-triplet
splitting and break Hunds rule, which explains the apparent
singlet-character of the $p$-shell electrons in this QD. It is
interesting to note that in this experiment, the QD nuclear spins
can be seen as a probe for the properties of the QD and allow for an
efficient test of Hunds rule in an individual QD.

To conclude this section, we would like to point out certain
subtleties, originally discussed in Chapt. 11 of
\textcite{Dyakonov:2008a}, regarding the interpretation of the
previously described measurements of DNP dynamics at low magnetic
fields. For external magnetic fields smaller than the typical
nuclear dipolar coupling (characterized by a dipolar field $B_{\rm
L}\approx0.1~$mT), nuclear spin is in general not a conserved
quantity, as dipolar interactions include ``non-secular'', spin
non-conserving processes. Therefore, a nuclear polarization at zero
magnetic field should completely relax within the nuclear coherence
time $T_\text{Dipole}\propto1/B_{\rm L}\approx100~\mu$s. The long nuclear spin
lifetimes shown in Fig.\,\ref{FigVgSwitch}, seem to contradict this
statement as DNP persists for a few minutes even at zero magnetic
field\,\cite{Maletinsky:2007b}. Two mechanisms could explain this
discrepancy:
\begin{itemize}
\item If the relevant energy scale for QD nuclei is set by an interaction other than nuclear dipolar interaction, the above argument is invalid as non-secular dipolar coupling terms will be suppressed. This is the case if an external magnetic field is applied to the nuclei, but can also be induced at zero external field by quadrupolar interactions (see Sec.,\,\ref{sec:quadlong}).
\item As was already discussed in Sec.\,\ref{sec:zerofield}, the Knight field of the electron can ensure nuclear spin polarization and thereby nuclear spin cooling, even at zero magnetic field \footnote{In the absence of any external magnetic field the natural spin quantization axis is normal to the QD plane, set by the stronger confinement along growth ($z$) direction. The combination of circularly polarized excitation and the large spin-orbit interaction in the valence band then ensures that high-purity electron spin and heavy hole pseudo spin eigenstates can be prepared using laser pulses propagating along the $z$ direction. The average optically prepared electron spin is oriented parallel to the $z$ direction in this case.}. In contrast to spin polarization, nuclear spin temperature is known to have a slow relaxation time constant\,\cite{Goldman:1970a} (characterized by $T_1$) and can therefore persist over timescales much longer than $T_\text{Dipole}$.
\end{itemize}

These two scenarios can be distinguished experimentally by changing
the light-polarization of the probe pulse used in nuclear pump-probe
experiments: In the case where nuclear spins are stabilized by
quadrupolar interactions, the detected nuclear spin polarization
should be independent of this polarization. Conversely, in the case
of Knight-field stabilized nuclear spins, the sign of the nuclear
spin polarization measured in the probe-pulse should follow the
helicity of the probe laser-light. In particular, DNP should be zero
for a linearly polarized probe pulse and parallel (antiparallel) to
the initially created nuclear polarization if the probe pulse
helicity is equal (opposite) to the pump pulse helicity.

These additional tests have been performed
by\,\textcite{Maletinsky:2008a} and showed that in the case of
self-assembled InAs QDs long nuclear spin lifetimes at zero
magnetic field are indeed enabled by the strong nuclear quadrupolar
shifts introduced in Sec.\,\ref{sec:quadintro} \footnote{We note that there is still a substantial degree
of Knight-field stabilization in these systems as evidenced by the
discussion in Sec.\,\ref{sec:zerofield}. These Knight field related
effects are likely to be caused by nuclei occupying spin
states $m_I=\pm1/2$.}.

\subsection{Nonlinear nuclear spin dynamics in high magnetic fields}
\label{sec:DynamicsBext}

\begin{figure}[t]
\centering
\includegraphics*[scale=1]{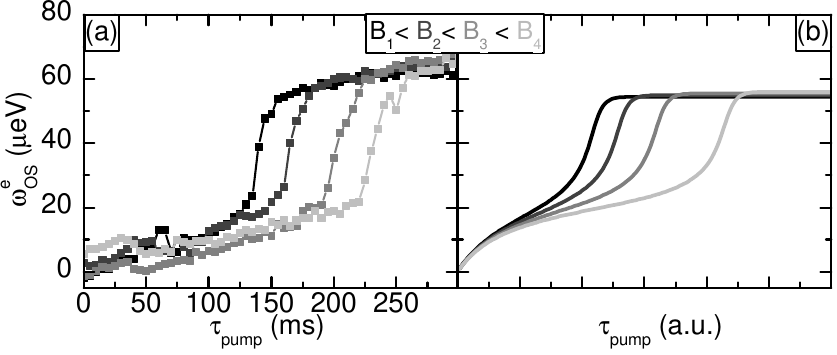}
\caption[Buildup of nuclear spin polarization in high magnetic
fields]{\label{FigUpBext} (a) Buildup of DNP in external magnetic
fields on the order of the final Overhauser field. Experiments were
performed with the procedure and external parameters described in
the main text at magnetic fields $B_1=1.1~$T, $B_2=1.2~$T,
$B_3=1.3~$T and $B_4=1.4~$T. (b) Simulations according to the
classical nonlinear rate equation\,\eqref{eq:DNP}. The magnetic
fields used for the simulation are $B_1=1.22~$T, $B_2=1.24~$T,
$B_3=1.26~$T and $B_4=1.28~$T. Adapted
from\,\textcite{Maletinsky:2008a}}
\end{figure}

In view of the nonlinear coupling between the electron and the
nuclear spin system that was demonstrated in
Sec.\,\ref{sec:nonlinear}, the purely exponential buildup and decay
curves presented in Sec.\,\ref{sec:Dynamics} might come as a
surprise. Since the nuclear spin relaxation rate $T_{\rm 1e}$ due to
the QD electron depends on electron spin detuning, the buildup and
decay rates of DNP should depend on the degree of nuclear spin
polarization and therefore change during the time traces presented
in Fig.\,\ref{FigUpDown}. These nonlinear effects should be most
prominent at the moment where the external and nuclear magnetic
fields cancel and are therefore more easily observed at elevated
external magnetic fields.

Nonlinearities in buildup and decay of DNP have been studied
by \textcite{Chekhovich:2010b} and \textcite{Maletinsky:2008a}
using pump-probe techniques similar to the ones described in the
previous section. Figure\,\ref{FigUpBext}(a) shows DNP buildup
curves measured at various external magnetic fields and clearly
demonstrates the discussed nonlinear buildup dynamics. A numerical
simulation of the dynamics described by the nonlinear equation of
motion Eq.\,\eqref{eq:DNP} at the corresponding magnetic fields is
presented in Fig.\,\ref{FigUpBext}(b) and shows qualitative
agreement with the experimental data.

A much more interesting situation arises for the decay of DNP in
sizable external magnetic fields. Since the nuclear spin decay rate
depends strongly on the electronic environment of the nuclei, the
dependence of the electron-mediated DNP decay rate on $\omega^e_Z$
can have various forms, depending on the relative importance of the
different possible mechanisms discussed in
Sec.\,\ref{sec:ElMediatedDecay}.

A good picture of the different decay characteristics at various QD
gate voltages in high magnetic fields can be obtained by measuring
DNP simultaneously as a function of gate voltage and time. The
nuclei are first initialized in a state of maximally achievable DNP. The gate
voltage is then switched to a value $V_2$ and DNP is measured after
a waiting time $\tau_{\rm wait}$. The measurement result as a
function of  $V_2$ and $\tau_{\rm wait}$ is shown in
Fig.\,\ref{FigDecBext}(a), where the final degree of DNP is encoded
in gray-scale. In accordance with the discussion in
Sec.\,\ref{sec:ElMediatedDecay}, significant nuclear spin
relaxation is only observed if the QD is occupied with a single
electron. There, the decay rate shows a marked increase when $V_2$
approaches the edge of the $1e^-$-plateau, where co-tunneling rates (introduced in section \ref{sec:orientation})
increase substantially\,\cite{Smith:2005a}.
This illustrates the importance of co-tunneling in electron-mediated
DNP decay, which is twofold: Co-tunneling ensures that the mean
electron spin polarization is zero due to the coupling to the
(unpolarized) electron reservoir and therefore sets the equilibrium nuclear
spin polarization. Furthermore, co-tunneling shortens the electron
spin correlation time $\tau^e_c$\, which broadens the electron spin
states and therefore allows for electron-nuclear spin flips to
happen.

\begin{figure}[t]
\centering
\includegraphics*[scale=1]{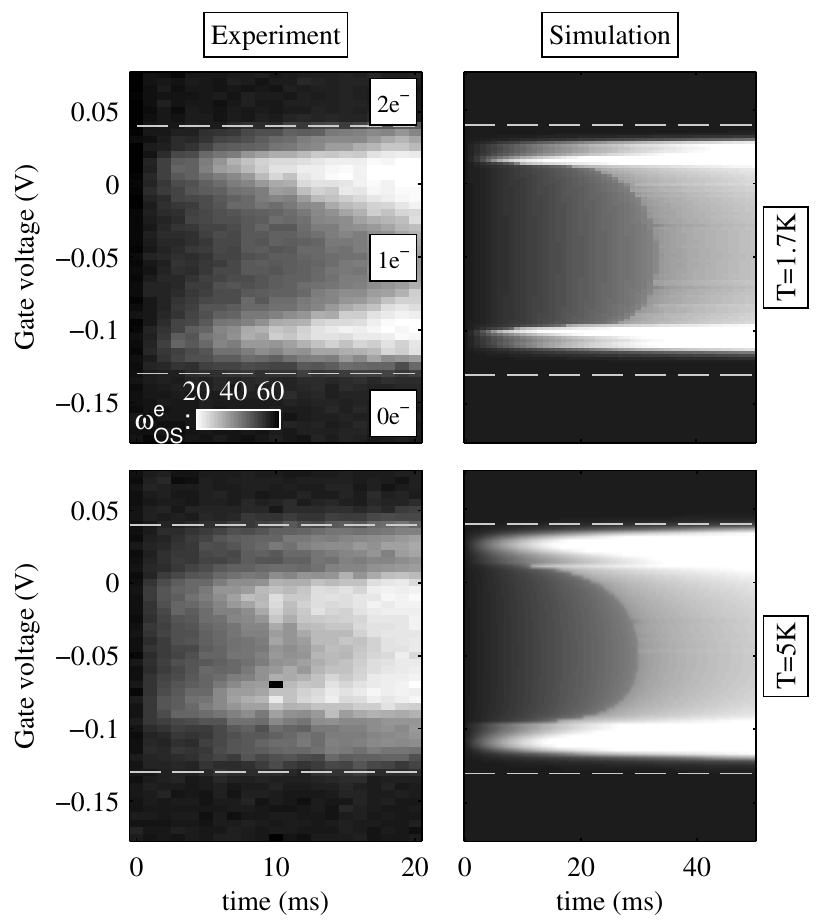}
\caption[Decay of nuclear spin polarization in high magnetic
fields]{\label{FigDecBext} Decay of DNP in an external magnetic
field of $B_{\rm ext}=1~$T. The nuclear spin polarization was
initialized with a $100~$ms, $\sigma^+$-polarized pump pulse at a
gate voltage $V_1$ corresponding to the center of the
$1e^-$-plateau, resulting in an initial Overhauser shift
$\omega^e_{OS}(\tau_{\rm wait}=0)\approx 55~\mu$eV. Immediately
after this nuclear spin initialization, the gate voltage was
switched to a value $V_2$. (Left) Measurement of $\omega^e_{OS}$ in $\mu$eV as a
function of waiting time $\tau_{\rm wait}$ and gate voltage $V_2$ at $T=1.7$~K (upper panel) and $T=5$~K (lower panel). The
dashed, gray lines indicate the transition between the QD charging
states (Right) Simulations according to Eq.\,\eqref{eq:Te}. Adapted
from\,\textcite{Maletinsky:2008a}}

\end{figure}

Some features in the data shown in Fig.\,\ref{FigDecBext}(a) seem to
contradict the picture of electron-mediated nuclear spin decay. When
approaching the $1e^-$-$2e^-$ transition point ($0.02V<V_2<0.04V$),
nuclear spin lifetime increases again, even though the stable
configuration of the QD is still singly charged. This observation is
a signature of motional narrowing of the nuclear spins: While a
finite $\tau^e_c$ is necessary to overcome the energy mismatch of
the initial and final states of an electron-nuclear spin flip-flop,
the nuclei cannot undergo such a transition if the electron spin
fluctuations become too fast\,\cite{Abragam:1961a,Meier:1984a}. This
becomes apparent by inspecting Eq.\,\eqref{eq:Te} which shows that
$T_{1e}$ has a maximum for
$\tau^e_c=1/\omega^e$, after which $T^{-1}_{1e}$ drops for decreasing
$\tau^e_c$. We observe the maximal electron-nuclear spin relaxation
rate at a gate voltage $V_2=0.02~$V. Since at $\tau_{\rm wait}=0$
the total electron Zeeman splitting,
$\hbar\omega^e$, amounts to $\sim20~\mu$eV, the corresponding
electron co-tunneling rate at this gate voltage should be on the
order of $30~$GHz, according to Eq.\,\eqref{eq:Te}. This is in reasonable
agreement with independent calculations of the electron
co-tunneling rate in QD structures similar to the one studied
here\,\cite{Smith:2005a}. Motional narrowing is not observed on the
$0e^-$-$1e^-$ transition, where one would at first expect a similar
behavior as in the $1e^-$-$2e^-$ transition. However, the electron
tunneling rate is growing with increasing gate voltage
which leads to slower co-tunneling rates for the $0e^-$-$1e^-$
transition compared to the $1e^-$-$2e^-$ transition. $\tau^e_c$
might therefore never reach the value $1/\omega^e$ on the
low-voltage side of the $1e^-$-plateau.

\subsection{Nuclear Spin dynamics at milliKelvin temperatures}
\label{sec:DecmK}

The experiments we described earlier in charge tunable InAs QDs have identified the role of
exchange coupling to an electron gas in determining the nuclear spin
relaxation in single-electron charged QDs. It is then natural to ask
if elimination of the ensuing co-tunneling processes by
either increasing the tunnel barrier thickness or by reducing the temperature would
reveal other, intrinsic Overhauser field decay mechanisms \footnote{The optical investigation of nuclear spin dynamics is carried out almost exclusively in the regime where $kT > \omega_Z^e$ (for the singular exception of this section \ref{sec:DecmK}). In this regime, the spin relaxation and excitation rates are both proportional to kT. }. Motivated
by this question, \textcite{Latta:2011a} have studied the Overhauser
field dynamics on a 35~nm tunnel barrier sample (compared to the more conventional 25~nm studied in the previous section) in a regime where
the coupling to the degenerate electron gas is vanishingly small and
the Overhauser field dynamics is determined solely by the coupling of each
nucleus to a confined electron spin. These experiments revealed that in
an external magnetic field of B$_z=5$~T in Faraday geometry and
temperatures of $\sim 200$~mK there are two distinct mechanisms for
the Overhauser field decay: \\
\textbf{(1)} a spatially limited,
temperature-independent, nuclear spin diffusion \textit{within the dot} originating from
electron mediated nuclear spin interactions\\
\textbf{(2)} a co-tunneling
mediated, temperature dependent \cite{Dreiser:2008a}, decay of the Overhauser field approaching
$\tau_\text{decay}\approx 10^5$~s. Remarkably, the diffusion induced reduction in the Overhauser
field taking place on $\sim100$~s timescale can be strongly
suppressed by repeating the preparation cycle consisting of
polarization (pump) and free-evolution (wait). In these experiments
an Overhauser field was established by resonant dragging of the blue-shifted
Zeeman line of the $X^0$ transition, as described in Sec.\,\ref{sec:locking}, followed by a waiting period
$\tau_{\rm wait}$ at a gate voltage where the QD contained a single
electron and the laser field is completely off-resonant with all QD
transitions. Finally the Overhauser field remaining after
$\tau_{\rm wait}$ was determined by first ejecting the resident electron
from the QD and then scanning the laser field across the $X^0$ resonance quickly so as to measure but not destroy the QD nuclear polarization.

When the gate voltage is chosen such that the QD is singly charged with a co-tunneling rate ($\kappa\ge 10^7
s^{-1}$) that does not allow for appreciable electron spin
pumping~\cite{Atature:2006a}, the Overhauser field exhibits a fast
decay~\cite{Maletinsky:2007a,Latta:2009a} on the order of a few
seconds. The observed decay is temperature dependent, consistent
with the predictions of the co-tunneling mediated process discussed in Sect.\,\ref{sec:ElMediatedDecay}.

When the exchange coupling between the QD electron and the Fermi reservoir is
minimized by choosing a gate voltage during $\tau_{\rm wait}$
corresponding to the center for the $X^-$ plateau, the temperature
dependence of the long-time decay rate becomes even more prominent:
Figure\,\ref{latta:diffusion01}(a) shows two measurements at 4 K and
200 mK of the Overhauser field magnitude as a function of waiting time
$\tau_{\rm wait}$. In both cases, there is an initial partial decay
taking place at $\sim100$~s time-scale which saturates after the
Overhauser field decays to half of its value. The slow decay at 4 K
takes place on a time scale of $5 \times 10^3$~s, whereas at 200 mK
the corresponding decay time exceeds $10^4$~s.

The temperature-dependent decay of the Overhauser field can be
explained by a second order process originating from an effective
non-collinear dipolar hyperfine interaction $\hat{H}_{\rm hf}^{\rm
nc}$ of Eq.\,\ref{eq:nc-hyperfine}: as discussed in Sec.\,\ref{sec:locking} such an effective coupling appears when the
quadrupolar axes $z'$ of the QD nuclear spins are not parallel to the
external field $B_z$. The energy
conservation in this irreversible nuclear spin flip process is
ensured by the exchange coupling of the QD electron to the
degenerate electron gas leading to a co-tunneling rate
$\kappa=1/\tau_{\rm cot}$; the corresponding Overhauser field
decay rate can then be shown to be
$\left(A_\mathrm{nc}^i/\hbar\omega_Z^n\right)^2/\tau_{\rm cot}$. In
the studied sample $\tau_{\rm cot}$ is estimated to be on the order
of $10~$ms at 4 K in the plateau center which allows to obtain the
value of $A_\mathrm{nc}^i\simeq 0.03 \tilde{A}_i$ for the effective
non-collinear dipolar hyperfine coupling constant between the
electron and the nuclei. It should be emphasized that the energy
difference of the states coupled by $\hat{H}_{\rm hf}^{\mathrm{nc}}$
is given by $\omega_Z^n < 10$~mK which is in turn much smaller than
the electron temperatures that can be reached. This observation
suggests that the relevant co-tunneling rate and hence the
Overhauser field decay rate will simply be linearly proportional to
the electron temperature. Finally, even though $1/\tau_{\rm cot} \ll
\omega_Z^n$, the fact that $T \gg \omega_Z^n$ ensures the validity
of the Markov approximation in describing the nuclear spin flips
associated with $\hat{H}_{\rm hf}^{\rm nc}$.

\begin{figure}[t]
\centering
\includegraphics*[scale=1]{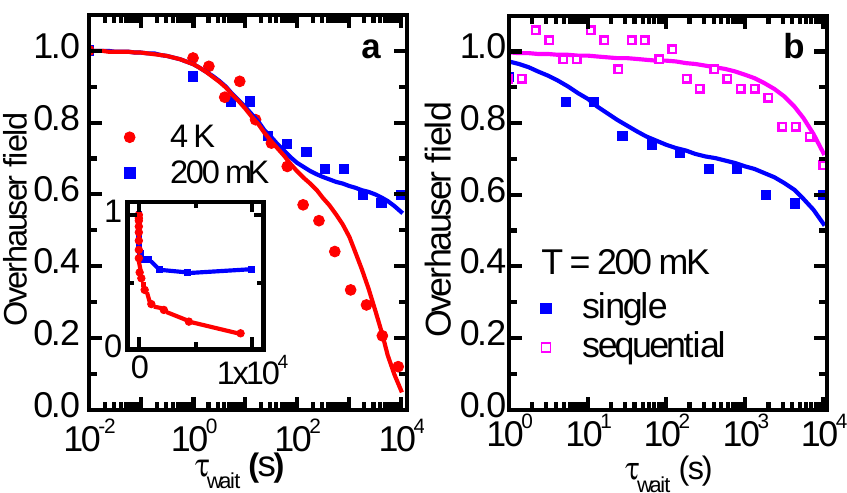}
\caption{ a) Decay of the Overhauser field with negligible co-tunneling for
the case of a resident
electron ($V_\text{wait}=530$~mV) at 200~mK ($\blacksquare$) and 4 K
(\textbullet). The inset shows the same data in a linear-linear
plot. b) Demonstration of the spatially limited nuclear spin
diffusion: By sequential polarization of the nuclear spins, the
polarization can be saturated, suppressing further nuclear spin
diffusion \label{latta:diffusion01} \cite{Latta:2011a}}\end{figure}

As is seen in Fig.\,\ref{latta:diffusion01}(a), the observed nuclear
spin dynamics is much richer than a simple exponential decay curve.
To explain the initial partial decay of DNP taking place on a
$\sim100$~s time scale, we consider the nuclear spin spatial
diffusion mediated by an indirect interaction through the electron present in the dot already introduced in Sec.\,\ref{sec:dnpgeneral}:

\begin{equation}
\hat{H}_{\rm ind}=\sum_{i,j}\frac{A^i A^j}{\omega^e_Z}\hat{S}_z\hat{I}^i_+\hat{I}^j_-
\tag{\ref{eq:hfind}} \;.
\end{equation}

This Hamiltonian
ensures the conservation of the total nuclear spin polarization and
leads to diffusion within the region where the electron
wave-function is non-vanishing. Although the total magnitude of QD
nuclear spin polarization does not decrease due to this diffusion
process, the Overhauser field seen by the electron decays partially
due to a redistribution of the nuclear spin polarization within the
QD.

A strong evidence that the initial partial decay of the
experienced Overhauser field stems from $\hat{H}_{\rm ind}$ is
provided by repeating the {\sl polarization-wait-measure} cycle and
observing its effect on the decay dynamics. In these experiments, it
was ensured that the initial value of the Overhauser field
(immediately after the polarization cycle) is identical for all
repetitions. The total waiting time in each cycle was chosen to be
200 s, which is longer than the time scale over which the initial
limited decay takes place. The corresponding decay of the Overhauser
field is shown in Fig.\,\ref{latta:diffusion01}(b). Clearly, the
initial limited decay is suppressed in this case, indicating a
saturation of the nuclear spin polarization within the QD. The
experimental data are in excellent agreement with numerical
calculations based on semi-classical rate equations taking into
account electron mediated diffusion and the pure decay of nuclear
spin polarization \cite{Latta:2011a}. The model assumes that the coherent evolution
described by $\hat{H}_{\rm ind}$ is interrupted by a dephasing
process; the absence of temperature dependence of the saturable
decay and hence the dephasing process suggests that the relevant
noise is channelled through electrical wiring. Finally, the finite
bandwidth of the measurement set-up ensured that this electrical
noise has vanishing contribution at $\omega_Z^n$, explaining why
there is no temperature-independent contribution to the decay
processes stemming from $\hat{H}_{\rm hf}^{\rm nc}$.


 \subsection{Dynamic nuclear polarization in a transverse field: the anomalous Hanle effect}
\label{sec:ahanle}

Since a finite  nuclear polarization can be achieved in InAs/GaAs
QDs along the optical axis $z$ in \emph{zero} external
magnetic field (see Sec.\,\ref{sec:zerofield}), it is obviously
interesting to study its evolution against a transverse magnetic
field $\bm{B}=B_x \bm{u}_x$. In particular, the  quadrupolar
interaction with principal axis $\parallel z$  is expected to
inhibit the alignment of the nuclear polarization parallel or
anti-parallel to $\bm{B}$, as it could be expected from
Eq.\,\eqref{eq:BnBulk}, up to fields of the order of a few
$B_\text{Q}$'s. The splitting $n\times\hbar\omega_\text{Q}$ between
the states  $|n\pm1/2\rangle$ reduces  indeed drastically the effect
of the magnetic coupling, except for the  states
$\left|\pm1/2\right>$ ($n=0$) which still split linearly in low
field according to $\Delta_{\pm1/2}=2\hbar \gamma_n B_x$. Their spin
polarization $\parallel z$ should thus be cancelled in a very small
transverse field.  In contrast,  the pairs of levels  $\left|\pm
m\right>$ with $|m|>1/2$   have  a  vanishingly small splitting
$\propto \hbar \gamma_n B_x(B_x/B_\text{Q})^{2m-1}$ as long as $B_x<
B_\text{Q}$. One may thus infer   that their nuclear polarization
should be substantially preserved  in fields below $B_\text{Q}$,
maintaining a  nuclear field  $\parallel z$ of a few 100~mT's.

Interestingly, the evolution of the  nuclear field  can be
investigated through the decrease of  electron spin polarization in
the transverse field $B_x$ (Hanle effect) which is directly
accessible from the photoluminescence of $X^+$ trion photo-created
with a circularly polarized excitation.
In the absence of nuclear effects, such Hanle depolarization curves assume a Lorentzian profile with half width $B_{1/2}=\hbar/(|g_e|\mu_B\tau_{r})$. Deviations from this well understood behavior are usually caused by nuclear effects and can thus be used to study DNP in QDs.
Such measurements of the
Hanle effect have been very successful  in the past to demonstrate
nuclear spin cooling in the electron Knight field at very low external
fields for bulk semiconductor, as well as the magnetic  anisotropy
of nuclei due to the quadrupolar interaction in alloys like
AlGaAs~\cite{Meier:1984a}. More recently,  the Hanle effect has been
measured in  an ensemble of p-type doped InP/InGaP QDs
\cite{Dzhioev:2008a}:   the half-width $B_{1/2}$ of the Lorentzian
decrease  was found to be 3 times larger when a nuclear field was
created (under constant $\sigma^+$ excitation) than in  absence of
nuclear polarization (excitation with 50~kHz modulated
$\sigma^+/\sigma^-$ polarization). From this observation it was
concluded that a finite nuclear field is  maintained perpendicular
to the external field thanks to the quadrupolar splitting of the
levels as described above. However, a direct measurement of the
Overhauser field to confirm this hypothesis is not possible in
experiments on ensembles of QDs.

The study of \textit{individual} InAs/GaAs QDs allows
to measure both the Hanle effect and the Overhauser field to
further refine this interpretation, while revealing more pronounced
anomalies of the Hanle depolarization curves.

\begin{figure}[h]
\centering
\includegraphics*[width=0.48 \textwidth,angle=0]{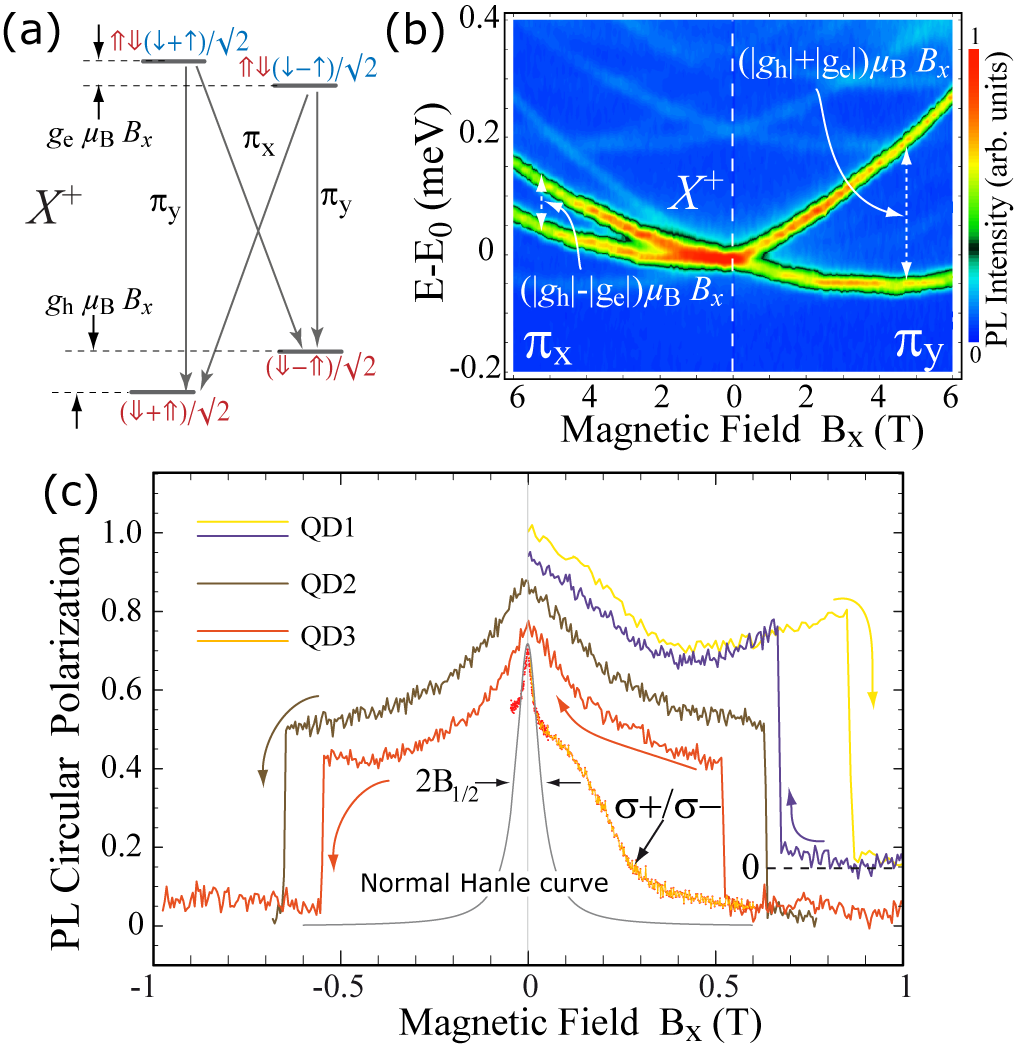}
\caption{(a) Schematics of optical transitions and polarization rules
for an  $X^+$ trion in a transverse magnetic field $B_x$. (b)
Typical density plot of $X^+$ PL intensity from a single QD against
magnetic field $B_x$ and  detection energy around $E_0=1.355$~eV.
The excitation is linearly polarized  and the detection either
$\pi_x$ or $\pi_y$ as indicated. (c) Depolarization curves  (Hanle
effect) of $X^+$ lines from 3 different QDs under $\sigma^+$
quasi-resonant excitation. QD1 measurements are vertically shifted
for clarity.  (after \textcite{Krebs:2010a})}\label{LPN-AHE1}
\end{figure}

Hanle effect in single QDs has been first investigated in
interfacial GaAs  QDs \cite{Bracker:2005a}, yet  in conditions where
no nuclear field was generated. Besides, for these unstrained and
almost pure GaAs QDs the  electric field gradient  should cancel out
on nuclear sites making the quadrupolar interaction vanishingly
small. Depolarization curves following a Lorentzian profile were
indeed obtained by Bracker \textit{et al.} with  a $B_{1/2}$ field in good agreement with the
expected estimate. In contrast, a strongly anomalous Hanle effect is
reported for InGaAs/GaAs QDs under excitation conditions
producing   DNP in zero  field \cite{Krebs:2010a}. For these QDs the
theoretical $B_{1/2}$ field (in absence of nuclear polarization)
amounts to $\sim$30~mT. Note that the value of $g_e$ which
determines $B_{1/2}$ can be  precisely inferred from the Zeeman
effect in a magnetic field $B_x\agt 1$~T, since it produces 4
distinct linearly polarized transitions with splittings given by
$(|g_e|\pm|g_h|)\mu_\text{B} B_x$, see  Fig.\,\ref{LPN-AHE1}(a),(b).
As shown in Fig.\,\ref{LPN-AHE1}(c), the experimental Hanle curves
of $X^+$ trions reveal that a sizeable electron spin polarization
$\sim$50\% is maintained in  fields as high as $\sim$1~T, up to a
critical field $B_x^c$ where it abruptly collapses. Moreover, this
evolution is symmetrical in magnetic field which means that it does
not depend on the  specific helicity  $\sigma^+$ or $\sigma^-$ of
the illumination while keeping the field direction constant.
However, under excitation with  50~kHz-modulated $\sigma^+/\sigma^-$
polarization such that no nuclear field $\parallel z$ is created,
the electron spin stabilization is significantly reduced.

These observations agree qualitatively  well with the interpretation
that a  nuclear field $\parallel z$ would be maintained thanks to
the quadrupolar interaction. Yet, the magnitude of this nuclear
field should be as high as $B_x^c$ in order to keep the electron
spin polarization above  50\% up to $B_x^c$. The direct measurement
of this nuclear field component $B_{n,z}$, deduced as in Faraday configuration
from the  splitting of the $\sigma^\pm$ lines, indicates in contrast
that $B_{n,z}$ monotonically decreases with $B_x$ from a maximum value
around $0.35$~T, down to $\sim0$~T at 0.5~T \cite{Krebs:2010a}, which obviously
invalidates the above interpretation. If a nuclear field is
responsible for the electron spin stabilization, it must be nearly
opposite to the external field ($B_{n,x}\approx-B_x$), such that the
in-plane component of the total field remains smaller than $B_{1/2}$
up to $B_x^c$.

Determining the actual magnetic field experienced by the electron in
the field range of interest, namely  around $B_x^c$ requires in
principle a very high spectral resolution to separate the four
different lines of $X^+$. However, if the hole $g$-factor is
sufficiently large,  this requirement can be circumvented because
the splittings of the $\pi_x$-polarized   lines and
$\pi_y$-polarized lines can be measured separately for fields as
small as $\sim$0.5~T, with a   $\pi_x$- or $\pi_y$-polarized
detection while keeping a $\sigma^+$ excitation,  see
Fig.\,\ref{LPN-AHE2}(a). For both polarizations  the $X^+$ splitting
turns out to be  reduced solely to the hole Zeeman splitting
$g_h\mu_\text{B}B_x$ for fields below $B_x^c$ and present jumps at
$B_x^c$ where they recover their normal values $(g_h\pm
g_e)\mu_\text{B}B_x$, see Fig.\,\ref{LPN-AHE2}(c). This behavior
proves that the anomalous Hanle effect is due to a nuclear field
which essentially cancels out the applied field, as  depicted
in Fig.\,\ref{LPN-AHE2}(b). The detailed mechanism leading to the
establishment of this in-plane nuclear field is yet to be
elucidated, the possible role of the non-collinear hyperfine interaction (see Sec.\,\ref{sec:locking}) deserves further investigation. In this context, we remark that a mean field approach
such as the one proposed in \textcite{Dzhioev:2008a} predicts in
contrast the build-up of an in-plane nuclear field pointing in the
same direction as the transverse field.

\begin{figure}[h]
\centering
\includegraphics*[width=0.48 \textwidth,angle=0]{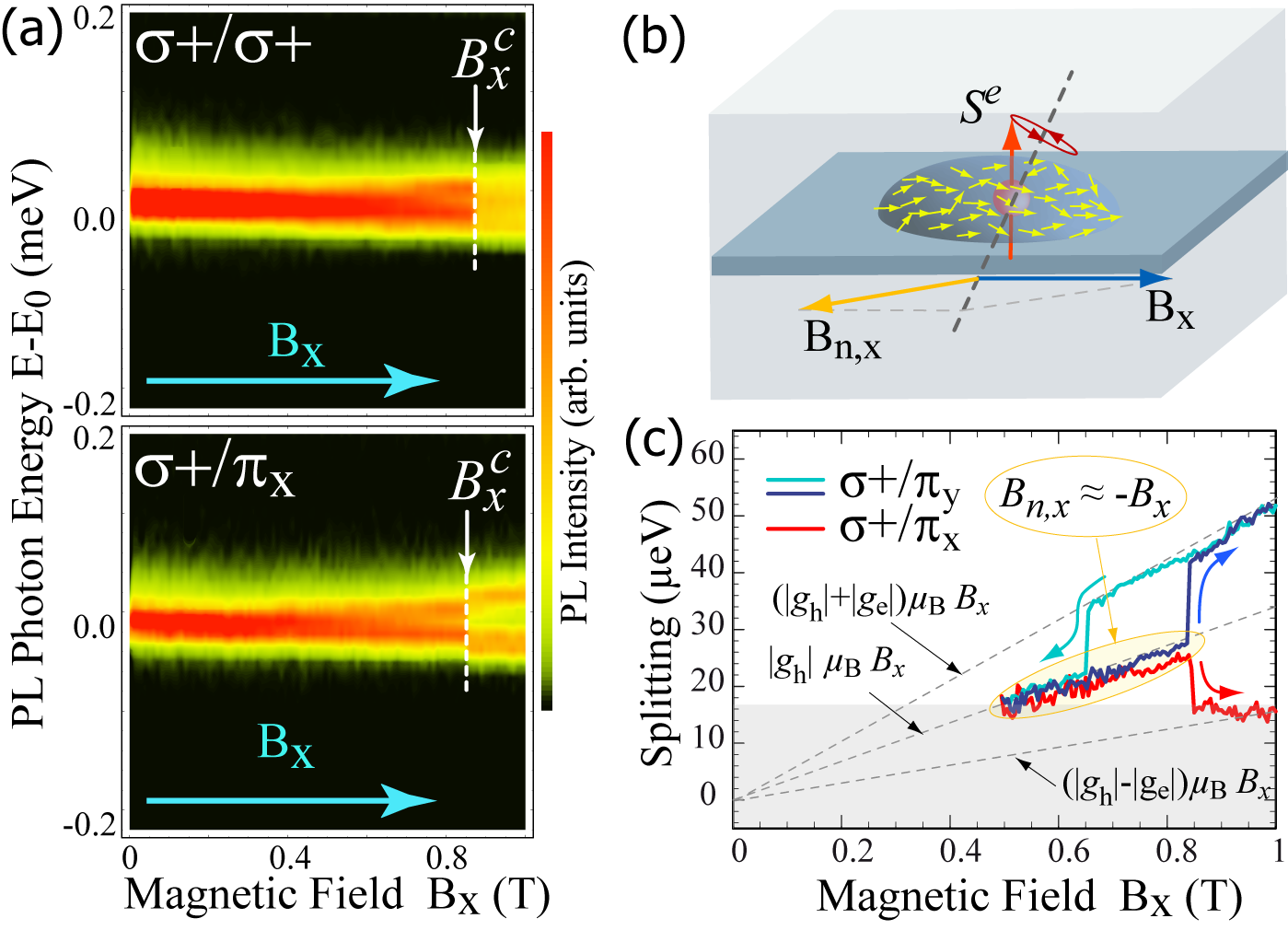}
\caption{ (a) Comparison of $X^+$ PL density plots measured under
$\sigma^+/\sigma^+$ or $\sigma^+/\pi_x$ configurations of excitation/detection.
polarizations. The latter enables one to resolve the  transition
splittings for fields above $\sim0.5$~T and  thus to infer the
nuclear field. (b) Sketch of the nuclear polarization generated
almost parallel to $B_x$ giving rise to a nuclear field
$B_{n,x}\approx-B_x$ responsible for the anomalous Hanle effect. (c)
Splitting of $X^+$ $\pi_x$- and $\pi_y$-polarized transitions
evidencing the cancellation of the external field  for $B_x<B_x^c$.
Normal splittings  are recovered above $B_x^c$  when the nuclear
polarization  is eventually destroyed. (adapted from
\textcite{Krebs:2010a})}\label{LPN-AHE2}
\end{figure}

\subsection{Optically detected NMR on semiconductor quantum dots}
\label{sec:odnmr}

The Overhauser shift of the emission spectrum of a single dot
allows for a direct measurement of the steady state nuclear
spin polarization. Further information about the QD composition and
the QD nuclear spin system can in principle be gained through
optically detected nuclear magnetic resonance (NMR) experiments.
Under optical pumping conditions, the spin polarization of the
nuclei can be decreased by applying an oscillating magnetic field
orthogonal to the nuclear spin quantization axis at a frequency
which matches the nuclear level splitting in the RF range. This
drives transitions between nuclear spin states and the corresponding
change in nuclear spin polarization can be detected via changes of
the Overhauser shift. The first such experiment has been
demonstrated by \textcite{Gammon:1997a} in GaAs interface
fluctuation QDs and has provided the first direct proof of
nuclear spin pumping in these QD systems. Figure\,\ref{FigNMR}(a)
shows the observed NMR spectra of $^{75}$As and $^{69}$Ga; the
corresponding resonance lines are remarkably narrow, indicating that
the nuclear spin system in interfacial QDs is rather unperturbed by
inhomogeneous strain or other line-broadening
mechanisms linked to nuclear quardrupole effects (see Sec.\,\ref{sec:quadintro})\footnote{\textcite{Gammon:1997a} note however a certain
variability of NMR frequencies between different QDs as well as
occasional lines which are anomalously broad.}. Another
line-broadening mechanism can stem from (inhomogeneous)
Knight-shifts of NMR lines which are caused by the effective
magnetic field that a spin-polarized electron exerts on the QD
nuclei (see Sec.\,\ref{sec:hyperfine}). NMR line-broadening and
line-shifts due to this Knight-field were recently observed by
\textcite{Makhonin:2010a} in NMR experiments in individual
interfacial QDs.

Performing optically detected NMR experiments on the well isolated
nuclear spin system in self assembled QDs, such as InAs QDs in GaAs,
has remained an open challenge for many years. Only very recently
first optically detected NMR experiments on large ensembles of self
assembled InAs QDs were reported by
\textcite{Flisinski:2010a,Cherbunin:2011a}. The difficulty in
observing NMR in self-assembled QDs lies in the significant
inhomogeneity of the nuclear spin system due to the strongly
strained lattice of these QDs. Strain in the QD lattice
results in strong, inhomogeneous quadrupolar splittings of the
nuclear spin states. As a result, NMR lines become strongly
broadened and are difficult to observe.

\textcite{Flisinski:2010a} partly circumvent these difficulties in
two ways: The splitting between nuclear $m_s=\pm 1/2$-states is
invariant under nuclear quadrupole interactions \cite{Abragam:1961a}
and the corresponding NMR lines are thus expected to be
insensitive to inhomogeneous quadrupolar fields.
Alternatively, by sweeping the NMR driving frequency over a broad
range, all possible nuclear spin transitions can be addressed
simultaneously. \textcite{Flisinski:2010a} employ both these
techniques to detect NMR in Hanle depolarization experiments
performed on a large ensemble of self-assembled InGaAs QDs.
Fig.\,\ref{FigNMR}(b) shows modifications of Hanle curves due to the
presence of a RF field which depolarizes the nuclei. This data
clearly shows that nuclear spins in self-assembled QDs can be
addressed using RF driving fields together with optical detection of
DNP. By keeping the frequency of the NMR driving field fixed,
\textcite{Flisinski:2010a} were able to observe resonant NMR
features in the low-field region of the Hanle depolarization curves
(Fig.\,\ref{FigNMR}(b), inset). These resonances correspond to
transitions between  $m_I=\pm1/2$ states of $^{71}$Ga and $^{75}$As
and provide an additional fingerprint of the strong quadrupolar
interactions present in the QDs studied in this work: For 3/2-spins
(such as $^{71}$Ga and $^{75}$As) the gyromagnetic ratio for the
$m_I=\pm 1/2$ is enhanced by a factor of two in the presence of
strong quadrupolar interactions and a magnetic field applied
perpendicular to the quantization axis of the spins. The data
shown in Fig.\,\ref{FigNMR}(c) confirms this picture and
demonstrates that the strain in the QDs under study is predominantly
oriented along the QD growth direction. The broadening of the NMR
lines in Fig.\,\ref{FigNMR}(b) can then be interpreted as a
variation of strain-axis, within individual QDs or between different
QDs in the ensemble.

Finally we note that recent advances in optically detected NMR
spectroscopy in single dots have allowed an isotope sensitive
determination of hyperfine constants for holes coupled to nuclear
spins in both strained and unstrained dot systems
\cite{Chekhovich:2011c}.

\begin{figure}
\epsfysize=5in
\epsfbox{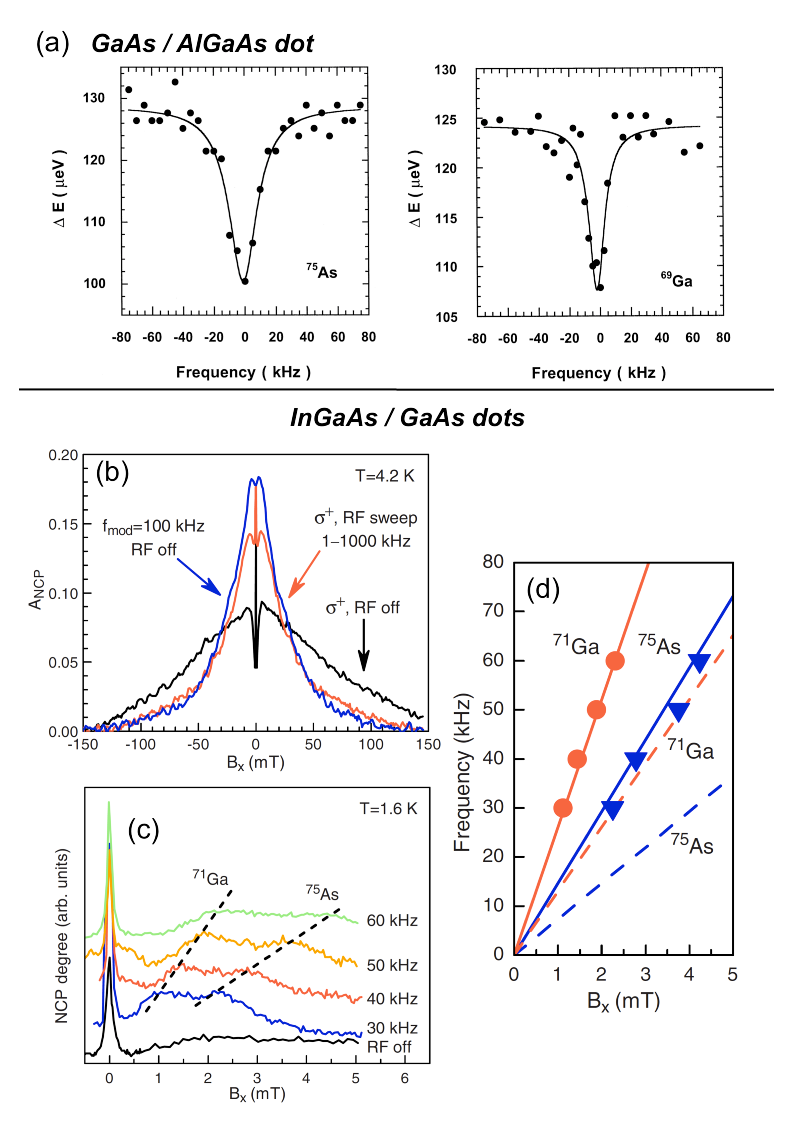}
\caption[ODNMR on semiconductor QDs]{\label{FigNMR} Optically
detected nuclear magnetic resonance in semiconductor QDs (a) NMR
spectra of a single interfacial, GaAs QD: The Zeeman splitting of
the QD in a longitudinal magnetic field of $1~$T changes as a
function of the frequency of an applied transverse RF magnetic field
when the frequency matches the nuclear Larmor precession frequency
(adapted from \textcite{Brown:1998a}). Frequency offset for $^{75}$As: 7.274~MHz, for $^{69}$Ga:10.193~MHz.
(b) Hanle curves measured on
an ensemble of self-assembled InAs QDs for $\sigma^+$-polarized
excitation in absence and presence of an RF field applied along the
z-axis. The radio-frequency is scanned over all the In and As
nuclear magnetic resonances for the relevant magnetic field range.
For comparison, the Hanle curve detected for excitation with
polarization modulation at $f_{\rm mod}=100~$kHz is shown.
(c) Effect of fixed radio-frequency irradiation (with RF field
along the y axis), which reveals distinct nuclear resonances for the
low-field range indicated by the dashed rectangle (curves are
shifted vertically for clarity). (d) Magnetic field dependencies of
resonant frequencies extracted from the inset in (a). The
dependencies of NMR for $m_I=\pm 1/2$ states in $^{71}$Ga and
$^{75}$As nuclei calculated with and without quadrupole interaction
are shown by the solid and dashed lines, respectively (adapted from
\textcite{Flisinski:2010a}). }\end{figure}

\subsection{Irreversibility and hysteresis in demagnetization experiments}
\label{sec:demag}

A remarkable feature of the nuclear-spin system of InAs/GaAs QD is
the excellent isolation from its environment if the QD is charged
with an even number of electrons and kept in the dark (see
Sec.\,\ref{sec:Dynamics} and \,\ref{sec:DynamicsBext}). In the specific case where all charge carriers are removed from the QD
after a sequence of optical DNP, the relaxation of the nuclear polarization turns
out to be extremely slow. Its characteristic time  is shown to
exceed one hour when the QD is subject to a longitudinal magnetic
field $B_\text{ext}$ in the 0-2~T  range, see Fig.\,\ref{FigVgSwitch} and \textcite{Maletinsky:2009a}.
This clearly  indicates that not only the dipolar relaxation but
also the nuclear spin diffusion towards the surrounding material are
strongly suppressed. Under this circumstance,  manipulating the
nuclear polarization  by slowly  varying  an external parameter like
the magnetic field is technically possible. Besides its intrinsic
interest it may also provide  valuable information on the
thermodynamics  of the nuclear-spin system in one QD.

While the nuclear polarization achieved under constant circularly
polarized optical excitation depends drastically on the magnetic
field (see Sec.\,\ref{sec:DNPvsBz}), if the QD is  isolated, under
conditions where relaxation is suppressed, its polarization should
evolve adiabatically as a function of a slowly varying  magnetic
field. In the absence of quadrupolar interaction, the nuclear spin
component $\parallel z$ is a good quantum number so that the change
of the external field should not produce any change of the nuclear
level  populations and the initial nuclear polarization should
remain constant. Only in magnetic fields  $\alt B_{L}$, the local
field induced by the dipolar interaction between nuclei, should the
polarization decrease and eventually vanish in strictly zero field \footnote{Note that for an empty dot (neither conduction electron nor valence hole present) the depolarization due to the indirect interaction of Eq.\,\ref{eq:hfind} as used to describe nuclear spin depolarization in fields of 5T in Sec.\,\ref{sec:DecmK} is not applicable as it requires the presence of charge carriers.}.
Yet, if the system is well isolated (no heat flow) its entropy does
not  change during the adiabatic demagnetization, so that applying
again a magnetic field  above $B_{L}$  restores in principle the
initial polarization. This thermodynamic approach corresponds to a
description of the system by a spin temperature $T_\text{s}$ which
is related to the nuclear polarization according to Curie's
law. Since the entropy of the QD spin ensemble depends only on the
ratio $(B_\text{ext}^2+ B_\text{L}^2)/T_\text{s}^2$
\cite{Slichter:1990a}, the adiabatic demagnetization leads to a
reduction (in absolute value) of $T_\text{s}$  proportional to the
field until $B_\text{ext}\approx B_\text{L}$.  Remarkably, when a
magnetic field above $B_\text{L}$ is re-applied   Curie's law
predicts that the polarization should be recovered and aligned
parallel to the field, independent of  its orientation,  in the same
or opposite direction depending only on the sign of $T_\text{s}$.
Such  reorientation of an initially created nuclear polarization
$\parallel z$ to a transverse direction was nicely evidenced in bulk
GaAs \cite{Meier:1984a}, demonstrating convincingly  the validity of
the spin temperature concept in this case. Clearly, in a
self-assembled QD with large quadrupolar interaction the
validity of the spin temperature concept requires further
consideration.

To address this issue in the case of InAs/GaAs QDs,
adiabatic demagnetization experiments have been performed in a
longitudinal magnetic field \cite{Maletinsky:2009a}. The  procedure
is sketched in Fig.\,\ref{LPN-Demag}(a). The nuclear spins of a
single QD are first polarized by optical pumping for a duration
$\tau_\text{pump}\approx600$~ms and in an initial field
$B_\text{i}=$2~T parallel to $z$. When the excitation is switched
off, the gate voltage of the sample is immediately changed to
uncharge the dot (to assure long nuclear spin relaxation times, as in Fig.\,\ref{FigVgSwitch}), and the magnetic field
is slowly varied  at a rate $\gamma_\text{B}=10$~mT~s$^{-1}$ to a
final value $B_\text{f}$. To read out the remaining polarization, an
electron is re-injected in the dot by applying the required  voltage
and a linearly-polarized optical pulse, short enough to not destroy
the DNP, is  used to measure the Overhauser shift from  the
$X^-$ trion PL. The normalized polarization
$P_\text{nuc}(B_\text{f})/P_\text{nuc}(B_\text{i})$ following the
demagnetization process $B_\text{i}\rightarrow B_\text{f}$ is
plotted in Fig.\,\ref{LPN-Demag}(b) with  $B_\text{f}$ going from
2~T ($B_\text{f}=B_\text{i}$) to -1~T. These measurements are
completed by re-magnetization experiments where the field is ramped
down and up according to $B_\text{i}\rightarrow
B'_\text{f}\rightarrow B_\text{f}$ with  $B'_\text{f}=-$1~T, in
order to probe the reversibility of the whole process. At first
glance, the system obeys satisfactorily the Curie's law: the
nuclear polarization is almost constant when the field decreases
from $B_\text{i}$ down to $\sim$0.3~T,  then decreases rapidly  and
changes sign  when the magnetic field passes through zero, and
eventually recover to a constant level for $B_\text{f}<-$0.3~T.
Significant discrepancies to the spin temperature model are yet
noticeable: there is a finite remnant nuclear polarization
$P_\text{nuc}^\text{rem}\simeq$0.2 at $B_\text{f}=0$~T, the field at
which the polarization starts decreasing or recovering  is much
higher than the dipolar field $B_\text{L}$, and the final
polarization recovered in negative fields amounts to only  60\%  of
its  initial magnitude. When the QD is re-magnetized towards
positive field a similar behavior is observed so that  a  complete
cycle evidences a drastic irreversibility.
Figure\,\ref{LPN-Demag}(d) shows that the irreversibility for a
round trip field ramp  $B_\text{i}\rightarrow B'_\text{f}\rightarrow
B_\text{i}$ develops  for  $|B'_\text{f}|<0.3$~T.

The strain-induced quadrupolar interaction of self-assembled QDs is
most certainly responsible for the observed disagreements to the
spin temperature description. On the one hand, the inhomogeneous
dispersion of the  principal  axis angle $\theta$ produces
anti-crossings of the nuclear levels  in fields of a few 100~mT, see
Fig.\,\ref{LPN-QI}(a,b). When the field  ramp passes through
anti-crossings of  energy splitting
$\hbar\omega_\text{S}\approx\hbar\theta\omega_\text{Q}$ between
states with $\Delta m=\pm1$  and such  that
$\omega_\text{S}\gg\sqrt{\gamma_n\gamma_\text{B}}\sim300$~Hz
(Landau-Zener criterion for adiabatic anti-crossing), the population
of each level is  conserved while the corresponding eigenstates are
progressively exchanged. One ends up  with an effective exchange of
the actual populations and therefore an inversion of their relative
polarization, e.g. $p_{+3/2}\rightleftarrows p_{+1/2}$ for $B_z$
around $B_\text{Q}$ when the $\left|+3/2\right>$ and
$\left|+1/2\right>$ states anticross. This explains why the
polarization starts decreasing in fields around 0.3~T. On the other
hand, the quadrupolar splitting quenches the effect of the dipolar
interaction in zero field for states $|m|>1/2$. The typical dipolar
splitting of $\left|\pm1/2\right>$ states $\hbar\gamma_n B_\text{L}$
is reduced by factors $(B_\text{L}/B_\text{Q})^{2m-1}\ll1$ for those
states, so that the randomization of their polarization is
suppressed yielding a finite  $P_\text{nuc}^\text{rem}$ in zero
field. The latter  is noticeably larger than that  directly created
in zero field (see Sec.\,\ref{sec:zerofield}), but obviously
decreases with the magnitude of the initial  polarization when the
initial field $B_\text{i}$  decreases (Fig.\,\ref{LPN-Demag}(c)).
Actually the pairs of  levels $\left|\pm m\right>$ with $|m|>1/2$ do
cross in zero field in the sense that their splitting $\ll
\sqrt{\gamma_n\gamma_\text{B}}$.
Therefore and in contrast to the $\vert \pm 1/2 \rangle $ states, their polarization is conserved without changing sign when the field passes through zero.

\begin{figure}[h]
\centering
\includegraphics*[width=0.48 \textwidth,angle=0]{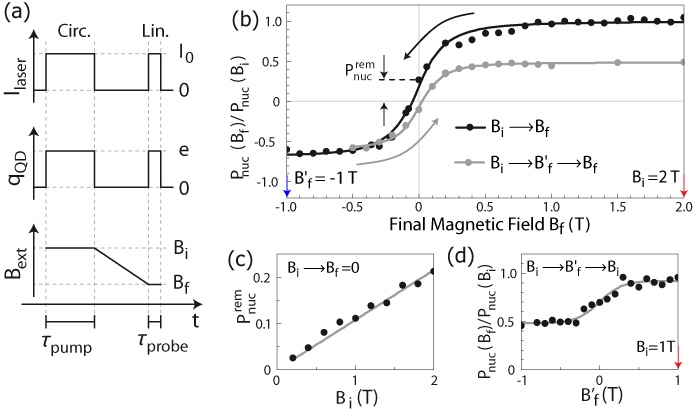}
\caption{(a). Schematic diagram of the experimental procedure for
adiabatic demagnetization of QD nuclear  spins in an charge tuneable
InGaAs/GaAs QD. The nuclei are optically pumped at
$B_\text{ext}=B_\text{i}$ via  quasi-resonant excitation of $X^-$
trion while the QD is in   charge state $q_\text{QD}=1 e$.
Immediately after the pumping pulse, the electron is ejected from
the QD. $B_\text{ext}$ is then linearly ramped at a rate
$\gamma_\text{B}$  to a final value $B_\text{f}$, at which the
nuclear polarization $P_\text{nuc}$ is measured  with a short
linearly polarized optical pulse. (b) Demagnetization as a function
of the final magnetic field $B_\text{f}$, for a monotonic decrease
from $B_\text{i}=$2~T (dark) and for a return from an intermediate
field $B'_\text{f}=$-1~T (gray). (c) Remanent polarization at zero
field  normalized to the polarization generated at 2~T, when
$B_\text{i}$ is varied. (d) Nuclear polarization following a
complete round trip  as indicated. (adapted from
\textcite{Maletinsky:2009a})}\label{LPN-Demag}
\end{figure}

The combination of these polarization-conserving crossings and
polarization-inverting anti-crossings is mostly  responsible for
the   reduction  of the  polarization which is recovered  in fields
$<$-0.3~T. However, the pronounced irreversibility observed when the
QD is   remagnetized  can not be explained by these processes which
are perfectly reversible. A specific mechanism leading to the
increase of the system entropy is required to explain this
observation. In \textcite{Maletinsky:2009a}, it has been suggested that
cross-relaxation between pairs of nuclear spins should be
dramatically enhanced at specific fields where a harmonic
configuration of any three nuclear levels is created \cite{Abragam:1961a}. Efficient
thermal relaxation involving a coupling to the environment (in
particular to the unpolarized nuclei surrounding the QD) would then
make these crossover transitions irreversible.  This interpretation
was shown to agree well with a numerical simulation of the demagnetization  and
re-magnetization curves relying on the  adiabatic evolution of the
populations as discussed above, but  including this irreversible
process \cite{Maletinsky:2009a}.

In addition, one could question if all 2-level crossings or anti-crossings
of a given nucleus are truly adiabatic. Assuming an average  dipolar local field
$B_L\sim0.15$~mT,  the zero field splitting of $\left|\pm m\right>$
states, $\gamma_n B_\text{L}\sim$1~kHz,  is roughly of the same
order as  $\sqrt{\gamma_n\gamma_\text{B}}$ at the ramping rate
$\gamma_\text{B}$ of 10mT~s$^{-1}$. As a result, the
adiabaticity condition is probably not  verified for all the
nuclei in a magnetic field ramp through zero.  Similarly, for Indium
nuclei ($I$=9/2) the quadrupolar interaction gives rise to many
different anti-crossings at  different magnetic fields; their
typical splittings cover several orders of magnitude because for
small $\theta$ they scale as $\hbar\omega_\text{Q}\theta^{|\Delta
m|}$ where $|\Delta m|\in\{1,\ldots,8\}$  is the difference of
angular momentum. Taking into account in addition the  dispersion of
$\theta$ in a QD, it seems very likely that  for a large
fraction of nuclei in a QD there is at least one pair of
anti-crossing levels such that $\omega_\text{Q}\theta^{|\Delta
m|}\sim \sqrt{\gamma_n\gamma_\text{B}}$, which would result in
irreversibility. The fact that the observed irreversibility  (see
Fig.\,\ref{LPN-Demag}(d)) develops essentially in the field region
where these quadrupole-induced anti-crossings take place, rather
supports the above idea, although we note that in these experiments changes in the magnetic field ramp speed did not cause any noticeable effect on the experimental results.


\section{HOLE SPINS COUPLED TO NUCLEAR SPINS}
\label{sec:hole}

The interaction between valence hole spins and nuclear spins has usually
been ignored in semiconductors for two main reasons.
First, the $p$-symmetry of the
periodic part of the valence Bloch wave function results in
negligible overlap with the nuclear spins yielding vanishing Fermi
contact interaction. Therefore $\hat{H}_{hf}^{fc}$ of Eq\,\ref{eq:eqHf}, that is at the origin of the hyperfine interaction discussed for conduction electrons in sections \ref{sec:elecdynamics}, \ref{sec:dnp} and \ref{sec:tdnp} does not apply. Second, the hole
spin in bulk or quantum well structures is very fragile: the hole
spin relaxation time is of the order of 10 ps or less in bulk
GaAs due to strong heavy-light hole mixing in the valence bands
\cite{Meier:1984a}, which leads to a correlation time for the
hyperfine interaction that is too short to achieve a significant
dynamic nuclear polarization.  In QDs the hole spin is much more
robust than in bulk or quantum well structures because of the
discrete energy states
\cite{Heiss:2007a,Flissikowski:2003a,Laurent:2005a} and significant
effects linked to the hole-nuclear spin interaction have been
revealed.

The hyperfine interaction of nuclear spins with an electron in the
valence band  is primarily dipolar. For a given nucleus, the
Hamiltonian of this interaction reads \cite{Abragam:1961a}:
\begin{equation}
\label{eq:dipabra}
\hat{H}_\text{hf}^\text{dip}=2\mu_B\frac{\mu_I}{I}\hat{\bm{I}}.\left[\frac{\hat{\bm{L}}}{\rho^3}-\frac{\hat{\bm{S}}}{\rho^3}+3\frac{\bm{\rho}(\hat{\bm{S}}.\bm{\rho})}{\rho^5}\right]
\end{equation}
where $\mu_B$ is the Bohr magneton and $\mu_I$ is the nuclear magnetic moment; $\hat{I}$ is the nuclear spin operator; $\bm{\rho}$ is the electron position vector with respect to the nucleus; $\hat{L}$ and $\hat{S}$ are the electron orbital momentum and spin operators, respectively. The dipolar hyperfine interaction depends on the valence band mixing $\beta$, which quantifies the deviation of the confined hole from the ideal, pure heavy hole case. In a simple picture, for a heavy hole $J_z=3/2$ to do a flip-flop with a nucleus ($\Delta J_z=-1$), it would have to access the $J_z=1/2$ light hole state. This is energetically forbidden as heavy and light holes are separated in energy by $\Delta_{HL}$. Valence band mixing is thus required to circumvent this blockade. The valence band mixing $\beta$ may arise from anisotropic strain fields within the QD and/or shape and interface anisotropy \cite{Krizhanovskii:2005a,Belhadj:2010a,Krebs:1996a}. This makes it necessary to consider the mixed hole states by including the light hole component as in:
\begin{equation}
\label{valencemixing}
|\widetilde{\pm 3/2}\rangle=\frac{1}{\sqrt{1+|\beta|^{2}}}\left( |\pm 3/2\rangle+\beta |\mp 1/2\rangle\right)
\end{equation}
where $|\beta|$ should be much smaller than unity, as a direct
consequence of the energy splitting $\Delta_{HL}$ between  heavy and
light hole states (tens of meV in InAs/GaAs dots). The dipolar hole
spin-nuclear spin hamiltonian can then be written as:
\begin{equation}
\label{dipprecise}
\hat{H}_\text{hf}^\text{dip}=\nu_0 \sum_{j}  \frac{A^h_j}{1+\beta^2}  \vert\psi(\bm{r}_j)\vert^2 \left(\hat{I}_z^j\hat{S}_z^h+ \frac{\alpha}{2} [\hat{I}_+^j\hat{S}_-^h+\hat{I}_-^j\hat{S}_+^h]\right)
\end{equation}
where $\alpha=\frac{2|\beta|}{\sqrt{3}}$ is the anisotropy factor
\cite{Fischer:2008a,Testelin:2009a}. In the case of moderate heavy-hole light-hole  mixing and assuming a
constant wavefunction $\psi(\bm{r})=\sqrt{2/N\nu_0}$ an approximate,
simplified hamiltonian is obtained:
\begin{equation}
\label{dipolarhamiltonian}
\hat{H}_\text{hf}^\text{dip}=\frac{2\tilde{A}^h}{N}\left( \hat{I }_{z}\hat{S}_{z}^{h}+\alpha\left[ \frac{\hat{I}_{+}\hat{S}_{-}^{h} +\hat{I}_{-}\hat{S}_{+}^{h}  }{2}  \right]     \right).
\end{equation}
Here $\tilde{A}^h$ is the average value of the dipole-dipole
hyperfine constant \cite{Gryncharova:1977a,Testelin:2009a},
$\hat{S}_{\pm}^{h}$ denote the heavy hole pseudo-spin $1/2$
operators acting on $\tilde{|\frac{3}{2}}\rangle$.

The Hamiltonians $\hat{H}_\text{hf}^\text{fc}$ for electrons and
$\hat{H}_\text{hf}^\text{dip}$ for holes have a similar form (compare
Eq.\,\eqref{eq:eqHf1} and Eq.\,\eqref{dipolarhamiltonian}), but
there are important differences: \\
\textbf{(i)} the ratio $|\tilde{A}^h| /
\tilde{A}$ is about $\simeq 0.1$, as has been theoretically predicted by \cite{Fischer:2008a} followed by experimental demostrations as in  
Fig.\,\ref{decoherence}(c) and
Fig.\,\ref{holeDNP}
\textcite{Eble:2009a,Fallahi:2010a,Chekhovich:2011a,Testelin:2009a};\\
\textbf{(ii)} the amplitude of the flip-flop term is proportional to
heavy-hole light-hole mixing, which varies from dot to dot.

\begin{figure}
\epsfxsize=2.5in
\epsfbox{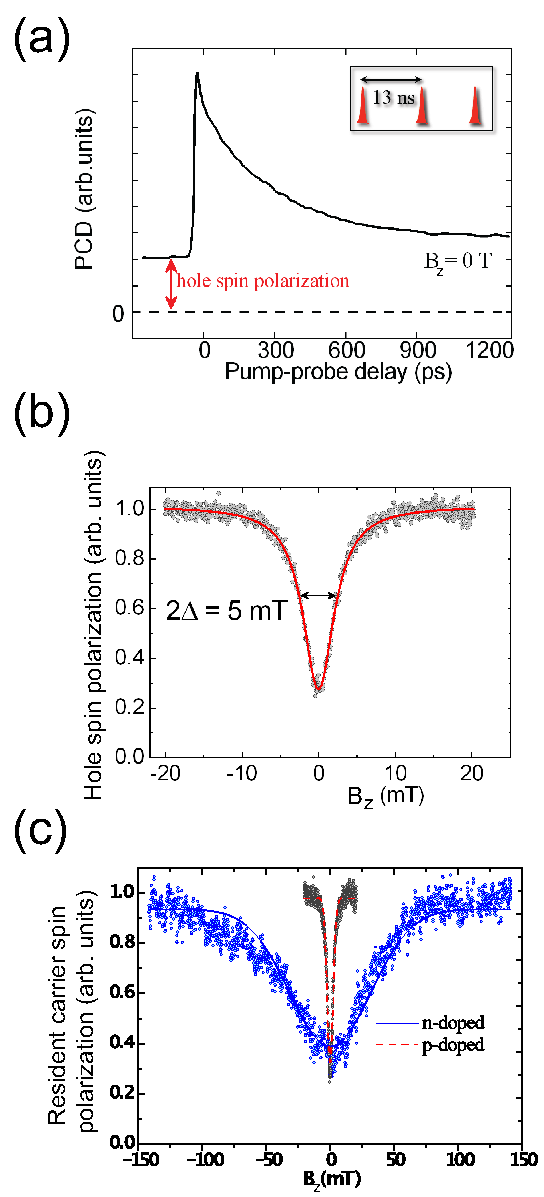}
\caption{(a) photo-induced circular dichroism (PCD) signal as a function of pump-probe delay, when the p-doped QDs are excited by a periodic train of ultra-short pulses. (b) normalized PCD amplitude at negative pump-probe delay $t=-130$~ps (i.e. reflecting the \textbf{hole} spin polarization) versus the applied longitudinal magnetic field $B_z$, performed on a QD sample with \textbf{one hole per dot}. The HWHM equals 2.5~mT. The solid line is a fit using the model developed in \textcite{Eble:2010a} .  (c) blue curve: normalized PCD  amplitude at negative pump-probe delay $t=-130$~ps (i.e. reflecting the \textbf{electron} spin polarization) versus the applied longitudinal magnetic field, performed on QDs \textbf{charged with one electron}. The HWHM equals 47~mT. Similar measurement on QDs charged with one hole (red curve) is added on the same figure to directly compare the efficiency of hyperfine induced dephasing for electrons and holes \cite{Desfonds:2010a}.
 \label{decoherence}}
\end{figure}

In the framework of the simple quantum box model, the strength of
the flip-flop process $\tilde{A}$ for electrons directly depends on
the materials that form the QD, whereas the dipolar
flip-flop term for holes not only depends on the chemical
composition of the dot, but also on the valence band mixing $\beta$,
which takes into account fine details of the real QD
spatial strain distribution and shape anisotropy. In recent
measurements, $\tilde{A}^h$ was found to be negative for In and Ga,  and
positive for As in different dot systems made out of III-V
compounds \cite{Chekhovich:2011c}, the origin of this surprising finding is still under discussion.

\subsection{Hole spin dephasing due to hyperfine interaction}
\label{sec:holedeco}

As the hyperfine interaction with the fluctuating nuclear field is
severely limiting electron spin coherence, the hole spin with
negligible Fermi contact hyperfine interaction seems a more
promising candidate for achieving long spin coherence times. With
this motivation in mind, detailed investigations led to promising
reports \cite{Gerardot:2008a,Brunner:2009a} under certain
experimental conditions. On the other hand,  surprisingly
short hole spin dephasing times due to nuclear spin fluctuations
were also reported \cite{Eble:2009a}. To clarify the limitations for
hole spin coherence, an the of hole spin dynamics in the model
system of singly positively charged InAs QDs is discussed
below, revealing that dipole-dipole hyperfine interaction does play
an important role in decoherence of the resident hole spin at cryogenic
temperatures.

Figure\,\ref{decoherence}(a) shows  the photoinduced circular
dichroism (PCD) signals versus pump-probe delay, obtained when no
magnetic field is applied in an ensemble of p-doped QDs containing a
resident hole. The experiments are performed under modulated
$\sigma^+/\sigma^-$ helicity of the excitation laser pulse
($f_\text{mod}=50$~KHz) in order to avoid any dynamic nuclear spin
polarization (here $\langle\hat{I}_z\rangle=0$). A nonzero PCD
signal at negative pump-probe delays is observed, indicating that
the spin polarization is not fully relaxed within the $T_{L}=13$~ns
repetition period of the laser pulses. This long-living component of
the PCD signal is directly associated with the net spin polarization
of the \textit{resident} holes, the only species present in the
sample after the radiative recombination of $X^+$ trions
($\tau_{r}\sim800$~ps). From this experimental observation, one can
understand how the spin polarization of the resident holes in the
dot is built-up by a resonant pulsed excitation, in the following
sequential way : (i) the $\sigma^{+}$ circularly polarized pump beam
photo-creates spin oriented trions $X^+$ with electron spin
$\downarrow$. (ii) During the lifetime of the excited states, the
efficient electron-nuclear hyperfine interaction leads to a coherent
coupling of their spin projections along the light direction. (iii)
Finally, the spontaneous decay of the trion states by emission of
polarized photons leads to an unbalanced hole-spin population with
$\rho^{h}_{\Uparrow}>\rho^{h}_{\Downarrow}$ (as in the hole spin pumping scheme of Fig.\,\ref{xmfig10}(a)).

Figure\,\ref{decoherence}(b) shows the experimental data for the PCD
signal at  negative delays, PCD$(0^{-})$, i.e. the hole-spin
polarization at $13~$ns, as a function of the applied magnetic
field. The application of a small magnetic field $B_z$ in the milli-Tesla
range has a dramatic impact on the hole spin polarization. The
Lorentzian -like dependence  with a half width at half maximum
(HWHM) of only $2.5~$mT is interpreted as the progressive magnetic
field quenching of the hyperfine-induced hole spin relaxation, just
as in the case of the electron (Fig.\,\ref{xmfig2}(a)), but at much
lower field. For B=0 the hole spin dephasing time due to the
interaction with nuclear spins is of the order of $10~$ns
\cite{Eble:2009a}.

To give a first interpretation that only  involves hyperfine-induced
dephasing it is convenient to treat, $\hat{H}_\text{hf}^\text{fc}$ and
$\hat{H}_\text{hf}^\text{dip}$ from Eq.\,\eqref{dipolarhamiltonian} (see
\textcite{Merkulov:2002a} and \textcite{Testelin:2009a}), as
semi-classical magnetic fields randomly distributed from dot to dot.
The orientation of the nuclear field fluctuations $\delta B_n$
responsible for electron spin dephasing is \textit{isotropic}.
In stark contrast, $\delta B_n^h$ is highly \textit{anisotropic} and
the corresponding Gaussian distribution of the nuclear field $B_n^h$
acting on holes at zero average nuclear field is:
\begin{equation}
\label{eq:bnhgauss}
W(B_{n}^{h})\propto\exp\left(-\frac{\left(B^{h}_{n,z}\right)^{2}}{2(\delta B_{n,{\parallel}}^h)^2}\right)\exp\left(-\frac{\left(B^{h}_{n,x}\right)^{2}+\left(B^{h}_{n,y}\right)^{2}}{ 2(\delta B_{n,{\perp}}^h)^2 }\right)
\end{equation}
where $\delta B_{n,{\perp}}^h=\alpha \delta B_{n,{\parallel}}^h$ and
$\delta B_{n,{\perp}}^h=\hbar/(g_{h}^{z}\mu_{B}T_{\Delta}^{h})$, directions $\perp$ and $\parallel$ with respect to the $z$-axis (i.e. quantization and light propagation axis).
$T_\Delta^h$ is the ensemble spin dephasing time, arising from the
random hole precession directions and frequencies in the randomly
distributed frozen nuclear field. Following the same approach as
\textcite{Merkulov:2002a}, adapted for the anisotropic interaction of hole spins with nuclei by 
\textcite{Testelin:2009a}, the decay of the z-component of the hole pseudo-spin is given by \footnote{Note that in our simple approach this time diverges for $\beta\rightarrow0$, but the dephasing time is still finite in this limit, see theory of \textcite{Fischer:2008a} and recent experiments \cite{Fras:2011a}.}:
\begin{equation}
\label{eq:tdeltah}
T_\Delta^h=\hbar\frac{1+\beta^2}{2\beta/\sqrt{3}}\left(\frac{3N}{4\sum_{\varepsilon} I^{\varepsilon}(I^{\varepsilon}+1)(A_j^h)^2}\right)^{1/2}
\end{equation}
where $\varepsilon=$ In, As or Ga. Using a heavy-light hole
mixing characterized by $|\beta|=0.4$, this equation yields a hole
spin dephasing time of the order of 10 ns, comparable to the one
which has been measured in the PCD experiment. As the electron
-nuclear spin dephasing time can be measured in the same p-doped
sample by photoluminescence dynamics (see Fig.\,\ref{xmfig2}(a)),
the ratio between the electron and hole nuclear spin dephasing time
can be extracted: it is found for these dots that the dephasing
time of the hole with nuclear spins is about ten times longer than
the dephasing time of the electron spins, but it is not negligible.
Note that the exact values of $|\beta|$ are extremely dot and sample
dependent. The heavy hole - light hole mixing depends on the exact
dot shape and strain which can both vary significantly from dot to
dot, even in the same sample. $|\beta|=0.4$ used to fit experiments
as in Fig.\,\ref{decoherence}(b)  corresponds to a very strong
mixing in the sample investigated in \textcite{Eble:2009a}, whereas
for dots of the same material used in other studies
\cite{Gerardot:2008a,Fallahi:2010a} the mixing deduced from
analysing optical pumping experiments is significantly
smaller.\\
\indent PCD measurements also allow revisiting the study performed on
electron spin dynamics by PL experiments.
Figure\,\ref{decoherence}(c) presents the PCD evolution versus
longitudinal magnetic field performed on an n-doped QD
ensemble. The interaction between nuclei and the \textit{electron}
takes place in the ground state of the QDs \textit{after}
$X^-$ recombination. The external magnetic field can efficiently
cancel the effect of the hyperfine interaction on the carrier-spin
dephasing time if its magnitude becomes larger than the
corresponding effective nuclear field fluctuation. This field is of
the order of several tens of mT for electrons (see
Fig.\,\ref{xmfig3} and Fig.\,\ref{xmfig4} of section \ref{sec:dephfara}), as can be seen in the comparison with the
values for holes  shown in Fig.\,\ref{decoherence}(c)
\cite{Desfonds:2010a}: the relative strength of the hole to electron
hyperfine coupling deduced from these measurements is on the
order of $10\%$. In summary, the hole spin confined to a QD
\textit{does} interact with the fluctuating nuclear spins through
dipole-dipole coupling. The strength of this interaction depends on
the degree of  heavy hole-light hole mixing and is a significant
source of decoherence. \\
\indent There are still open questions concerning the exact link between valence band mixing and the transverse components of the hyperfine interaction: A strong in-plane hyperfine field for holes and the resulting short hole spin coherence time is a priori linked to strong heavy hole-light hole mixing \cite{Testelin:2009a}. Conversely, measuring strong heavy hole-light hole mixing by analysing the transverse hole g-factor, for example, does \textit{not} automatically result in short hole spin decoherence times for the QDs investigated \cite{Brunner:2009a}. This underlines that hole spin dephasing due to the hyperfine interaction is a very recent research topic \cite{Fischer:2008a,Eble:2009a}, that will surely stimulate further innovative theoretical and experimental investigations \cite{Godden:2012a,Greilich:2011a,DeGreve:2011a}.

\subsection{Overhauser effect for holes}
\label{sec:holeover}

\begin{figure}
\epsfxsize=3in
\epsfbox{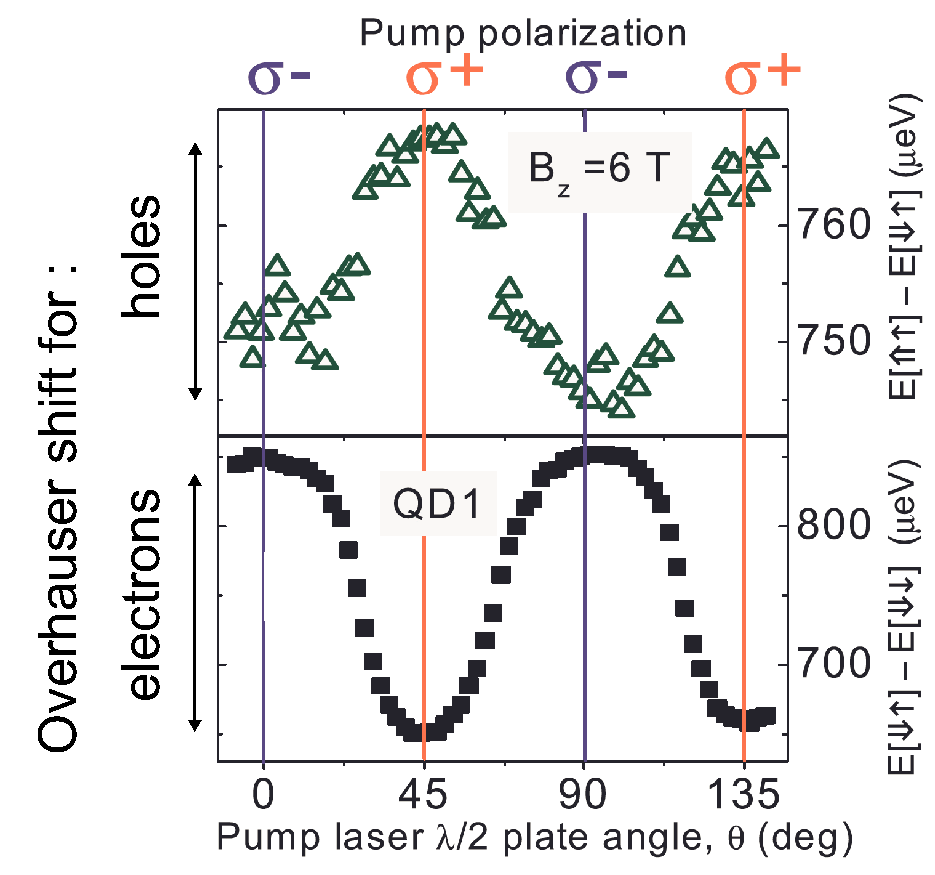}
\caption{Measurement of the electron- and hole nuclear spin interaction in a neutral InP dot at $B_z = 6~$T.  The angle  of a $\lambda/2$ plate is varied to change the polarization of the pump laser resulting in a change of nuclear spin polarization \cite{Chekhovich:2011a}.
 \label{holeDNP}}
\end{figure}

In addition to a spin dephasing contribution,  the non-zero
interaction between hole and nuclear spins can also lead to the
observation of Overhauser effect for holes. An existing nuclear
polarization can split pure heavy hole spin states (Overhauser
effect) due to the Ising term
$\frac{2\tilde{A}^h}{N}\hat{I}_{z}\hat{S}_{z}^{h}$  in
Eq.\,\eqref{dipolarhamiltonian} (see
\textcite{Fischer:2008a,Fallahi:2010a,Chekhovich:2011a}).

The dynamic nuclear polarization created through  non-resonant
excitation of an InP dot was monitored by \citealp{Chekhovich:2011a}
via the emission of the neutral exciton bright states $\left|
\Uparrow \downarrow \right>$ and $\left|  \Downarrow \uparrow
\right>$ and the dark states $\left|  \Uparrow \uparrow \right>$ and
$\left|  \Downarrow \downarrow \right>$. Here, the dark
exciton states are weakly optically active due to heavy hole - light
hole mixing. It is therefore possible to compare for the same dot
the effect of a finite nuclear polarization on (i) the electron spin
by measuring the energy difference between the $\left|  \Downarrow
\uparrow \right>$ and $\left|  \Downarrow \downarrow \right>$ state
and (ii) very interestingly, the hole spin by measuring the energy
difference between the $\left|  \Uparrow \uparrow \right>$ and
$\left| \Downarrow \uparrow \right>$. Figure\,\ref{holeDNP} shows
directly, that the hyperfine constant in the investigated InP QDs is about a factor of 10
stronger for electrons than for holes, and the signs are opposite
(note the different vertical scales). The relative strength of the
hyperfine interactions as well as the relative sign of the electron
and hole Overhauser effect can vary for different nuclear spin
species \cite{Chekhovich:2011c}. In InAs dots the strength and the
sign of hyperfine interaction of the heavy hole with nuclear spins
has also been measured \cite{Fallahi:2010a}). By using an experiment
based on the locking of the QD resonance to the incident
laser frequency (see Sec.\,\ref{sec:locking}), it has been possible
to measure very accurately the Overhauser shift due to electron or
exciton (including both electron and hole contributions). A ratio
$\simeq -0.1$ of the heavy-hole and electron hyperfine interaction
has been deduced a value very similar to the one measured in InP
dots \cite{Chekhovich:2011a}. In both of these experiments the hole
feels the nuclear field which has been dynamically created. The
relative contribution of electron and hole spins to DNP is an open
issue \cite{Xu:2009a,Yang:2010a}, as in most optical experiments a
single spin of either species is present in the dot at some stage of
the absorption-emission cycle. As $\tilde{A}\gg\tilde{A}^h$, DNP due
to the electron will most likely dominate when electrons are
present during some stage in the dot.

\section{PERSPECTIVES}

The physics reviewed in this article has progressed immensely over
the past years due to the fruitful exchange of ideas from distinct
scientific communities working on nuclear magnetism, electron spin
physics in nanostructures, quantum optics and quantum dot photonics.
For the latter, experiments without applied magnetic fields are
important for potential applications. Here for example the impact of
the nuclear spin bath on the carrier spin polarization and hence
emitted photon polarization via the optical selection rules
has recently been shown to play an important role in the
context of entangled photon pair emission \cite{Stevenson:2011a}.
Even at $B_z=0$ the screening of the nuclear field fluctuations
$\delta B_n$ by an optically created Overhauser field $B_n$ has been
shown to allow for tuning of the polarization states of both charged
\cite{Lai:2006a} and neutral excitons \cite{Belhadj:2009a}. Here the
exact interplay between the Knight field $B_K$ and  quadrupolar
effects due to strain and alloy disorder that allows the creation of
nuclear polarization is yet to be clarified. A promising route will
be the investigation of strain free systems such as GaAs/AlGas
droplet dots for which quadrupolar effects are far less important.
This latter system has the additional advantage of containing dots
truly isolated from each other as no wetting layer is formed under
certain growth conditions \cite{Sallen:2011a} permitting studies of
nuclear spin diffusion across the AlGaAs barrier
\cite{Malinowski:2001a} and comparison with the physics of GaAs dots
formed due to interface fluctuations in GaAs/AlGaAs quantum wells
\cite{Gammon:2001a}.

One challenge is to prolong carrier spin dephasing times in quantum
dots  by eliminating the effects of the fluctuating nuclear field
$\delta B_n$ without changing its mean value. Here sophisticated
spin echo techniques in both quantum dot optics \cite{Press:2010a}
and transport \cite{Bluhm:2010b} are promising routes to decouple
the electron spin from the nuclear spin bath. In electron spin
resonance experiments again the interplay between electron spin
dephasing and dynamic nuclear polarization can be expected to give
rise to surprising, non-linear behavior and memory effects
\cite{Kroner:2008a}.

An alternative approach is to try to polarize as many nuclear spins
as possible to finally achieve a net reduction in the nuclear spin
fluctuations $\delta B_n$ once the nuclear polarization is
approaching 100\%. So far the maximum nuclear polarization achieved
in different material systems is about 60\%  \cite{Bracker:2005a,
Urbaszek:2007a,Chekhovich:2010a}. Future experiments should clarify
if this nuclear polarization presents some fundamental limit as dark
nuclear states might form \cite{Imamoglu:2003a} or if complete
nuclear spin polarization is accessible in experiments in dots where
each lattice nucleus has non-zero nuclear spin \cite{Feng:2007a}. In
the regime of complete nuclear spin polarization, the possibility of
using the long lived, well isolated nuclear spin system as a quantum
memory \cite{Taylor:2003a,Witzel:2007a}) can be explored.

Another exciting approach is based  on the recent observation of the
locking of the QD resonance to an optical
\cite{Latta:2009a,Xu:2009a} or microwave driving
\cite{Vink:2009a} field (see Sec.\,\ref{sec:beyond}). Here the
achieved dynamical polarization is far below $100$\% but the
amplitude of the nuclear field fluctuations $\delta B_n$ can be
reduced significantly under certain conditions. Further reduction
of the nuclear field fluctuations are the motivation of further
experimental and theoretical work
\cite{Yang:2010a,Hogele:2012a,Issler:2010a} that aims to clarify the
exact nature of the interplay between Fermi-contact and
dipole-dipole type interactions for electrons and holes with nuclear
spins.

The mesoscopic nuclear spin system  of one QD coupled to a
single electron spin may be used to study exciting cooperative effects such
as phase transitions \cite{Kessler:2010a,Kessler:2012a}. The reason for the abrupt
collapse of the nuclear polarization experienced by an electron in a
transverse magnetic field shown in Fig.\,\ref{LPN-AHE1} is not
understood \cite{Krebs:2010a} and one could speculate that
collective phenomena  play a role. If this is realistic for quantum
dots with strong quadrupolar effects has to be clarified, ideally
trying to do similar experiments in transverse fields in dots with
vanishing quadrupolar effects
\cite{Gammon:2001a,Belhadj:2008a}.

The emission of polarized photons as a  result of the radiative
recombination of electrons with well defined spin and (unpolarized)
holes is at the heart of an important type of device called
Spin-LEDs, often with QDs in the active region
\cite{Asshoff:2011a,Lombez:2007b,Li:2005a}. The carriers are
injected electrically, the electrons spin orientation is assured by the passage through a ferromagnetic
contact. Due to the robustness of the electron spin even this highly
non-resonant carrier injection leads to substantial electron spin
polarization in the QD ground state and first promising
results show that this carrier polarization leads in turn to
measurable Overhauser fields \cite{Asshoff:2011b}, paving an
alternative way for electrical control of nuclear spin polarization.

The fascinating physics of one carrier spin coupled  to a mesoscopic
nuclear spin ensemble has been revealed through optical
investigations of semiconductor QDs. Very sophisticated
transport measurements on gate defined QDs have provided an
alternative approach to study the coupled spin systems, resulting in
an impressive level of control of spin coherence and relaxation (for
example
\textcite{Hanson:2007a,Bluhm:2010b,Petta:2005a,Takahashi:2011a}). A
promising approach for the future might be to combine fast and
convenient optical techniques to manipulate spins in high quality
gate defined QDs.

\section*{Acknowledgments}

We would like to thank ANR QUAMOS, ITN SPINOPTRONICS and NCCR
Nanoscience for support. We thank Mete Atature, Manfred Bayer,
Benoit Eble, Khaled Karrai, Martin Kroner, Chih-Wei Lai, Alexander
Tartakovksii and Richard Warburton for many fruitful discussions.

\bibliographystyle{apsrmp}

\begin{thebibliography}{203}
\expandafter\ifx\csname natexlab\endcsname\relax\def\natexlab#1{#1}\fi
\expandafter\ifx\csname bibnamefont\endcsname\relax
  \def\bibnamefont#1{#1}\fi
\expandafter\ifx\csname bibfnamefont\endcsname\relax
  \def\bibfnamefont#1{#1}\fi
\expandafter\ifx\csname citenamefont\endcsname\relax
  \def\citenamefont#1{#1}\fi
\expandafter\ifx\csname url\endcsname\relax
  \def\url#1{\texttt{#1}}\fi
\expandafter\ifx\csname urlprefix\endcsname\relax\def\urlprefix{URL }\fi
\providecommand{\bibinfo}[2]{#2}
\providecommand{\eprint}[2][]{\url{#2}}

\bibitem[{\citenamefont{Abragam}(1961)}]{Abragam:1961a}
\bibinfo{author}{\bibnamefont{Abragam}, \bibfnamefont{A.}},
  \bibinfo{year}{1961}, \bibinfo{journal}{Oxford University Press} .

\bibitem[{\citenamefont{Akimov} \emph{et~al.}(2006)\citenamefont{Akimov, Feng,
  and Henneberger}}]{Akimov:2006a}
\bibinfo{author}{\bibnamefont{Akimov}, \bibfnamefont{I.~A.}},
  \bibinfo{author}{\bibfnamefont{D.~H.} \bibnamefont{Feng}}, and
  \bibinfo{author}{\bibfnamefont{F.}~\bibnamefont{Henneberger}},
  \bibinfo{year}{2006}, \bibinfo{journal}{Phys. Rev. Lett.}
  \textbf{\bibinfo{volume}{97}}(\bibinfo{number}{5}), \bibinfo{pages}{056602}.

\bibitem[{\citenamefont{Akimov} \emph{et~al.}(2002)\citenamefont{Akimov, Hundt,
  Flissikowski, and Henneberger}}]{Akimov:2002a}
\bibinfo{author}{\bibnamefont{Akimov}, \bibfnamefont{I.~A.}},
  \bibinfo{author}{\bibfnamefont{A.}~\bibnamefont{Hundt}},
  \bibinfo{author}{\bibfnamefont{T.}~\bibnamefont{Flissikowski}}, and
  \bibinfo{author}{\bibfnamefont{F.}~\bibnamefont{Henneberger}},
  \bibinfo{year}{2002}, \bibinfo{journal}{Applied Physics Letters}
  \textbf{\bibinfo{volume}{81}}(\bibinfo{number}{25}), \bibinfo{pages}{4730}.

\bibitem[{\citenamefont{Akopian} \emph{et~al.}(2006)\citenamefont{Akopian,
  Lindner, Poem, Berlatzky, Avron, Gershoni, Gerardot, and
  Petroff}}]{Akopian:2006a}
\bibinfo{author}{\bibnamefont{Akopian}, \bibfnamefont{N.}},
  \bibinfo{author}{\bibfnamefont{N.~H.} \bibnamefont{Lindner}},
  \bibinfo{author}{\bibfnamefont{E.}~\bibnamefont{Poem}},
  \bibinfo{author}{\bibfnamefont{Y.}~\bibnamefont{Berlatzky}},
  \bibinfo{author}{\bibfnamefont{J.}~\bibnamefont{Avron}},
  \bibinfo{author}{\bibfnamefont{D.}~\bibnamefont{Gershoni}},
  \bibinfo{author}{\bibfnamefont{B.~D.} \bibnamefont{Gerardot}}, and
  \bibinfo{author}{\bibfnamefont{P.~M.} \bibnamefont{Petroff}},
  \bibinfo{year}{2006}, \bibinfo{journal}{Phys. Rev. Lett.}
  \textbf{\bibinfo{volume}{96}}(\bibinfo{number}{13}), \bibinfo{pages}{130501}.

\bibitem[{\citenamefont{Alen} \emph{et~al.}(2006)\citenamefont{Alen, Hogele,
  Kroner, Seidl, Karrai, Warburton, Badolato, Medeiros-Ribeiro, and
  Petroff}}]{Alen:2006a}
\bibinfo{author}{\bibnamefont{Alen}, \bibfnamefont{B.}},
  \bibinfo{author}{\bibfnamefont{A.}~\bibnamefont{Hogele}},
  \bibinfo{author}{\bibfnamefont{M.}~\bibnamefont{Kroner}},
  \bibinfo{author}{\bibfnamefont{S.}~\bibnamefont{Seidl}},
  \bibinfo{author}{\bibfnamefont{K.}~\bibnamefont{Karrai}},
  \bibinfo{author}{\bibfnamefont{R.~J.} \bibnamefont{Warburton}},
  \bibinfo{author}{\bibfnamefont{A.}~\bibnamefont{Badolato}},
  \bibinfo{author}{\bibfnamefont{G.}~\bibnamefont{Medeiros-Ribeiro}}, and
  \bibinfo{author}{\bibfnamefont{P.~M.} \bibnamefont{Petroff}},
  \bibinfo{year}{2006}, \bibinfo{journal}{Applied Physics Letters}
  \textbf{\bibinfo{volume}{89}}(\bibinfo{number}{12}), \bibinfo{eid}{123124}.

\bibitem[{\citenamefont{Amasha} \emph{et~al.}(2008)\citenamefont{Amasha,
  MacLean, Radu, Zumb\"uhl, Kastner, Hanson, and Gossard}}]{Amasha:2008a}
\bibinfo{author}{\bibnamefont{Amasha}, \bibfnamefont{S.}},
  \bibinfo{author}{\bibfnamefont{K.}~\bibnamefont{MacLean}},
  \bibinfo{author}{\bibfnamefont{I.~P.} \bibnamefont{Radu}},
  \bibinfo{author}{\bibfnamefont{D.~M.} \bibnamefont{Zumb\"uhl}},
  \bibinfo{author}{\bibfnamefont{M.~A.} \bibnamefont{Kastner}},
  \bibinfo{author}{\bibfnamefont{M.~P.} \bibnamefont{Hanson}}, and
  \bibinfo{author}{\bibfnamefont{A.~C.} \bibnamefont{Gossard}},
  \bibinfo{year}{2008}, \bibinfo{journal}{Phys. Rev. Lett.}
  \textbf{\bibinfo{volume}{100}}, \bibinfo{pages}{046803}.

\bibitem[{\citenamefont{Asshoff}
  \emph{et~al.}(2011{\natexlab{a}})\citenamefont{Asshoff, Merz, Kalt, and
  Hetterich}}]{Asshoff:2011a}
\bibinfo{author}{\bibnamefont{Asshoff}, \bibfnamefont{P.}},
  \bibinfo{author}{\bibfnamefont{A.}~\bibnamefont{Merz}},
  \bibinfo{author}{\bibfnamefont{H.}~\bibnamefont{Kalt}}, and
  \bibinfo{author}{\bibfnamefont{M.}~\bibnamefont{Hetterich}},
  \bibinfo{year}{2011}{\natexlab{a}}, \bibinfo{journal}{Appl. Phys. Lett.}
  \textbf{\bibinfo{volume}{98}}(\bibinfo{number}{11}), \bibinfo{pages}{112106}.

\bibitem[{\citenamefont{Asshoff}
  \emph{et~al.}(2011{\natexlab{b}})\citenamefont{Asshoff, W\"ust, Merz,
  Litvinov, Gerthsen, Kalt, and Hetterich}}]{Asshoff:2011b}
\bibinfo{author}{\bibnamefont{Asshoff}, \bibfnamefont{P.}},
  \bibinfo{author}{\bibfnamefont{G.}~\bibnamefont{W\"ust}},
  \bibinfo{author}{\bibfnamefont{A.}~\bibnamefont{Merz}},
  \bibinfo{author}{\bibfnamefont{D.}~\bibnamefont{Litvinov}},
  \bibinfo{author}{\bibfnamefont{D.}~\bibnamefont{Gerthsen}},
  \bibinfo{author}{\bibfnamefont{H.}~\bibnamefont{Kalt}}, and
  \bibinfo{author}{\bibfnamefont{M.}~\bibnamefont{Hetterich}},
  \bibinfo{year}{2011}{\natexlab{b}}, \bibinfo{journal}{Phys. Rev. B}
  \textbf{\bibinfo{volume}{84}}, \bibinfo{pages}{125302}.

\bibitem[{\citenamefont{Atature} \emph{et~al.}(2006)\citenamefont{Atature,
  Dreiser, Badolato, Hogele, Karrai, and Imamoglu}}]{Atature:2006a}
\bibinfo{author}{\bibnamefont{Atature}, \bibfnamefont{M.}},
  \bibinfo{author}{\bibfnamefont{J.}~\bibnamefont{Dreiser}},
  \bibinfo{author}{\bibfnamefont{A.}~\bibnamefont{Badolato}},
  \bibinfo{author}{\bibfnamefont{A.}~\bibnamefont{Hogele}},
  \bibinfo{author}{\bibfnamefont{K.}~\bibnamefont{Karrai}}, and
  \bibinfo{author}{\bibfnamefont{A.}~\bibnamefont{Imamoglu}},
  \bibinfo{year}{2006}, \bibinfo{journal}{Science}
  \textbf{\bibinfo{volume}{312}}(\bibinfo{number}{5773}), \bibinfo{pages}{551}.

\bibitem[{\citenamefont{Auer} \emph{et~al.}(2009)\citenamefont{Auer, Oulton,
  Bauschulte, Yakovlev, Bayer, Verbin, Cherbunin, Reuter, and
  Wieck}}]{Auer:2009a}
\bibinfo{author}{\bibnamefont{Auer}, \bibfnamefont{T.}},
  \bibinfo{author}{\bibfnamefont{R.}~\bibnamefont{Oulton}},
  \bibinfo{author}{\bibfnamefont{A.}~\bibnamefont{Bauschulte}},
  \bibinfo{author}{\bibfnamefont{D.~R.} \bibnamefont{Yakovlev}},
  \bibinfo{author}{\bibfnamefont{M.}~\bibnamefont{Bayer}},
  \bibinfo{author}{\bibfnamefont{S.~Y.} \bibnamefont{Verbin}},
  \bibinfo{author}{\bibfnamefont{R.~V.} \bibnamefont{Cherbunin}},
  \bibinfo{author}{\bibfnamefont{D.}~\bibnamefont{Reuter}}, and
  \bibinfo{author}{\bibfnamefont{A.~D.} \bibnamefont{Wieck}},
  \bibinfo{year}{2009}, \bibinfo{journal}{Phys. Rev. B}
  \textbf{\bibinfo{volume}{80}}(\bibinfo{number}{20}), \bibinfo{pages}{205303}.

\bibitem[{\citenamefont{Balasubramanian}
  \emph{et~al.}(2009)\citenamefont{Balasubramanian, Neumann, Twitchen, Markham,
  Mizuochi, Isoya, Achard, Beck, Tissler, Jacques, Hemmer, Jelezko}
  \emph{et~al.}}]{Balasubramanian:2009a}
\bibinfo{author}{\bibnamefont{Balasubramanian}, \bibfnamefont{G.}},
  \bibinfo{author}{\bibfnamefont{P.}~\bibnamefont{Neumann}},
  \bibinfo{author}{\bibfnamefont{D.}~\bibnamefont{Twitchen}},
  \bibinfo{author}{\bibfnamefont{R.}~\bibnamefont{Markham},
  \bibfnamefont{M.~Kolesov}},
  \bibinfo{author}{\bibfnamefont{N.}~\bibnamefont{Mizuochi}},
  \bibinfo{author}{\bibfnamefont{J.}~\bibnamefont{Isoya}},
  \bibinfo{author}{\bibfnamefont{J.}~\bibnamefont{Achard}},
  \bibinfo{author}{\bibfnamefont{J.}~\bibnamefont{Beck}},
  \bibinfo{author}{\bibfnamefont{J.}~\bibnamefont{Tissler}},
  \bibinfo{author}{\bibfnamefont{V.}~\bibnamefont{Jacques}},
  \bibinfo{author}{\bibfnamefont{P.~R.} \bibnamefont{Hemmer}},
  \bibinfo{author}{\bibfnamefont{F.}~\bibnamefont{Jelezko}}, \emph{et~al.},
  \bibinfo{year}{2009}, \bibinfo{journal}{Nature Materials}
  \textbf{\bibinfo{volume}{8}}, \bibinfo{pages}{383}.

\bibitem[{\citenamefont{Barnes and Economou}(2011)}]{Barnes:2011a}
\bibinfo{author}{\bibnamefont{Barnes}, \bibfnamefont{E.}}, and
  \bibinfo{author}{\bibfnamefont{S.~E.} \bibnamefont{Economou}},
  \bibinfo{year}{2011}, \bibinfo{journal}{Phys. Rev. Lett.}
  \textbf{\bibinfo{volume}{107}}, \bibinfo{pages}{047601}.

\bibitem[{\citenamefont{Baugh} \emph{et~al.}(2007)\citenamefont{Baugh,
  Kitamura, Ono, and Tarucha}}]{Baugh:2007a}
\bibinfo{author}{\bibnamefont{Baugh}, \bibfnamefont{J.}},
  \bibinfo{author}{\bibfnamefont{Y.}~\bibnamefont{Kitamura}},
  \bibinfo{author}{\bibfnamefont{K.}~\bibnamefont{Ono}}, and
  \bibinfo{author}{\bibfnamefont{S.}~\bibnamefont{Tarucha}},
  \bibinfo{year}{2007}, \bibinfo{journal}{Phys. Rev. Lett.}
  \textbf{\bibinfo{volume}{99}}, \bibinfo{pages}{096804}.

\bibitem[{\citenamefont{Bayer} \emph{et~al.}(2002)\citenamefont{Bayer, Ortner,
  Stern, Kuther, Gorbunov, Forchel, Hawrylak, Fafard, Hinzer, Reinecke, Walck,
  Reithmaier} \emph{et~al.}}]{Bayer:2002a}
\bibinfo{author}{\bibnamefont{Bayer}, \bibfnamefont{M.}},
  \bibinfo{author}{\bibfnamefont{G.}~\bibnamefont{Ortner}},
  \bibinfo{author}{\bibfnamefont{O.}~\bibnamefont{Stern}},
  \bibinfo{author}{\bibfnamefont{A.}~\bibnamefont{Kuther}},
  \bibinfo{author}{\bibfnamefont{A.~A.} \bibnamefont{Gorbunov}},
  \bibinfo{author}{\bibfnamefont{A.}~\bibnamefont{Forchel}},
  \bibinfo{author}{\bibfnamefont{P.}~\bibnamefont{Hawrylak}},
  \bibinfo{author}{\bibfnamefont{S.}~\bibnamefont{Fafard}},
  \bibinfo{author}{\bibfnamefont{K.}~\bibnamefont{Hinzer}},
  \bibinfo{author}{\bibfnamefont{T.~L.} \bibnamefont{Reinecke}},
  \bibinfo{author}{\bibfnamefont{S.~N.} \bibnamefont{Walck}},
  \bibinfo{author}{\bibfnamefont{J.~P.} \bibnamefont{Reithmaier}},
  \emph{et~al.}, \bibinfo{year}{2002}, \bibinfo{journal}{Phys. Rev. B}
  \textbf{\bibinfo{volume}{65}}(\bibinfo{number}{19}), \bibinfo{pages}{195315}.

\bibitem[{\citenamefont{Belhadj} \emph{et~al.}(2010)\citenamefont{Belhadj,
  Amand, Kunold, Simon, Kuroda, Abbarchi, Mano, Sakoda, Kunz, Marie, and
  Urbaszek}}]{Belhadj:2010a}
\bibinfo{author}{\bibnamefont{Belhadj}, \bibfnamefont{T.}},
  \bibinfo{author}{\bibfnamefont{T.}~\bibnamefont{Amand}},
  \bibinfo{author}{\bibfnamefont{A.}~\bibnamefont{Kunold}},
  \bibinfo{author}{\bibfnamefont{C.-M.} \bibnamefont{Simon}},
  \bibinfo{author}{\bibfnamefont{T.}~\bibnamefont{Kuroda}},
  \bibinfo{author}{\bibfnamefont{M.}~\bibnamefont{Abbarchi}},
  \bibinfo{author}{\bibfnamefont{T.}~\bibnamefont{Mano}},
  \bibinfo{author}{\bibfnamefont{K.}~\bibnamefont{Sakoda}},
  \bibinfo{author}{\bibfnamefont{S.}~\bibnamefont{Kunz}},
  \bibinfo{author}{\bibfnamefont{X.}~\bibnamefont{Marie}}, and
  \bibinfo{author}{\bibfnamefont{B.}~\bibnamefont{Urbaszek}},
  \bibinfo{year}{2010}, \bibinfo{journal}{Applied Physics Letters}
  \textbf{\bibinfo{volume}{97}}(\bibinfo{number}{5}), \bibinfo{eid}{051111}
  (pages~\bibinfo{numpages}{3}).

\bibitem[{\citenamefont{Belhadj} \emph{et~al.}(2008)\citenamefont{Belhadj,
  Kuroda, Simon, Amand, Mano, Sakoda, Koguchi, Marie, and
  Urbaszek}}]{Belhadj:2008a}
\bibinfo{author}{\bibnamefont{Belhadj}, \bibfnamefont{T.}},
  \bibinfo{author}{\bibfnamefont{T.}~\bibnamefont{Kuroda}},
  \bibinfo{author}{\bibfnamefont{C.-M.} \bibnamefont{Simon}},
  \bibinfo{author}{\bibfnamefont{T.}~\bibnamefont{Amand}},
  \bibinfo{author}{\bibfnamefont{T.}~\bibnamefont{Mano}},
  \bibinfo{author}{\bibfnamefont{K.}~\bibnamefont{Sakoda}},
  \bibinfo{author}{\bibfnamefont{N.}~\bibnamefont{Koguchi}},
  \bibinfo{author}{\bibfnamefont{X.}~\bibnamefont{Marie}}, and
  \bibinfo{author}{\bibfnamefont{B.}~\bibnamefont{Urbaszek}},
  \bibinfo{year}{2008}, \bibinfo{journal}{Phys. Rev. B}
  \textbf{\bibinfo{volume}{78}}(\bibinfo{number}{20}), \bibinfo{pages}{205325}.

\bibitem[{\citenamefont{Belhadj} \emph{et~al.}(2009)\citenamefont{Belhadj,
  Simon, Amand, Renucci, Chatel, Krebs, Lema\^\i{}tre, Voisin, Marie, and
  Urbaszek}}]{Belhadj:2009a}
\bibinfo{author}{\bibnamefont{Belhadj}, \bibfnamefont{T.}},
  \bibinfo{author}{\bibfnamefont{C.-M.} \bibnamefont{Simon}},
  \bibinfo{author}{\bibfnamefont{T.}~\bibnamefont{Amand}},
  \bibinfo{author}{\bibfnamefont{P.}~\bibnamefont{Renucci}},
  \bibinfo{author}{\bibfnamefont{B.}~\bibnamefont{Chatel}},
  \bibinfo{author}{\bibfnamefont{O.}~\bibnamefont{Krebs}},
  \bibinfo{author}{\bibfnamefont{A.}~\bibnamefont{Lema\^\i{}tre}},
  \bibinfo{author}{\bibfnamefont{P.}~\bibnamefont{Voisin}},
  \bibinfo{author}{\bibfnamefont{X.}~\bibnamefont{Marie}}, and
  \bibinfo{author}{\bibfnamefont{B.}~\bibnamefont{Urbaszek}},
  \bibinfo{year}{2009}, \bibinfo{journal}{Phys. Rev. Lett.}
  \textbf{\bibinfo{volume}{103}}(\bibinfo{number}{8}), \bibinfo{pages}{086601}.

\bibitem[{\citenamefont{Besombes} \emph{et~al.}(2000)\citenamefont{Besombes,
  Kheng, and Martrou}}]{Besombes:2000a}
\bibinfo{author}{\bibnamefont{Besombes}, \bibfnamefont{L.}},
  \bibinfo{author}{\bibfnamefont{K.}~\bibnamefont{Kheng}}, and
  \bibinfo{author}{\bibfnamefont{D.}~\bibnamefont{Martrou}},
  \bibinfo{year}{2000}, \bibinfo{journal}{Phys. Rev. Lett.}
  \textbf{\bibinfo{volume}{85}}(\bibinfo{number}{2}), \bibinfo{pages}{425}.

\bibitem[{\citenamefont{Bester} \emph{et~al.}(2007)\citenamefont{Bester,
  Reuter, He, Zunger, Kailuweit, Wieck, Zeitler, Maan, Wibbelhoff, and
  Lorke}}]{Bester:2007a}
\bibinfo{author}{\bibnamefont{Bester}, \bibfnamefont{G.}},
  \bibinfo{author}{\bibfnamefont{D.}~\bibnamefont{Reuter}},
  \bibinfo{author}{\bibfnamefont{L.}~\bibnamefont{He}},
  \bibinfo{author}{\bibfnamefont{A.}~\bibnamefont{Zunger}},
  \bibinfo{author}{\bibfnamefont{P.}~\bibnamefont{Kailuweit}},
  \bibinfo{author}{\bibfnamefont{A.~D.} \bibnamefont{Wieck}},
  \bibinfo{author}{\bibfnamefont{U.}~\bibnamefont{Zeitler}},
  \bibinfo{author}{\bibfnamefont{J.~C.} \bibnamefont{Maan}},
  \bibinfo{author}{\bibfnamefont{O.}~\bibnamefont{Wibbelhoff}}, and
  \bibinfo{author}{\bibfnamefont{A.}~\bibnamefont{Lorke}},
  \bibinfo{year}{2007}, \bibinfo{journal}{Phys. Rev. B}
  \textbf{\bibinfo{volume}{76}}, \bibinfo{pages}{075338}.

\bibitem[{\citenamefont{Bester and Zunger}(2005)}]{Bester:2005a}
\bibinfo{author}{\bibnamefont{Bester}, \bibfnamefont{G.}}, and
  \bibinfo{author}{\bibfnamefont{A.}~\bibnamefont{Zunger}},
  \bibinfo{year}{2005}, \bibinfo{journal}{Phys. Rev. B}
  \textbf{\bibinfo{volume}{71}}, \bibinfo{pages}{045318}.

\bibitem[{\citenamefont{Bluhm} \emph{et~al.}(2010)\citenamefont{Bluhm, Foletti,
  Neder, Rudner, Mahalu, Umansky, and Yacoby}}]{Bluhm:2010b}
\bibinfo{author}{\bibnamefont{Bluhm}, \bibfnamefont{H.}},
  \bibinfo{author}{\bibfnamefont{S.}~\bibnamefont{Foletti}},
  \bibinfo{author}{\bibfnamefont{I.}~\bibnamefont{Neder}},
  \bibinfo{author}{\bibfnamefont{M.}~\bibnamefont{Rudner}},
  \bibinfo{author}{\bibfnamefont{D.}~\bibnamefont{Mahalu}},
  \bibinfo{author}{\bibfnamefont{V.}~\bibnamefont{Umansky}}, and
  \bibinfo{author}{\bibfnamefont{A.}~\bibnamefont{Yacoby}},
  \bibinfo{year}{2010}, \bibinfo{journal}{Nature Physics}
  \textbf{\bibinfo{volume}{7}}, \bibinfo{pages}{109}.

\bibitem[{\citenamefont{Bracker} \emph{et~al.}(2008)\citenamefont{Bracker,
  Gammon, and Korenev}}]{Bracker:2008a}
\bibinfo{author}{\bibnamefont{Bracker}, \bibfnamefont{A.~S.}},
  \bibinfo{author}{\bibfnamefont{D.}~\bibnamefont{Gammon}}, and
  \bibinfo{author}{\bibfnamefont{V.~L.} \bibnamefont{Korenev}},
  \bibinfo{year}{2008}, \bibinfo{journal}{Semiconductor Science and Technology}
  \textbf{\bibinfo{volume}{23}}(\bibinfo{number}{11}), \bibinfo{pages}{114004}.

\bibitem[{\citenamefont{Bracker} \emph{et~al.}(2005)\citenamefont{Bracker,
  Stinaff, Gammon, Ware, Tischler, Shabaev, Efros, Park, Gershoni, Korenev, and
  Merkulov}}]{Bracker:2005a}
\bibinfo{author}{\bibnamefont{Bracker}, \bibfnamefont{A.~S.}},
  \bibinfo{author}{\bibfnamefont{E.~A.} \bibnamefont{Stinaff}},
  \bibinfo{author}{\bibfnamefont{D.}~\bibnamefont{Gammon}},
  \bibinfo{author}{\bibfnamefont{M.~E.} \bibnamefont{Ware}},
  \bibinfo{author}{\bibfnamefont{J.~G.} \bibnamefont{Tischler}},
  \bibinfo{author}{\bibfnamefont{A.}~\bibnamefont{Shabaev}},
  \bibinfo{author}{\bibfnamefont{A.~L.} \bibnamefont{Efros}},
  \bibinfo{author}{\bibfnamefont{D.}~\bibnamefont{Park}},
  \bibinfo{author}{\bibfnamefont{D.}~\bibnamefont{Gershoni}},
  \bibinfo{author}{\bibfnamefont{V.~L.} \bibnamefont{Korenev}}, and
  \bibinfo{author}{\bibfnamefont{I.~A.} \bibnamefont{Merkulov}},
  \bibinfo{year}{2005}, \bibinfo{journal}{Phys. Rev. Lett.}
  \textbf{\bibinfo{volume}{94}}(\bibinfo{number}{4}), \bibinfo{pages}{047402}.

\bibitem[{\citenamefont{Braun}
  \emph{et~al.}(2006{\natexlab{a}})\citenamefont{Braun, , Eble, Lombez,
  Urbaszek, Amand, Marie, Renucci, Krebs, Lemaitre, Voisin, Kalevich}
  \emph{et~al.}}]{Braun:2006b}
\bibinfo{author}{\bibnamefont{Braun}, \bibfnamefont{P.-F.}}, ,
  \bibinfo{author}{\bibfnamefont{B.}~\bibnamefont{Eble}},
  \bibinfo{author}{\bibfnamefont{L.}~\bibnamefont{Lombez}},
  \bibinfo{author}{\bibfnamefont{B.}~\bibnamefont{Urbaszek}},
  \bibinfo{author}{\bibfnamefont{T.}~\bibnamefont{Amand}},
  \bibinfo{author}{\bibfnamefont{X.}~\bibnamefont{Marie}},
  \bibinfo{author}{\bibfnamefont{P.}~\bibnamefont{Renucci}},
  \bibinfo{author}{\bibfnamefont{O.}~\bibnamefont{Krebs}},
  \bibinfo{author}{\bibfnamefont{A.}~\bibnamefont{Lemaitre}},
  \bibinfo{author}{\bibfnamefont{P.}~\bibnamefont{Voisin}},
  \bibinfo{author}{\bibfnamefont{V.~K.} \bibnamefont{Kalevich}}, \emph{et~al.},
  \bibinfo{year}{2006}{\natexlab{a}}, \bibinfo{journal}{phys. stat. sol. (b)}
  \textbf{\bibinfo{volume}{243}}(\bibinfo{number}{15}), \bibinfo{pages}{3917}.

\bibitem[{\citenamefont{Braun} \emph{et~al.}(2005)\citenamefont{Braun, Marie,
  Lombez, Urbaszek, Amand, Renucci, Kalevich, Kavokin, Krebs, Voisin, and
  Masumoto}}]{Braun:2005a}
\bibinfo{author}{\bibnamefont{Braun}, \bibfnamefont{P.-F.}},
  \bibinfo{author}{\bibfnamefont{X.}~\bibnamefont{Marie}},
  \bibinfo{author}{\bibfnamefont{L.}~\bibnamefont{Lombez}},
  \bibinfo{author}{\bibfnamefont{B.}~\bibnamefont{Urbaszek}},
  \bibinfo{author}{\bibfnamefont{T.}~\bibnamefont{Amand}},
  \bibinfo{author}{\bibfnamefont{P.}~\bibnamefont{Renucci}},
  \bibinfo{author}{\bibfnamefont{V.~K.} \bibnamefont{Kalevich}},
  \bibinfo{author}{\bibfnamefont{K.~V.} \bibnamefont{Kavokin}},
  \bibinfo{author}{\bibfnamefont{O.}~\bibnamefont{Krebs}},
  \bibinfo{author}{\bibfnamefont{P.}~\bibnamefont{Voisin}}, and
  \bibinfo{author}{\bibfnamefont{Y.}~\bibnamefont{Masumoto}},
  \bibinfo{year}{2005}, \bibinfo{journal}{Phys. Rev. Lett.}
  \textbf{\bibinfo{volume}{94}}(\bibinfo{number}{11}), \bibinfo{pages}{116601}.

\bibitem[{\citenamefont{Braun}
  \emph{et~al.}(2006{\natexlab{b}})\citenamefont{Braun, Urbaszek, Amand, Marie,
  Krebs, Eble, Lemaitre, and Voisin}}]{Braun:2006a}
\bibinfo{author}{\bibnamefont{Braun}, \bibfnamefont{P.-F.}},
  \bibinfo{author}{\bibfnamefont{B.}~\bibnamefont{Urbaszek}},
  \bibinfo{author}{\bibfnamefont{T.}~\bibnamefont{Amand}},
  \bibinfo{author}{\bibfnamefont{X.}~\bibnamefont{Marie}},
  \bibinfo{author}{\bibfnamefont{O.}~\bibnamefont{Krebs}},
  \bibinfo{author}{\bibfnamefont{B.}~\bibnamefont{Eble}},
  \bibinfo{author}{\bibfnamefont{A.}~\bibnamefont{Lemaitre}}, and
  \bibinfo{author}{\bibfnamefont{P.}~\bibnamefont{Voisin}},
  \bibinfo{year}{2006}{\natexlab{b}}, \bibinfo{journal}{Phys. Rev. B}
  \textbf{\bibinfo{volume}{74}}(\bibinfo{number}{24}), \bibinfo{pages}{245306}.

\bibitem[{\citenamefont{Brossel} \emph{et~al.}(1952)\citenamefont{Brossel,
  Kastler, and Winter}}]{Kastler:1952a}
\bibinfo{author}{\bibnamefont{Brossel}, \bibfnamefont{J.}},
  \bibinfo{author}{\bibfnamefont{A.}~\bibnamefont{Kastler}}, and
  \bibinfo{author}{\bibfnamefont{J.}~\bibnamefont{Winter}},
  \bibinfo{year}{1952}, \bibinfo{journal}{J. de Physique et le Radium}
  \textbf{\bibinfo{volume}{13}}, \bibinfo{pages}{668}.

\bibitem[{\citenamefont{Brown} \emph{et~al.}(1998)\citenamefont{Brown, Kennedy,
  and Gammon}}]{Brown:1998a}
\bibinfo{author}{\bibnamefont{Brown}, \bibfnamefont{S.~W.}},
  \bibinfo{author}{\bibfnamefont{T.~A.} \bibnamefont{Kennedy}}, and
  \bibinfo{author}{\bibfnamefont{D.}~\bibnamefont{Gammon}},
  \bibinfo{year}{1998}, \bibinfo{journal}{Solid State Nuclear Magnetic
  Resonance} \textbf{\bibinfo{volume}{11}}(\bibinfo{number}{1-2}),
  \bibinfo{pages}{49 }.

\bibitem[{\citenamefont{Brunner} \emph{et~al.}(2009)\citenamefont{Brunner,
  Gerardot, Dalgarno, Wost, Karrai, Stoltz, Petroff, and
  Warburton}}]{Brunner:2009a}
\bibinfo{author}{\bibnamefont{Brunner}, \bibfnamefont{D.}},
  \bibinfo{author}{\bibfnamefont{B.~D.} \bibnamefont{Gerardot}},
  \bibinfo{author}{\bibfnamefont{P.~A.} \bibnamefont{Dalgarno}},
  \bibinfo{author}{\bibfnamefont{G.}~\bibnamefont{Wost}},
  \bibinfo{author}{\bibfnamefont{K.}~\bibnamefont{Karrai}},
  \bibinfo{author}{\bibfnamefont{N.~G.} \bibnamefont{Stoltz}},
  \bibinfo{author}{\bibfnamefont{P.~M.} \bibnamefont{Petroff}}, and
  \bibinfo{author}{\bibfnamefont{R.~J.} \bibnamefont{Warburton}},
  \bibinfo{year}{2009}, \bibinfo{journal}{Science}
  \textbf{\bibinfo{volume}{325}}(\bibinfo{number}{5936}), \bibinfo{pages}{70}.

\bibitem[{\citenamefont{Bulutay}(2012)}]{Bulutay:2012a}
\bibinfo{author}{\bibnamefont{Bulutay}, \bibfnamefont{C.}},
  \bibinfo{year}{2012}, \bibinfo{journal}{Phys. Rev. B}
  \textbf{\bibinfo{volume}{85}}, \bibinfo{pages}{115313}.

\bibitem[{\citenamefont{Burkard} \emph{et~al.}(1999)\citenamefont{Burkard,
  Loss, and DiVincenzo}}]{Burkard:1999a}
\bibinfo{author}{\bibnamefont{Burkard}, \bibfnamefont{G.}},
  \bibinfo{author}{\bibfnamefont{D.}~\bibnamefont{Loss}}, and
  \bibinfo{author}{\bibfnamefont{D.~P.} \bibnamefont{DiVincenzo}},
  \bibinfo{year}{1999}, \bibinfo{journal}{Phys. Rev. B}
  \textbf{\bibinfo{volume}{59}}(\bibinfo{number}{3}), \bibinfo{pages}{2070}.

\bibitem[{\citenamefont{Calarco} \emph{et~al.}(2003)\citenamefont{Calarco,
  Datta, Fedichev, Pazy, and Zoller}}]{Calarco:2003a}
\bibinfo{author}{\bibnamefont{Calarco}, \bibfnamefont{T.}},
  \bibinfo{author}{\bibfnamefont{A.}~\bibnamefont{Datta}},
  \bibinfo{author}{\bibfnamefont{P.}~\bibnamefont{Fedichev}},
  \bibinfo{author}{\bibfnamefont{E.}~\bibnamefont{Pazy}}, and
  \bibinfo{author}{\bibfnamefont{P.}~\bibnamefont{Zoller}},
  \bibinfo{year}{2003}, \bibinfo{journal}{Phys. Rev. A}
  \textbf{\bibinfo{volume}{68}}(\bibinfo{number}{1}), \bibinfo{pages}{012310}.

\bibitem[{\citenamefont{{Chekhovich}}
  \emph{et~al.}(2011)\citenamefont{{Chekhovich}, {Krysa}, {Hopkinson},
  {Senellart}, {Lemaitre}, {Skolnick}, and {Tartakovskii}}}]{Chekhovich:2011c}
\bibinfo{author}{\bibnamefont{{Chekhovich}}, \bibfnamefont{E.~A.}},
  \bibinfo{author}{\bibfnamefont{A.~B.} \bibnamefont{{Krysa}}},
  \bibinfo{author}{\bibfnamefont{M.}~\bibnamefont{{Hopkinson}}},
  \bibinfo{author}{\bibfnamefont{P.}~\bibnamefont{{Senellart}}},
  \bibinfo{author}{\bibfnamefont{A.}~\bibnamefont{{Lemaitre}}},
  \bibinfo{author}{\bibfnamefont{M.~S.} \bibnamefont{{Skolnick}}}, and
  \bibinfo{author}{\bibfnamefont{A.~I.} \bibnamefont{{Tartakovskii}}},
  \bibinfo{year}{2011}, \bibinfo{journal}{ArXiv} \eprint{1109.0733}.

\bibitem[{\citenamefont{Chekhovich}
  \emph{et~al.}(2011{\natexlab{a}})\citenamefont{Chekhovich, Krysa, Skolnick,
  and Tartakovskii}}]{Chekhovich:2011a}
\bibinfo{author}{\bibnamefont{Chekhovich}, \bibfnamefont{E.~A.}},
  \bibinfo{author}{\bibfnamefont{A.~B.} \bibnamefont{Krysa}},
  \bibinfo{author}{\bibfnamefont{M.~S.} \bibnamefont{Skolnick}}, and
  \bibinfo{author}{\bibfnamefont{A.~I.} \bibnamefont{Tartakovskii}},
  \bibinfo{year}{2011}{\natexlab{a}}, \bibinfo{journal}{Phys. Rev. Lett.}
  \textbf{\bibinfo{volume}{106}}(\bibinfo{number}{2}), \bibinfo{pages}{027402}.

\bibitem[{\citenamefont{Chekhovich}
  \emph{et~al.}(2011{\natexlab{b}})\citenamefont{Chekhovich, Krysa, Skolnick,
  and Tartakovskii}}]{Chekhovich:2011b}
\bibinfo{author}{\bibnamefont{Chekhovich}, \bibfnamefont{E.~A.}},
  \bibinfo{author}{\bibfnamefont{A.~B.} \bibnamefont{Krysa}},
  \bibinfo{author}{\bibfnamefont{M.~S.} \bibnamefont{Skolnick}}, and
  \bibinfo{author}{\bibfnamefont{A.~I.} \bibnamefont{Tartakovskii}},
  \bibinfo{year}{2011}{\natexlab{b}}, \bibinfo{journal}{Phys. Rev. B}
  \textbf{\bibinfo{volume}{83}}(\bibinfo{number}{12}), \bibinfo{pages}{125318}.

\bibitem[{\citenamefont{Chekhovich}
  \emph{et~al.}(2010{\natexlab{a}})\citenamefont{Chekhovich, Makhonin, Kavokin,
  Krysa, Skolnick, and Tartakovskii}}]{Chekhovich:2010a}
\bibinfo{author}{\bibnamefont{Chekhovich}, \bibfnamefont{E.~A.}},
  \bibinfo{author}{\bibfnamefont{M.~N.} \bibnamefont{Makhonin}},
  \bibinfo{author}{\bibfnamefont{K.~V.} \bibnamefont{Kavokin}},
  \bibinfo{author}{\bibfnamefont{A.~B.} \bibnamefont{Krysa}},
  \bibinfo{author}{\bibfnamefont{M.~S.} \bibnamefont{Skolnick}}, and
  \bibinfo{author}{\bibfnamefont{A.~I.} \bibnamefont{Tartakovskii}},
  \bibinfo{year}{2010}{\natexlab{a}}, \bibinfo{journal}{Phys. Rev. Lett.}
  \textbf{\bibinfo{volume}{104}}(\bibinfo{number}{6}), \bibinfo{pages}{066804}.

\bibitem[{\citenamefont{Chekhovich}
  \emph{et~al.}(2010{\natexlab{b}})\citenamefont{Chekhovich, Makhonin,
  Skiba-Szymanska, Krysa, Kulakovskii, Skolnick, and
  Tartakovskii}}]{Chekhovich:2010b}
\bibinfo{author}{\bibnamefont{Chekhovich}, \bibfnamefont{E.~A.}},
  \bibinfo{author}{\bibfnamefont{M.~N.} \bibnamefont{Makhonin}},
  \bibinfo{author}{\bibfnamefont{J.}~\bibnamefont{Skiba-Szymanska}},
  \bibinfo{author}{\bibfnamefont{A.~B.} \bibnamefont{Krysa}},
  \bibinfo{author}{\bibfnamefont{V.~D.} \bibnamefont{Kulakovskii}},
  \bibinfo{author}{\bibfnamefont{M.~S.} \bibnamefont{Skolnick}}, and
  \bibinfo{author}{\bibfnamefont{A.~I.} \bibnamefont{Tartakovskii}},
  \bibinfo{year}{2010}{\natexlab{b}}, \bibinfo{journal}{Phys. Rev. B}
  \textbf{\bibinfo{volume}{81}}(\bibinfo{number}{24}), \bibinfo{pages}{245308}.

\bibitem[{\citenamefont{Cherbunin} \emph{et~al.}(2011)\citenamefont{Cherbunin,
  Flisinski, Gerlovin, Ignatiev, Kuznetsova, Petrov, Yakovlev, Reuter, Wieck,
  and Bayer}}]{Cherbunin:2011a}
\bibinfo{author}{\bibnamefont{Cherbunin}, \bibfnamefont{R.~V.}},
  \bibinfo{author}{\bibfnamefont{K.}~\bibnamefont{Flisinski}},
  \bibinfo{author}{\bibfnamefont{I.~Y.} \bibnamefont{Gerlovin}},
  \bibinfo{author}{\bibfnamefont{I.~V.} \bibnamefont{Ignatiev}},
  \bibinfo{author}{\bibfnamefont{M.~S.} \bibnamefont{Kuznetsova}},
  \bibinfo{author}{\bibfnamefont{M.~Y.} \bibnamefont{Petrov}},
  \bibinfo{author}{\bibfnamefont{D.~R.} \bibnamefont{Yakovlev}},
  \bibinfo{author}{\bibfnamefont{D.}~\bibnamefont{Reuter}},
  \bibinfo{author}{\bibfnamefont{A.~D.} \bibnamefont{Wieck}}, and
  \bibinfo{author}{\bibfnamefont{M.}~\bibnamefont{Bayer}},
  \bibinfo{year}{2011}, \bibinfo{journal}{Phys. Rev. B}
  \textbf{\bibinfo{volume}{84}}(\bibinfo{number}{4}), \bibinfo{pages}{041304}.

\bibitem[{\citenamefont{Cherbunin} \emph{et~al.}(2009)\citenamefont{Cherbunin,
  Verbin, Auer, Yakovlev, Reuter, Wieck, Gerlovin, Ignatiev, Vishnevsky, and
  Bayer}}]{Cherbunin:2009a}
\bibinfo{author}{\bibnamefont{Cherbunin}, \bibfnamefont{R.~V.}},
  \bibinfo{author}{\bibfnamefont{S.~Y.} \bibnamefont{Verbin}},
  \bibinfo{author}{\bibfnamefont{T.}~\bibnamefont{Auer}},
  \bibinfo{author}{\bibfnamefont{D.~R.} \bibnamefont{Yakovlev}},
  \bibinfo{author}{\bibfnamefont{D.}~\bibnamefont{Reuter}},
  \bibinfo{author}{\bibfnamefont{A.~D.} \bibnamefont{Wieck}},
  \bibinfo{author}{\bibfnamefont{I.~Y.} \bibnamefont{Gerlovin}},
  \bibinfo{author}{\bibfnamefont{I.~V.} \bibnamefont{Ignatiev}},
  \bibinfo{author}{\bibfnamefont{D.~V.} \bibnamefont{Vishnevsky}}, and
  \bibinfo{author}{\bibfnamefont{M.}~\bibnamefont{Bayer}},
  \bibinfo{year}{2009}, \bibinfo{journal}{Phys. Rev. B}
  \textbf{\bibinfo{volume}{80}}(\bibinfo{number}{3}), \bibinfo{pages}{035326}.

\bibitem[{\citenamefont{Childress} \emph{et~al.}(2006)\citenamefont{Childress,
  Gurudev~Dutt, Taylor, Zibrov, Jelezko, Wrachtrup, Hemmer, and
  Lukin}}]{Childress:2006a}
\bibinfo{author}{\bibnamefont{Childress}, \bibfnamefont{L.}},
  \bibinfo{author}{\bibfnamefont{M.~V.} \bibnamefont{Gurudev~Dutt}},
  \bibinfo{author}{\bibfnamefont{J.~M.} \bibnamefont{Taylor}},
  \bibinfo{author}{\bibfnamefont{A.~S.} \bibnamefont{Zibrov}},
  \bibinfo{author}{\bibfnamefont{F.}~\bibnamefont{Jelezko}},
  \bibinfo{author}{\bibfnamefont{J.}~\bibnamefont{Wrachtrup}},
  \bibinfo{author}{\bibfnamefont{P.~R.} \bibnamefont{Hemmer}}, and
  \bibinfo{author}{\bibfnamefont{M.~D.} \bibnamefont{Lukin}},
  \bibinfo{year}{2006}, \bibinfo{journal}{Science}
  \textbf{\bibinfo{volume}{314}}(\bibinfo{number}{5797}), \bibinfo{pages}{281}.

\bibitem[{\citenamefont{Colton} \emph{et~al.}(2004)\citenamefont{Colton,
  Kennedy, Bracker, and Gammon}}]{Colton:2004a}
\bibinfo{author}{\bibnamefont{Colton}, \bibfnamefont{J.~S.}},
  \bibinfo{author}{\bibfnamefont{T.~A.} \bibnamefont{Kennedy}},
  \bibinfo{author}{\bibfnamefont{A.~S.} \bibnamefont{Bracker}}, and
  \bibinfo{author}{\bibfnamefont{D.}~\bibnamefont{Gammon}},
  \bibinfo{year}{2004}, \bibinfo{journal}{Phys. Rev. B}
  \textbf{\bibinfo{volume}{69}}(\bibinfo{number}{12}), \bibinfo{pages}{121307}.

\bibitem[{\citenamefont{Cortez} \emph{et~al.}(2002)\citenamefont{Cortez, Krebs,
  Laurent, Senes, Marie, Voisin, Ferreira, Bastard, G\'erard, and
  Amand}}]{Cortez:2002a}
\bibinfo{author}{\bibnamefont{Cortez}, \bibfnamefont{S.}},
  \bibinfo{author}{\bibfnamefont{O.}~\bibnamefont{Krebs}},
  \bibinfo{author}{\bibfnamefont{S.}~\bibnamefont{Laurent}},
  \bibinfo{author}{\bibfnamefont{M.}~\bibnamefont{Senes}},
  \bibinfo{author}{\bibfnamefont{X.}~\bibnamefont{Marie}},
  \bibinfo{author}{\bibfnamefont{P.}~\bibnamefont{Voisin}},
  \bibinfo{author}{\bibfnamefont{R.}~\bibnamefont{Ferreira}},
  \bibinfo{author}{\bibfnamefont{G.}~\bibnamefont{Bastard}},
  \bibinfo{author}{\bibfnamefont{J.-M.} \bibnamefont{G\'erard}}, and
  \bibinfo{author}{\bibfnamefont{T.}~\bibnamefont{Amand}},
  \bibinfo{year}{2002}, \bibinfo{journal}{Phys. Rev. Lett.}
  \textbf{\bibinfo{volume}{89}}(\bibinfo{number}{20}), \bibinfo{pages}{207401}.

\bibitem[{\citenamefont{Crooker} \emph{et~al.}(2010)\citenamefont{Crooker,
  Brandt, Sandfort, Greilich, Yakovlev, Reuter, Wieck, and
  Bayer}}]{Crooker:2010a}
\bibinfo{author}{\bibnamefont{Crooker}, \bibfnamefont{S.~A.}},
  \bibinfo{author}{\bibfnamefont{J.}~\bibnamefont{Brandt}},
  \bibinfo{author}{\bibfnamefont{C.}~\bibnamefont{Sandfort}},
  \bibinfo{author}{\bibfnamefont{A.}~\bibnamefont{Greilich}},
  \bibinfo{author}{\bibfnamefont{D.~R.} \bibnamefont{Yakovlev}},
  \bibinfo{author}{\bibfnamefont{D.}~\bibnamefont{Reuter}},
  \bibinfo{author}{\bibfnamefont{A.~D.} \bibnamefont{Wieck}}, and
  \bibinfo{author}{\bibfnamefont{M.}~\bibnamefont{Bayer}},
  \bibinfo{year}{2010}, \bibinfo{journal}{Phys. Rev. Lett.}
  \textbf{\bibinfo{volume}{104}}, \bibinfo{pages}{036601}.

\bibitem[{\citenamefont{Cywinski} \emph{et~al.}(2009)\citenamefont{Cywinski,
  Witzel, and Das~Sarma}}]{Cywinski:2009a}
\bibinfo{author}{\bibnamefont{Cywinski}, \bibfnamefont{L.}},
  \bibinfo{author}{\bibfnamefont{W.~M.} \bibnamefont{Witzel}}, and
  \bibinfo{author}{\bibfnamefont{S.}~\bibnamefont{Das~Sarma}},
  \bibinfo{year}{2009}, \bibinfo{journal}{Phys. Rev. Lett.}
  \textbf{\bibinfo{volume}{102}}(\bibinfo{number}{5}), \bibinfo{pages}{057601}.

\bibitem[{\citenamefont{Dahbashi} \emph{et~al.}(2012)\citenamefont{Dahbashi,
  H\"{u}bner, Berski, Wiegand, Marie, Pierz, Schumacher, and
  Oestreich}}]{Dahbashi:2012a}
\bibinfo{author}{\bibnamefont{Dahbashi}, \bibfnamefont{R.}},
  \bibinfo{author}{\bibfnamefont{J.}~\bibnamefont{H\"{u}bner}},
  \bibinfo{author}{\bibfnamefont{F.}~\bibnamefont{Berski}},
  \bibinfo{author}{\bibfnamefont{J.}~\bibnamefont{Wiegand}},
  \bibinfo{author}{\bibfnamefont{X.}~\bibnamefont{Marie}},
  \bibinfo{author}{\bibfnamefont{K.}~\bibnamefont{Pierz}},
  \bibinfo{author}{\bibfnamefont{H.~W.} \bibnamefont{Schumacher}}, and
  \bibinfo{author}{\bibfnamefont{M.}~\bibnamefont{Oestreich}},
  \bibinfo{year}{2012}, \bibinfo{journal}{Applied Physics Letters}
  \textbf{\bibinfo{volume}{100}}(\bibinfo{number}{3}), \bibinfo{eid}{031906}.

\bibitem[{\citenamefont{Damen} \emph{et~al.}(1991)\citenamefont{Damen, Via,
  Cunningham, Shah, and Sham}}]{Damen:1991a}
\bibinfo{author}{\bibnamefont{Damen}, \bibfnamefont{T.~C.}},
  \bibinfo{author}{\bibfnamefont{L.}~\bibnamefont{Via}},
  \bibinfo{author}{\bibfnamefont{J.~E.} \bibnamefont{Cunningham}},
  \bibinfo{author}{\bibfnamefont{J.}~\bibnamefont{Shah}}, and
  \bibinfo{author}{\bibfnamefont{L.~J.} \bibnamefont{Sham}},
  \bibinfo{year}{1991}, \bibinfo{journal}{Phys. Rev. Lett.}
  \textbf{\bibinfo{volume}{67}}(\bibinfo{number}{24}), \bibinfo{pages}{3432}.

\bibitem[{\citenamefont{Desfonds} \emph{et~al.}(2010)\citenamefont{Desfonds,
  Eble, Fras, Testelin, Bernardot, Chamarro, Urbaszek, Amand, Marie, Gerard,
  Thierry-Mieg, Miard} \emph{et~al.}}]{Desfonds:2010a}
\bibinfo{author}{\bibnamefont{Desfonds}, \bibfnamefont{P.}},
  \bibinfo{author}{\bibfnamefont{B.}~\bibnamefont{Eble}},
  \bibinfo{author}{\bibfnamefont{F.}~\bibnamefont{Fras}},
  \bibinfo{author}{\bibfnamefont{C.}~\bibnamefont{Testelin}},
  \bibinfo{author}{\bibfnamefont{F.}~\bibnamefont{Bernardot}},
  \bibinfo{author}{\bibfnamefont{M.}~\bibnamefont{Chamarro}},
  \bibinfo{author}{\bibfnamefont{B.}~\bibnamefont{Urbaszek}},
  \bibinfo{author}{\bibfnamefont{T.}~\bibnamefont{Amand}},
  \bibinfo{author}{\bibfnamefont{X.}~\bibnamefont{Marie}},
  \bibinfo{author}{\bibfnamefont{J.~M.} \bibnamefont{Gerard}},
  \bibinfo{author}{\bibfnamefont{V.}~\bibnamefont{Thierry-Mieg}},
  \bibinfo{author}{\bibfnamefont{A.}~\bibnamefont{Miard}}, \emph{et~al.},
  \bibinfo{year}{2010}, \bibinfo{journal}{Appl. Phys. Lett.}
  \textbf{\bibinfo{volume}{96}}(\bibinfo{number}{{17}}).

\bibitem[{\citenamefont{Dou} \emph{et~al.}(2011)\citenamefont{Dou, Sun, Jiang,
  Ni, and Niu}}]{Dou:2011a}
\bibinfo{author}{\bibnamefont{Dou}, \bibfnamefont{X.~M.}},
  \bibinfo{author}{\bibfnamefont{B.~Q.} \bibnamefont{Sun}},
  \bibinfo{author}{\bibfnamefont{D.~S.} \bibnamefont{Jiang}},
  \bibinfo{author}{\bibfnamefont{H.~Q.} \bibnamefont{Ni}}, and
  \bibinfo{author}{\bibfnamefont{Z.~C.} \bibnamefont{Niu}},
  \bibinfo{year}{2011}, \bibinfo{journal}{Phys. Rev. B}
  \textbf{\bibinfo{volume}{84}}(\bibinfo{number}{3}), \bibinfo{pages}{033302}.

\bibitem[{\citenamefont{Dousse} \emph{et~al.}(2010)\citenamefont{Dousse,
  Suffczynski, Beveratos, Krebs, Lemaitre, Sagnes, Bloch, Voisin, and
  Senellart}}]{Dousse:2010a}
\bibinfo{author}{\bibnamefont{Dousse}, \bibfnamefont{A.}},
  \bibinfo{author}{\bibfnamefont{J.}~\bibnamefont{Suffczynski}},
  \bibinfo{author}{\bibfnamefont{A.}~\bibnamefont{Beveratos}},
  \bibinfo{author}{\bibfnamefont{O.}~\bibnamefont{Krebs}},
  \bibinfo{author}{\bibfnamefont{A.}~\bibnamefont{Lemaitre}},
  \bibinfo{author}{\bibfnamefont{I.}~\bibnamefont{Sagnes}},
  \bibinfo{author}{\bibfnamefont{J.}~\bibnamefont{Bloch}},
  \bibinfo{author}{\bibfnamefont{P.}~\bibnamefont{Voisin}}, and
  \bibinfo{author}{\bibfnamefont{P.}~\bibnamefont{Senellart}},
  \bibinfo{year}{2010}, \bibinfo{journal}{Nature}
  \textbf{\bibinfo{volume}{466}}, \bibinfo{pages}{217}.

\bibitem[{\citenamefont{Dreiser} \emph{et~al.}(2008)\citenamefont{Dreiser,
  Atat\"ure, Galland, M\"uller, Badolato, and Imamoglu}}]{Dreiser:2008a}
\bibinfo{author}{\bibnamefont{Dreiser}, \bibfnamefont{J.}},
  \bibinfo{author}{\bibfnamefont{M.}~\bibnamefont{Atat\"ure}},
  \bibinfo{author}{\bibfnamefont{C.}~\bibnamefont{Galland}},
  \bibinfo{author}{\bibfnamefont{T.}~\bibnamefont{M\"uller}},
  \bibinfo{author}{\bibfnamefont{A.}~\bibnamefont{Badolato}}, and
  \bibinfo{author}{\bibfnamefont{A.}~\bibnamefont{Imamoglu}},
  \bibinfo{year}{2008}, \bibinfo{journal}{Phys. Rev. B}
  \textbf{\bibinfo{volume}{77}}(\bibinfo{number}{7}), \bibinfo{pages}{075317}.

\bibitem[{\citenamefont{Drexler} \emph{et~al.}(1994)\citenamefont{Drexler,
  Leonard, Hansen, Kotthaus, and Petroff}}]{Drexler:1994a}
\bibinfo{author}{\bibnamefont{Drexler}, \bibfnamefont{H.}},
  \bibinfo{author}{\bibfnamefont{D.}~\bibnamefont{Leonard}},
  \bibinfo{author}{\bibfnamefont{W.}~\bibnamefont{Hansen}},
  \bibinfo{author}{\bibfnamefont{J.~P.} \bibnamefont{Kotthaus}}, and
  \bibinfo{author}{\bibfnamefont{P.~M.} \bibnamefont{Petroff}},
  \bibinfo{year}{1994}, \bibinfo{journal}{Phys. Rev. Lett.}
  \textbf{\bibinfo{volume}{73}}(\bibinfo{number}{16}), \bibinfo{pages}{2252}.

\bibitem[{\citenamefont{Dutt} \emph{et~al.}(2005)\citenamefont{Dutt, Cheng, Li,
  Xu, Li, Berman, Steel, Bracker, Gammon, Economou, Liu, and
  Sham}}]{Dutt:2005a}
\bibinfo{author}{\bibnamefont{Dutt}, \bibfnamefont{M.~V.~G.}},
  \bibinfo{author}{\bibfnamefont{J.}~\bibnamefont{Cheng}},
  \bibinfo{author}{\bibfnamefont{B.}~\bibnamefont{Li}},
  \bibinfo{author}{\bibfnamefont{X.}~\bibnamefont{Xu}},
  \bibinfo{author}{\bibfnamefont{X.}~\bibnamefont{Li}},
  \bibinfo{author}{\bibfnamefont{P.~R.} \bibnamefont{Berman}},
  \bibinfo{author}{\bibfnamefont{D.~G.} \bibnamefont{Steel}},
  \bibinfo{author}{\bibfnamefont{A.~S.} \bibnamefont{Bracker}},
  \bibinfo{author}{\bibfnamefont{D.}~\bibnamefont{Gammon}},
  \bibinfo{author}{\bibfnamefont{S.~E.} \bibnamefont{Economou}},
  \bibinfo{author}{\bibfnamefont{R.-B.} \bibnamefont{Liu}}, and
  \bibinfo{author}{\bibfnamefont{L.~J.} \bibnamefont{Sham}},
  \bibinfo{year}{2005}, \bibinfo{journal}{Phys. Rev. Lett.}
  \textbf{\bibinfo{volume}{94}}(\bibinfo{number}{22}), \bibinfo{pages}{227403}.

\bibitem[{\citenamefont{Dyakonov}(2008)}]{Dyakonov:2008a}
\bibinfo{author}{\bibnamefont{Dyakonov}, \bibfnamefont{M.}},
  \bibinfo{year}{2008}, \bibinfo{journal}{Springer Series in Solid-State
  Science, Springer-Verlag Berlin} \textbf{\bibinfo{volume}{157}}.

\bibitem[{\citenamefont{Dyakonov and Perel.}({1973})}]{Dyakonov:1973a}
\bibinfo{author}{\bibnamefont{Dyakonov}, \bibfnamefont{M.}}, and
  \bibinfo{author}{\bibfnamefont{V.}~\bibnamefont{Perel.}},
  \bibinfo{year}{{1973}}, \bibinfo{journal}{{Zh.E ksp. Teor. Fiz.}}
  \textbf{\bibinfo{volume}{{65}}}(\bibinfo{number}{{362}}).

\bibitem[{\citenamefont{Dyakonov and Perel.}({1974})}]{Dyakonov:1974a}
\bibinfo{author}{\bibnamefont{Dyakonov}, \bibfnamefont{M.}}, and
  \bibinfo{author}{\bibfnamefont{V.}~\bibnamefont{Perel.}},
  \bibinfo{year}{{1974}}, \bibinfo{journal}{{Sov. Phys. JETP}}
  \textbf{\bibinfo{volume}{{38}}}(\bibinfo{number}{{177}}).

\bibitem[{\citenamefont{Dzhioev and Korenev}(2007)}]{Dzhioev:2008a}
\bibinfo{author}{\bibnamefont{Dzhioev}, \bibfnamefont{R.~I.}}, and
  \bibinfo{author}{\bibfnamefont{V.~L.} \bibnamefont{Korenev}},
  \bibinfo{year}{2007}, \bibinfo{journal}{Phys. Rev. Lett.}
  \textbf{\bibinfo{volume}{99}}, \bibinfo{pages}{037401}.

\bibitem[{\citenamefont{Dzhioev} \emph{et~al.}(2002)\citenamefont{Dzhioev,
  Korenev, Merkulov, Zakharchenya, Gammon, Efros, and Katzer}}]{Dzhioev:2002a}
\bibinfo{author}{\bibnamefont{Dzhioev}, \bibfnamefont{R.~I.}},
  \bibinfo{author}{\bibfnamefont{V.~L.} \bibnamefont{Korenev}},
  \bibinfo{author}{\bibfnamefont{I.~A.} \bibnamefont{Merkulov}},
  \bibinfo{author}{\bibfnamefont{B.~P.} \bibnamefont{Zakharchenya}},
  \bibinfo{author}{\bibfnamefont{D.}~\bibnamefont{Gammon}},
  \bibinfo{author}{\bibfnamefont{A.~L.} \bibnamefont{Efros}}, and
  \bibinfo{author}{\bibfnamefont{D.~S.} \bibnamefont{Katzer}},
  \bibinfo{year}{2002}, \bibinfo{journal}{Phys. Rev. Lett.}
  \textbf{\bibinfo{volume}{88}}(\bibinfo{number}{25}), \bibinfo{pages}{256801}.

\bibitem[{\citenamefont{Dzhioev}
  \emph{et~al.}(1998{\natexlab{a}})\citenamefont{Dzhioev, Zakharchenya,
  Ivchenko, Korenev, Kusraev, Ledentsov, Ustinov, Zhukov, and
  Tsatsul�nikov}}]{Dzhioev:1998a}
\bibinfo{author}{\bibnamefont{Dzhioev}, \bibfnamefont{R.~I.}},
  \bibinfo{author}{\bibfnamefont{B.~P.} \bibnamefont{Zakharchenya}},
  \bibinfo{author}{\bibfnamefont{E.~L.} \bibnamefont{Ivchenko}},
  \bibinfo{author}{\bibfnamefont{V.~L.} \bibnamefont{Korenev}},
  \bibinfo{author}{\bibfnamefont{Y.~G.} \bibnamefont{Kusraev}},
  \bibinfo{author}{\bibfnamefont{N.~N.} \bibnamefont{Ledentsov}},
  \bibinfo{author}{\bibfnamefont{V.~M.} \bibnamefont{Ustinov}},
  \bibinfo{author}{\bibfnamefont{A.~E.} \bibnamefont{Zhukov}}, and
  \bibinfo{author}{\bibfnamefont{A.~F.} \bibnamefont{Tsatsul�nikov}},
  \bibinfo{year}{1998}{\natexlab{a}}, \bibinfo{journal}{Phys. Sol. Sta.}
  \textbf{\bibinfo{volume}{40}}, \bibinfo{pages}{790}.

\bibitem[{\citenamefont{Dzhioev}
  \emph{et~al.}(1998{\natexlab{b}})\citenamefont{Dzhioev, Zakharchenya,
  Korenev, Pak, Vinokurov, Kovalenkov, and Tarasov}}]{Dzhioev:1998b}
\bibinfo{author}{\bibnamefont{Dzhioev}, \bibfnamefont{R.~I.}},
  \bibinfo{author}{\bibfnamefont{B.~P.} \bibnamefont{Zakharchenya}},
  \bibinfo{author}{\bibfnamefont{V.~L.} \bibnamefont{Korenev}},
  \bibinfo{author}{\bibfnamefont{P.~E.} \bibnamefont{Pak}},
  \bibinfo{author}{\bibfnamefont{D.~A.} \bibnamefont{Vinokurov}},
  \bibinfo{author}{\bibfnamefont{O.~V.} \bibnamefont{Kovalenkov}}, and
  \bibinfo{author}{\bibfnamefont{I.~S.} \bibnamefont{Tarasov}},
  \bibinfo{year}{1998}{\natexlab{b}}, \bibinfo{journal}{Physics of the Solid
  State} \textbf{\bibinfo{volume}{40}}(\bibinfo{number}{24}),
  \bibinfo{pages}{1587}.

\bibitem[{\citenamefont{Ebbens} \emph{et~al.}(2005)\citenamefont{Ebbens,
  Krizhanovskii, Tartakovskii, Pulizzi, Wright, Savelyev, Skolnick, and
  Hopkinson}}]{Ebbens:2005a}
\bibinfo{author}{\bibnamefont{Ebbens}, \bibfnamefont{A.}},
  \bibinfo{author}{\bibfnamefont{D.~N.} \bibnamefont{Krizhanovskii}},
  \bibinfo{author}{\bibfnamefont{A.~I.} \bibnamefont{Tartakovskii}},
  \bibinfo{author}{\bibfnamefont{F.}~\bibnamefont{Pulizzi}},
  \bibinfo{author}{\bibfnamefont{T.}~\bibnamefont{Wright}},
  \bibinfo{author}{\bibfnamefont{A.~V.} \bibnamefont{Savelyev}},
  \bibinfo{author}{\bibfnamefont{M.~S.} \bibnamefont{Skolnick}}, and
  \bibinfo{author}{\bibfnamefont{M.}~\bibnamefont{Hopkinson}},
  \bibinfo{year}{2005}, \bibinfo{journal}{Phys. Rev. B}
  \textbf{\bibinfo{volume}{72}}, \bibinfo{pages}{73307}.

\bibitem[{\citenamefont{Eble} \emph{et~al.}(2010)\citenamefont{Eble, Desfonds,
  Bernardot, Testelin, and Chamarro}}]{Eble:2010a}
\bibinfo{author}{\bibnamefont{Eble}, \bibfnamefont{B.}},
  \bibinfo{author}{\bibfnamefont{P.}~\bibnamefont{Desfonds}},
  \bibinfo{author}{\bibfnamefont{F.}~\bibnamefont{Bernardot}},
  \bibinfo{author}{\bibfnamefont{C.}~\bibnamefont{Testelin}}, and
  \bibinfo{author}{\bibfnamefont{M.}~\bibnamefont{Chamarro}},
  \bibinfo{year}{2010}, \bibinfo{journal}{Phys. Rev. B}
  \textbf{\bibinfo{volume}{81}}(\bibinfo{number}{045322}), \bibinfo{pages}{8}.

\bibitem[{\citenamefont{Eble} \emph{et~al.}(2006)\citenamefont{Eble, Krebs,
  Lema\^\i{}tre, Kowalik, Kudelski, Voisin, Urbaszek, Marie, and
  Amand}}]{Eble:2006a}
\bibinfo{author}{\bibnamefont{Eble}, \bibfnamefont{B.}},
  \bibinfo{author}{\bibfnamefont{O.}~\bibnamefont{Krebs}},
  \bibinfo{author}{\bibfnamefont{A.}~\bibnamefont{Lema\^\i{}tre}},
  \bibinfo{author}{\bibfnamefont{K.}~\bibnamefont{Kowalik}},
  \bibinfo{author}{\bibfnamefont{A.}~\bibnamefont{Kudelski}},
  \bibinfo{author}{\bibfnamefont{P.}~\bibnamefont{Voisin}},
  \bibinfo{author}{\bibfnamefont{B.}~\bibnamefont{Urbaszek}},
  \bibinfo{author}{\bibfnamefont{X.}~\bibnamefont{Marie}}, and
  \bibinfo{author}{\bibfnamefont{T.}~\bibnamefont{Amand}},
  \bibinfo{year}{2006}, \bibinfo{journal}{Phys. Rev. B}
  \textbf{\bibinfo{volume}{74}}(\bibinfo{number}{8}), \bibinfo{pages}{081306}.

\bibitem[{\citenamefont{Eble} \emph{et~al.}(2009)\citenamefont{Eble, Testelin,
  Desfonds, Bernardot, Balocchi, Amand, Miard, Lema\^\i{}tre, Marie, and
  Chamarro}}]{Eble:2009a}
\bibinfo{author}{\bibnamefont{Eble}, \bibfnamefont{B.}},
  \bibinfo{author}{\bibfnamefont{C.}~\bibnamefont{Testelin}},
  \bibinfo{author}{\bibfnamefont{P.}~\bibnamefont{Desfonds}},
  \bibinfo{author}{\bibfnamefont{F.}~\bibnamefont{Bernardot}},
  \bibinfo{author}{\bibfnamefont{A.}~\bibnamefont{Balocchi}},
  \bibinfo{author}{\bibfnamefont{T.}~\bibnamefont{Amand}},
  \bibinfo{author}{\bibfnamefont{A.}~\bibnamefont{Miard}},
  \bibinfo{author}{\bibfnamefont{A.}~\bibnamefont{Lema\^\i{}tre}},
  \bibinfo{author}{\bibfnamefont{X.}~\bibnamefont{Marie}}, and
  \bibinfo{author}{\bibfnamefont{M.}~\bibnamefont{Chamarro}},
  \bibinfo{year}{2009}, \bibinfo{journal}{Phys. Rev. Lett.}
  \textbf{\bibinfo{volume}{102}}(\bibinfo{number}{14}),
  \bibinfo{pages}{146601}.

\bibitem[{\citenamefont{Elzerman} \emph{et~al.}(2004)\citenamefont{Elzerman,
  Hanson, van Beveren, Witkamp, Vandersypen, and Kouwenhoven}}]{Elzerman:2004a}
\bibinfo{author}{\bibnamefont{Elzerman}, \bibfnamefont{J.~M.}},
  \bibinfo{author}{\bibfnamefont{R.}~\bibnamefont{Hanson}},
  \bibinfo{author}{\bibfnamefont{L.~H.~W.} \bibnamefont{van Beveren}},
  \bibinfo{author}{\bibfnamefont{B.}~\bibnamefont{Witkamp}},
  \bibinfo{author}{\bibfnamefont{L.~M.~K.} \bibnamefont{Vandersypen}}, and
  \bibinfo{author}{\bibfnamefont{L.~P.} \bibnamefont{Kouwenhoven}},
  \bibinfo{year}{2004}, \bibinfo{journal}{Nature}
  \textbf{\bibinfo{volume}{430}}, \bibinfo{pages}{431}.

\bibitem[{\citenamefont{Erlingsson}
  \emph{et~al.}(2001)\citenamefont{Erlingsson, Nazarov, and
  Fal'ko}}]{Erlingsson:2001a}
\bibinfo{author}{\bibnamefont{Erlingsson}, \bibfnamefont{S.~I.}},
  \bibinfo{author}{\bibfnamefont{Y.~V.} \bibnamefont{Nazarov}}, and
  \bibinfo{author}{\bibfnamefont{V.~I.} \bibnamefont{Fal'ko}},
  \bibinfo{year}{2001}, \bibinfo{journal}{Phys. Rev. B}
  \textbf{\bibinfo{volume}{64}}(\bibinfo{number}{19}), \bibinfo{pages}{195306}.

\bibitem[{\citenamefont{Fallahi} \emph{et~al.}(2010)\citenamefont{Fallahi,
  Y\ifmmode \imath \else~\i \fi{}lmaz, and Imamo\ifmmode~\breve{g}\else
  \u{g}\fi{}lu}}]{Fallahi:2010a}
\bibinfo{author}{\bibnamefont{Fallahi}, \bibfnamefont{P.}},
  \bibinfo{author}{\bibfnamefont{S.~T.} \bibnamefont{Y\ifmmode \imath \else~\i
  \fi{}lmaz}}, and
  \bibinfo{author}{\bibfnamefont{A.}~\bibnamefont{Imamo\ifmmode~\breve{g}\else
  \u{g}\fi{}lu}}, \bibinfo{year}{2010}, \bibinfo{journal}{Phys. Rev. Lett.}
  \textbf{\bibinfo{volume}{105}}(\bibinfo{number}{25}),
  \bibinfo{pages}{257402}.

\bibitem[{\citenamefont{Feng} \emph{et~al.}(2007)\citenamefont{Feng, Akimov,
  and Henneberger}}]{Feng:2007a}
\bibinfo{author}{\bibnamefont{Feng}, \bibfnamefont{D.~H.}},
  \bibinfo{author}{\bibfnamefont{I.~A.} \bibnamefont{Akimov}}, and
  \bibinfo{author}{\bibfnamefont{F.}~\bibnamefont{Henneberger}},
  \bibinfo{year}{2007}, \bibinfo{journal}{Phys. Rev. Lett.}
  \textbf{\bibinfo{volume}{99}}(\bibinfo{number}{3}), \bibinfo{pages}{036604}.

\bibitem[{\citenamefont{Fischer} \emph{et~al.}(2008)\citenamefont{Fischer,
  Coish, Bulaev, and Loss}}]{Fischer:2008a}
\bibinfo{author}{\bibnamefont{Fischer}, \bibfnamefont{J.}},
  \bibinfo{author}{\bibfnamefont{W.~A.} \bibnamefont{Coish}},
  \bibinfo{author}{\bibfnamefont{D.~V.} \bibnamefont{Bulaev}}, and
  \bibinfo{author}{\bibfnamefont{D.}~\bibnamefont{Loss}}, \bibinfo{year}{2008},
  \bibinfo{journal}{Phys. Rev. B}
  \textbf{\bibinfo{volume}{78}}(\bibinfo{number}{15}), \bibinfo{pages}{155329}.

\bibitem[{\citenamefont{Flisinski} \emph{et~al.}(2010)\citenamefont{Flisinski,
  Gerlovin, Ignatiev, Petrov, Verbin, Yakovlev, Reuter, Wieck, and
  Bayer}}]{Flisinski:2010a}
\bibinfo{author}{\bibnamefont{Flisinski}, \bibfnamefont{K.}},
  \bibinfo{author}{\bibfnamefont{I.~Y.} \bibnamefont{Gerlovin}},
  \bibinfo{author}{\bibfnamefont{I.~V.} \bibnamefont{Ignatiev}},
  \bibinfo{author}{\bibfnamefont{M.~Y.} \bibnamefont{Petrov}},
  \bibinfo{author}{\bibfnamefont{S.~Y.} \bibnamefont{Verbin}},
  \bibinfo{author}{\bibfnamefont{D.~R.} \bibnamefont{Yakovlev}},
  \bibinfo{author}{\bibfnamefont{D.}~\bibnamefont{Reuter}},
  \bibinfo{author}{\bibfnamefont{A.~D.} \bibnamefont{Wieck}}, and
  \bibinfo{author}{\bibfnamefont{M.}~\bibnamefont{Bayer}},
  \bibinfo{year}{2010}, \bibinfo{journal}{Phys. Rev. B}
  \textbf{\bibinfo{volume}{82}}(\bibinfo{number}{8}), \bibinfo{pages}{081308}.

\bibitem[{\citenamefont{Flissikowski}
  \emph{et~al.}(2003)\citenamefont{Flissikowski, Akimov, Hundt, and
  Henneberger}}]{Flissikowski:2003a}
\bibinfo{author}{\bibnamefont{Flissikowski}, \bibfnamefont{T.}},
  \bibinfo{author}{\bibfnamefont{I.~A.} \bibnamefont{Akimov}},
  \bibinfo{author}{\bibfnamefont{A.}~\bibnamefont{Hundt}}, and
  \bibinfo{author}{\bibfnamefont{F.}~\bibnamefont{Henneberger}},
  \bibinfo{year}{2003}, \bibinfo{journal}{Phys. Rev. B}
  \textbf{\bibinfo{volume}{68}}(\bibinfo{number}{16}), \bibinfo{pages}{161309}.

\bibitem[{\citenamefont{Flissikowski}
  \emph{et~al.}(2001)\citenamefont{Flissikowski, Hundt, Lowisch, Rabe, and
  Henneberger}}]{Flissikowski:2001a}
\bibinfo{author}{\bibnamefont{Flissikowski}, \bibfnamefont{T.}},
  \bibinfo{author}{\bibfnamefont{A.}~\bibnamefont{Hundt}},
  \bibinfo{author}{\bibfnamefont{M.}~\bibnamefont{Lowisch}},
  \bibinfo{author}{\bibfnamefont{M.}~\bibnamefont{Rabe}}, and
  \bibinfo{author}{\bibfnamefont{F.}~\bibnamefont{Henneberger}},
  \bibinfo{year}{2001}, \bibinfo{journal}{Phys. Rev. Lett.}
  \textbf{\bibinfo{volume}{86}}(\bibinfo{number}{14}), \bibinfo{pages}{3172}.

\bibitem[{\citenamefont{Fras} \emph{et~al.}(2011)\citenamefont{Fras, Eble,
  Desfonds, Bernardot, Testelin, Chamarro, Miard, and Lemaitre}}]{Fras:2011a}
\bibinfo{author}{\bibnamefont{Fras}, \bibfnamefont{F.}},
  \bibinfo{author}{\bibfnamefont{B.}~\bibnamefont{Eble}},
  \bibinfo{author}{\bibfnamefont{P.}~\bibnamefont{Desfonds}},
  \bibinfo{author}{\bibfnamefont{F.}~\bibnamefont{Bernardot}},
  \bibinfo{author}{\bibfnamefont{C.}~\bibnamefont{Testelin}},
  \bibinfo{author}{\bibfnamefont{M.}~\bibnamefont{Chamarro}},
  \bibinfo{author}{\bibfnamefont{A.}~\bibnamefont{Miard}}, and
  \bibinfo{author}{\bibfnamefont{A.}~\bibnamefont{Lemaitre}},
  \bibinfo{year}{2011}, \bibinfo{journal}{Phys. Rev. B}
  \textbf{\bibinfo{volume}{84}}, \bibinfo{pages}{125431}.

\bibitem[{\citenamefont{Gammon} \emph{et~al.}(1997)\citenamefont{Gammon, Brown,
  Snow, Kennedy, Katzer, and Park}}]{Gammon:1997a}
\bibinfo{author}{\bibnamefont{Gammon}, \bibfnamefont{D.}},
  \bibinfo{author}{\bibfnamefont{S.}~\bibnamefont{Brown}},
  \bibinfo{author}{\bibfnamefont{E.}~\bibnamefont{Snow}},
  \bibinfo{author}{\bibfnamefont{T.}~\bibnamefont{Kennedy}},
  \bibinfo{author}{\bibfnamefont{D.}~\bibnamefont{Katzer}}, and
  \bibinfo{author}{\bibfnamefont{D.}~\bibnamefont{Park}}, \bibinfo{year}{1997},
  \bibinfo{journal}{Science} \textbf{\bibinfo{volume}{277}},
  \bibinfo{pages}{85}.

\bibitem[{\citenamefont{Gammon} \emph{et~al.}(2001)\citenamefont{Gammon, Efros,
  Kennedy, Rosen, Katzer, Park, Brown, Korenev, and Merkulov}}]{Gammon:2001a}
\bibinfo{author}{\bibnamefont{Gammon}, \bibfnamefont{D.}},
  \bibinfo{author}{\bibfnamefont{A.~L.} \bibnamefont{Efros}},
  \bibinfo{author}{\bibfnamefont{T.~A.} \bibnamefont{Kennedy}},
  \bibinfo{author}{\bibfnamefont{M.}~\bibnamefont{Rosen}},
  \bibinfo{author}{\bibfnamefont{D.~S.} \bibnamefont{Katzer}},
  \bibinfo{author}{\bibfnamefont{D.}~\bibnamefont{Park}},
  \bibinfo{author}{\bibfnamefont{S.~W.} \bibnamefont{Brown}},
  \bibinfo{author}{\bibfnamefont{V.~L.} \bibnamefont{Korenev}}, and
  \bibinfo{author}{\bibfnamefont{I.~A.} \bibnamefont{Merkulov}},
  \bibinfo{year}{2001}, \bibinfo{journal}{Phys. Rev. Lett.}
  \textbf{\bibinfo{volume}{86}}(\bibinfo{number}{22}), \bibinfo{pages}{5176}.

\bibitem[{\citenamefont{Gammon} \emph{et~al.}(1996)\citenamefont{Gammon, Snow,
  Shanabrook, Katzer, and Park}}]{Gammon:1996a}
\bibinfo{author}{\bibnamefont{Gammon}, \bibfnamefont{D.}},
  \bibinfo{author}{\bibfnamefont{E.~S.} \bibnamefont{Snow}},
  \bibinfo{author}{\bibfnamefont{B.~V.} \bibnamefont{Shanabrook}},
  \bibinfo{author}{\bibfnamefont{D.~S.} \bibnamefont{Katzer}}, and
  \bibinfo{author}{\bibfnamefont{D.}~\bibnamefont{Park}}, \bibinfo{year}{1996},
  \bibinfo{journal}{Phys. Rev. Lett.}
  \textbf{\bibinfo{volume}{76}}(\bibinfo{number}{16}), \bibinfo{pages}{3005}.

\bibitem[{\citenamefont{Gerardot} \emph{et~al.}(2008)\citenamefont{Gerardot,
  Brunner, Dalgarno, Ohberg, Seidl, Kroner, Karrai, G.~Stoltz, Petroff, and
  Warburton}}]{Gerardot:2008a}
\bibinfo{author}{\bibnamefont{Gerardot}, \bibfnamefont{B.~D.}},
  \bibinfo{author}{\bibfnamefont{D.}~\bibnamefont{Brunner}},
  \bibinfo{author}{\bibfnamefont{P.~A.} \bibnamefont{Dalgarno}},
  \bibinfo{author}{\bibfnamefont{P.}~\bibnamefont{Ohberg}},
  \bibinfo{author}{\bibfnamefont{S.}~\bibnamefont{Seidl}},
  \bibinfo{author}{\bibfnamefont{M.}~\bibnamefont{Kroner}},
  \bibinfo{author}{\bibfnamefont{K.}~\bibnamefont{Karrai}},
  \bibinfo{author}{\bibfnamefont{N.}~\bibnamefont{G.~Stoltz}},
  \bibinfo{author}{\bibfnamefont{P.~M.} \bibnamefont{Petroff}}, and
  \bibinfo{author}{\bibfnamefont{R.~J.} \bibnamefont{Warburton}},
  \bibinfo{year}{2008}, \bibinfo{journal}{Nature}
  \textbf{\bibinfo{volume}{451}}, \bibinfo{pages}{441}.

\bibitem[{\citenamefont{Girard} \emph{et~al.}(2009)\citenamefont{Girard,
  Lema\^{i}tre, Miard, David, and Wang}}]{Girard:2009a}
\bibinfo{author}{\bibnamefont{Girard}, \bibfnamefont{J.~C.}},
  \bibinfo{author}{\bibfnamefont{A.}~\bibnamefont{Lema\^{i}tre}},
  \bibinfo{author}{\bibfnamefont{A.}~\bibnamefont{Miard}},
  \bibinfo{author}{\bibfnamefont{C.}~\bibnamefont{David}}, and
  \bibinfo{author}{\bibfnamefont{Z.~Z.} \bibnamefont{Wang}},
  \bibinfo{year}{2009} (\bibinfo{publisher}{AVS}), volume~\bibinfo{volume}{27},
  pp. \bibinfo{pages}{891--894}.

\bibitem[{\citenamefont{Glazov} \emph{et~al.}(2012)\citenamefont{Glazov,
  Yugova, and Efros}}]{Glazov:2012a}
\bibinfo{author}{\bibnamefont{Glazov}, \bibfnamefont{M.~M.}},
  \bibinfo{author}{\bibfnamefont{I.~A.} \bibnamefont{Yugova}}, and
  \bibinfo{author}{\bibfnamefont{A.~L.} \bibnamefont{Efros}},
  \bibinfo{year}{2012}, \bibinfo{journal}{Phys. Rev. B}
  \textbf{\bibinfo{volume}{85}}, \bibinfo{pages}{041303}.

\bibitem[{\citenamefont{Godden} \emph{et~al.}(2012)\citenamefont{Godden,
  Quilter, Ramsay, Wu, Brereton, Boyle, Luxmoore, Puebla-Nunez, Fox, and
  Skolnick}}]{Godden:2012a}
\bibinfo{author}{\bibnamefont{Godden}, \bibfnamefont{T.~M.}},
  \bibinfo{author}{\bibfnamefont{J.~H.} \bibnamefont{Quilter}},
  \bibinfo{author}{\bibfnamefont{A.~J.} \bibnamefont{Ramsay}},
  \bibinfo{author}{\bibfnamefont{Y.}~\bibnamefont{Wu}},
  \bibinfo{author}{\bibfnamefont{P.}~\bibnamefont{Brereton}},
  \bibinfo{author}{\bibfnamefont{S.~J.} \bibnamefont{Boyle}},
  \bibinfo{author}{\bibfnamefont{I.~J.} \bibnamefont{Luxmoore}},
  \bibinfo{author}{\bibfnamefont{J.}~\bibnamefont{Puebla-Nunez}},
  \bibinfo{author}{\bibfnamefont{A.~M.} \bibnamefont{Fox}}, and
  \bibinfo{author}{\bibfnamefont{M.~S.} \bibnamefont{Skolnick}},
  \bibinfo{year}{2012}, \bibinfo{journal}{Phys. Rev. Lett.}
  \textbf{\bibinfo{volume}{108}}, \bibinfo{pages}{017402}.

\bibitem[{\citenamefont{Goldman}(1970)}]{Goldman:1970a}
\bibinfo{author}{\bibnamefont{Goldman}, \bibfnamefont{M.}},
  \bibinfo{year}{1970}, \emph{\bibinfo{title}{Spin Temperature and nuclear
  Magnetic Resonance in Solids}} (\bibinfo{publisher}{Oxford University
  Press}).

\bibitem[{\citenamefont{Goldstein} \emph{et~al.}(1985)\citenamefont{Goldstein,
  Glas, Marzin, Charasse, and Roux}}]{Goldstein:1985a}
\bibinfo{author}{\bibnamefont{Goldstein}, \bibfnamefont{L.}},
  \bibinfo{author}{\bibfnamefont{F.}~\bibnamefont{Glas}},
  \bibinfo{author}{\bibfnamefont{J.~Y.} \bibnamefont{Marzin}},
  \bibinfo{author}{\bibfnamefont{M.~N.} \bibnamefont{Charasse}}, and
  \bibinfo{author}{\bibfnamefont{G.~L.} \bibnamefont{Roux}},
  \bibinfo{year}{1985}, \bibinfo{journal}{Applied Physics Letters}
  \textbf{\bibinfo{volume}{47}}(\bibinfo{number}{10}), \bibinfo{pages}{1099}.

\bibitem[{\citenamefont{Greilich} \emph{et~al.}(2011)\citenamefont{Greilich,
  Carter, Kim, Bracker, and Gammon}}]{Greilich:2011a}
\bibinfo{author}{\bibnamefont{Greilich}, \bibfnamefont{A.}},
  \bibinfo{author}{\bibfnamefont{S.~G.} \bibnamefont{Carter}},
  \bibinfo{author}{\bibfnamefont{D.}~\bibnamefont{Kim}},
  \bibinfo{author}{\bibfnamefont{A.~S.} \bibnamefont{Bracker}}, and
  \bibinfo{author}{\bibfnamefont{D.}~\bibnamefont{Gammon}},
  \bibinfo{year}{2011}, \bibinfo{journal}{Nature Photonics}
  \textbf{\bibinfo{volume}{5}}, \bibinfo{pages}{702–708}.

\bibitem[{\citenamefont{Greilich}
  \emph{et~al.}(2006{\natexlab{a}})\citenamefont{Greilich, Oulton, Zhukov,
  Yugova, Yakovlev, Bayer, Shabaev, Efros, Merkulov, Stavarache, Reuter, and
  Wieck}}]{Greilich:2006b}
\bibinfo{author}{\bibnamefont{Greilich}, \bibfnamefont{A.}},
  \bibinfo{author}{\bibfnamefont{R.}~\bibnamefont{Oulton}},
  \bibinfo{author}{\bibfnamefont{E.~A.} \bibnamefont{Zhukov}},
  \bibinfo{author}{\bibfnamefont{I.~A.} \bibnamefont{Yugova}},
  \bibinfo{author}{\bibfnamefont{D.~R.} \bibnamefont{Yakovlev}},
  \bibinfo{author}{\bibfnamefont{M.}~\bibnamefont{Bayer}},
  \bibinfo{author}{\bibfnamefont{A.}~\bibnamefont{Shabaev}},
  \bibinfo{author}{\bibfnamefont{A.~L.} \bibnamefont{Efros}},
  \bibinfo{author}{\bibfnamefont{I.~A.} \bibnamefont{Merkulov}},
  \bibinfo{author}{\bibfnamefont{V.}~\bibnamefont{Stavarache}},
  \bibinfo{author}{\bibfnamefont{D.}~\bibnamefont{Reuter}}, and
  \bibinfo{author}{\bibfnamefont{A.}~\bibnamefont{Wieck}},
  \bibinfo{year}{2006}{\natexlab{a}}, \bibinfo{journal}{Phys. Rev. Lett.}
  \textbf{\bibinfo{volume}{96}}(\bibinfo{number}{22}), \bibinfo{pages}{227401}.

\bibitem[{\citenamefont{Greilich} \emph{et~al.}(2007)\citenamefont{Greilich,
  Shabaev, Yakovlev, Efros, Yugova, Reuter, Wieck, and Bayer}}]{Greilich:2007a}
\bibinfo{author}{\bibnamefont{Greilich}, \bibfnamefont{A.}},
  \bibinfo{author}{\bibfnamefont{A.}~\bibnamefont{Shabaev}},
  \bibinfo{author}{\bibfnamefont{D.~R.} \bibnamefont{Yakovlev}},
  \bibinfo{author}{\bibfnamefont{A.~L.} \bibnamefont{Efros}},
  \bibinfo{author}{\bibfnamefont{I.~A.} \bibnamefont{Yugova}},
  \bibinfo{author}{\bibfnamefont{D.}~\bibnamefont{Reuter}},
  \bibinfo{author}{\bibfnamefont{A.~D.} \bibnamefont{Wieck}}, and
  \bibinfo{author}{\bibfnamefont{M.}~\bibnamefont{Bayer}},
  \bibinfo{year}{2007}, \bibinfo{journal}{Science}
  \textbf{\bibinfo{volume}{317}}(\bibinfo{number}{5846}),
  \bibinfo{pages}{1896}.

\bibitem[{\citenamefont{Greilich}
  \emph{et~al.}(2006{\natexlab{b}})\citenamefont{Greilich, Yakovlev, Shabaev,
  Efros, Yugova, Oulton, Stavarache, Reuter, Wieck, and
  Bayer}}]{Greilich:2006a}
\bibinfo{author}{\bibnamefont{Greilich}, \bibfnamefont{A.}},
  \bibinfo{author}{\bibfnamefont{D.~R.} \bibnamefont{Yakovlev}},
  \bibinfo{author}{\bibfnamefont{A.}~\bibnamefont{Shabaev}},
  \bibinfo{author}{\bibfnamefont{A.~L.} \bibnamefont{Efros}},
  \bibinfo{author}{\bibfnamefont{I.~A.} \bibnamefont{Yugova}},
  \bibinfo{author}{\bibfnamefont{R.}~\bibnamefont{Oulton}},
  \bibinfo{author}{\bibfnamefont{V.}~\bibnamefont{Stavarache}},
  \bibinfo{author}{\bibfnamefont{D.}~\bibnamefont{Reuter}},
  \bibinfo{author}{\bibfnamefont{A.}~\bibnamefont{Wieck}}, and
  \bibinfo{author}{\bibfnamefont{M.}~\bibnamefont{Bayer}},
  \bibinfo{year}{2006}{\natexlab{b}}, \bibinfo{journal}{Science}
  \textbf{\bibinfo{volume}{313}}(\bibinfo{number}{5785}), \bibinfo{pages}{341}.

\bibitem[{\citenamefont{Greve} \emph{et~al.}(2011)\citenamefont{Greve, McMahon,
  Press, Bisping, Schneider, Kamp, Worschech, Höfling, Forchel, and
  Yamamoto}}]{DeGreve:2011a}
\bibinfo{author}{\bibnamefont{Greve}, \bibfnamefont{K.~D.}},
  \bibinfo{author}{\bibfnamefont{P.~L.} \bibnamefont{McMahon}},
  \bibinfo{author}{\bibfnamefont{D.}~\bibnamefont{Press}},
  \bibinfo{author}{\bibfnamefont{T.~D. L.~D.} \bibnamefont{Bisping}},
  \bibinfo{author}{\bibfnamefont{C.}~\bibnamefont{Schneider}},
  \bibinfo{author}{\bibfnamefont{M.}~\bibnamefont{Kamp}},
  \bibinfo{author}{\bibfnamefont{L.}~\bibnamefont{Worschech}},
  \bibinfo{author}{\bibfnamefont{S.}~\bibnamefont{Höfling}},
  \bibinfo{author}{\bibfnamefont{A.}~\bibnamefont{Forchel}}, and
  \bibinfo{author}{\bibfnamefont{Y.}~\bibnamefont{Yamamoto}},
  \bibinfo{year}{2011}, \bibinfo{journal}{Nature Physics}
  \textbf{\bibinfo{volume}{7}}, \bibinfo{pages}{872–878}.

\bibitem[{\citenamefont{Grundmann} \emph{et~al.}(1995)\citenamefont{Grundmann,
  Stier, and Bimberg}}]{Grundmann:1995a}
\bibinfo{author}{\bibnamefont{Grundmann}, \bibfnamefont{M.}},
  \bibinfo{author}{\bibfnamefont{O.}~\bibnamefont{Stier}}, and
  \bibinfo{author}{\bibfnamefont{D.}~\bibnamefont{Bimberg}},
  \bibinfo{year}{1995}, \bibinfo{journal}{Phys. Rev. B}
  \textbf{\bibinfo{volume}{52}}, \bibinfo{pages}{11969}.

\bibitem[{\citenamefont{Gryncharova and Perel}(1977)}]{Gryncharova:1977a}
\bibinfo{author}{\bibnamefont{Gryncharova}, \bibfnamefont{E.}}, and
  \bibinfo{author}{\bibfnamefont{V.}~\bibnamefont{Perel}},
  \bibinfo{year}{1977}, \bibinfo{journal}{Sov. Phys. Semicond.}
  \textbf{\bibinfo{volume}{11}}(\bibinfo{number}{997}).

\bibitem[{\citenamefont{Hanson} \emph{et~al.}(2007)\citenamefont{Hanson,
  Kouwenhoven, Petta, Tarucha, and Vandersypen}}]{Hanson:2007a}
\bibinfo{author}{\bibnamefont{Hanson}, \bibfnamefont{R.}},
  \bibinfo{author}{\bibfnamefont{L.~P.} \bibnamefont{Kouwenhoven}},
  \bibinfo{author}{\bibfnamefont{J.~R.} \bibnamefont{Petta}},
  \bibinfo{author}{\bibfnamefont{S.}~\bibnamefont{Tarucha}}, and
  \bibinfo{author}{\bibfnamefont{L.~M.~K.} \bibnamefont{Vandersypen}},
  \bibinfo{year}{2007}, \bibinfo{journal}{Rev. Mod. Phys.}
  \textbf{\bibinfo{volume}{79}}(\bibinfo{number}{4}), \bibinfo{pages}{1217}.

\bibitem[{\citenamefont{Heiss} \emph{et~al.}(2007)\citenamefont{Heiss, Schaeck,
  Huebl, Bichler, Abstreiter, Finley, Bulaev, and Loss}}]{Heiss:2007a}
\bibinfo{author}{\bibnamefont{Heiss}, \bibfnamefont{D.}},
  \bibinfo{author}{\bibfnamefont{S.}~\bibnamefont{Schaeck}},
  \bibinfo{author}{\bibfnamefont{H.}~\bibnamefont{Huebl}},
  \bibinfo{author}{\bibfnamefont{M.}~\bibnamefont{Bichler}},
  \bibinfo{author}{\bibfnamefont{G.}~\bibnamefont{Abstreiter}},
  \bibinfo{author}{\bibfnamefont{J.~J.} \bibnamefont{Finley}},
  \bibinfo{author}{\bibfnamefont{D.~V.} \bibnamefont{Bulaev}}, and
  \bibinfo{author}{\bibfnamefont{D.}~\bibnamefont{Loss}}, \bibinfo{year}{2007},
  \bibinfo{journal}{Phys. Rev. B} \textbf{\bibinfo{volume}{76}},
  \bibinfo{pages}{241306}.

\bibitem[{\citenamefont{Henneberger and Benson}(2008)}]{Henneberger:2008a}
\bibinfo{author}{\bibnamefont{Henneberger}, \bibfnamefont{F.}}, and
  \bibinfo{author}{\bibfnamefont{O.}~\bibnamefont{Benson}},
  \bibinfo{year}{2008}, \bibinfo{journal}{Pan Stanford Publishing} .

\bibitem[{\citenamefont{H\"ogele} \emph{et~al.}(2012)\citenamefont{H\"ogele,
  Kroner, Latta, Claassen, Carusotto, Bulutay, and Imamoglu}}]{Hogele:2012a}
\bibinfo{author}{\bibnamefont{H\"ogele}, \bibfnamefont{A.}},
  \bibinfo{author}{\bibfnamefont{M.}~\bibnamefont{Kroner}},
  \bibinfo{author}{\bibfnamefont{C.}~\bibnamefont{Latta}},
  \bibinfo{author}{\bibfnamefont{M.}~\bibnamefont{Claassen}},
  \bibinfo{author}{\bibfnamefont{I.}~\bibnamefont{Carusotto}},
  \bibinfo{author}{\bibfnamefont{C.}~\bibnamefont{Bulutay}}, and
  \bibinfo{author}{\bibfnamefont{A.}~\bibnamefont{Imamoglu}},
  \bibinfo{year}{2012}, \bibinfo{journal}{Phys. Rev. Lett.}
  \textbf{\bibinfo{volume}{108}}, \bibinfo{pages}{197403}.

\bibitem[{\citenamefont{H\"ogele} \emph{et~al.}(2004)\citenamefont{H\"ogele,
  Seidl, Kroner, Karrai, Warburton, Gerardot, and Petroff}}]{Hogele:2004a}
\bibinfo{author}{\bibnamefont{H\"ogele}, \bibfnamefont{A.}},
  \bibinfo{author}{\bibfnamefont{S.}~\bibnamefont{Seidl}},
  \bibinfo{author}{\bibfnamefont{M.}~\bibnamefont{Kroner}},
  \bibinfo{author}{\bibfnamefont{K.}~\bibnamefont{Karrai}},
  \bibinfo{author}{\bibfnamefont{R.~J.} \bibnamefont{Warburton}},
  \bibinfo{author}{\bibfnamefont{B.~D.} \bibnamefont{Gerardot}}, and
  \bibinfo{author}{\bibfnamefont{P.~M.} \bibnamefont{Petroff}},
  \bibinfo{year}{2004}, \bibinfo{journal}{Phys. Rev. Lett.}
  \textbf{\bibinfo{volume}{93}}(\bibinfo{number}{21}), \bibinfo{pages}{217401}.

\bibitem[{\citenamefont{Hours} \emph{et~al.}(2005)\citenamefont{Hours,
  Senellart, Peter, Cavanna, and Bloch}}]{Hours:2005a}
\bibinfo{author}{\bibnamefont{Hours}, \bibfnamefont{J.}},
  \bibinfo{author}{\bibfnamefont{P.}~\bibnamefont{Senellart}},
  \bibinfo{author}{\bibfnamefont{E.}~\bibnamefont{Peter}},
  \bibinfo{author}{\bibfnamefont{A.}~\bibnamefont{Cavanna}}, and
  \bibinfo{author}{\bibfnamefont{J.}~\bibnamefont{Bloch}},
  \bibinfo{year}{2005}, \bibinfo{journal}{Phys. Rev. B}
  \textbf{\bibinfo{volume}{71}}(\bibinfo{number}{16}), \bibinfo{pages}{161306}.

\bibitem[{\citenamefont{Huang and Hu}(2010)}]{Huang:2010a}
\bibinfo{author}{\bibnamefont{Huang}, \bibfnamefont{C.-W.}}, and
  \bibinfo{author}{\bibfnamefont{X.}~\bibnamefont{Hu}}, \bibinfo{year}{2010},
  \bibinfo{journal}{Phys. Rev. B} \textbf{\bibinfo{volume}{81}},
  \bibinfo{pages}{205304}.

\bibitem[{\citenamefont{Imamoglu} \emph{et~al.}(1999)\citenamefont{Imamoglu,
  Awschalom, Burkard, DiVincenzo, Loss, Sherwin, and Small}}]{Imamoglu:1999a}
\bibinfo{author}{\bibnamefont{Imamoglu}, \bibfnamefont{A.}},
  \bibinfo{author}{\bibfnamefont{D.~D.} \bibnamefont{Awschalom}},
  \bibinfo{author}{\bibfnamefont{G.}~\bibnamefont{Burkard}},
  \bibinfo{author}{\bibfnamefont{D.~P.} \bibnamefont{DiVincenzo}},
  \bibinfo{author}{\bibfnamefont{D.}~\bibnamefont{Loss}},
  \bibinfo{author}{\bibfnamefont{M.}~\bibnamefont{Sherwin}}, and
  \bibinfo{author}{\bibfnamefont{A.}~\bibnamefont{Small}},
  \bibinfo{year}{1999}, \bibinfo{journal}{Phys. Rev. Lett.}
  \textbf{\bibinfo{volume}{83}}(\bibinfo{number}{20}), \bibinfo{pages}{4204}.

\bibitem[{\citenamefont{Imamoglu} \emph{et~al.}(2003)\citenamefont{Imamoglu,
  Knill, Tian, and Zoller}}]{Imamoglu:2003a}
\bibinfo{author}{\bibnamefont{Imamoglu}, \bibfnamefont{A.}},
  \bibinfo{author}{\bibfnamefont{E.}~\bibnamefont{Knill}},
  \bibinfo{author}{\bibfnamefont{L.}~\bibnamefont{Tian}}, and
  \bibinfo{author}{\bibfnamefont{P.}~\bibnamefont{Zoller}},
  \bibinfo{year}{2003}, \bibinfo{journal}{Phys. Rev. Lett.}
  \textbf{\bibinfo{volume}{91}}(\bibinfo{number}{1}), \bibinfo{pages}{017402}.

\bibitem[{\citenamefont{Issler} \emph{et~al.}(2010)\citenamefont{Issler,
  Kessler, Giedke, Yelin, Cirac, Lukin, and Imamoglu}}]{Issler:2010a}
\bibinfo{author}{\bibnamefont{Issler}, \bibfnamefont{M.}},
  \bibinfo{author}{\bibfnamefont{E.~M.} \bibnamefont{Kessler}},
  \bibinfo{author}{\bibfnamefont{G.}~\bibnamefont{Giedke}},
  \bibinfo{author}{\bibfnamefont{S.}~\bibnamefont{Yelin}},
  \bibinfo{author}{\bibfnamefont{I.}~\bibnamefont{Cirac}},
  \bibinfo{author}{\bibfnamefont{M.~D.} \bibnamefont{Lukin}}, and
  \bibinfo{author}{\bibfnamefont{A.}~\bibnamefont{Imamoglu}},
  \bibinfo{year}{2010}, \bibinfo{journal}{Phys. Rev. Lett.}
  \textbf{\bibinfo{volume}{105}}(\bibinfo{number}{26}),
  \bibinfo{pages}{267202}.

\bibitem[{\citenamefont{Kaji} \emph{et~al.}(2008)\citenamefont{Kaji, Adachi,
  Sasakura, and Muto}}]{Kaji:2008a}
\bibinfo{author}{\bibnamefont{Kaji}, \bibfnamefont{R.}},
  \bibinfo{author}{\bibfnamefont{S.}~\bibnamefont{Adachi}},
  \bibinfo{author}{\bibfnamefont{H.}~\bibnamefont{Sasakura}}, and
  \bibinfo{author}{\bibfnamefont{S.}~\bibnamefont{Muto}}, \bibinfo{year}{2008},
  \bibinfo{journal}{Phys. Rev. B}
  \textbf{\bibinfo{volume}{77}}(\bibinfo{number}{11}), \bibinfo{pages}{115345}.

\bibitem[{\citenamefont{Kalevich} \emph{et~al.}(2001)\citenamefont{Kalevich,
  Paillard, Kavokin, Marie, Kovsh, Amand, Zhukov, Musikhin, Ustinov, Vanelle,
  and Zakharchenya}}]{Kalevich:2001a}
\bibinfo{author}{\bibnamefont{Kalevich}, \bibfnamefont{V.~K.}},
  \bibinfo{author}{\bibfnamefont{M.}~\bibnamefont{Paillard}},
  \bibinfo{author}{\bibfnamefont{K.~V.} \bibnamefont{Kavokin}},
  \bibinfo{author}{\bibfnamefont{X.}~\bibnamefont{Marie}},
  \bibinfo{author}{\bibfnamefont{A.~R.} \bibnamefont{Kovsh}},
  \bibinfo{author}{\bibfnamefont{T.}~\bibnamefont{Amand}},
  \bibinfo{author}{\bibfnamefont{A.~E.} \bibnamefont{Zhukov}},
  \bibinfo{author}{\bibfnamefont{Y.~G.} \bibnamefont{Musikhin}},
  \bibinfo{author}{\bibfnamefont{V.~M.} \bibnamefont{Ustinov}},
  \bibinfo{author}{\bibfnamefont{E.}~\bibnamefont{Vanelle}}, and
  \bibinfo{author}{\bibfnamefont{B.~P.} \bibnamefont{Zakharchenya}},
  \bibinfo{year}{2001}, \bibinfo{journal}{Phys. Rev. B}
  \textbf{\bibinfo{volume}{64}}(\bibinfo{number}{4}), \bibinfo{pages}{045309}.

\bibitem[{\citenamefont{Karrai} \emph{et~al.}(2004)\citenamefont{Karrai,
  Warburton, Schulhauser, H\"ogele, Urbaszek, McGhee, Govorov, Garcia,
  Gerardot, and Petroff}}]{Karrai:2004a}
\bibinfo{author}{\bibnamefont{Karrai}, \bibfnamefont{K.}},
  \bibinfo{author}{\bibfnamefont{R.~J.} \bibnamefont{Warburton}},
  \bibinfo{author}{\bibfnamefont{C.}~\bibnamefont{Schulhauser}},
  \bibinfo{author}{\bibfnamefont{A.}~\bibnamefont{H\"ogele}},
  \bibinfo{author}{\bibfnamefont{B.}~\bibnamefont{Urbaszek}},
  \bibinfo{author}{\bibfnamefont{E.~J.} \bibnamefont{McGhee}},
  \bibinfo{author}{\bibfnamefont{A.~O.} \bibnamefont{Govorov}},
  \bibinfo{author}{\bibfnamefont{J.~M.} \bibnamefont{Garcia}},
  \bibinfo{author}{\bibfnamefont{B.~D.} \bibnamefont{Gerardot}}, and
  \bibinfo{author}{\bibfnamefont{P.~M.} \bibnamefont{Petroff}},
  \bibinfo{year}{2004}, \bibinfo{journal}{Nature}
  \textbf{\bibinfo{volume}{427}}, \bibinfo{pages}{135}.

\bibitem[{\citenamefont{Keizer} \emph{et~al.}(2010)\citenamefont{Keizer,
  Bocquel, Koenraad, Mano, Noda, and Sakoda}}]{Keizer:2010a}
\bibinfo{author}{\bibnamefont{Keizer}, \bibfnamefont{J.~G.}},
  \bibinfo{author}{\bibfnamefont{J.}~\bibnamefont{Bocquel}},
  \bibinfo{author}{\bibfnamefont{P.~M.} \bibnamefont{Koenraad}},
  \bibinfo{author}{\bibfnamefont{T.}~\bibnamefont{Mano}},
  \bibinfo{author}{\bibfnamefont{T.}~\bibnamefont{Noda}}, and
  \bibinfo{author}{\bibfnamefont{K.}~\bibnamefont{Sakoda}},
  \bibinfo{year}{2010}, \bibinfo{journal}{Applied Physics Letters}
  \textbf{\bibinfo{volume}{96}}(\bibinfo{number}{6}), \bibinfo{eid}{062101}
  (pages~\bibinfo{numpages}{3}).

\bibitem[{\citenamefont{Kessler} \emph{et~al.}(2012)\citenamefont{Kessler,
  Giedke, Imamoglu, Yelin, Lukin, and Cirac}}]{Kessler:2012a}
\bibinfo{author}{\bibnamefont{Kessler}, \bibfnamefont{E.~M.}},
  \bibinfo{author}{\bibfnamefont{G.}~\bibnamefont{Giedke}},
  \bibinfo{author}{\bibfnamefont{A.}~\bibnamefont{Imamoglu}},
  \bibinfo{author}{\bibfnamefont{S.~F.} \bibnamefont{Yelin}},
  \bibinfo{author}{\bibfnamefont{M.~D.} \bibnamefont{Lukin}}, and
  \bibinfo{author}{\bibfnamefont{J.~I.} \bibnamefont{Cirac}},
  \bibinfo{year}{2012}, \bibinfo{journal}{ArXiv e-prints} \eprint{1205.3341}.

\bibitem[{\citenamefont{Kessler} \emph{et~al.}(2010)\citenamefont{Kessler,
  Yelin, Lukin, Cirac, and Giedke}}]{Kessler:2010a}
\bibinfo{author}{\bibnamefont{Kessler}, \bibfnamefont{E.~M.}},
  \bibinfo{author}{\bibfnamefont{S.}~\bibnamefont{Yelin}},
  \bibinfo{author}{\bibfnamefont{M.~D.} \bibnamefont{Lukin}},
  \bibinfo{author}{\bibfnamefont{J.~I.} \bibnamefont{Cirac}}, and
  \bibinfo{author}{\bibfnamefont{G.}~\bibnamefont{Giedke}},
  \bibinfo{year}{2010}, \bibinfo{journal}{Phys. Rev. Lett.}
  \textbf{\bibinfo{volume}{104}}(\bibinfo{number}{14}),
  \bibinfo{pages}{143601}.

\bibitem[{\citenamefont{Khaetskii} \emph{et~al.}(2003)\citenamefont{Khaetskii,
  Loss, and Glazman}}]{Khaetskii:2003a}
\bibinfo{author}{\bibnamefont{Khaetskii}, \bibfnamefont{A.}},
  \bibinfo{author}{\bibfnamefont{D.}~\bibnamefont{Loss}}, and
  \bibinfo{author}{\bibfnamefont{L.}~\bibnamefont{Glazman}},
  \bibinfo{year}{2003}, \bibinfo{journal}{Phys. Rev. B}
  \textbf{\bibinfo{volume}{67}}, \bibinfo{pages}{195329}.

\bibitem[{\citenamefont{Khaetskii} \emph{et~al.}(2002)\citenamefont{Khaetskii,
  Loss, and Glazman}}]{Khaetskii:2002a}
\bibinfo{author}{\bibnamefont{Khaetskii}, \bibfnamefont{A.~V.}},
  \bibinfo{author}{\bibfnamefont{D.}~\bibnamefont{Loss}}, and
  \bibinfo{author}{\bibfnamefont{L.}~\bibnamefont{Glazman}},
  \bibinfo{year}{2002}, \bibinfo{journal}{Phys. Rev. Lett.}
  \textbf{\bibinfo{volume}{88}}(\bibinfo{number}{18}), \bibinfo{pages}{186802}.

\bibitem[{\citenamefont{Khaetskii and Nazarov}(2000)}]{Khaetskii:2000a}
\bibinfo{author}{\bibnamefont{Khaetskii}, \bibfnamefont{A.~V.}}, and
  \bibinfo{author}{\bibfnamefont{Y.~V.} \bibnamefont{Nazarov}},
  \bibinfo{year}{2000}, \bibinfo{journal}{Phys. Rev. B}
  \textbf{\bibinfo{volume}{61}}(\bibinfo{number}{19}), \bibinfo{pages}{12639}.

\bibitem[{\citenamefont{Klauser} \emph{et~al.}(2006)\citenamefont{Klauser,
  Coish, and Loss}}]{Klauser:2006a}
\bibinfo{author}{\bibnamefont{Klauser}, \bibfnamefont{D.}},
  \bibinfo{author}{\bibfnamefont{W.~A.} \bibnamefont{Coish}}, and
  \bibinfo{author}{\bibfnamefont{D.}~\bibnamefont{Loss}}, \bibinfo{year}{2006},
  \bibinfo{journal}{Phys. Rev. B} \textbf{\bibinfo{volume}{73}},
  \bibinfo{pages}{205302}.

\bibitem[{\citenamefont{Kloeffel} \emph{et~al.}(2011)\citenamefont{Kloeffel,
  Dalgarno, Urbaszek, Gerardot, Brunner, Petroff, Loss, and
  Warburton}}]{Kloeffel:2011a}
\bibinfo{author}{\bibnamefont{Kloeffel}, \bibfnamefont{C.}},
  \bibinfo{author}{\bibfnamefont{P.~A.} \bibnamefont{Dalgarno}},
  \bibinfo{author}{\bibfnamefont{B.}~\bibnamefont{Urbaszek}},
  \bibinfo{author}{\bibfnamefont{B.~D.} \bibnamefont{Gerardot}},
  \bibinfo{author}{\bibfnamefont{D.}~\bibnamefont{Brunner}},
  \bibinfo{author}{\bibfnamefont{P.~M.} \bibnamefont{Petroff}},
  \bibinfo{author}{\bibfnamefont{D.}~\bibnamefont{Loss}}, and
  \bibinfo{author}{\bibfnamefont{R.~J.} \bibnamefont{Warburton}},
  \bibinfo{year}{2011}, \bibinfo{journal}{Phys. Rev. Lett.}
  \textbf{\bibinfo{volume}{106}}(\bibinfo{number}{4}), \bibinfo{pages}{046802}.

\bibitem[{\citenamefont{Klotz} \emph{et~al.}(2010)\citenamefont{Klotz, Jovanov,
  Kierig, Clark, Bichler, Abstreiter, Brandt, Finley, Schwager, and
  Giedke}}]{Klotz:2010a}
\bibinfo{author}{\bibnamefont{Klotz}, \bibfnamefont{F.}},
  \bibinfo{author}{\bibfnamefont{V.}~\bibnamefont{Jovanov}},
  \bibinfo{author}{\bibfnamefont{J.}~\bibnamefont{Kierig}},
  \bibinfo{author}{\bibfnamefont{E.~C.} \bibnamefont{Clark}},
  \bibinfo{author}{\bibfnamefont{M.}~\bibnamefont{Bichler}},
  \bibinfo{author}{\bibfnamefont{G.}~\bibnamefont{Abstreiter}},
  \bibinfo{author}{\bibfnamefont{M.~S.} \bibnamefont{Brandt}},
  \bibinfo{author}{\bibfnamefont{J.~J.} \bibnamefont{Finley}},
  \bibinfo{author}{\bibfnamefont{H.}~\bibnamefont{Schwager}}, and
  \bibinfo{author}{\bibfnamefont{G.}~\bibnamefont{Giedke}},
  \bibinfo{year}{2010}, \bibinfo{journal}{Phys. Rev. B}
  \textbf{\bibinfo{volume}{82}}(\bibinfo{number}{12}), \bibinfo{pages}{121307}.

\bibitem[{\citenamefont{Knight}(1949)}]{Knight:1949a}
\bibinfo{author}{\bibnamefont{Knight}, \bibfnamefont{W.~D.}},
  \bibinfo{year}{1949}, \bibinfo{journal}{Phys. Rev.}
  \textbf{\bibinfo{volume}{76}}(\bibinfo{number}{8}), \bibinfo{pages}{1259}.

\bibitem[{\citenamefont{Koguchi} \emph{et~al.}(1991)\citenamefont{Koguchi,
  Takahashi, and Chikyow}}]{Koguchi:1991a}
\bibinfo{author}{\bibnamefont{Koguchi}, \bibfnamefont{N.}},
  \bibinfo{author}{\bibfnamefont{S.}~\bibnamefont{Takahashi}}, and
  \bibinfo{author}{\bibfnamefont{T.}~\bibnamefont{Chikyow}},
  \bibinfo{year}{1991}, \bibinfo{journal}{J. Cryst. Growth}
  \textbf{\bibinfo{volume}{111}}, \bibinfo{pages}{688}.

\bibitem[{\citenamefont{Korenev}(2007)}]{Korenev:2007a}
\bibinfo{author}{\bibnamefont{Korenev}, \bibfnamefont{V.~L.}},
  \bibinfo{year}{2007}, \bibinfo{journal}{Phys. Rev. Lett.}
  \textbf{\bibinfo{volume}{99}}(\bibinfo{number}{25}), \bibinfo{pages}{256405}.

\bibitem[{\citenamefont{Koudinov} \emph{et~al.}(2004)\citenamefont{Koudinov,
  Akimov, Kusrayev, and Henneberger}}]{Koudinov:2004a}
\bibinfo{author}{\bibnamefont{Koudinov}, \bibfnamefont{A.~V.}},
  \bibinfo{author}{\bibfnamefont{I.~A.} \bibnamefont{Akimov}},
  \bibinfo{author}{\bibfnamefont{Y.~G.} \bibnamefont{Kusrayev}}, and
  \bibinfo{author}{\bibfnamefont{F.}~\bibnamefont{Henneberger}},
  \bibinfo{year}{2004}, \bibinfo{journal}{Phys. Rev. B}
  \textbf{\bibinfo{volume}{70}}(\bibinfo{number}{24}), \bibinfo{pages}{241305}.

\bibitem[{\citenamefont{Kozlov}(2007)}]{Kozlov:2007a}
\bibinfo{author}{\bibnamefont{Kozlov}, \bibfnamefont{G.~G.}},
  \bibinfo{year}{2007}, \bibinfo{journal}{JETP} \textbf{\bibinfo{volume}{105}},
  \bibinfo{pages}{803}.

\bibitem[{\citenamefont{Krebs} \emph{et~al.}(2008)\citenamefont{Krebs, Eble,
  Lema\^\i{}tre, Voisin, Urbaszek, Amand, and Marie}}]{Krebs:2008a}
\bibinfo{author}{\bibnamefont{Krebs}, \bibfnamefont{O.}},
  \bibinfo{author}{\bibfnamefont{B.}~\bibnamefont{Eble}},
  \bibinfo{author}{\bibfnamefont{A.}~\bibnamefont{Lema\^\i{}tre}},
  \bibinfo{author}{\bibfnamefont{P.}~\bibnamefont{Voisin}},
  \bibinfo{author}{\bibfnamefont{B.}~\bibnamefont{Urbaszek}},
  \bibinfo{author}{\bibfnamefont{T.}~\bibnamefont{Amand}}, and
  \bibinfo{author}{\bibfnamefont{X.}~\bibnamefont{Marie}},
  \bibinfo{year}{2008}, \bibinfo{journal}{C.R. Physique}
  \textbf{\bibinfo{volume}{9}}, \bibinfo{pages}{874}.

\bibitem[{\citenamefont{Krebs} \emph{et~al.}(2010)\citenamefont{Krebs,
  Maletinsky, Amand, Urbaszek, Lema\^\i{}tre, Voisin, Marie, and
  Imamoglu}}]{Krebs:2010a}
\bibinfo{author}{\bibnamefont{Krebs}, \bibfnamefont{O.}},
  \bibinfo{author}{\bibfnamefont{P.}~\bibnamefont{Maletinsky}},
  \bibinfo{author}{\bibfnamefont{T.}~\bibnamefont{Amand}},
  \bibinfo{author}{\bibfnamefont{B.}~\bibnamefont{Urbaszek}},
  \bibinfo{author}{\bibfnamefont{A.}~\bibnamefont{Lema\^\i{}tre}},
  \bibinfo{author}{\bibfnamefont{P.}~\bibnamefont{Voisin}},
  \bibinfo{author}{\bibfnamefont{X.}~\bibnamefont{Marie}}, and
  \bibinfo{author}{\bibfnamefont{A.}~\bibnamefont{Imamoglu}},
  \bibinfo{year}{2010}, \bibinfo{journal}{Phys. Rev. Lett.}
  \textbf{\bibinfo{volume}{104}}(\bibinfo{number}{5}), \bibinfo{pages}{056603}.

\bibitem[{\citenamefont{Krebs and Voisin}(1996)}]{Krebs:1996a}
\bibinfo{author}{\bibnamefont{Krebs}, \bibfnamefont{O.}}, and
  \bibinfo{author}{\bibfnamefont{P.}~\bibnamefont{Voisin}},
  \bibinfo{year}{1996}, \bibinfo{journal}{Phys. Rev. Lett.}
  \textbf{\bibinfo{volume}{77}}, \bibinfo{pages}{1829}.

\bibitem[{\citenamefont{Krizhanovskii}
  \emph{et~al.}(2005)\citenamefont{Krizhanovskii, Ebbens, Tartakovskii,
  Pulizzi, Wright, Skolnick, and Hopkinson}}]{Krizhanovskii:2005a}
\bibinfo{author}{\bibnamefont{Krizhanovskii}, \bibfnamefont{D.~N.}},
  \bibinfo{author}{\bibfnamefont{A.}~\bibnamefont{Ebbens}},
  \bibinfo{author}{\bibfnamefont{A.~I.} \bibnamefont{Tartakovskii}},
  \bibinfo{author}{\bibfnamefont{F.}~\bibnamefont{Pulizzi}},
  \bibinfo{author}{\bibfnamefont{T.}~\bibnamefont{Wright}},
  \bibinfo{author}{\bibfnamefont{M.~S.} \bibnamefont{Skolnick}}, and
  \bibinfo{author}{\bibfnamefont{M.}~\bibnamefont{Hopkinson}},
  \bibinfo{year}{2005}, \bibinfo{journal}{Phys. Rev. B}
  \textbf{\bibinfo{volume}{72}}(\bibinfo{number}{16}), \bibinfo{pages}{161312}.

\bibitem[{\citenamefont{Kroner} \emph{et~al.}(2008)\citenamefont{Kroner, Weiss,
  Biedermann, Seidl, Manus, Holleitner, Badolato, Petroff, Gerardot, Warburton,
  and Karrai}}]{Kroner:2008a}
\bibinfo{author}{\bibnamefont{Kroner}, \bibfnamefont{M.}},
  \bibinfo{author}{\bibfnamefont{K.~M.} \bibnamefont{Weiss}},
  \bibinfo{author}{\bibfnamefont{B.}~\bibnamefont{Biedermann}},
  \bibinfo{author}{\bibfnamefont{S.}~\bibnamefont{Seidl}},
  \bibinfo{author}{\bibfnamefont{S.}~\bibnamefont{Manus}},
  \bibinfo{author}{\bibfnamefont{A.~W.} \bibnamefont{Holleitner}},
  \bibinfo{author}{\bibfnamefont{A.}~\bibnamefont{Badolato}},
  \bibinfo{author}{\bibfnamefont{P.~M.} \bibnamefont{Petroff}},
  \bibinfo{author}{\bibfnamefont{B.~D.} \bibnamefont{Gerardot}},
  \bibinfo{author}{\bibfnamefont{R.~J.} \bibnamefont{Warburton}}, and
  \bibinfo{author}{\bibfnamefont{K.}~\bibnamefont{Karrai}},
  \bibinfo{year}{2008}, \bibinfo{journal}{Phys. Rev. Lett.}
  \textbf{\bibinfo{volume}{100}}(\bibinfo{number}{15}),
  \bibinfo{pages}{156803}.

\bibitem[{\citenamefont{Kroutvar} \emph{et~al.}(2004)\citenamefont{Kroutvar,
  Ducommun, Heiss, Bichler, Schuh, Abstreiter, and Finley}}]{Kroutvar:2004a}
\bibinfo{author}{\bibnamefont{Kroutvar}, \bibfnamefont{M.}},
  \bibinfo{author}{\bibfnamefont{Y.}~\bibnamefont{Ducommun}},
  \bibinfo{author}{\bibfnamefont{D.}~\bibnamefont{Heiss}},
  \bibinfo{author}{\bibfnamefont{M.}~\bibnamefont{Bichler}},
  \bibinfo{author}{\bibfnamefont{D.}~\bibnamefont{Schuh}},
  \bibinfo{author}{\bibfnamefont{G.}~\bibnamefont{Abstreiter}}, and
  \bibinfo{author}{\bibfnamefont{J.~J.} \bibnamefont{Finley}},
  \bibinfo{year}{2004}, \bibinfo{journal}{Nature}
  \textbf{\bibinfo{volume}{432}}, \bibinfo{pages}{81}.

\bibitem[{\citenamefont{Lai} \emph{et~al.}(2006)\citenamefont{Lai, Maletinsky,
  Badolato, and Imamoglu}}]{Lai:2006a}
\bibinfo{author}{\bibnamefont{Lai}, \bibfnamefont{C.~W.}},
  \bibinfo{author}{\bibfnamefont{P.}~\bibnamefont{Maletinsky}},
  \bibinfo{author}{\bibfnamefont{A.}~\bibnamefont{Badolato}}, and
  \bibinfo{author}{\bibfnamefont{A.}~\bibnamefont{Imamoglu}},
  \bibinfo{year}{2006}, \bibinfo{journal}{Phys. Rev. Lett.}
  \textbf{\bibinfo{volume}{96}}(\bibinfo{number}{16}), \bibinfo{pages}{167403}.

\bibitem[{\citenamefont{Lampel}(1968)}]{Lampel:1968a}
\bibinfo{author}{\bibnamefont{Lampel}, \bibfnamefont{G.}},
  \bibinfo{year}{1968}, \bibinfo{journal}{Phys. Rev. Lett.}
  \textbf{\bibinfo{volume}{20}}(\bibinfo{number}{10}), \bibinfo{pages}{491}.

\bibitem[{\citenamefont{Larsson} \emph{et~al.}(2011)\citenamefont{Larsson,
  Moskalenko, and Holtz}}]{Larsson:2011a}
\bibinfo{author}{\bibnamefont{Larsson}, \bibfnamefont{L.~A.}},
  \bibinfo{author}{\bibfnamefont{E.~S.} \bibnamefont{Moskalenko}}, and
  \bibinfo{author}{\bibfnamefont{P.~O.} \bibnamefont{Holtz}},
  \bibinfo{year}{2011}, \bibinfo{journal}{Applied Physics Letters}
  \textbf{\bibinfo{volume}{98}}(\bibinfo{number}{7}), \bibinfo{eid}{071906}
  (pages~\bibinfo{numpages}{3}).

\bibitem[{\citenamefont{Latta}
  \emph{et~al.}(2011{\natexlab{a}})\citenamefont{Latta, Haupt, Hanl,
  Weichselbaum, Claassen, Wuester, Fallahi, Faelt, Glazman, von Delft, Tureci,
  and Imamoglu}}]{Latta:2011b}
\bibinfo{author}{\bibnamefont{Latta}, \bibfnamefont{C.}},
  \bibinfo{author}{\bibfnamefont{F.}~\bibnamefont{Haupt}},
  \bibinfo{author}{\bibfnamefont{M.}~\bibnamefont{Hanl}},
  \bibinfo{author}{\bibfnamefont{A.}~\bibnamefont{Weichselbaum}},
  \bibinfo{author}{\bibfnamefont{M.}~\bibnamefont{Claassen}},
  \bibinfo{author}{\bibfnamefont{W.}~\bibnamefont{Wuester}},
  \bibinfo{author}{\bibfnamefont{P.}~\bibnamefont{Fallahi}},
  \bibinfo{author}{\bibfnamefont{S.}~\bibnamefont{Faelt}},
  \bibinfo{author}{\bibfnamefont{L.}~\bibnamefont{Glazman}},
  \bibinfo{author}{\bibfnamefont{J.}~\bibnamefont{von Delft}},
  \bibinfo{author}{\bibfnamefont{H.~E.} \bibnamefont{Tureci}}, and
  \bibinfo{author}{\bibfnamefont{A.}~\bibnamefont{Imamoglu}},
  \bibinfo{year}{2011}{\natexlab{a}}, \bibinfo{journal}{Nature}
  \textbf{\bibinfo{volume}{474}}, \bibinfo{pages}{627}.

\bibitem[{\citenamefont{Latta} \emph{et~al.}(2009)\citenamefont{Latta, Hogele,
  Zhao, Vamivakas, Maletinsky, Kroner, Dreiser, Carusotto, Badolato, Schuh,
  Wegscheider, Atature} \emph{et~al.}}]{Latta:2009a}
\bibinfo{author}{\bibnamefont{Latta}, \bibfnamefont{C.}},
  \bibinfo{author}{\bibfnamefont{A.}~\bibnamefont{Hogele}},
  \bibinfo{author}{\bibfnamefont{Y.}~\bibnamefont{Zhao}},
  \bibinfo{author}{\bibfnamefont{A.~N.} \bibnamefont{Vamivakas}},
  \bibinfo{author}{\bibfnamefont{P.}~\bibnamefont{Maletinsky}},
  \bibinfo{author}{\bibfnamefont{M.}~\bibnamefont{Kroner}},
  \bibinfo{author}{\bibfnamefont{J.}~\bibnamefont{Dreiser}},
  \bibinfo{author}{\bibfnamefont{I.}~\bibnamefont{Carusotto}},
  \bibinfo{author}{\bibfnamefont{A.}~\bibnamefont{Badolato}},
  \bibinfo{author}{\bibfnamefont{D.}~\bibnamefont{Schuh}},
  \bibinfo{author}{\bibfnamefont{W.}~\bibnamefont{Wegscheider}},
  \bibinfo{author}{\bibfnamefont{M.}~\bibnamefont{Atature}}, \emph{et~al.},
  \bibinfo{year}{2009}, \bibinfo{journal}{Nature Phys.}
  \textbf{\bibinfo{volume}{5}}(\bibinfo{number}{8}), \bibinfo{pages}{758}.

\bibitem[{\citenamefont{Latta}
  \emph{et~al.}(2011{\natexlab{b}})\citenamefont{Latta, Srivastava, and
  Imamo\ifmmode~\breve{g}\else \u{g}\fi{}lu}}]{Latta:2011a}
\bibinfo{author}{\bibnamefont{Latta}, \bibfnamefont{C.}},
  \bibinfo{author}{\bibfnamefont{A.}~\bibnamefont{Srivastava}}, and
  \bibinfo{author}{\bibfnamefont{A.}~\bibnamefont{Imamo\ifmmode~\breve{g}\else
  \u{g}\fi{}lu}}, \bibinfo{year}{2011}{\natexlab{b}}, \bibinfo{journal}{Phys.
  Rev. Lett.} \textbf{\bibinfo{volume}{107}}, \bibinfo{pages}{167401}.

\bibitem[{\citenamefont{Laurent} \emph{et~al.}(2005)\citenamefont{Laurent,
  Eble, Krebs, Lema\^\i{}tre, Urbaszek, Marie, Amand, and
  Voisin}}]{Laurent:2005a}
\bibinfo{author}{\bibnamefont{Laurent}, \bibfnamefont{S.}},
  \bibinfo{author}{\bibfnamefont{B.}~\bibnamefont{Eble}},
  \bibinfo{author}{\bibfnamefont{O.}~\bibnamefont{Krebs}},
  \bibinfo{author}{\bibfnamefont{A.}~\bibnamefont{Lema\^\i{}tre}},
  \bibinfo{author}{\bibfnamefont{B.}~\bibnamefont{Urbaszek}},
  \bibinfo{author}{\bibfnamefont{X.}~\bibnamefont{Marie}},
  \bibinfo{author}{\bibfnamefont{T.}~\bibnamefont{Amand}}, and
  \bibinfo{author}{\bibfnamefont{P.}~\bibnamefont{Voisin}},
  \bibinfo{year}{2005}, \bibinfo{journal}{Phys. Rev. Lett.}
  \textbf{\bibinfo{volume}{94}}(\bibinfo{number}{14}), \bibinfo{pages}{147401}.

\bibitem[{\citenamefont{Laurent} \emph{et~al.}(2006)\citenamefont{Laurent,
  Senes, Krebs, Kalevich, Urbaszek, Marie, Amand, and Voisin}}]{Laurent:2006a}
\bibinfo{author}{\bibnamefont{Laurent}, \bibfnamefont{S.}},
  \bibinfo{author}{\bibfnamefont{M.}~\bibnamefont{Senes}},
  \bibinfo{author}{\bibfnamefont{O.}~\bibnamefont{Krebs}},
  \bibinfo{author}{\bibfnamefont{V.~K.} \bibnamefont{Kalevich}},
  \bibinfo{author}{\bibfnamefont{B.}~\bibnamefont{Urbaszek}},
  \bibinfo{author}{\bibfnamefont{X.}~\bibnamefont{Marie}},
  \bibinfo{author}{\bibfnamefont{T.}~\bibnamefont{Amand}}, and
  \bibinfo{author}{\bibfnamefont{P.}~\bibnamefont{Voisin}},
  \bibinfo{year}{2006}, \bibinfo{journal}{Phys. Rev. B}
  \textbf{\bibinfo{volume}{73}}(\bibinfo{number}{23}), \bibinfo{pages}{235302}.

\bibitem[{\citenamefont{L\'eger} \emph{et~al.}(2007)\citenamefont{L\'eger,
  Besombes, Maingault, and Mariette}}]{Leger:2007a}
\bibinfo{author}{\bibnamefont{L\'eger}, \bibfnamefont{Y.}},
  \bibinfo{author}{\bibfnamefont{L.}~\bibnamefont{Besombes}},
  \bibinfo{author}{\bibfnamefont{L.}~\bibnamefont{Maingault}}, and
  \bibinfo{author}{\bibfnamefont{H.}~\bibnamefont{Mariette}},
  \bibinfo{year}{2007}, \bibinfo{journal}{Phys. Rev. B}
  \textbf{\bibinfo{volume}{76}}(\bibinfo{number}{4}), \bibinfo{pages}{045331}.

\bibitem[{\citenamefont{Leonard} \emph{et~al.}(1994)\citenamefont{Leonard,
  Pond, and Petroff}}]{Leonard:1994a}
\bibinfo{author}{\bibnamefont{Leonard}, \bibfnamefont{D.}},
  \bibinfo{author}{\bibfnamefont{K.}~\bibnamefont{Pond}}, and
  \bibinfo{author}{\bibfnamefont{P.~M.} \bibnamefont{Petroff}},
  \bibinfo{year}{1994}, \bibinfo{journal}{Phys. Rev. B}
  \textbf{\bibinfo{volume}{50}}(\bibinfo{number}{16}), \bibinfo{pages}{11687}.

\bibitem[{\citenamefont{Li} \emph{et~al.}(2005)\citenamefont{Li, Kioseoglou,
  van Erve, Ware, Gammon, Stroud, Jonker, Mallory, Yasar, and
  Petrou}}]{Li:2005a}
\bibinfo{author}{\bibnamefont{Li}, \bibfnamefont{C.~H.}},
  \bibinfo{author}{\bibfnamefont{G.}~\bibnamefont{Kioseoglou}},
  \bibinfo{author}{\bibfnamefont{O.~M.~J.} \bibnamefont{van Erve}},
  \bibinfo{author}{\bibfnamefont{M.~E.} \bibnamefont{Ware}},
  \bibinfo{author}{\bibfnamefont{D.}~\bibnamefont{Gammon}},
  \bibinfo{author}{\bibfnamefont{R.~M.} \bibnamefont{Stroud}},
  \bibinfo{author}{\bibfnamefont{B.~T.} \bibnamefont{Jonker}},
  \bibinfo{author}{\bibfnamefont{R.}~\bibnamefont{Mallory}},
  \bibinfo{author}{\bibfnamefont{M.}~\bibnamefont{Yasar}}, and
  \bibinfo{author}{\bibfnamefont{A.}~\bibnamefont{Petrou}},
  \bibinfo{year}{2005}, \bibinfo{journal}{Appl. Phys. Lett.}
  \textbf{\bibinfo{volume}{86}}(\bibinfo{number}{13}), \bibinfo{pages}{132503}.

\bibitem[{\citenamefont{Li} \emph{et~al.}(2012)\citenamefont{Li, Sinitsyn,
  Smith, Reuter, Wieck, Yakovlev, Bayer, and Crooker}}]{Li:2012a}
\bibinfo{author}{\bibnamefont{Li}, \bibfnamefont{Y.}},
  \bibinfo{author}{\bibfnamefont{N.}~\bibnamefont{Sinitsyn}},
  \bibinfo{author}{\bibfnamefont{D.~L.} \bibnamefont{Smith}},
  \bibinfo{author}{\bibfnamefont{D.}~\bibnamefont{Reuter}},
  \bibinfo{author}{\bibfnamefont{A.~D.} \bibnamefont{Wieck}},
  \bibinfo{author}{\bibfnamefont{D.~R.} \bibnamefont{Yakovlev}},
  \bibinfo{author}{\bibfnamefont{M.}~\bibnamefont{Bayer}}, and
  \bibinfo{author}{\bibfnamefont{S.~A.} \bibnamefont{Crooker}},
  \bibinfo{year}{2012}, \bibinfo{journal}{Phys. Rev. Lett.}
  \textbf{\bibinfo{volume}{108}}, \bibinfo{pages}{186603}.

\bibitem[{\citenamefont{Liu} \emph{et~al.}(2007)\citenamefont{Liu, Whitaker,
  Smith, Kittilstved, Robinson, and Gamelin}}]{Liu:2007a}
\bibinfo{author}{\bibnamefont{Liu}, \bibfnamefont{W.~K.}},
  \bibinfo{author}{\bibfnamefont{K.~M.} \bibnamefont{Whitaker}},
  \bibinfo{author}{\bibfnamefont{A.~L.} \bibnamefont{Smith}},
  \bibinfo{author}{\bibfnamefont{K.~R.} \bibnamefont{Kittilstved}},
  \bibinfo{author}{\bibfnamefont{B.~H.} \bibnamefont{Robinson}}, and
  \bibinfo{author}{\bibfnamefont{D.~R.} \bibnamefont{Gamelin}},
  \bibinfo{year}{2007}, \bibinfo{journal}{Phys. Rev. Lett.}
  \textbf{\bibinfo{volume}{98}}(\bibinfo{number}{18}), \bibinfo{pages}{186804}.

\bibitem[{\citenamefont{Lombez}
  \emph{et~al.}(2007{\natexlab{a}})\citenamefont{Lombez, Braun, Marie, Renucci,
  Urbaszek, Amand, Krebs, and Voisin}}]{Lombez:2007a}
\bibinfo{author}{\bibnamefont{Lombez}, \bibfnamefont{L.}},
  \bibinfo{author}{\bibfnamefont{P.-F.} \bibnamefont{Braun}},
  \bibinfo{author}{\bibfnamefont{X.}~\bibnamefont{Marie}},
  \bibinfo{author}{\bibfnamefont{P.}~\bibnamefont{Renucci}},
  \bibinfo{author}{\bibfnamefont{B.}~\bibnamefont{Urbaszek}},
  \bibinfo{author}{\bibfnamefont{T.}~\bibnamefont{Amand}},
  \bibinfo{author}{\bibfnamefont{O.}~\bibnamefont{Krebs}}, and
  \bibinfo{author}{\bibfnamefont{P.}~\bibnamefont{Voisin}},
  \bibinfo{year}{2007}{\natexlab{a}}, \bibinfo{journal}{Phys. Rev. B}
  \textbf{\bibinfo{volume}{75}}(\bibinfo{number}{19}), \bibinfo{pages}{195314}.

\bibitem[{\citenamefont{Lombez}
  \emph{et~al.}(2007{\natexlab{b}})\citenamefont{Lombez, Renucci, Braun,
  Carrere, Marie, Amand, Urbaszek, Gauffier, Gallo, Camps, Arnoult, Fontaine}
  \emph{et~al.}}]{Lombez:2007b}
\bibinfo{author}{\bibnamefont{Lombez}, \bibfnamefont{L.}},
  \bibinfo{author}{\bibfnamefont{P.}~\bibnamefont{Renucci}},
  \bibinfo{author}{\bibfnamefont{P.~F.} \bibnamefont{Braun}},
  \bibinfo{author}{\bibfnamefont{H.}~\bibnamefont{Carrere}},
  \bibinfo{author}{\bibfnamefont{X.}~\bibnamefont{Marie}},
  \bibinfo{author}{\bibfnamefont{T.}~\bibnamefont{Amand}},
  \bibinfo{author}{\bibfnamefont{B.}~\bibnamefont{Urbaszek}},
  \bibinfo{author}{\bibfnamefont{J.~L.} \bibnamefont{Gauffier}},
  \bibinfo{author}{\bibfnamefont{P.}~\bibnamefont{Gallo}},
  \bibinfo{author}{\bibfnamefont{T.}~\bibnamefont{Camps}},
  \bibinfo{author}{\bibfnamefont{A.}~\bibnamefont{Arnoult}},
  \bibinfo{author}{\bibfnamefont{C.}~\bibnamefont{Fontaine}}, \emph{et~al.},
  \bibinfo{year}{2007}{\natexlab{b}}, \bibinfo{journal}{Appl. Phys. Lett.}
  \textbf{\bibinfo{volume}{90}}(\bibinfo{number}{8}), \bibinfo{pages}{081111}.

\bibitem[{\citenamefont{Lu} \emph{et~al.}(2010)\citenamefont{Lu, Zhao,
  Vamivakas, Matthiesen, F\"alt, Badolato, and Atat\"ure}}]{Lu:2010a}
\bibinfo{author}{\bibnamefont{Lu}, \bibfnamefont{C.-Y.}},
  \bibinfo{author}{\bibfnamefont{Y.}~\bibnamefont{Zhao}},
  \bibinfo{author}{\bibfnamefont{A.~N.} \bibnamefont{Vamivakas}},
  \bibinfo{author}{\bibfnamefont{C.}~\bibnamefont{Matthiesen}},
  \bibinfo{author}{\bibfnamefont{S.}~\bibnamefont{F\"alt}},
  \bibinfo{author}{\bibfnamefont{A.}~\bibnamefont{Badolato}}, and
  \bibinfo{author}{\bibfnamefont{M.}~\bibnamefont{Atat\"ure}},
  \bibinfo{year}{2010}, \bibinfo{journal}{Phys. Rev. B}
  \textbf{\bibinfo{volume}{81}}(\bibinfo{number}{3}), \bibinfo{pages}{035332}.

\bibitem[{\citenamefont{Lu} \emph{et~al.}(2006)\citenamefont{Lu, Hoch, Kuhns,
  Moulton, Gan, and Reyes}}]{Lu:2006a}
\bibinfo{author}{\bibnamefont{Lu}, \bibfnamefont{J.}},
  \bibinfo{author}{\bibfnamefont{M.~J.~R.} \bibnamefont{Hoch}},
  \bibinfo{author}{\bibfnamefont{P.~L.} \bibnamefont{Kuhns}},
  \bibinfo{author}{\bibfnamefont{W.~G.} \bibnamefont{Moulton}},
  \bibinfo{author}{\bibfnamefont{Z.}~\bibnamefont{Gan}}, and
  \bibinfo{author}{\bibfnamefont{A.~P.} \bibnamefont{Reyes}},
  \bibinfo{year}{2006}, \bibinfo{journal}{Phys. Rev. B}
  \textbf{\bibinfo{volume}{74}}(\bibinfo{number}{12}), \bibinfo{pages}{125208}.

\bibitem[{\citenamefont{Makhonin} \emph{et~al.}(2010)\citenamefont{Makhonin,
  Chekhovich, Senellart, Lema\^\i{}tre, Skolnick, and
  Tartakovskii}}]{Makhonin:2010a}
\bibinfo{author}{\bibnamefont{Makhonin}, \bibfnamefont{M.~N.}},
  \bibinfo{author}{\bibfnamefont{E.~A.} \bibnamefont{Chekhovich}},
  \bibinfo{author}{\bibfnamefont{P.}~\bibnamefont{Senellart}},
  \bibinfo{author}{\bibfnamefont{A.}~\bibnamefont{Lema\^\i{}tre}},
  \bibinfo{author}{\bibfnamefont{M.~S.} \bibnamefont{Skolnick}}, and
  \bibinfo{author}{\bibfnamefont{A.~I.} \bibnamefont{Tartakovskii}},
  \bibinfo{year}{2010}, \bibinfo{journal}{Phys. Rev. B}
  \textbf{\bibinfo{volume}{82}}(\bibinfo{number}{16}), \bibinfo{pages}{161309}.

\bibitem[{\citenamefont{Maletinsky}(2008)}]{Maletinsky:2008a}
\bibinfo{author}{\bibnamefont{Maletinsky}, \bibfnamefont{P.}},
  \bibinfo{year}{2008}, \emph{\bibinfo{title}{Polarization and Manipulation of
  a Mesoscopic Nuclear Spin Ensemble Using a Single Confined Electron Spin}},
  Ph.D. thesis, \bibinfo{school}{ETH Z\"urich},
  \urlprefix\url{http://e-collection.ethbib.ethz.ch/view/eth:30788}.

\bibitem[{\citenamefont{Maletinsky}
  \emph{et~al.}(2007{\natexlab{a}})\citenamefont{Maletinsky, Badolato, and
  Imamoglu}}]{Maletinsky:2007b}
\bibinfo{author}{\bibnamefont{Maletinsky}, \bibfnamefont{P.}},
  \bibinfo{author}{\bibfnamefont{A.}~\bibnamefont{Badolato}}, and
  \bibinfo{author}{\bibfnamefont{A.}~\bibnamefont{Imamoglu}},
  \bibinfo{year}{2007}{\natexlab{a}}, \bibinfo{journal}{Phys. Rev. Lett.}
  \textbf{\bibinfo{volume}{99}}(\bibinfo{number}{5}), \bibinfo{pages}{056804}.

\bibitem[{\citenamefont{Maletinsky}
  \emph{et~al.}(2009)\citenamefont{Maletinsky, Kroner, and
  Imamoglu}}]{Maletinsky:2009a}
\bibinfo{author}{\bibnamefont{Maletinsky}, \bibfnamefont{P.}},
  \bibinfo{author}{\bibfnamefont{M.}~\bibnamefont{Kroner}}, and
  \bibinfo{author}{\bibfnamefont{A.}~\bibnamefont{Imamoglu}},
  \bibinfo{year}{2009}, \bibinfo{journal}{Nature Phys.}
  \textbf{\bibinfo{volume}{5}}(\bibinfo{number}{6}), \bibinfo{pages}{407}.

\bibitem[{\citenamefont{Maletinsky}
  \emph{et~al.}(2007{\natexlab{b}})\citenamefont{Maletinsky, Lai, Badolato, and
  Imamoglu}}]{Maletinsky:2007a}
\bibinfo{author}{\bibnamefont{Maletinsky}, \bibfnamefont{P.}},
  \bibinfo{author}{\bibfnamefont{C.~W.} \bibnamefont{Lai}},
  \bibinfo{author}{\bibfnamefont{A.}~\bibnamefont{Badolato}}, and
  \bibinfo{author}{\bibfnamefont{A.}~\bibnamefont{Imamoglu}},
  \bibinfo{year}{2007}{\natexlab{b}}, \bibinfo{journal}{Phys. Rev. B}
  \textbf{\bibinfo{volume}{75}}(\bibinfo{number}{3}), \bibinfo{pages}{035409}.

\bibitem[{\citenamefont{Malinowski}
  \emph{et~al.}(2001)\citenamefont{Malinowski, Brand, and
  Harley}}]{Malinowski:2001a}
\bibinfo{author}{\bibnamefont{Malinowski}, \bibfnamefont{A.}},
  \bibinfo{author}{\bibfnamefont{M.~A.} \bibnamefont{Brand}}, and
  \bibinfo{author}{\bibfnamefont{R.~T.} \bibnamefont{Harley}},
  \bibinfo{year}{2001}, \bibinfo{journal}{Physica E}
  \textbf{\bibinfo{volume}{10}}, \bibinfo{pages}{13}.

\bibitem[{\citenamefont{Marzin} \emph{et~al.}(1994)\citenamefont{Marzin,
  G\'erard, Izra\"el, Barrier, and Bastard}}]{Marzin:1994a}
\bibinfo{author}{\bibnamefont{Marzin}, \bibfnamefont{J.~Y.}},
  \bibinfo{author}{\bibfnamefont{J.~M.} \bibnamefont{G\'erard}},
  \bibinfo{author}{\bibfnamefont{A.}~\bibnamefont{Izra\"el}},
  \bibinfo{author}{\bibfnamefont{D.}~\bibnamefont{Barrier}}, and
  \bibinfo{author}{\bibfnamefont{G.}~\bibnamefont{Bastard}},
  \bibinfo{year}{1994}, \bibinfo{journal}{Phys. Rev. Lett.}
  \textbf{\bibinfo{volume}{73}}(\bibinfo{number}{5}), \bibinfo{pages}{716}.

\bibitem[{\citenamefont{McNeil and Clark}(1976)}]{McNeil:1976a}
\bibinfo{author}{\bibnamefont{McNeil}, \bibfnamefont{J.~A.}}, and
  \bibinfo{author}{\bibfnamefont{W.~G.} \bibnamefont{Clark}},
  \bibinfo{year}{1976}, \bibinfo{journal}{Phys. Rev. B}
  \textbf{\bibinfo{volume}{13}}, \bibinfo{pages}{4705}.

\bibitem[{\citenamefont{Meier and Zakharchenya}(1984)}]{Meier:1984a}
\bibinfo{author}{\bibnamefont{Meier}, \bibfnamefont{F.}}, and
  \bibinfo{author}{\bibfnamefont{B.}~\bibnamefont{Zakharchenya}},
  \bibinfo{year}{1984}, \bibinfo{journal}{Modern Problems in Condensed Matter
  Sciences (North-Holland, Amsterdam).} \textbf{\bibinfo{volume}{8}}.

\bibitem[{\citenamefont{Merkulov} \emph{et~al.}(2002)\citenamefont{Merkulov,
  Efros, and Rosen}}]{Merkulov:2002a}
\bibinfo{author}{\bibnamefont{Merkulov}, \bibfnamefont{I.~A.}},
  \bibinfo{author}{\bibfnamefont{A.~L.} \bibnamefont{Efros}}, and
  \bibinfo{author}{\bibfnamefont{M.}~\bibnamefont{Rosen}},
  \bibinfo{year}{2002}, \bibinfo{journal}{Phys. Rev. B}
  \textbf{\bibinfo{volume}{65}}(\bibinfo{number}{20}), \bibinfo{pages}{205309}.

\bibitem[{\citenamefont{Michler} \emph{et~al.}(2000)\citenamefont{Michler,
  Kiraz, Becher, Schoenfeld, Petroff, Zhang, Hu, and Imamoglu}}]{Michler:2000a}
\bibinfo{author}{\bibnamefont{Michler}, \bibfnamefont{P.}},
  \bibinfo{author}{\bibfnamefont{A.}~\bibnamefont{Kiraz}},
  \bibinfo{author}{\bibfnamefont{C.}~\bibnamefont{Becher}},
  \bibinfo{author}{\bibfnamefont{W.~V.} \bibnamefont{Schoenfeld}},
  \bibinfo{author}{\bibfnamefont{P.~M.} \bibnamefont{Petroff}},
  \bibinfo{author}{\bibfnamefont{L.}~\bibnamefont{Zhang}},
  \bibinfo{author}{\bibfnamefont{E.}~\bibnamefont{Hu}}, and
  \bibinfo{author}{\bibfnamefont{A.}~\bibnamefont{Imamoglu}},
  \bibinfo{year}{2000}, \bibinfo{journal}{Science}
  \textbf{\bibinfo{volume}{290}}(\bibinfo{number}{5500}),
  \bibinfo{pages}{2282}.

\bibitem[{\citenamefont{Moskalenko}
  \emph{et~al.}(2009)\citenamefont{Moskalenko, Larsson, and
  Holtz}}]{Moskalenko:2009a}
\bibinfo{author}{\bibnamefont{Moskalenko}, \bibfnamefont{E.~S.}},
  \bibinfo{author}{\bibfnamefont{L.~A.} \bibnamefont{Larsson}}, and
  \bibinfo{author}{\bibfnamefont{P.~O.} \bibnamefont{Holtz}},
  \bibinfo{year}{2009}, \bibinfo{journal}{Phys. Rev. B}
  \textbf{\bibinfo{volume}{80}}(\bibinfo{number}{19}), \bibinfo{pages}{193413}.

\bibitem[{\citenamefont{Muller} \emph{et~al.}(2007)\citenamefont{Muller, Flagg,
  Bianucci, Wang, Deppe, Ma, Zhang, Salamo, Xiao, and Shih}}]{Muller:2007a}
\bibinfo{author}{\bibnamefont{Muller}, \bibfnamefont{A.}},
  \bibinfo{author}{\bibfnamefont{E.~B.} \bibnamefont{Flagg}},
  \bibinfo{author}{\bibfnamefont{P.}~\bibnamefont{Bianucci}},
  \bibinfo{author}{\bibfnamefont{X.~Y.} \bibnamefont{Wang}},
  \bibinfo{author}{\bibfnamefont{D.~G.} \bibnamefont{Deppe}},
  \bibinfo{author}{\bibfnamefont{W.}~\bibnamefont{Ma}},
  \bibinfo{author}{\bibfnamefont{J.}~\bibnamefont{Zhang}},
  \bibinfo{author}{\bibfnamefont{G.~J.} \bibnamefont{Salamo}},
  \bibinfo{author}{\bibfnamefont{M.}~\bibnamefont{Xiao}}, and
  \bibinfo{author}{\bibfnamefont{C.~K.} \bibnamefont{Shih}},
  \bibinfo{year}{2007}, \bibinfo{journal}{Phys. Rev. Lett.}
  \textbf{\bibinfo{volume}{99}}(\bibinfo{number}{18}), \bibinfo{pages}{187402}.

\bibitem[{\citenamefont{Nikolaenko}
  \emph{et~al.}(2009)\citenamefont{Nikolaenko, Chekhovich, Makhonin, Drouzas,
  Van'kov, Skiba-Szymanska, Skolnick, Senellart, Martrou, Lema\^\i{}tre, and
  Tartakovskii}}]{Nikolaenko:2009a}
\bibinfo{author}{\bibnamefont{Nikolaenko}, \bibfnamefont{A.~E.}},
  \bibinfo{author}{\bibfnamefont{E.~A.} \bibnamefont{Chekhovich}},
  \bibinfo{author}{\bibfnamefont{M.~N.} \bibnamefont{Makhonin}},
  \bibinfo{author}{\bibfnamefont{I.~W.} \bibnamefont{Drouzas}},
  \bibinfo{author}{\bibfnamefont{A.~B.} \bibnamefont{Van'kov}},
  \bibinfo{author}{\bibfnamefont{J.}~\bibnamefont{Skiba-Szymanska}},
  \bibinfo{author}{\bibfnamefont{M.~S.} \bibnamefont{Skolnick}},
  \bibinfo{author}{\bibfnamefont{P.}~\bibnamefont{Senellart}},
  \bibinfo{author}{\bibfnamefont{D.}~\bibnamefont{Martrou}},
  \bibinfo{author}{\bibfnamefont{A.}~\bibnamefont{Lema\^\i{}tre}}, and
  \bibinfo{author}{\bibfnamefont{A.~I.} \bibnamefont{Tartakovskii}},
  \bibinfo{year}{2009}, \bibinfo{journal}{Phys. Rev. B}
  \textbf{\bibinfo{volume}{79}}(\bibinfo{number}{8}), \bibinfo{pages}{081303}.

\bibitem[{\citenamefont{Offermans} \emph{et~al.}(2005)\citenamefont{Offermans,
  Koenraad, Wolter, Pierz, Roy, and Maksym}}]{Offermans:2005a}
\bibinfo{author}{\bibnamefont{Offermans}, \bibfnamefont{P.}},
  \bibinfo{author}{\bibfnamefont{P.}~\bibnamefont{Koenraad}},
  \bibinfo{author}{\bibfnamefont{J.}~\bibnamefont{Wolter}},
  \bibinfo{author}{\bibfnamefont{K.}~\bibnamefont{Pierz}},
  \bibinfo{author}{\bibfnamefont{M.}~\bibnamefont{Roy}}, and
  \bibinfo{author}{\bibfnamefont{P.}~\bibnamefont{Maksym}},
  \bibinfo{year}{2005}, \bibinfo{journal}{Physica E}
  \textbf{\bibinfo{volume}{26}}(\bibinfo{number}{1-4}), \bibinfo{pages}{236 }.

\bibitem[{\citenamefont{Oulton} \emph{et~al.}(2007)\citenamefont{Oulton,
  Greilich, Verbin, Cherbunin, Auer, Yakovlev, Bayer, Merkulov, Stavarache,
  Reuter, and Wieck}}]{Oulton:2007a}
\bibinfo{author}{\bibnamefont{Oulton}, \bibfnamefont{R.}},
  \bibinfo{author}{\bibfnamefont{A.}~\bibnamefont{Greilich}},
  \bibinfo{author}{\bibfnamefont{S.~Y.} \bibnamefont{Verbin}},
  \bibinfo{author}{\bibfnamefont{R.~V.} \bibnamefont{Cherbunin}},
  \bibinfo{author}{\bibfnamefont{T.}~\bibnamefont{Auer}},
  \bibinfo{author}{\bibfnamefont{D.~R.} \bibnamefont{Yakovlev}},
  \bibinfo{author}{\bibfnamefont{M.}~\bibnamefont{Bayer}},
  \bibinfo{author}{\bibfnamefont{I.~A.} \bibnamefont{Merkulov}},
  \bibinfo{author}{\bibfnamefont{V.}~\bibnamefont{Stavarache}},
  \bibinfo{author}{\bibfnamefont{D.}~\bibnamefont{Reuter}}, and
  \bibinfo{author}{\bibfnamefont{A.~D.} \bibnamefont{Wieck}},
  \bibinfo{year}{2007}, \bibinfo{journal}{Phys. Rev. Lett.}
  \textbf{\bibinfo{volume}{98}}(\bibinfo{number}{10}), \bibinfo{pages}{107401}.

\bibitem[{\citenamefont{Overhauser}(1953)}]{Overhauser:1953a}
\bibinfo{author}{\bibnamefont{Overhauser}, \bibfnamefont{A.~W.}},
  \bibinfo{year}{1953}, \bibinfo{journal}{Phys. Rev.}
  \textbf{\bibinfo{volume}{92}}(\bibinfo{number}{2}), \bibinfo{pages}{411}.

\bibitem[{\citenamefont{Paget}(1982)}]{Paget:1982a}
\bibinfo{author}{\bibnamefont{Paget}, \bibfnamefont{D.}}, \bibinfo{year}{1982},
  \bibinfo{journal}{Phys. Rev. B} \textbf{\bibinfo{volume}{25}},
  \bibinfo{pages}{4444}.

\bibitem[{\citenamefont{Paget} \emph{et~al.}(2008)\citenamefont{Paget, Amand,
  and Korb}}]{Paget:2008a}
\bibinfo{author}{\bibnamefont{Paget}, \bibfnamefont{D.}},
  \bibinfo{author}{\bibfnamefont{T.}~\bibnamefont{Amand}}, and
  \bibinfo{author}{\bibfnamefont{J.-P.} \bibnamefont{Korb}},
  \bibinfo{year}{2008}, \bibinfo{journal}{Phys. Rev. B}
  \textbf{\bibinfo{volume}{77}}(\bibinfo{number}{24}), \bibinfo{pages}{245201}.

\bibitem[{\citenamefont{Paget} \emph{et~al.}(1977)\citenamefont{Paget, Lampel,
  Sapoval, and Safarov}}]{Paget:1977a}
\bibinfo{author}{\bibnamefont{Paget}, \bibfnamefont{D.}},
  \bibinfo{author}{\bibfnamefont{G.}~\bibnamefont{Lampel}},
  \bibinfo{author}{\bibfnamefont{B.}~\bibnamefont{Sapoval}}, and
  \bibinfo{author}{\bibfnamefont{V.~I.} \bibnamefont{Safarov}},
  \bibinfo{year}{1977}, \bibinfo{journal}{Phys. Rev. B}
  \textbf{\bibinfo{volume}{15}}, \bibinfo{pages}{5780}.

\bibitem[{\citenamefont{Paillard} \emph{et~al.}(2001)\citenamefont{Paillard,
  Marie, Renucci, Amand, Jbeli, and G\'erard}}]{Paillard:2001a}
\bibinfo{author}{\bibnamefont{Paillard}, \bibfnamefont{M.}},
  \bibinfo{author}{\bibfnamefont{X.}~\bibnamefont{Marie}},
  \bibinfo{author}{\bibfnamefont{P.}~\bibnamefont{Renucci}},
  \bibinfo{author}{\bibfnamefont{T.}~\bibnamefont{Amand}},
  \bibinfo{author}{\bibfnamefont{A.}~\bibnamefont{Jbeli}}, and
  \bibinfo{author}{\bibfnamefont{J.~M.} \bibnamefont{G\'erard}},
  \bibinfo{year}{2001}, \bibinfo{journal}{Phys. Rev. Lett.}
  \textbf{\bibinfo{volume}{86}}(\bibinfo{number}{8}), \bibinfo{pages}{1634}.

\bibitem[{\citenamefont{Paillard} \emph{et~al.}(2000)\citenamefont{Paillard,
  Marie, Vanelle, Amand, Kalevich, Kovsh, Zhukov, and
  Ustinov}}]{Paillard:2000a}
\bibinfo{author}{\bibnamefont{Paillard}, \bibfnamefont{M.}},
  \bibinfo{author}{\bibfnamefont{X.}~\bibnamefont{Marie}},
  \bibinfo{author}{\bibfnamefont{E.}~\bibnamefont{Vanelle}},
  \bibinfo{author}{\bibfnamefont{T.}~\bibnamefont{Amand}},
  \bibinfo{author}{\bibfnamefont{V.~K.} \bibnamefont{Kalevich}},
  \bibinfo{author}{\bibfnamefont{A.~R.} \bibnamefont{Kovsh}},
  \bibinfo{author}{\bibfnamefont{A.~E.} \bibnamefont{Zhukov}}, and
  \bibinfo{author}{\bibfnamefont{V.~M.} \bibnamefont{Ustinov}},
  \bibinfo{year}{2000}, \bibinfo{journal}{Applied Physics Letters}
  \textbf{\bibinfo{volume}{76}}(\bibinfo{number}{1}), \bibinfo{pages}{76}.

\bibitem[{\citenamefont{Pal} \emph{et~al.}(2007)\citenamefont{Pal, Verbin,
  Ignatiev, Ikezawa, and Masumoto}}]{Pal:2007a}
\bibinfo{author}{\bibnamefont{Pal}, \bibfnamefont{B.}},
  \bibinfo{author}{\bibfnamefont{S.~Y.} \bibnamefont{Verbin}},
  \bibinfo{author}{\bibfnamefont{I.~V.} \bibnamefont{Ignatiev}},
  \bibinfo{author}{\bibfnamefont{M.}~\bibnamefont{Ikezawa}}, and
  \bibinfo{author}{\bibfnamefont{Y.}~\bibnamefont{Masumoto}},
  \bibinfo{year}{2007}, \bibinfo{journal}{Phys. Rev. B}
  \textbf{\bibinfo{volume}{75}}(\bibinfo{number}{12}), \bibinfo{pages}{125322}.

\bibitem[{\citenamefont{Petrov} \emph{et~al.}(2009)\citenamefont{Petrov,
  Kozlov, Ignatiev, Cherbunin, Yakovlev, and Bayer}}]{Petrov:2009a}
\bibinfo{author}{\bibnamefont{Petrov}, \bibfnamefont{M.~Y.}},
  \bibinfo{author}{\bibfnamefont{G.~G.} \bibnamefont{Kozlov}},
  \bibinfo{author}{\bibfnamefont{I.~V.} \bibnamefont{Ignatiev}},
  \bibinfo{author}{\bibfnamefont{R.~V.} \bibnamefont{Cherbunin}},
  \bibinfo{author}{\bibfnamefont{D.~R.} \bibnamefont{Yakovlev}}, and
  \bibinfo{author}{\bibfnamefont{M.}~\bibnamefont{Bayer}},
  \bibinfo{year}{2009}, \bibinfo{journal}{Phys. Rev. B}
  \textbf{\bibinfo{volume}{80}}, \bibinfo{pages}{125318}.

\bibitem[{\citenamefont{Petta} \emph{et~al.}(2005)\citenamefont{Petta, Johnson,
  Taylor, Laird, Yacoby, Lukin, Marcus, Hanson, and Gossard}}]{Petta:2005a}
\bibinfo{author}{\bibnamefont{Petta}, \bibfnamefont{J.~R.}},
  \bibinfo{author}{\bibfnamefont{A.~C.} \bibnamefont{Johnson}},
  \bibinfo{author}{\bibfnamefont{J.~M.} \bibnamefont{Taylor}},
  \bibinfo{author}{\bibfnamefont{E.~A.} \bibnamefont{Laird}},
  \bibinfo{author}{\bibfnamefont{A.}~\bibnamefont{Yacoby}},
  \bibinfo{author}{\bibfnamefont{M.~D.} \bibnamefont{Lukin}},
  \bibinfo{author}{\bibfnamefont{C.~M.} \bibnamefont{Marcus}},
  \bibinfo{author}{\bibfnamefont{M.~P.} \bibnamefont{Hanson}}, and
  \bibinfo{author}{\bibfnamefont{A.~C.} \bibnamefont{Gossard}},
  \bibinfo{year}{2005}, \bibinfo{journal}{Science}
  \textbf{\bibinfo{volume}{309}}(\bibinfo{number}{5744}),
  \bibinfo{pages}{2180}.

\bibitem[{\citenamefont{Pines} \emph{et~al.}(1957)\citenamefont{Pines, Bardeen,
  and Slichter}}]{Pines:1957a}
\bibinfo{author}{\bibnamefont{Pines}, \bibfnamefont{D.}},
  \bibinfo{author}{\bibfnamefont{J.}~\bibnamefont{Bardeen}}, and
  \bibinfo{author}{\bibfnamefont{C.~P.} \bibnamefont{Slichter}},
  \bibinfo{year}{1957}, \bibinfo{journal}{Phys. Rev.}
  \textbf{\bibinfo{volume}{106}}, \bibinfo{pages}{489}.

\bibitem[{\citenamefont{Press} \emph{et~al.}(2010)\citenamefont{Press,
  De~Greve, McMahon, Ladd, Friess, Schneider, Kamp, Hoefling, Forchel, and
  Yamamoto}}]{Press:2010a}
\bibinfo{author}{\bibnamefont{Press}, \bibfnamefont{D.}},
  \bibinfo{author}{\bibfnamefont{K.}~\bibnamefont{De~Greve}},
  \bibinfo{author}{\bibfnamefont{P.~L.} \bibnamefont{McMahon}},
  \bibinfo{author}{\bibfnamefont{T.~D.} \bibnamefont{Ladd}},
  \bibinfo{author}{\bibfnamefont{B.}~\bibnamefont{Friess}},
  \bibinfo{author}{\bibfnamefont{C.}~\bibnamefont{Schneider}},
  \bibinfo{author}{\bibfnamefont{M.}~\bibnamefont{Kamp}},
  \bibinfo{author}{\bibfnamefont{S.}~\bibnamefont{Hoefling}},
  \bibinfo{author}{\bibfnamefont{A.}~\bibnamefont{Forchel}}, and
  \bibinfo{author}{\bibfnamefont{Y.}~\bibnamefont{Yamamoto}},
  \bibinfo{year}{2010}, \bibinfo{journal}{Nature Photonics}
  \textbf{\bibinfo{volume}{4}}, \bibinfo{pages}{367}.

\bibitem[{\citenamefont{Rudner} \emph{et~al.}(2011)\citenamefont{Rudner,
  Vandersypen, Vuleti\ifmmode~\acute{c}\else \'{c}\fi{}, and
  Levitov}}]{Rudner:2011a}
\bibinfo{author}{\bibnamefont{Rudner}, \bibfnamefont{M.~S.}},
  \bibinfo{author}{\bibfnamefont{L.~M.~K.} \bibnamefont{Vandersypen}},
  \bibinfo{author}{\bibfnamefont{V.}~\bibnamefont{Vuleti\ifmmode~\acute{c}\else
  \'{c}\fi{}}}, and \bibinfo{author}{\bibfnamefont{L.~S.}
  \bibnamefont{Levitov}}, \bibinfo{year}{2011}, \bibinfo{journal}{Phys. Rev.
  Lett.} \textbf{\bibinfo{volume}{107}}, \bibinfo{pages}{206806}.

\bibitem[{\citenamefont{Sallen} \emph{et~al.}(2011)\citenamefont{Sallen,
  Urbaszek, Glazov, Ivchenko, Kuroda, Mano, Kunz, Abbarchi, Sakoda, Lagarde,
  Balocchi, Marie} \emph{et~al.}}]{Sallen:2011a}
\bibinfo{author}{\bibnamefont{Sallen}, \bibfnamefont{G.}},
  \bibinfo{author}{\bibfnamefont{B.}~\bibnamefont{Urbaszek}},
  \bibinfo{author}{\bibfnamefont{M.~M.} \bibnamefont{Glazov}},
  \bibinfo{author}{\bibfnamefont{E.~L.} \bibnamefont{Ivchenko}},
  \bibinfo{author}{\bibfnamefont{T.}~\bibnamefont{Kuroda}},
  \bibinfo{author}{\bibfnamefont{T.}~\bibnamefont{Mano}},
  \bibinfo{author}{\bibfnamefont{S.}~\bibnamefont{Kunz}},
  \bibinfo{author}{\bibfnamefont{M.}~\bibnamefont{Abbarchi}},
  \bibinfo{author}{\bibfnamefont{K.}~\bibnamefont{Sakoda}},
  \bibinfo{author}{\bibfnamefont{D.}~\bibnamefont{Lagarde}},
  \bibinfo{author}{\bibfnamefont{A.}~\bibnamefont{Balocchi}},
  \bibinfo{author}{\bibfnamefont{X.}~\bibnamefont{Marie}}, \emph{et~al.},
  \bibinfo{year}{2011}, \bibinfo{journal}{Phys. Rev. Lett.}
  \textbf{\bibinfo{volume}{107}}, \bibinfo{pages}{166604}.

\bibitem[{\citenamefont{Scheibner} \emph{et~al.}(2003)\citenamefont{Scheibner,
  Bacher, Weber, Forchel, Passow, and Hommel}}]{Scheibner:2003a}
\bibinfo{author}{\bibnamefont{Scheibner}, \bibfnamefont{M.}},
  \bibinfo{author}{\bibfnamefont{G.}~\bibnamefont{Bacher}},
  \bibinfo{author}{\bibfnamefont{S.}~\bibnamefont{Weber}},
  \bibinfo{author}{\bibfnamefont{A.}~\bibnamefont{Forchel}},
  \bibinfo{author}{\bibfnamefont{T.}~\bibnamefont{Passow}}, and
  \bibinfo{author}{\bibfnamefont{D.}~\bibnamefont{Hommel}},
  \bibinfo{year}{2003}, \bibinfo{journal}{Phys. Rev. B}
  \textbf{\bibinfo{volume}{67}}(\bibinfo{number}{15}), \bibinfo{pages}{153302}.

\bibitem[{\citenamefont{Schrieffer and Wolff}(1966)}]{Schrieffer:1966a}
\bibinfo{author}{\bibnamefont{Schrieffer}, \bibfnamefont{J.~R.}}, and
  \bibinfo{author}{\bibfnamefont{P.~A.} \bibnamefont{Wolff}},
  \bibinfo{year}{1966}, \bibinfo{journal}{Phys. Rev.}
  \textbf{\bibinfo{volume}{149}}, \bibinfo{pages}{491}.

\bibitem[{\citenamefont{Schulhauser}(2004)}]{Schulhauser:2004a}
\bibinfo{author}{\bibnamefont{Schulhauser}, \bibfnamefont{C.}},
  \bibinfo{year}{2004}, \emph{\bibinfo{title}{Elektronische
  Quantenpunktzust{\"a}nde induziert durch Photoemission}}, Ph.D. thesis,
  \bibinfo{school}{LMU M{\"u}nchen},
  \urlprefix\url{http://edoc.ub.uni-muenchen.de/2685/}.

\bibitem[{\citenamefont{Schulten and Wolynes}(1978)}]{Schulten:1978a}
\bibinfo{author}{\bibnamefont{Schulten}, \bibfnamefont{K.}}, and
  \bibinfo{author}{\bibfnamefont{P.~G.} \bibnamefont{Wolynes}},
  \bibinfo{year}{1978}, \bibinfo{journal}{J. Chem. Phys.}
  \textbf{\bibinfo{volume}{68}}, \bibinfo{pages}{3292}.

\bibitem[{\citenamefont{Semenov and Kim}(2003)}]{Semenov:2003a}
\bibinfo{author}{\bibnamefont{Semenov}, \bibfnamefont{Y.~G.}}, and
  \bibinfo{author}{\bibfnamefont{K.~W.} \bibnamefont{Kim}},
  \bibinfo{year}{2003}, \bibinfo{journal}{Phys. Rev. B}
  \textbf{\bibinfo{volume}{67}}(\bibinfo{number}{7}), \bibinfo{pages}{073301}.

\bibitem[{\citenamefont{S\'en\`es} \emph{et~al.}(2004)\citenamefont{S\'en\`es,
  Marie, Urbaszek, Renucci, Amand, and G\'erard}}]{Senes:2004a}
\bibinfo{author}{\bibnamefont{S\'en\`es}, \bibfnamefont{M.}},
  \bibinfo{author}{\bibfnamefont{X.}~\bibnamefont{Marie}},
  \bibinfo{author}{\bibfnamefont{B.}~\bibnamefont{Urbaszek}},
  \bibinfo{author}{\bibfnamefont{P.}~\bibnamefont{Renucci}},
  \bibinfo{author}{\bibfnamefont{T.}~\bibnamefont{Amand}}, and
  \bibinfo{author}{\bibfnamefont{J.-M.} \bibnamefont{G\'erard}},
  \bibinfo{year}{2004}, \bibinfo{journal}{phys. stat. sol.(c)}
  \textbf{\bibinfo{volume}{1}}, \bibinfo{pages}{594}.

\bibitem[{\citenamefont{S\'en\`es} \emph{et~al.}(2005)\citenamefont{S\'en\`es,
  Urbaszek, Marie, Amand, Tribollet, Bernardot, Testelin, Chamarro, and
  G\'erard}}]{Senes:2005a}
\bibinfo{author}{\bibnamefont{S\'en\`es}, \bibfnamefont{M.}},
  \bibinfo{author}{\bibfnamefont{B.}~\bibnamefont{Urbaszek}},
  \bibinfo{author}{\bibfnamefont{X.}~\bibnamefont{Marie}},
  \bibinfo{author}{\bibfnamefont{T.}~\bibnamefont{Amand}},
  \bibinfo{author}{\bibfnamefont{J.}~\bibnamefont{Tribollet}},
  \bibinfo{author}{\bibfnamefont{F.}~\bibnamefont{Bernardot}},
  \bibinfo{author}{\bibfnamefont{C.}~\bibnamefont{Testelin}},
  \bibinfo{author}{\bibfnamefont{M.}~\bibnamefont{Chamarro}}, and
  \bibinfo{author}{\bibfnamefont{J.-M.} \bibnamefont{G\'erard}},
  \bibinfo{year}{2005}, \bibinfo{journal}{Phys. Rev. B}
  \textbf{\bibinfo{volume}{71}}(\bibinfo{number}{11}), \bibinfo{pages}{115334}.

\bibitem[{\citenamefont{Shabaev} \emph{et~al.}(2009)\citenamefont{Shabaev,
  Stinaff, Bracker, Gammon, Efros, Korenev, and Merkulov}}]{Shabaev:2009a}
\bibinfo{author}{\bibnamefont{Shabaev}, \bibfnamefont{A.}},
  \bibinfo{author}{\bibfnamefont{E.~A.} \bibnamefont{Stinaff}},
  \bibinfo{author}{\bibfnamefont{A.~S.} \bibnamefont{Bracker}},
  \bibinfo{author}{\bibfnamefont{D.}~\bibnamefont{Gammon}},
  \bibinfo{author}{\bibfnamefont{A.~L.} \bibnamefont{Efros}},
  \bibinfo{author}{\bibfnamefont{V.~L.} \bibnamefont{Korenev}}, and
  \bibinfo{author}{\bibfnamefont{I.}~\bibnamefont{Merkulov}},
  \bibinfo{year}{2009}, \bibinfo{journal}{Phys. Rev. B}
  \textbf{\bibinfo{volume}{79}}(\bibinfo{number}{3}), \bibinfo{pages}{035322}.

\bibitem[{\citenamefont{Simon} \emph{et~al.}(2011)\citenamefont{Simon, Belhadj,
  Chatel, Amand, Renucci, Lemaitre, Krebs, Dalgarno, Warburton, Marie, and
  Urbaszek}}]{Simon:2011a}
\bibinfo{author}{\bibnamefont{Simon}, \bibfnamefont{C.-M.}},
  \bibinfo{author}{\bibfnamefont{T.}~\bibnamefont{Belhadj}},
  \bibinfo{author}{\bibfnamefont{B.}~\bibnamefont{Chatel}},
  \bibinfo{author}{\bibfnamefont{T.}~\bibnamefont{Amand}},
  \bibinfo{author}{\bibfnamefont{P.}~\bibnamefont{Renucci}},
  \bibinfo{author}{\bibfnamefont{A.}~\bibnamefont{Lemaitre}},
  \bibinfo{author}{\bibfnamefont{O.}~\bibnamefont{Krebs}},
  \bibinfo{author}{\bibfnamefont{P.~A.} \bibnamefont{Dalgarno}},
  \bibinfo{author}{\bibfnamefont{R.~J.} \bibnamefont{Warburton}},
  \bibinfo{author}{\bibfnamefont{X.}~\bibnamefont{Marie}}, and
  \bibinfo{author}{\bibfnamefont{B.}~\bibnamefont{Urbaszek}},
  \bibinfo{year}{2011}, \bibinfo{journal}{Phys. Rev. Lett.}
  \textbf{\bibinfo{volume}{106}}, \bibinfo{pages}{166801}.

\bibitem[{\citenamefont{Skiba-Szymanska}
  \emph{et~al.}(2008)\citenamefont{Skiba-Szymanska, Chekhovich, Nikolaenko,
  Tartakovskii, Makhonin, Drouzas, Skolnick, and
  Krysa}}]{Skiba-Szymanska:2008a}
\bibinfo{author}{\bibnamefont{Skiba-Szymanska}, \bibfnamefont{J.}},
  \bibinfo{author}{\bibfnamefont{E.~A.} \bibnamefont{Chekhovich}},
  \bibinfo{author}{\bibfnamefont{A.~E.} \bibnamefont{Nikolaenko}},
  \bibinfo{author}{\bibfnamefont{A.~I.} \bibnamefont{Tartakovskii}},
  \bibinfo{author}{\bibfnamefont{M.~N.} \bibnamefont{Makhonin}},
  \bibinfo{author}{\bibfnamefont{I.}~\bibnamefont{Drouzas}},
  \bibinfo{author}{\bibfnamefont{M.~S.} \bibnamefont{Skolnick}}, and
  \bibinfo{author}{\bibfnamefont{A.~B.} \bibnamefont{Krysa}},
  \bibinfo{year}{2008}, \bibinfo{journal}{Phys. Rev. B}
  \textbf{\bibinfo{volume}{77}}(\bibinfo{number}{16}), \bibinfo{pages}{165338}.

\bibitem[{\citenamefont{Slichter}(1990)}]{Slichter:1990a}
\bibinfo{author}{\bibnamefont{Slichter}, \bibfnamefont{C.~P.}},
  \bibinfo{year}{1990}, \bibinfo{journal}{Springer-Verlag} .

\bibitem[{\citenamefont{Smith} \emph{et~al.}(2005)\citenamefont{Smith,
  Dalgarno, Warburton, Govorov, Karrai, Gerardot, and Petroff}}]{Smith:2005a}
\bibinfo{author}{\bibnamefont{Smith}, \bibfnamefont{J.~M.}},
  \bibinfo{author}{\bibfnamefont{P.~A.} \bibnamefont{Dalgarno}},
  \bibinfo{author}{\bibfnamefont{R.~J.} \bibnamefont{Warburton}},
  \bibinfo{author}{\bibfnamefont{A.~O.} \bibnamefont{Govorov}},
  \bibinfo{author}{\bibfnamefont{K.}~\bibnamefont{Karrai}},
  \bibinfo{author}{\bibfnamefont{B.~D.} \bibnamefont{Gerardot}}, and
  \bibinfo{author}{\bibfnamefont{P.~M.} \bibnamefont{Petroff}},
  \bibinfo{year}{2005}, \bibinfo{journal}{Phys. Rev. Lett.}
  \textbf{\bibinfo{volume}{94}}, \bibinfo{pages}{197402}.

\bibitem[{\citenamefont{Stevenson} \emph{et~al.}(2011)\citenamefont{Stevenson,
  Salter, Boyer de~la Giroday, Farrer, Nicoll, Ritchie, and
  Shields}}]{Stevenson:2011a}
\bibinfo{author}{\bibnamefont{Stevenson}, \bibfnamefont{R.~M.}},
  \bibinfo{author}{\bibfnamefont{C.~L.} \bibnamefont{Salter}},
  \bibinfo{author}{\bibfnamefont{A.}~\bibnamefont{Boyer de~la Giroday}},
  \bibinfo{author}{\bibfnamefont{I.~A.} \bibnamefont{Farrer}},
  \bibinfo{author}{\bibfnamefont{C.~A.} \bibnamefont{Nicoll}},
  \bibinfo{author}{\bibfnamefont{D.~A.} \bibnamefont{Ritchie}}, and
  \bibinfo{author}{\bibfnamefont{A.~J.} \bibnamefont{Shields}},
  \bibinfo{year}{2011}, \bibinfo{journal}{ArXiv e-prints} \eprint{1103.2969}.

\bibitem[{\citenamefont{Takahashi} \emph{et~al.}(2011)\citenamefont{Takahashi,
  Kono, Tarucha, and Ono}}]{Takahashi:2011a}
\bibinfo{author}{\bibnamefont{Takahashi}, \bibfnamefont{R.}},
  \bibinfo{author}{\bibfnamefont{K.}~\bibnamefont{Kono}},
  \bibinfo{author}{\bibfnamefont{S.}~\bibnamefont{Tarucha}}, and
  \bibinfo{author}{\bibfnamefont{K.}~\bibnamefont{Ono}}, \bibinfo{year}{2011},
  \bibinfo{journal}{Phys. Rev. Lett.}
  \textbf{\bibinfo{volume}{107}}(\bibinfo{number}{2}), \bibinfo{pages}{026602}.

\bibitem[{\citenamefont{Tartakovskii}
  \emph{et~al.}(2007)\citenamefont{Tartakovskii, Wright, Russell, Fal'ko,
  Van'kov, Skiba-Szymanska, Drouzas, Kolodka, Skolnick, Fry, Tahraoui, Liu}
  \emph{et~al.}}]{Tartakovskii:2007a}
\bibinfo{author}{\bibnamefont{Tartakovskii}, \bibfnamefont{A.~I.}},
  \bibinfo{author}{\bibfnamefont{T.}~\bibnamefont{Wright}},
  \bibinfo{author}{\bibfnamefont{A.}~\bibnamefont{Russell}},
  \bibinfo{author}{\bibfnamefont{V.~I.} \bibnamefont{Fal'ko}},
  \bibinfo{author}{\bibfnamefont{A.~B.} \bibnamefont{Van'kov}},
  \bibinfo{author}{\bibfnamefont{J.}~\bibnamefont{Skiba-Szymanska}},
  \bibinfo{author}{\bibfnamefont{I.}~\bibnamefont{Drouzas}},
  \bibinfo{author}{\bibfnamefont{R.~S.} \bibnamefont{Kolodka}},
  \bibinfo{author}{\bibfnamefont{M.~S.} \bibnamefont{Skolnick}},
  \bibinfo{author}{\bibfnamefont{P.~W.} \bibnamefont{Fry}},
  \bibinfo{author}{\bibfnamefont{A.}~\bibnamefont{Tahraoui}},
  \bibinfo{author}{\bibfnamefont{H.-Y.} \bibnamefont{Liu}}, \emph{et~al.},
  \bibinfo{year}{2007}, \bibinfo{journal}{Phys. Rev. Lett.}
  \textbf{\bibinfo{volume}{98}}(\bibinfo{number}{2}), \bibinfo{pages}{026806}.

\bibitem[{\citenamefont{Taylor} \emph{et~al.}(2003)\citenamefont{Taylor,
  Marcus, and Lukin}}]{Taylor:2003a}
\bibinfo{author}{\bibnamefont{Taylor}, \bibfnamefont{J.~M.}},
  \bibinfo{author}{\bibfnamefont{C.~M.} \bibnamefont{Marcus}}, and
  \bibinfo{author}{\bibfnamefont{M.~D.} \bibnamefont{Lukin}},
  \bibinfo{year}{2003}, \bibinfo{journal}{Phys. Rev. Lett.}
  \textbf{\bibinfo{volume}{90}}(\bibinfo{number}{20}), \bibinfo{pages}{206803}.

\bibitem[{\citenamefont{Testelin} \emph{et~al.}(2009)\citenamefont{Testelin,
  Bernardot, Eble, and Chamarro}}]{Testelin:2009a}
\bibinfo{author}{\bibnamefont{Testelin}, \bibfnamefont{C.}},
  \bibinfo{author}{\bibfnamefont{F.}~\bibnamefont{Bernardot}},
  \bibinfo{author}{\bibfnamefont{B.}~\bibnamefont{Eble}}, and
  \bibinfo{author}{\bibfnamefont{M.}~\bibnamefont{Chamarro}},
  \bibinfo{year}{2009}, \bibinfo{journal}{Phys. Rev. B}
  \textbf{\bibinfo{volume}{79}}(\bibinfo{number}{19}), \bibinfo{pages}{195440}.

\bibitem[{\citenamefont{Tong and Wu}(2011)}]{Tong:2011a}
\bibinfo{author}{\bibnamefont{Tong}, \bibfnamefont{H.}}, and
  \bibinfo{author}{\bibfnamefont{M.~W.} \bibnamefont{Wu}},
  \bibinfo{year}{2011}, \bibinfo{journal}{Phys. Rev. B}
  \textbf{\bibinfo{volume}{83}}, \bibinfo{pages}{235323}.

\bibitem[{\citenamefont{Urbaszek} \emph{et~al.}(2007)\citenamefont{Urbaszek,
  Braun, Amand, Krebs, Belhadj, Lema\'\i{}tre, Voisin, and
  Marie}}]{Urbaszek:2007a}
\bibinfo{author}{\bibnamefont{Urbaszek}, \bibfnamefont{B.}},
  \bibinfo{author}{\bibfnamefont{P.-F.} \bibnamefont{Braun}},
  \bibinfo{author}{\bibfnamefont{T.}~\bibnamefont{Amand}},
  \bibinfo{author}{\bibfnamefont{O.}~\bibnamefont{Krebs}},
  \bibinfo{author}{\bibfnamefont{T.}~\bibnamefont{Belhadj}},
  \bibinfo{author}{\bibfnamefont{A.}~\bibnamefont{Lema\'\i{}tre}},
  \bibinfo{author}{\bibfnamefont{P.}~\bibnamefont{Voisin}}, and
  \bibinfo{author}{\bibfnamefont{X.}~\bibnamefont{Marie}},
  \bibinfo{year}{2007}, \bibinfo{journal}{Phys. Rev. B}
  \textbf{\bibinfo{volume}{76}}(\bibinfo{number}{20}), \bibinfo{pages}{201301}.

\bibitem[{\citenamefont{Urbaszek} \emph{et~al.}(2003)\citenamefont{Urbaszek,
  Warburton, Karrai, Gerardot, Petroff, and Garcia}}]{Urbaszek:2003a}
\bibinfo{author}{\bibnamefont{Urbaszek}, \bibfnamefont{B.}},
  \bibinfo{author}{\bibfnamefont{R.~J.} \bibnamefont{Warburton}},
  \bibinfo{author}{\bibfnamefont{K.}~\bibnamefont{Karrai}},
  \bibinfo{author}{\bibfnamefont{B.~D.} \bibnamefont{Gerardot}},
  \bibinfo{author}{\bibfnamefont{P.~M.} \bibnamefont{Petroff}}, and
  \bibinfo{author}{\bibfnamefont{J.~M.} \bibnamefont{Garcia}},
  \bibinfo{year}{2003}, \bibinfo{journal}{Phys. Rev. Lett.}
  \textbf{\bibinfo{volume}{90}}, \bibinfo{pages}{247403}.

\bibitem[{\citenamefont{Vamivakas} \emph{et~al.}(2009)\citenamefont{Vamivakas,
  Zhao, Lu, and Atat\"ure}}]{Vamivakas:2009a}
\bibinfo{author}{\bibnamefont{Vamivakas}, \bibfnamefont{N.~A.}},
  \bibinfo{author}{\bibfnamefont{Y.}~\bibnamefont{Zhao}},
  \bibinfo{author}{\bibfnamefont{C.-Y.} \bibnamefont{Lu}}, and
  \bibinfo{author}{\bibfnamefont{M.}~\bibnamefont{Atat\"ure}},
  \bibinfo{year}{2009}, \bibinfo{journal}{Nature Phys.}
  \textbf{\bibinfo{volume}{5}}, \bibinfo{pages}{198}.

\bibitem[{\citenamefont{Verzelen} \emph{et~al.}(2002)\citenamefont{Verzelen,
  Bastard, and Ferreira}}]{Verzelen:2002a}
\bibinfo{author}{\bibnamefont{Verzelen}, \bibfnamefont{O.}},
  \bibinfo{author}{\bibfnamefont{G.}~\bibnamefont{Bastard}}, and
  \bibinfo{author}{\bibfnamefont{R.}~\bibnamefont{Ferreira}},
  \bibinfo{year}{2002}, \bibinfo{journal}{Phys. Rev. B}
  \textbf{\bibinfo{volume}{66}}, \bibinfo{pages}{081308}.

\bibitem[{\citenamefont{Vink} \emph{et~al.}(2009)\citenamefont{Vink, Nowack,
  Koppens, Danon, Nazarov, and Vandersypen}}]{Vink:2009a}
\bibinfo{author}{\bibnamefont{Vink}, \bibfnamefont{I.~T.}},
  \bibinfo{author}{\bibfnamefont{K.~C.} \bibnamefont{Nowack}},
  \bibinfo{author}{\bibfnamefont{F.~H.~L.} \bibnamefont{Koppens}},
  \bibinfo{author}{\bibfnamefont{J.}~\bibnamefont{Danon}},
  \bibinfo{author}{\bibfnamefont{Y.~V.} \bibnamefont{Nazarov}}, and
  \bibinfo{author}{\bibfnamefont{L.~M.~K.} \bibnamefont{Vandersypen}},
  \bibinfo{year}{2009}, \bibinfo{journal}{Nature Phys.}
  \textbf{\bibinfo{volume}{5}}, \bibinfo{pages}{764}.

\bibitem[{\citenamefont{Warburton} \emph{et~al.}(1998)\citenamefont{Warburton,
  Miller, D\"urr, B\"odefeld, Karrai, Kotthaus, Medeiros-Ribeiro, Petroff, and
  Huant}}]{Warburton:1998a}
\bibinfo{author}{\bibnamefont{Warburton}, \bibfnamefont{R.~J.}},
  \bibinfo{author}{\bibfnamefont{B.~T.} \bibnamefont{Miller}},
  \bibinfo{author}{\bibfnamefont{C.~S.} \bibnamefont{D\"urr}},
  \bibinfo{author}{\bibfnamefont{C.}~\bibnamefont{B\"odefeld}},
  \bibinfo{author}{\bibfnamefont{K.}~\bibnamefont{Karrai}},
  \bibinfo{author}{\bibfnamefont{J.~P.} \bibnamefont{Kotthaus}},
  \bibinfo{author}{\bibfnamefont{G.}~\bibnamefont{Medeiros-Ribeiro}},
  \bibinfo{author}{\bibfnamefont{P.~M.} \bibnamefont{Petroff}}, and
  \bibinfo{author}{\bibfnamefont{S.}~\bibnamefont{Huant}},
  \bibinfo{year}{1998}, \bibinfo{journal}{Phys. Rev. B}
  \textbf{\bibinfo{volume}{58}}, \bibinfo{pages}{16221}.

\bibitem[{\citenamefont{Warburton} \emph{et~al.}(2000)\citenamefont{Warburton,
  Sch{\"a}flein, Haft, Bickel, Lorke, K.Karrai, Garcia, Schoenfeld, and
  Petroff}}]{Warburton:2000a}
\bibinfo{author}{\bibnamefont{Warburton}, \bibfnamefont{R.~J.}},
  \bibinfo{author}{\bibfnamefont{C.}~\bibnamefont{Sch{\"a}flein}},
  \bibinfo{author}{\bibfnamefont{D.}~\bibnamefont{Haft}},
  \bibinfo{author}{\bibfnamefont{F.}~\bibnamefont{Bickel}},
  \bibinfo{author}{\bibfnamefont{A.}~\bibnamefont{Lorke}},
  \bibinfo{author}{\bibnamefont{K.Karrai}},
  \bibinfo{author}{\bibfnamefont{J.}~\bibnamefont{Garcia}},
  \bibinfo{author}{\bibfnamefont{W.}~\bibnamefont{Schoenfeld}}, and
  \bibinfo{author}{\bibfnamefont{P.~M.} \bibnamefont{Petroff}},
  \bibinfo{year}{2000}, \bibinfo{journal}{Nature (London)}
  \textbf{\bibinfo{volume}{405}}, \bibinfo{pages}{926}.

\bibitem[{\citenamefont{Ware} \emph{et~al.}(2005)\citenamefont{Ware, Stinaff,
  Gammon, Doty, Bracker, Gershoni, Korenev, B\ifmmode~\u{a}\else
  \u{a}\fi{}descu, Lyanda-Geller, and Reinecke}}]{Ware:2005a}
\bibinfo{author}{\bibnamefont{Ware}, \bibfnamefont{M.~E.}},
  \bibinfo{author}{\bibfnamefont{E.~A.} \bibnamefont{Stinaff}},
  \bibinfo{author}{\bibfnamefont{D.}~\bibnamefont{Gammon}},
  \bibinfo{author}{\bibfnamefont{M.~F.} \bibnamefont{Doty}},
  \bibinfo{author}{\bibfnamefont{A.~S.} \bibnamefont{Bracker}},
  \bibinfo{author}{\bibfnamefont{D.}~\bibnamefont{Gershoni}},
  \bibinfo{author}{\bibfnamefont{V.~L.} \bibnamefont{Korenev}},
  \bibinfo{author}{\bibfnamefont{i.~m. c.~C.} \bibnamefont{B\ifmmode~\u{a}\else
  \u{a}\fi{}descu}},
  \bibinfo{author}{\bibfnamefont{Y.}~\bibnamefont{Lyanda-Geller}}, and
  \bibinfo{author}{\bibfnamefont{T.~L.} \bibnamefont{Reinecke}},
  \bibinfo{year}{2005}, \bibinfo{journal}{Phys. Rev. Lett.}
  \textbf{\bibinfo{volume}{95}}, \bibinfo{pages}{177403}.

\bibitem[{\citenamefont{Whitaker} \emph{et~al.}(2010)\citenamefont{Whitaker,
  Ochsenbein, Smith, Echodu, Robinson, and Gamelin}}]{Whitaker:2010a}
\bibinfo{author}{\bibnamefont{Whitaker}, \bibfnamefont{K.~M.}},
  \bibinfo{author}{\bibfnamefont{S.~T.} \bibnamefont{Ochsenbein}},
  \bibinfo{author}{\bibfnamefont{A.~L.} \bibnamefont{Smith}},
  \bibinfo{author}{\bibfnamefont{D.~C.} \bibnamefont{Echodu}},
  \bibinfo{author}{\bibfnamefont{B.~H.} \bibnamefont{Robinson}}, and
  \bibinfo{author}{\bibfnamefont{D.~R.} \bibnamefont{Gamelin}},
  \bibinfo{year}{2010}, \bibinfo{journal}{J. Phys. Chem. C}
  \textbf{\bibinfo{volume}{114}}(\bibinfo{number}{34}), \bibinfo{pages}{14467}.

\bibitem[{\citenamefont{Williams}(1991)}]{Williams:1991a}
\bibinfo{author}{\bibnamefont{Williams}, \bibfnamefont{W.}},
  \bibinfo{year}{1991}, \bibinfo{journal}{Clarendon Press, Oxford} .

\bibitem[{\citenamefont{Witzel and Das~Sarma}(2007)}]{Witzel:2007a}
\bibinfo{author}{\bibnamefont{Witzel}, \bibfnamefont{W.~M.}}, and
  \bibinfo{author}{\bibfnamefont{S.}~\bibnamefont{Das~Sarma}},
  \bibinfo{year}{2007}, \bibinfo{journal}{Phys. Rev. B}
  \textbf{\bibinfo{volume}{76}}(\bibinfo{number}{4}), \bibinfo{pages}{045218}.

\bibitem[{\citenamefont{Xu} \emph{et~al.}(2007)\citenamefont{Xu, Wu, Sun,
  Huang, Cheng, Steel, Bracker, Gammon, Emary, and Sham}}]{Xu:2007a}
\bibinfo{author}{\bibnamefont{Xu}, \bibfnamefont{X.}},
  \bibinfo{author}{\bibfnamefont{Y.}~\bibnamefont{Wu}},
  \bibinfo{author}{\bibfnamefont{B.}~\bibnamefont{Sun}},
  \bibinfo{author}{\bibfnamefont{Q.}~\bibnamefont{Huang}},
  \bibinfo{author}{\bibfnamefont{J.}~\bibnamefont{Cheng}},
  \bibinfo{author}{\bibfnamefont{D.~G.} \bibnamefont{Steel}},
  \bibinfo{author}{\bibfnamefont{A.~S.} \bibnamefont{Bracker}},
  \bibinfo{author}{\bibfnamefont{D.}~\bibnamefont{Gammon}},
  \bibinfo{author}{\bibfnamefont{C.}~\bibnamefont{Emary}}, and
  \bibinfo{author}{\bibfnamefont{L.~J.} \bibnamefont{Sham}},
  \bibinfo{year}{2007}, \bibinfo{journal}{Phys. Rev. Lett.}
  \textbf{\bibinfo{volume}{99}}, \bibinfo{pages}{097401}.

\bibitem[{\citenamefont{Xu} \emph{et~al.}(2009)\citenamefont{Xu, Yao, Sun,
  Steel, Bracker, Gammon, and Sham}}]{Xu:2009a}
\bibinfo{author}{\bibnamefont{Xu}, \bibfnamefont{X.}},
  \bibinfo{author}{\bibfnamefont{W.}~\bibnamefont{Yao}},
  \bibinfo{author}{\bibfnamefont{B.}~\bibnamefont{Sun}},
  \bibinfo{author}{\bibfnamefont{D.~G.} \bibnamefont{Steel}},
  \bibinfo{author}{\bibfnamefont{A.~S.} \bibnamefont{Bracker}},
  \bibinfo{author}{\bibfnamefont{D.}~\bibnamefont{Gammon}}, and
  \bibinfo{author}{\bibfnamefont{L.~J.} \bibnamefont{Sham}},
  \bibinfo{year}{2009}, \bibinfo{journal}{Nature (London)}
  \textbf{\bibinfo{volume}{459}}, \bibinfo{pages}{1105}.

\bibitem[{\citenamefont{{Yang} and {Sham}}(2010)}]{Yang:2010a}
\bibinfo{author}{\bibnamefont{{Yang}}, \bibfnamefont{W.}}, and
  \bibinfo{author}{\bibfnamefont{L.~J.} \bibnamefont{{Sham}}},
  \bibinfo{year}{2010}, \bibinfo{journal}{ArXiv e-prints} \eprint{1012.0060}.

\bibitem[{\citenamefont{Yao} \emph{et~al.}(2006)\citenamefont{Yao, Liu, and
  Sham}}]{Yao:2006a}
\bibinfo{author}{\bibnamefont{Yao}, \bibfnamefont{W.}},
  \bibinfo{author}{\bibfnamefont{R.-B.} \bibnamefont{Liu}}, and
  \bibinfo{author}{\bibfnamefont{L.~J.} \bibnamefont{Sham}},
  \bibinfo{year}{2006}, \bibinfo{journal}{Phys. Rev. B}
  \textbf{\bibinfo{volume}{74}}, \bibinfo{pages}{195301}.

\bibitem[{\citenamefont{Yilmaz} \emph{et~al.}(2010)\citenamefont{Yilmaz,
  Fallahi, and Imamoglu}}]{Yilmaz:2010a}
\bibinfo{author}{\bibnamefont{Yilmaz}, \bibfnamefont{S.~T.}},
  \bibinfo{author}{\bibfnamefont{P.}~\bibnamefont{Fallahi}}, and
  \bibinfo{author}{\bibfnamefont{A.}~\bibnamefont{Imamoglu}},
  \bibinfo{year}{2010}, \bibinfo{journal}{Phys. Rev. Lett.}
  \textbf{\bibinfo{volume}{105}}(\bibinfo{number}{3}), \bibinfo{pages}{033601}.

\bibitem[{\citenamefont{Yugova} \emph{et~al.}(2007)\citenamefont{Yugova,
  Greilich, Zhukov, Yakovlev, Bayer, Reuter, and Wieck}}]{Yugova:2007a}
\bibinfo{author}{\bibnamefont{Yugova}, \bibfnamefont{I.~A.}},
  \bibinfo{author}{\bibfnamefont{A.}~\bibnamefont{Greilich}},
  \bibinfo{author}{\bibfnamefont{E.~A.} \bibnamefont{Zhukov}},
  \bibinfo{author}{\bibfnamefont{D.~R.} \bibnamefont{Yakovlev}},
  \bibinfo{author}{\bibfnamefont{M.}~\bibnamefont{Bayer}},
  \bibinfo{author}{\bibfnamefont{D.}~\bibnamefont{Reuter}}, and
  \bibinfo{author}{\bibfnamefont{A.~D.} \bibnamefont{Wieck}},
  \bibinfo{year}{2007}, \bibinfo{journal}{Phys. Rev. B}
  \textbf{\bibinfo{volume}{75}}(\bibinfo{number}{19}), \bibinfo{pages}{195325}.

\bibitem[{\citenamefont{Zhang} \emph{et~al.}(2006)\citenamefont{Zhang,
  Dobrovitski, Al-Hassanieh, Dagotto, and Harmon}}]{Zhang:2006a}
\bibinfo{author}{\bibnamefont{Zhang}, \bibfnamefont{W.}},
  \bibinfo{author}{\bibfnamefont{V.~V.} \bibnamefont{Dobrovitski}},
  \bibinfo{author}{\bibfnamefont{K.~A.} \bibnamefont{Al-Hassanieh}},
  \bibinfo{author}{\bibfnamefont{E.}~\bibnamefont{Dagotto}}, and
  \bibinfo{author}{\bibfnamefont{B.~N.} \bibnamefont{Harmon}},
  \bibinfo{year}{2006}, \bibinfo{journal}{Phys. Rev. B}
  \textbf{\bibinfo{volume}{74}}, \bibinfo{pages}{205313}.

\end{thebibliography}


\newpage

\begin{table}
\caption{\label{tab:tabledef1} Definitions}
\begin{center}
\begin{longtable}{l p{12cm}}
symbol & meaning \\
\hline
$N$ & number of nuclei in the dot  \\
$\Gamma$ & spontaneous emission rate  \\
$\tau_r=1/\Gamma$ & radiative recombination time  \\
$\hbar\omega^e_Z$ & electron Zeeman splitting  \\
$\hbar\omega^h_Z$ & hole Zeeman splitting  \\
$\hbar\omega^X_Z$ & exciton Zeeman splitting  \\
$\hbar\omega^n_Z$ & nuclear Zeeman splitting  \\
$\hbar\omega^e_{OS}$ & electron Overhauser splitting \\
$\hbar\omega^h_{OS}$ & hole Overhauser splitting \\
$B_K$ & Knight field \\
$\bm{B}_n=(B_{n,x},B_{n,y},B_{n,z})$ & Overhauser field \\
$\delta B_n $ & Nuclear Field Fluctuations \\
$\bm{B}=(B_x,B_y,B_z)$ & external magnetic field \\
$B_{tot}$ & total magnetic field experienced by electron \\
$B_{tot}^h$ & total magnetic field experienced by hole \\
$A_j$ & hyperfine const. for nuclear species $j$  coupling to electrons \\
$\tilde{A}$ & effective hyperfine constant for coupling to electrons \\
$A_j^h$ & hyperfine constant for nuclear species $j$  coupling to holes \\
$\tilde{A}^h$ & effective hyperfine constant for coupling to holes \\
$A_i^{\rm nc}$ & non-collinear hyperfine coupling of $i$-th nucleus\\
$\uparrow$ or $\downarrow$ & electron spin state in ``$z$'' basis \\
$\Uparrow$ or $\Downarrow$ & hole pseudo-spin  state in ``$z$'' basis \\
write $\sum_{j} I_{j,z}$ as $\left| I_z \right>$ & nuclear spin $z$ projection \\
$S_z$ & electron spin $z$ projection \\
$S^h_z$ & hole 'pseudo' spin $z$ projection \\
$J$ & hole angular momentum \\
$J_z$ & hole angular momentum $z$ projection \\
$T_\Delta$ & electron spin dephasing time in randomly distributed frozen fluctuation of the hyperfine field\\
$T_\text{Dipole}$ & average precession time of a nuclear spin in the local field fluctuations $\delta B_L$\\
$T_\text{1e}$ & electron-induced of nuclear spins \\
$T_d$ & decay time of average nuclear spin \\
$T_\text{1d}$ & spin flip time experienced by electron due to flip-flops with nuclei \\
$\tau^e_c$ & electron correlation time \\
$\tau^h_c$ & hole correlation time \\
$\tau^n_c$ & correlation time of nuclei \\
$\tau^e_s$ & electron spin relaxation time \\
$\tau^h_s$ & hole spin relaxation time \\
$g^e_z$ & longitudinal electron g-factor \\
$g^h_z$ & longitudinal hole electron g-factor \\
$g^e_\bot$ & transverse electron g-factor \\
$g^h_\bot$ & transverse hole g-factor \\
DNP & Dynamic Nuclear (Spin) Polarization \\
$ X^0$ & neutral exciton \\
$X^-$ & negatively charged exciton \\
$X^+$ & positively charged exciton \\
$\delta_1$ & fine structure splitting of J=1 $X^0$ due to anisotropic e-h Coulomb exchange interaction\\
$\delta_2$ & fine structure splitting of J=2 $X^0$ due to anisotropic e-h Coulomb exchange interaction\\
$\delta_0$ & fine structure splitting of between J=2 and J=1 $X^0$ due to isotropic e-h Coulomb exchange interaction\\
$\kappa$ & electron co-tunelling rate\\
$\Delta\nu$ & QD absorption linewidth\\
$\omega_{\rm X}$ & QD exciton transition frequency\\
$\omega_{\rm L}$ & laser frequency\\
$\Omega_{\rm L}$ & Rabi frequency\\
$\Delta\omega=\omega_{\rm X}-\omega_{\rm L}$ & laser detuning from QD resonance\\
$\omega_{\rm X}$ & QD exciton transition frequency\\

\end{longtable}
\end{center}
\end{table}

\end{document}